\documentclass{aa}  

\usepackage{graphicx}
\usepackage{tikz}
\usepackage{booktabs}
\usepackage{footmisc}
\usetikzlibrary{shapes.geometric, arrows, calc}
\tikzset{every node/.style={inner sep=10pt,minimum height=1cm}}
\tikzstyle{startstop} = [rectangle, rounded corners, minimum width=1cm, minimum height=1cm,text centered, draw=black, fill=blue!50]
\tikzstyle{io} = [trapezium, trapezium left angle=70, trapezium right angle=110, minimum width=1cm, text width=5cm, minimum height=1cm, text centered, draw=black, fill=blue!30]
\tikzstyle{process} = [rectangle, minimum width=1cm, minimum height=1cm, text centered, text width=8cm, draw=black, fill=gray!20]
\tikzstyle{middle} = [trapezium, trapezium left angle=80, trapezium right angle=100, minimum width=0cm, text width=6cm, minimum height=1cm, text centered, draw=black, fill=green!10]
\tikzstyle{middle2} = [trapezium, trapezium left angle=80, trapezium right angle=100, minimum width=0cm, text width=6cm, minimum height=1cm, text centered, draw=black, fill=green!10]
\tikzstyle{subio} = [rectangle, minimum width=1cm, minimum height=1cm, text centered, draw=black, fill=blue!30]
\tikzstyle{arrow} = [thick,->,>=stealth]

\usepackage{txfonts}
\usepackage{bm}
\usepackage{amsmath}
\usepackage{hyperref}
\makeatletter
\renewcommand*\aa@pageof{, page \thepage{} of \pageref*{LastPage}}
\makeatother

\hypersetup{
    colorlinks=true,
    linkcolor=blue,
    filecolor=blue,      
    urlcolor=blue,
    citecolor=blue
}

\DeclareMathOperator{\yellm}{Y}
\newcommand\yellmarb[3]{\, _{#1}\!\!\yellm^{#2}_{#3}}

\defcitealias{lin2022a}{L23}
\begin{document}

   \title{{KiDS-SBI}: Simulation-based inference analysis of KiDS-1000 cosmic shear}

   \author{Maximilian von Wietersheim-Kramsta
          \inst{1,2,3}
          \and
          Kiyam Lin\inst{1}
          \and
          Nicolas Tessore\inst{1}
          \and
          Benjamin Joachimi\inst{1}
          \and
          Arthur Loureiro\inst{4, 5}
          \and
          Robert Reischke\inst{6,7}
          \and
          Angus H. Wright\inst{7}
          }

   \institute{Department of Physics and Astronomy, University College London, Gower Street, London, WC1E 6BT, UK
        \and
            Department of Physics, Institute for Computational  Cosmology, Durham University, South Road, Durham DH1 3LE, UK
        \and
             Department of Physics, Centre for Extragalactic Astronomy, Durham University, South Road, Durham DH1 3LE, UK
         \and
             The Oskar Klein Centre, Department of Physics, Stockholm University, AlbaNova University Centre, SE-106 91, Stockholm, Sweden
        \and
            Astrophysics Group and Imperial Centre for Inference and Cosmology (ICIC), Blackett Laboratory, Imperial College London, London SW7 2AZ, UK
        \and
            Argelander-Institut für Astronomie, Auf dem Hügel 71, D-53121 Bonn, Germany
        \and
            Ruhr University Bochum, Faculty of Physics and Astronomy, Astronomical Institute (AIRUB), German Centre for Cosmological Lensing, D-44780 Bochum, Germany
             }

   \date{Received 23 April 2024 / Accepted 10 December 2024}
 
  \abstract
  {We present a simulation-based inference (SBI) cosmological analysis of cosmic shear two-point statistics from the fourth weak gravitational lensing data release of the ESO Kilo-Degree Survey (KiDS-1000). KiDS-SBI efficiently performs non-Limber projection of the matter power spectrum via \texttt{Levin}'s method and constructs log-normal random matter fields on the curved sky for arbitrary cosmologies, including effective prescriptions for intrinsic alignments and baryonic feedback. The forward model samples realistic galaxy positions and shapes, based on the observational characteristics of KiDS-1000. It incorporates shear measurement and redshift calibration uncertainties, as well as angular anisotropies due to variable survey depth and point spread function (PSF) variations. To enable direct comparisons with standard inference, we limited our analysis to pseudo-angular power spectra as summary statistics. Here, the SBI is based on neural density estimation of the likelihood with active learning to infer the posterior distribution of spatially flat $\Lambda$CDM cosmological parameters from 18,000 realisations. We inferred a mean marginal for the growth of the structure parameter of $S_{8}  \equiv \sigma_8 (\Omega_\mathrm{m} / 0.3)^{0.5} = 0.731\pm 0.033$ ($68 \%$). We present a measurement of the goodness-of-fit for SBI, determining that the forward model fits the data well, with a probability-to-exceed of $0.42$. For a fixed cosmology, the learnt likelihood is approximately Gaussian, while its constraints are wider, compared to a Gaussian likelihood analysis due to the cosmology dependence in the covariance. Neglecting variable depth and anisotropies in the point spread function in the model can cause $S_{8}$ to be overestimated by ${\sim}5\%$. Our results are in agreement with previous analyses of KiDS-1000 and reinforce a $2.9 \sigma$ tension with early Universe constraints from cosmic microwave background measurements. This work highlights the importance of forward-modelling systematic effects in upcoming galaxy surveys, such as \textit{Euclid}, Rubin, and \textit{Roman}.}


   \maketitle
%
%-------------------------------------------------------------------

\section{Introduction}
The weak gravitational lensing effect on distant galaxies due to matter in the foreground, known as cosmic shear, is a powerful tool for studying the distribution of matter in the Universe and to probe its large-scale structure. By measuring the distortions in the shapes of galaxy images caused by the gravitational influence of intervening matter, we can infer a combination of the matter density and the amplitude of the matter power spectrum within the framework of a spatially flat cold dark matter cosmology with a cosmological constant, $\Lambda$CDM. Recent cosmic shear analyses of the data taken in stage-III galaxy surveys, such as the Kilo-Degree Survey\footnote{\url{https://kids.strw.leidenuniv.nl/}} (KiDS; \citealt{kuijken2019the, asgari2021kids, heymans2021kids, loureiro2021kids, busch2022kids, li2023kids}), the Hyper Suprime-Cam survey\footnote{\url{https://hsc.mtk.nao.ac.jp/ssp/}} (HSC; \citealt{sugiyama2022hsc, aihara2022third, li2023hyper}), and the Dark Energy Survey\footnote{\url{https://www.darkenergysurvey.org/}} (DES; \citealt{gatti2021dark, amon2022dark, secco2022dark}), have constrained these cosmological parameters with unprecedented precision. Upcoming stage-IV galaxy surveys, such as \emph{Euclid}\footnote{\url{https://www.euclid-ec.org/}} \citep{laureijs2011euclid}, Rubin\footnote{\url{https://www.lsst.org/}} \citep{lsst2009lsst}, and \emph{Roman}\footnote{\url{https://roman.gsfc.nasa.gov/}} \citep{spergel2015wide}, will further improve upon these constraints, as these surveys will be wider, deeper, and more precise. However, many analytical challenges still remain.

Current stage-III galaxy survey measurements of the root-mean square of the matter overdensity field at 8 Mpc $h^{-1}$, $\sigma_8$, or its analogue $S_8 \equiv \sigma_8 (\Omega_\mathrm{m} / 0.3)^{0.5}$, are in good agreement with each other \citep{hikage2019cosmology, asgari2021kids, busch2022kids, amon2022dark, secco2022dark, li2023hyper, dalal2023hyper}, despite the use of largely independent methodologies and survey volumes. However, the constraints from observations of the late Universe \citep{asgari2021kids, heymans2021kids, busch2022kids, amon2022dark, abbott2022dark, li2023kids, deskids2023} are in disagreement by up to ${\sim}$3.4$\sigma$ with the $\sigma_8$ value consistent with the early Universe observations based on the cosmic microwave background (CMB; \citealt{planck2020planck}). Recently, cluster abundances measured by SRG/eROSITA \citep{ghirardini2024the} have provided another measurement of the fluctuation amplitude even higher than those of the CMB, although similar recent cluster studies have obtained $S_{8}$ constraints more consistent with weak lensing and CMB analyses \citep{mantz2015weighing, garrel2022xxl, chiu2023cosmological, bocquet2024spt}. This discrepancy might be pointing at new physical phenomena, but it could also be caused by unconsidered systematic effects when modelling the low-redshift large-scale structure or the CMB signal and noise (see for example \citealt{handley2021quantifying, deskids2023}). Such systematic effects may be physical, such as baryonic feedback or the intrinsic alignments of galaxies \citep{mandelbaum2018weak, amon2022a, li2023hyper, miyatake2023hyper, dalal2023hyper}, or observational.

We aim to shed some light on this issue by conducting a full cosmic shear analysis using neural likelihood estimation simulation-based inference (SBI) characterised by the same complexity as current stage-III analyses. SBI, also known as 'likelihood-free inference' or implicit likelihood inference', is a Bayesian inference method that does not require an explicit formulation for the likelihood function of the data given the parameters of interest. Instead, the likelihood is implicitly calculated by evaluating the joint probability of data and parameters of interest from forward simulations, which map the parameters to the corresponding mock data vectors. This comes with multiple advantages with respect to other standard approaches that require an explicit form for the likelihood. Firstly, the likelihood is allowed to take an arbitrary form, so we can avoid the typical assumption of a Gaussian likelihood or avoid having to define a complex analytical expression for the likelihood. Secondly, for some models and measurements, it may not even be possible to define an analytical likelihood or it would be too resource-intensive; for instance, the case of covariance among n$^{\mathrm{th}}$-order statistics depending on up to 2n$^{\mathrm{th}}$-order correlation functions, which become geometrically more expensive to compute. In such cases, as long as the observables can be simulated, an effective or implicit likelihood may be found using SBI. Lastly, in standard analyses, it often is required to validate the signal and noise modelling with forward simulations, so that we can make use of existing tools to perform SBI.

When running a Markov chain Monte Carlo (MCMC) to sample the posterior distributions from a Gaussian likelihood, we either have to compute a numerical covariance matrix from forward simulations or create an analytical model for the covariance, which ought to be validated with a numerical one. In both cases, for a data vector, $\mathbf{d}$, we would require a number of simulations $>|\mathbf{d}|$ to get an accurate sample covariance matrix. This is already the approximate amount of forward simulations needed for an SBI analysis \citep{alsing2018massive, lin2022a}. Therefore, SBI allows us to perform a full Bayesian uncertainty propagation from the data to obtain parameters for any model that can be simulated, without substantial additional computational costs when compared to standard approaches at a comparable accuracy.

With respect to cosmic shear, we know that the commonly used two-point statistics to quantify the shear-shear correlations of galaxies on the sky have an approximately Gaussian likelihood, but this assumption has its limitations \citep{schneider2009constrained, sellentin2018on, sellentin2018the, taylor2019cosmic, upham2021sufficiency, hall2022non}. Additionally, when observational and physical systematics are included, the likelihood may become even less Gaussian \citep{jeffrey2021}. In such cases, like the one presented in this paper, SBI allows us to capture potential non-Gaussanities and propagate them to the posterior distributions.

The SBI approach can also enable us to compute effective likelihoods for summary statistics for which analytical methods may be intractable, such as higher order cosmic shear statistics \citep{fluri2022full, euclid2023euclid, lu2023cosmological}. In this work, we limit the scope to only angular power spectra of cosmic shear and leave higher order statistics or field-level statistics within SBI as an avenue for future work. This allows us to make a direct comparison to previous standard cosmic shear analyses (such as \citealt{asgari2021kids, loureiro2021kids}), while also harnessing the efficiency of log-normal forward models of matter density fields which are mostly accurate for two-point statistics \citep{hall2022non, piras2023fast}.

A potential cause for deviations from Gaussianity in the likelihood of two-point statistics/angular power spectra is the anisotropy in the observational depth of a galaxy survey across the sky \citep{guzik2005inhomogeneous, shirasaki2019mock}. As a survey such as KiDS images the sky, each pointing and each pixel are subject to variable environmental and background conditions that may affect the selection of galaxies (for example atmospheric conditions, thermal expansion of the telescope, moonlight, zodiacal light, Galactic extinction, etc.).

Due to the given survey strategy and/or dithering to compensate for gaps in the telescope's charge-coupled device (CCD), it is possible that there is spatial variation in the overlap between pointings. When the images of each pointing are subsequently combined, some galaxies are observed more often than others that can cause variations in the signal-to-noise ratio (S/N), which are spatially correlated.

In effect, all these variations lead to fluctuations in the effective galaxy density, which are observed across the footprint, while also selecting galaxies according to their shape. This has been found to be a percent-level effect on the cosmic shear signal measured by KiDS \citep{heydenreich2020the}, while also affecting the standard deviation of the KiDS-1000 cosmic shear signal (by $\lesssim 20\%$; \citealt{joachimi2021kids}).  For KiDS-1000, this is within the statistical uncertainty, however, it will become important for upcoming galaxy shear surveys such as \emph{Euclid}, Rubin, and \emph{Roman}. Additionally, the effect of spatial variability on the non-Gaussianity of the likelihood has not been explored in previous works. To address these needs, in this work, our forward model includes the effects of the observational depth variations in KiDS, so that we may propagate the effects of spatial variations in the depth to the uncertainty of inferred cosmological parameters.

 To conduct the SBI, we chose to make use of the density estimation likelihood-free inference (DELFI, \citealt{alsing2019fast}) method, as it offers good performance and scales better when increasing the dimensionality of the parameter space of interest than other SBI methods \citep{leclercq2018bayesian, alsing2018massive}, such as approximate Bayesian computation (ABC; \citealt{rubin1984bayesianly, pritchard1999population, lin2015a, lin2016a, beaumont2019approximate}). DELFI discards less information from the forward simulations than ABC, whereas the latter only keeps the forward simulations which have a certain level of agreement with the desired data vector. In contrast, DELFI learns a probability density distribution of the data vectors as a function of the model parameters based on all the forward simulations. To achieve this, DELFI employs ensembles of neural density estimators to learn the sampling distribution of the data and the parameters from the forward simulations.

 Of course, as with any inference method, we need to be mindful of any intrinsic biases that might come with SBI, as well as how to mitigate and avoid them. As the likelihood is not known analytically, it is important that the chosen SBI method, sampling scheme, and number of simulations are sufficient to ensure that the learnt likelihood has converged to the model's true likelihood function. To avoid biasing the inference, this requires careful robustness and convergence tests of the implicit likelihood, which may not always be necessary for a standard Gaussian likelihood analysis. Hence, we dedicate a considerable effort in this work to ensure the inference pipeline is validated, the neural density estimators of DELFI have converged, the inferred posterior is accurate, and the inferred parameters give a good fit to the data.

Applications of SBI are increasing in popularity, in major part because it requires about an order of magnitude fewer evaluations of the likelihood than a standard Gaussian likelihood MCMC analysis to constrain the posterior \citep{papamakarios2016fast, alsing2018massive, gerardi2021unbiased, jeffrey2021, lemos2021sum, legin2021simulation, mishra2022neural, mancini2022bayesian, lin2022a, chen2022lightweight, hu2022measuring, lemos2023robust, lemos2023simbig, gatti2023dark, moser2024simulation, abellan2024fast}. We highlight the methodology of \citet{gatti2023dark}, which used  a similar analysis to the one outlined in this paper, with the DES Year 3 weak lensing data. We refer to  \citet{jeffrey2024des} for a full description of the results. While the DES analysis focuses on SBI of two-point and higher-order shear statistics, along with cosmologies beyond $\Lambda$CDM based on a fixed number of N-body simulations, while incorporating standard systematics, the analysis in this paper exclusively considers two-point statistics within the context of $\Lambda$CDM. We highlight that in our analysis, the data likelihood is learned by actively sampling as many efficient forward simulations as needed. This allows us to capture many physical and observational systematics, including anisotropic systematics that are typically not considered in weak lensing analyses. With this, we aim to give more insight into the robustness against systematics of standard weak lensing analyses.

In a previous paper \citep[][L23 hereafter]{lin2022a}, we introduced a new inference pipeline for estimating the cosmological parameters from the cosmic shear data of KiDS-1000 \citep{kuijken2019the} using DELFI. In \citetalias{lin2022a}, the SBI pipeline for KiDS is shown to be robust, accurate and efficient, even when constraining a 12-dimensional (12D) posterior distribution. However, in that work, the simulated vectors were based on random samples from the covariance matrix used in the fiducial Gaussian KiDS-1000 analysis, so the likelihood was Gaussian by construction to allow for direct comparison with the standard formalism. In this paper, we present a novel suite of physically motivated and realistic forward simulations of the cosmic shear observables, as seen by KiDS. Thus, the neural density estimators can learn any non-Gaussianities in the likelihood induced by our model, plus any complex relations between the data and the model parameters. We then used these simulations to create two inference pipelines: a fiducial one which makes the same modelling choices as the KiDS-1000 analysis \citep{joachimi2021kids} and a more realistic one, which explicitly models systematic effects that would be prohibitively difficult to incorporate into a standard likelihood analysis. From each pipeline, we obtained a separate posterior estimate for $S_8$, which we then scrutinised with a novel measure of the goodness-of-fit for SBI posteriors.

This paper has the following structure. Section~\ref{data} presents the KiDS data used for this analysis. Section~\ref{fs} gives a detailed description of the forward simulations developed for the SBI analysis. Section~\ref{measurements} details how the galaxy shapes and summary statistics are measured in both the simulations and the data. Section~\ref{method:sbi} describes the setup of the simulation-based inference pipeline using DELFI based on \citetalias{lin2022a}. Section~\ref{method:gof} presents the novel measure of goodness-of-fit for SBI  we use in this paper. In Sect.~\ref{validation}, we describe our  consistency tests to ensure the SBI pipeline is unbiased and robust, while in Sect.~\ref{gaussian_likelihood}, we discusse the validity of a Gaussian likelihood assumption. Section~\ref{results:real_data} presents the cosmological inference results from the KiDS-1000 weak lensing data. Lastly, we give our concluding remarks in Sect.~\ref{conclusions}. We provide additional details about the analysis in the appendix.

\begin{figure*}
    \centering
    \includegraphics[trim=0 265 0 0,clip,width=15cm]{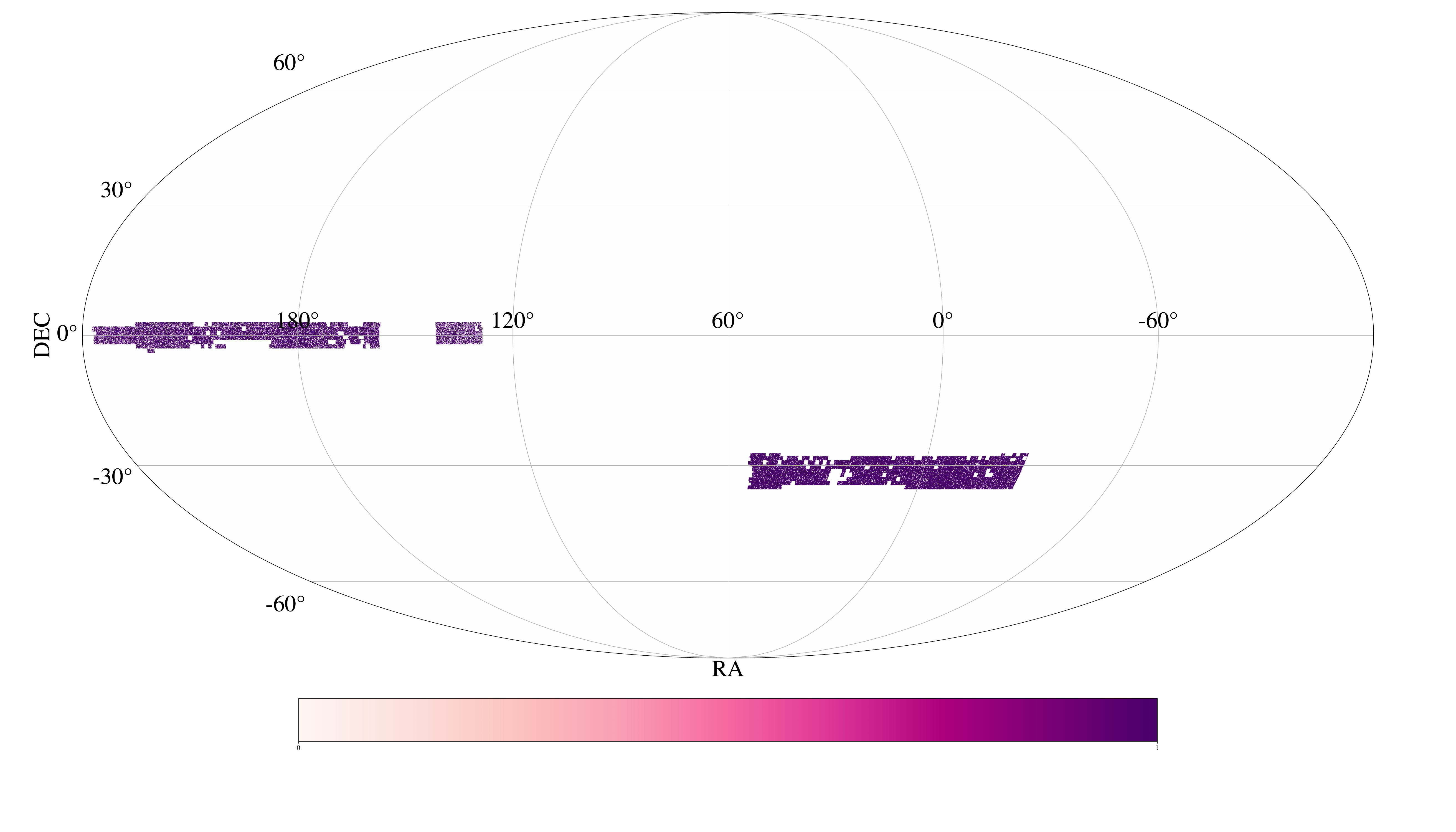}
    \includegraphics[width=17cm]{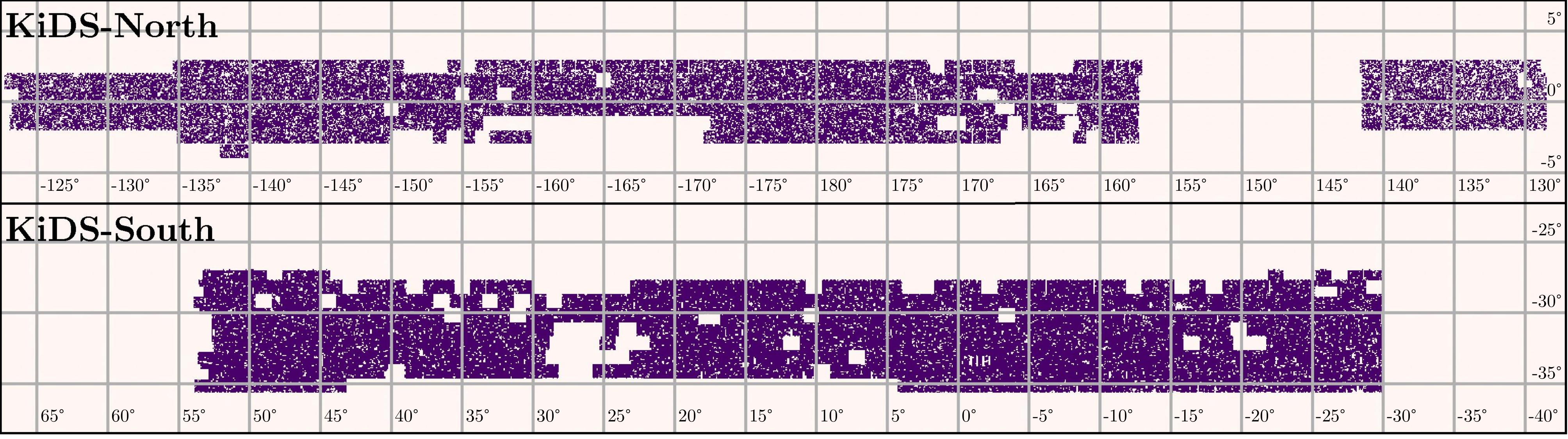}
    \caption{Spatial map of the KiDS-1000 footprint. The top panel shows a Mollweide projection of the full KiDS-1000 footprint, while the two panels at the bottom show zoomed-in Cartesian projections of KiDS-North and KiDS-South fields, respectively.}
    \label{fig:kids_footprint}
\end{figure*}

\begin{figure}
    \centering
    \includegraphics[width=9cm]{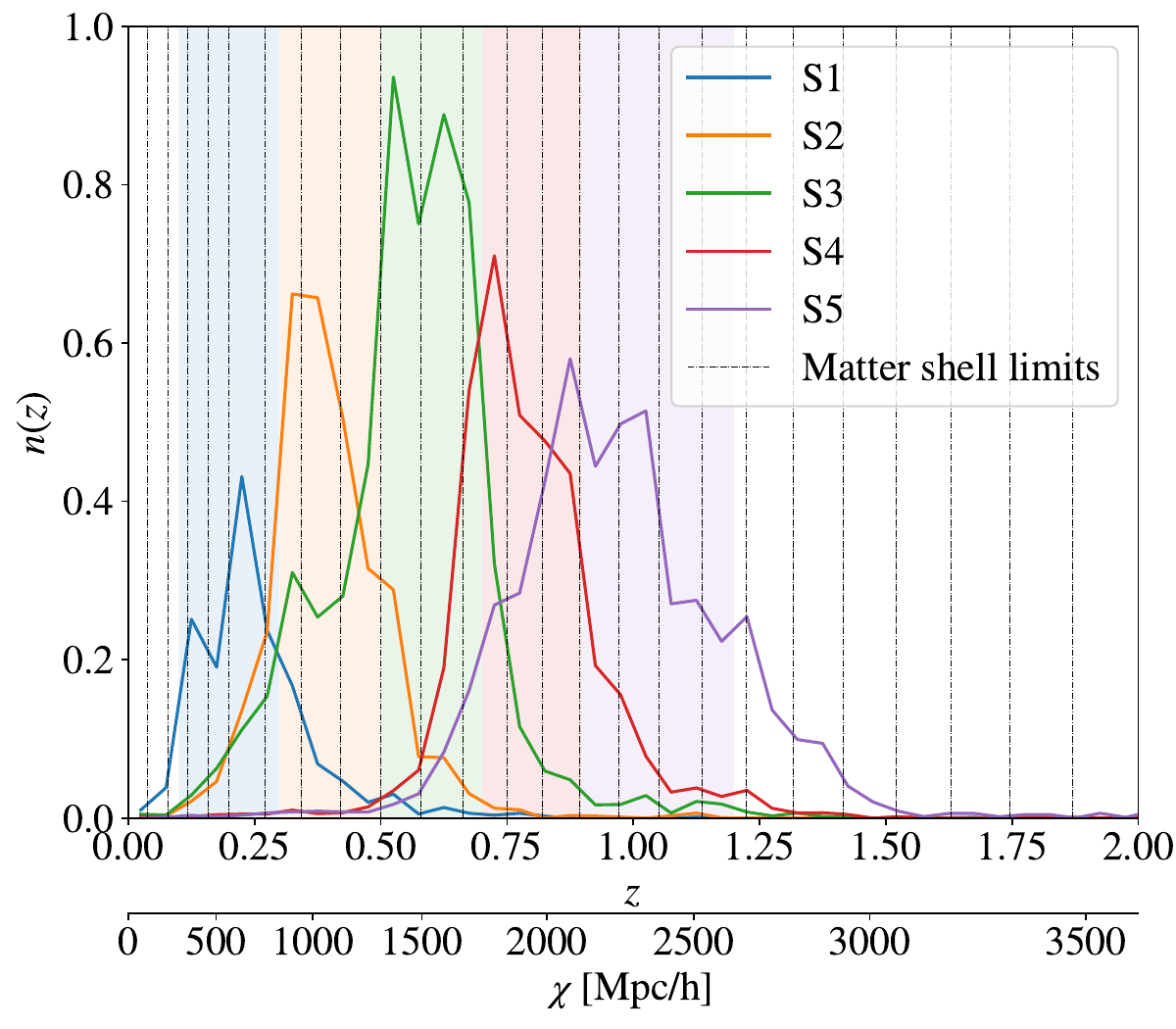}
    \caption{Redshift distributions of the five KiDS-1000 tomographic bins. The shaded areas show to limits of each tomographic bin, while the solid lines show the $n(z)$ of the source galaxies in each tomographic bin as a function of both redshift, $z$, and comoving distance, $\chi$ (the latter is derived assuming a Planck 2018 cosmology; \citealt{planck2020planck}). The black dashed lines show the limits of the spherical matter shells in our forward simulations.}
    \label{fig:kids_nz}
\end{figure}

\section{KiDS-1000 data} \label{data}
The Kilo-Degree Survey (KiDS) is a large public galaxy survey conducted by the European Southern Observatory using the OmegaCAM CCD mosaic camera \citep{kuijken2011omegacam},  attached to the 2.6~m VLT Survey Telescope (VST). The survey covers approximately 1,350 deg$^2$ between two distinct fields known as KiDS North and South, which are fields located in the northern and southern galactic caps (with DEC $\approx$ 0º and DEC $\approx$ -30º, respectively). This area is the same as the one covered by the VISTA Kilo-degree INfrared Galaxy survey (VIKING; \citealt{edge2013the}), which means that both surveys together observe every object with a total of nine photometric bands: $ugriZYJHK_{\rm s}$. The analysis presented in this paper makes use of DR4 \citep{kuijken2019the}, also known as KiDS-1000, which covers approximately 1,000 deg$^2$. The effective area covered by KiDS-1000 shape measurements (shown in Fig.~\ref{fig:kids_footprint}) extends over 773.3 deg$^2$ \citep{joachimi2021kids}. This area is calculated directly from the mosaic mask defined a the native OmegaCAM pixel scale of 0.213 arcsec. 

The KiDS images were processed using \texttt{Astro-WISE} \citep{mcfarland2013the} and the methodology outlined in \citet{wright2019kids} to combine the VST and VIKING photometric measurements. To obtain the shape measurements, \texttt{Theli} \citep{erben2005gabods} was used to process the $r$-band images, while the gravitational shear estimates are then obtained from \texttt{lensfit} \citep{miller2007bayesian, miller2013bayesian, fenech2017calibration}. Depending on the shape noise variance and the ellipticity measurement noise variance, \texttt{lensfit} assigns a weight to the shear measurement of each galaxy, $w_{i}$, which scales the shear signal such that the signal-to-noise ratio is optimal. Although {KiDS-SBI} currently does not model \texttt{lensfit} weights (all galaxies are weighted equally by default), the weights are carried through the measurement pipeline when measuring the cosmic shear summary statistics from the real KiDS-1000 data. If the \texttt{lensfit} weight is mischaracterised, this may lead to shear bias. For this purpose, the shape measurements in the KiDS-1000 gold sample are calibrated to determine any shear biases due to selection biases, noise, weight bias, point spread function (PSF) residuals, or otherwise \citep{giblin2021kids}. The resulting estimate of the linear multiplicative shear bias per tomographic bin is then used to unbias the galaxy shear measurements in KiDS-1000. All details on the reduction of the images and the shape calibration are given in \citet{kuijken2019the, wright2019kids, giblin2021kids}.

The photometric redshifts were estimated from the KiDS 9-band photometry using Bayesian template-fitting as incorporated within the \texttt{BPZ} code \citep{benitez2000bayesian, wright2019kids, wright2020photometric}. The photometric redshifts are calibrated with spectroscopic redshift measurements made of galaxies in the KiDS sample \citep{hildebrandt2021kids} based on a subsample of $99\%$ completeness. The mapping from photometric to spectroscopic redshift, $P(z|z_{\mathrm{ph}})$, is learnt through a self-organising map \citep{wright2020photometric, hildebrandt2021kids}. As is the case for previous KiDS-1000 analyses, we use the KiDS-1000 'gold-sample', which only considers galaxies with reliable redshift and shape measurements. The maximum posterior photometric redshift is subsequently used to split the source galaxy catalogue into five tomographic bins at the following boundaries: $(0.1, 0.3, 0.5, 0.7, 0.9, 1.2]$. These  give the redshift distributions shown in Fig.~\ref{fig:kids_nz}. As the cosmological constraints are sensitive to shifts in the redshift distributions, the KiDS-1000 analysis also allows the shift in the mean of each tomographic bin to vary freely in the cosmological inference as a nuisance parameter while incorporating for the correlations between them \citep{asgari2021kids, heymans2021kids, busch2022kids}. This not only avoids biases in the cosmological parameter estimates, but also provides an additional test on the redshift calibration.

We also note that KiDS-SBI treats the KiDS-1000 catalogue as a population of point sources unaffected by source blending, since the KiDS data already accounts for overlapping sources in multiple ways. Firstly, the KiDS photometry integrates Gaussian Aperture and PSF photometry (GAaP; \citealt{kuijiken2015gravitational, kuijken2019the, dejong2017third}), where a Gaussian aperture is applied to the images of all sources that is optimised for S/N in the fluxes while also avoiding source blending. An aperture size of 2” is found to be optimal for these purposes and applied to all sources \citep{kuijiken2015gravitational, kuijken2019the, dejong2017third}. Secondly, at the source detection step with  \texttt{SExtractor} \citep{bertin1996}, explicit flagging of overlapping sources occurs. Approximately 5\% of sources \citep{kuijken2019the, giblin2021kids} are excluded from the sample due to this selection. Additionally, \texttt{lensfit} weights also account for any biases due to irregular objects which are not affected by the selection. We note that deblending of sources is not attempted at any point. Lastly, at the calibration stage, the KiDS-1000 shape measurements are validated against image simulations from \citet{kannawadi2019towards}. The multiplicative shear bias values obtained in this analysis \citep{giblin2021kids} incorporate any residual biases due to correlated noise in the shape measurements, including source blending.

\section{Forward simulations}\label{fs}
The forward simulations for cosmic shear analyses described in this paper are fully implemented and available in the {KiDS-SBI}\footnote{Kilo-Degree Survey -- Simulation-Based Inference; \url{https://github.com/mwiet/kids_sbi}} module. This code is built within the Cosmological Survey Inference System (\texttt{CosmoSIS}\footnote{\url{https://github.com/joezuntz/cosmosis}}; \citealt{zuntz2015cosmosis}) and based upon the KiDS Cosmology Analysis Pipeline (\texttt{KCAP}\footnote{\url{https://github.com/KiDS-WL/kcap}}; \citealt{joachimi2021kids, asgari2021kids, heymans2021kids, troester2021kids}). In addition, the {KiDS-SBI} pipeline is generalisable to other cosmic shear analyses.

\begin{figure*}
    \centering
    \begin{tikzpicture}[node distance=1.4cm]
        \node (start) [startstop, text width=8cm] {Cosmological parameters, $\mathbf{\Theta}$};
        \node (camb) [middle, below of=start, text width=7cm] {3D matter power spectrum, $P_{\delta, \mathrm{nl}}(\mathbf{k}, z)$: Sect.~\ref{method:fs:cosm}};
        \node (camb_in) [io, right of=camb,text width=3cm, xshift=7cm] {Baryonic feedback};
        \node (shell) [process, below of=camb, yshift=0cm] {Split space into concentric shells: $i, j...$: Sect.~\ref{method:fs:shells}};
        \node (shell_in) [io, right of=shell,text width=4cm, xshift=7cm] {Resolution along the line of sight};
        \node (levin) [process, below of=shell] {Non-Limber 2D projection: Sect.~\ref{method:fs:nonlimber}};
        \node (cl) [middle, below of=levin, text width=5cm] {2D power spectra, $C_{\delta \delta}^{(i j)}(\ell)$};
         \node (random) [process, below of=cl] {Generate log-normal random fields: Sect.~\ref{method:fs:glass}};
        \node (glass) [middle, below of=random, text width=6.5cm] {Matter, $\delta^{(i)}(\bm{\theta})$, \& convergence fields, $\kappa^{(i)}(\bm{\theta})$};
        \node (glass_in) [io, right of=glass, xshift=7cm, text width=4cm] {Intrinsic alignments, $\kappa_{\mathrm{IA}}^{(i)}(\bm{\theta})$: Sect.~\ref{method:fs:nla}};
        \node (sampling) [process, below of=glass, yshift=-0.3cm] {Sample galaxies: Sect.~\ref{method:fs:salmo} $\&$ \ref{method:fs:salmo_shears}};
        \node (salmo) [middle, below of=sampling, yshift=-0.2cm, text width=8cm] {Angular galaxy positions, $\bm{\theta}$, redshifts, $z_{\mathrm{true}}$, observed tomographic bin, $p$, and observed ellipticity, $\epsilon$};
        \node (salmo_in) [io, right of=sampling, xshift=7cm, text width=4cm] {Survey \\ characteristics};
             \node (salmo_in_in) [subio, below of=salmo_in, text width=5cm, yshift=-0.8cm, xshift=0.2cm] {\begin{itemize}\itemsep0em \vspace{-\baselineskip}  \item Mask
             \vspace{-0.5\baselineskip}\item Redshift distributions
             \vspace{-0.5\baselineskip}\item Shear biases
             \vspace{-0.5\baselineskip}\item Point spread function
             \vspace{-0.5\baselineskip}\item Galaxy density
             \vspace{-0.5\baselineskip}\item Intrinsic galaxy shapes
             \vspace{-0.5\baselineskip} \item Depth variability: Sect.~\ref{method:fs:salmo_vd}
             \end{itemize}};
        \node (measure) [process, below of=salmo, yshift=-0.25cm] {Measure and correct shear: Sect.~\ref{method:fs:shapes}};
        \node (obs) [middle, below of=measure, yshift=0cm, text width=4cm] {Observed shear, $\epsilon^{\mathrm{corr}}$};
        \node (measurecl) [process, below of=obs, yshift=0cm] {Measure pseudo-Cls, subtract noise and binning: Sect.~\ref{method:fs:pcl}};
        \node (pcl) [startstop, below of=measurecl, yshift=0cm, text width=8cm] {Pseudo-Cl, $\Tilde{C}_{\epsilon \epsilon}^{(p q)}(\ell)$};
        
        \draw [arrow] (start) -- (camb);
        \draw [arrow] (camb) -- (shell);
        \draw [arrow] (shell) -- (levin);
        \draw [arrow] (levin) -- (cl);
        \draw [arrow] (cl) -- (random);
        \draw [arrow] (random) -- (glass);
        \draw [arrow] (glass) -- (sampling);
        \draw [arrow] (sampling) -- (salmo);
        \draw [arrow] (salmo) -- (measure);
        \draw [arrow] (measure) -- (obs);
        \draw [arrow] (obs) -- (measurecl);
        \draw [arrow] (measurecl) -- (pcl);
        
        \draw [arrow] (camb_in) -- (camb);
        \draw [arrow] (shell_in) -- (shell);
        \draw [arrow] (glass_in) -- (glass);
        \draw [arrow] (salmo_in) -- (sampling);
    \end{tikzpicture}
\caption{Flowchart describing the steps in a single forward simulation of cosmic shear observables from cosmological parameters. The dark blue rounded boxes represent the inputs and outputs which are given to the simulation-based inference pipeline. The green slanted boxes represent relevant quantities which are calculated during the simulation. The grey rectangular boxes show steps in the calculations, while the blue slanted boxes show any (systematic) effects which are included. All variables are defined within the respective sections quoted in the diagram.}
\label{fig:sim_diagram}
\end{figure*}
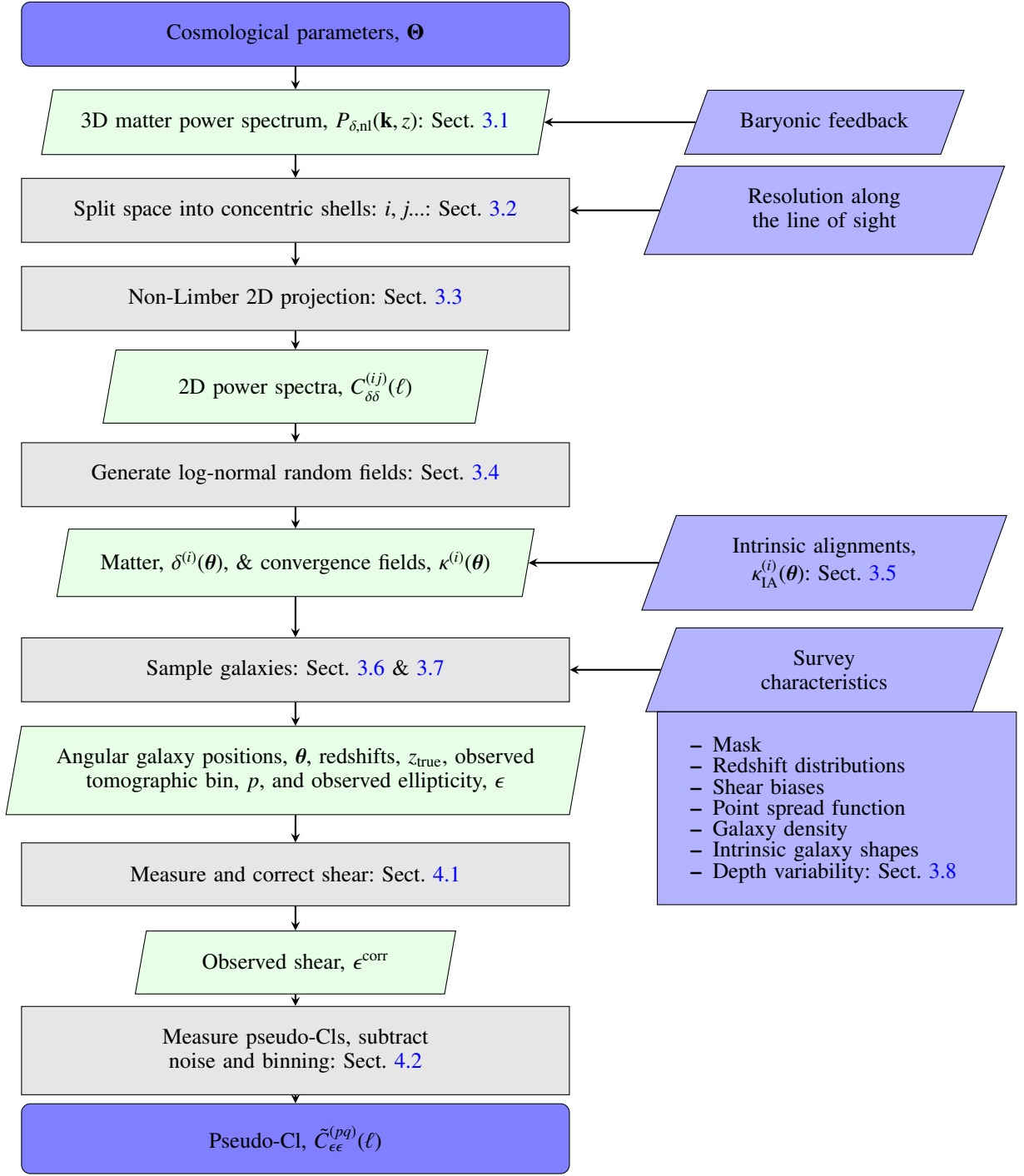

The overall outline of the forward simulations is shown in Fig.~\ref{fig:sim_diagram}. Firstly, the simulation is fed a set of cosmological and astrophysical parameters, $\mathbf{\Theta}$, from a sampler of our choice. All of these parameters are passed into \texttt{CAMB} \citep{lewis2000efficient, lewis2002evolution, howlett2012cmb} which calculates the 3D matter power spectrum, $P_{\delta, \mathrm{nl}}$, at a given cosmology (see also Sect.~\ref{method:fs:cosm} for more details). We incorporate baryonic feedback using \texttt{HMCode} \citep{mead2016accurate}. Next, the simulation splits the light cone along the line of sight into concentric spherical shells centred at the observer, as required by our large-scale structure simulator of choice, \texttt{GLASS} \citep{tessore2023glass}. Such an approach is computationally efficient as all observables are simulated with a coarse resolution along the line of sight. Within each shell, we sample log-normal random matter fields to efficiently model the two-point statistics with high accuracy, while also incorporating some higher-order fluctuations. As an added benefit, this simplifies the implementation of effects such as redshift evolution as well as survey characteristics, as they usually also follow a spherical geometry. On the other hand, the concentric shells also allow for the Universe to be split into shells, which are of a scale large enough such that a log-normal field becomes an accurate description of the matter over-densities within it (see Sect.~\ref{method:fs:shells} for details).

Once we know the geometry of our matter shells, we can project the 3D matter power spectrum using the redshift distributions of these shells in order to obtain their 2D angular power spectra, $C_{\delta \delta}$. The latter describes the correlations between the matter over-densities in each shell. Since we also require accurate cross-correlations between different shells to compute the log-normal fields, the Limber approximation is not applicable here, as the cross-correlations are zero in that approximation with the shells being non-overlapping along the line of sight. To avoid this, we project the 3D matter power spectrum with a non-Limber integral. This is usually an expensive computation, so we show here a novel implementation of the Levin method of integration \citep{levin1996fast}, which allows for the efficient integration of the spherical Bessel functions found in non-Limber integrals (more details are given in Sect.~\ref{method:fs:nonlimber}).

We can then construct log-normal matter fields which are consistent with these correlations and their associated cosmology. To do so, we used the Generator for Large Scale Structure (\texttt{GLASS}\footnote{Generator for Large Scale Structure; \url{https://github.com/glass-dev/glass}}, \citealt{tessore2023glass}), which efficiently generates correlated log-normal random matter fields, $\delta$, for each shell. Then, it accurately computes the respective convergence field, $\kappa$, (see Sect.~\ref{method:fs:glass}) while also incorporating an effective convergence field due to intrinsic galaxy alignments, $\kappa_{\mathrm{IA}}$ (see Sect.~\ref{method:fs:nla}).

Upon the construction of the matter, $\delta$, and convergence fields, $\kappa$, we used them to Poisson-sample galaxies that reside within the matter fields as well as their properties, namely, intrinsic ellipticities and shear. During this step, with the aid of the \texttt{SALMO} module \citep{joachimi2021kids}, we also consider survey characteristics, such as the survey footprint, along with its spatial variability in depth and in its redshift distributions, as well as in the shape and shot noise, as described in Sect.~\ref{method:fs:salmo}. Thus, we obtained a full galaxy catalogue with shape information consistent with the input cosmology as well as with the previously mentioned survey properties.

The last steps of the forward simulation pipeline involve the post-processing and compression of the simulated data catalogue. In principle, any data vector could be obtained here, such as two-point shear statistics \citep{asgari2021kids, busch2022kids}, the shear fields \citep{porqueres2022lifting, loureiro2023almanac}, or even the full catalogues. However, in order to reduce the dimensionality of the data vector efficiently, while still retaining most information about the cosmology, similar to previous KiDS-1000 analyses \citep{loureiro2021kids, troester2022joint}, we chose the observed angular power spectra, also known as 'pseudo-Cls' \citep{peebles1973statistical}, as our data vector of choice. This allowed us to compare our SBI analysis to the results from the standard analysis of KiDS-1000 data using pseudo-Cls in \cite{loureiro2021kids}. The {KiDS-SBI} pipeline is, in principle, set up to make cosmological inferences from cosmic shear based on any statistic which could be derived from a galaxy catalogue, but the underlying statistical random fields are currently only capable of achieving percent-level accuracy for two-point statistics. Any higher-order statistics measured from the current forward simulations are not necessarily accurate enough to conduct inference. The statistics and their post-processing are defined in Sect.~\ref{method:fs:pcl}.

The forward simulations are designed to model weak gravitational lensing observations on the level of galaxy catalogues. The galaxy populations are sampled such that they trace the underlying log-normal random matter fields, while the galaxy shapes are lensed in accordance with the lensing potential of the matter fields along the line of sight. As the simulations do not model images, any image-level systematic effects, such as shear biases or variable depth, are included as spatially varying probability density functions.

\subsection{Cosmology dependence: 3D matter power spectrum}\label{method:fs:cosm}
The basis of the forward simulations is the 3D matter power spectrum, $P_{\delta, \mathrm{nl}}(k, z; \mathbf{\Theta})$. In the linear regime, the equal-time 3D matter power spectrum's cosmology dependence comes from the following relation:
\begin{equation}
    P_{\delta, \mathrm{l}}(k, z; \mathbf{\Theta}) =  T^{2}(k; \mathbf{\Theta}) \, D^2(z; \mathbf{\Theta}) \, P_{\mathcal{R}}(k; \mathbf{\Theta})\,,
\end{equation}
\noindent where $k$ is a wavenumber, $z$ is redshift, $\mathbf{\Theta}$ is the set of cosmological parameters, $P_{\mathcal{R}}(k; \mathbf{\Theta})$ is the primordial density fluctuation power spectrum, $D$ is the growth factor of structure, $T$ is the transfer function between the matter overdensity field, $\mathbf{\delta}(\mathbf{k}, z)$, and the primordial curvature fluctuation field, $\mathcal{R}(\mathbf{k}; \mathbf{\Theta})$, such that $\mathbf{\delta}(\mathbf{k}, z) = T(\mathbf{k}; \mathbf{\Theta})  D(z; \mathbf{\Theta}) \mathcal{R}(\mathbf{k}; \mathbf{\Theta})$. To compute this, we make use of \texttt{CAMB} \citep{lewis2000efficient, lewis2002evolution, howlett2012cmb}. To stay in line with the main KiDS-1000 analysis \citep{joachimi2021kids, asgari2021kids, busch2022kids}, we assume a normal neutrino hierarchy, while also assuming a fixed sum of neutrino masses of $\sum m_{\nu} = 0.06$ eV/$c^2$.

We can compute the non-linear matter power spectrum via a non-perturbative model, \texttt{HMCode-2016} \citep{mead2015an, mead2016accurate}. Although updated iterations of this model exist \citep{mead2021hmcode}, we limited our study to \texttt{HMCode-2016}, so our model remains comparable to the KiDS-1000 cosmic shear analysis. \texttt{HMCode} uses a halo model approach to incorporate the effects of baryonic feedback on the matter distribution. The main driving factor in this is the baryonic matter expelled by active galactic nuclei (AGNs). This suppresses the power at small scales, namely, large values of $k$, as a function of the amplitude of the halo mass-concentration relation, $A_{\mathrm{bary}}$, and the halo bloating parameter, $\eta_{0}$. We also fixed the relation between these two parameters to $\eta_{0} = 0.98 - 0.12 A_{\mathrm{bary}}$ in accordance with \citet{joudaki2018kids} and \citet{joachimi2021kids}, and treated $A_{\mathrm{bary}}$ as the only free parameter related to baryonic feedback.

We note that this treatment may lead to systematic biases at small scales, as the galaxies will be sampled from a matter power spectrum which is already modified by baryonic feedback. It would be more physically accurate to sample the galaxies from a matter power spectrum without baryonic non-linearities and subsequently add a perturbation to the matter density contrast fields due to baryonic feedback (similar to the approach used for the intrinsic alignments described in Sect.~\ref{method:fs:nla}). Nevertheless, we do not expect this to cause a large discrepancy for the scales which we are probing in the cosmic shear signal ($\ell < 1500$). Field-level implementations of baryonic feedback already exist \citep{schneider2015a, schneider2019quantifying, schneider2022constraining, porqueres2023field} and they would constitute interesting avenues for future extensions to {KiDS-SBI}. 

\subsection{Working on the sphere}\label{method:fs:shells}

Astronomical observations can be viewed as images of a 3D space projected onto the inside of a 2D surface of a sphere at a given distance with the observer at its centre. Therefore, galaxy position measurements, most observational biases and some systematics are naturally expressed in spherical coordinates on the sky. For this reason, the forward model presented in this work is expressed in the same geometry.

The {KiDS-SBI} forward simulations use \texttt{GLASS} \citep{tessore2023glass} to model the underlying large-scale structure. Within \texttt{GLASS}, we chose to model the large-scale structure through log-normal random matter fields within non-overlapping concentric shells centred at the observer. Therefore, a given shell is a comoving volume which spans the full sky and has a finite width along the line of sight from one redshift, $z_{i}$, to another, $z_{i+1}$, such that $i \in \{1, 2, 3, ..., N_{\mathrm{shells}}\}$. We define the matter weight function $W^{(i)}(z)$ describing the distribution along the line of sight of the comoving volume spanned by a given shell as \citep{tessore2023glass},
\begin{equation}
    W^{(i)}(z; \mathbf{\Theta}) = \begin{cases} f_{k}^2(z; \mathbf{\Theta})/E(z; \mathbf{\Theta}), & \text{if $z_{i} \leq z < z_{i+1}$\,,} \\ 0, & \text{otherwise\,,} \end{cases}
    \label{eq:method:fs:shells:weight}
\end{equation}
\noindent where $E(z; \mathbf{\Theta})$ is the dimensionless Hubble function and $f_{k}(z; \mathbf{\Theta})$ is the transverse comoving distance for a given redshift and cosmology, which is defined as
\begin{equation}
    f_{k}(z; \mathbf{\Theta}) = \begin{cases}
        \frac{c}{H_{0} \sqrt{\Omega_{k}}} \mathrm{sinh}\bigg(\frac{\sqrt{\Omega_{k}} H_{0} }{c} \chi(z; \mathbf{\Theta}) \bigg), & \text{if $\Omega_{k} > 0$\,,} \\
        \chi(z; \mathbf{\Theta}), & \text{if $\Omega_{k} = 0$}\,, \\
        \frac{c}{H_{0} \sqrt{|\Omega_{k}|}} \mathrm{sin}\bigg(\frac{\sqrt{|\Omega_{k}|} H_{0} }{c} \chi(z; \mathbf{\Theta}) \bigg), & \text{if $\Omega_{k} < 0$\,,}
    \end{cases}
\end{equation}
\noindent where $H_{0}$ is the Hubble constant, $\Omega_{k}$ is the curvature density parameter and $c$ is the speed of light in a vacuum. As noted in \citet{tessore2023glass}, different choice of window function may induce percent-level biases in the resulting shear power spectrum. We test for this in Appendix~\ref{appendix:consistency}, and find sub-percent level agreement between the measured shear from the simulations with this choice of window function and the shear signal expected from theory (see Appendix~\ref{appendix:signal}). 

A given shell should not be thicker than a few hundred $\mathrm{Mpc}$ of comoving distance along the line of sight for log-normal random fields to be an accurate enough representation of the distribution of large-scale structure \citep{xavier2016improving, hall2022non, tessore2023glass}. Otherwise, discretisation effects would smooth away large-scale structure \citep{tessore2023glass}. Additionally, we need to be mindful of computational resources, because the number required of angular power spectra between the matter fields within each shell scales as $2N_{\mathrm{shells}}$. Simultaneously, the simulation run-time for a given set of cosmological parameters increases with $N_{\mathrm{shells}}$. To balance these factors, we aim to have as many shells along the line of sight as computationally feasible. 

For an accurate weak lensing simulation at the relevant redshifts, $0.6 \lesssim z \lesssim 1.2$, we require shells that finely sample the lowest redshifts. The lensing signal in a given matter shell depends on the weighted sum of the matter fields of the shells with lower redshifts. At redshifts where lensing becomes more important as the lensing efficiency increases, prioritising thinner shells helps to reduce discretisation effects. We find that the following set of redshift limits for 19 shells are sufficient for the resolution needed for KiDS-1000, while still being efficient: $\{0, \, 0.04, \, 0.08, \, 0.12, \, 0.16, \, 0.2, \, 0.27, \, 0.34, \, 0.42, \, 0.5, \, 0.58, \, 0.66,$ $ \, 0.75, \, 0.88, \, 1.03, \, 1.19, \, 1.36, \, 1.55, \, 1.76, \, 2 \}$ (see Fig.~\ref{fig:kids_nz}).

\subsection{Non-Limber projection}\label{method:fs:nonlimber}
Although the 3D matter power spectrum characterises all cosmological dependence of the matter fields in our simulation, for it to define the cosmology dependence in the simulated matter fields, it has to be projected according to the geometry of the shells considered in Sect.~\ref{method:fs:shells}. To determine the correlations between the matter fields of comoving spherical shells, we define the 2D angular matter power spectra for a given cosmology, $\mathbf{\Theta}$, and a given shell combination $(i j)$ as follows
\begin{equation}
    C_{\delta \delta, \ell}^{(i j)}(\mathbf{\Theta}) = \langle \Tilde{\delta}^{(i)}_{\ell m}(\mathbf{\Theta}) \, \Tilde{\delta}^{(j)*}_{\ell m}(\mathbf{\Theta}) \rangle\,,
    \label{eq:method:fs:nonlimber:cl}
\end{equation}
\noindent where $\Tilde{\delta}^{(i)}_{\ell m}(\mathbf{\Theta})$ represents the harmonic coefficients defined through a spherical harmonic transform of the projection of the spin-0 matter field, $\delta^{(i)}$, as given by \citep{zaldarriaga1997all, reinecke2011libpsht},
\begin{equation}
    \Tilde{\delta}^{(i)}_{\ell m}(\mathbf{\Theta}) = \int \mathrm{d}^{2}\bm{\theta} \,\delta^{(i)}(\bm{\theta}; \mathbf{\Theta})\, \yellmarb{0}{*}{\ell m}(\bm{\theta})\,,
    \label{eq:method:fs:nonlimber:delta_lm}
\end{equation}
\noindent where $\yellmarb{0}{}{\ell m} (\bm{\theta})$ are the spherical harmonics for a spin value of 0 as a function of the 2D spatial sky position $\bm{\theta}$. The projected matter field, $\delta^{(i)}(\bm{\theta}; \mathbf{\Theta})$, can in turn be defined with respect to the underlying 3D matter overdensity field, $\delta(\bm{\theta}, z; \mathbf{\Theta})$, as
\begin{equation}
     \delta^{(i)}(\bm{\theta}; \mathbf{\Theta}) = \sum_{\ell m} \Tilde{\delta}^{(i)}_{\ell m}(\mathbf{\Theta})\yellmarb{0}{}{\ell m}(\bm{\theta}) = \int \mathrm{d}z \, W^{(i)}(z; \mathbf{\Theta}) \, \delta(\bm{\theta}, z; \mathbf{\Theta})\,,
     \label{eq:method:fs:nonlimber:delta_theta}
\end{equation}
\noindent where $W^{(i)}(z; \mathbf{\Theta})$ is the weight function of the given shell as defined in Eq.~(\ref{eq:method:fs:shells:weight}).

The 3D matter overdensity field's cosmology dependence can be quantified as follows
\begin{equation}
     \langle \delta(\mathbf{k}, z; \mathbf{\Theta}) \, \delta^{*}(\mathbf{k}', z'; \mathbf{\Theta}) \rangle = (2\pi)^{3} \delta_{\mathrm{D}}(\mathbf{k}-\mathbf{k}') \, P_{\delta, \mathrm{nl}}(k, z, z'; \mathbf{\Theta})\,,
     \label{eq:method:fs:nonlimber:pk}
\end{equation}
\noindent where $\delta(\mathbf{k}, z; \mathbf{\Theta})$ is the Fourier transform of the 3D matter overdensity field at $z$, $\delta(\bm{\theta}, z,; \mathbf{\Theta})$, while $\delta_{\mathrm{D}}(\mathbf{k}-\mathbf{k}')$ is the Dirac delta function.

If we combine the relations in Eqs.~(\ref{eq:method:fs:nonlimber:cl}) to (\ref{eq:method:fs:nonlimber:pk}) and, for the sake of simplicity, assume that $\Omega_{k}=0$, specifically $f_k(z; \mathbf{\Theta}) = \chi(z; \mathbf{\Theta})$, we can characterise the cosmology dependence of $C_{\delta \delta}^{(i j)}(\ell; \mathbf{\Theta})$ directly with respect to the 3D matter power spectrum, $P_{\delta, \mathrm{nl}}(k, z, z'; \mathbf{\Theta})$, using the following relation
\begin{multline}
    C_{\delta \delta}^{(i j)}(\ell; \mathbf{\Theta}) = \frac{2}{\pi} \int \mathrm{d}\chi \, W^{(i)}(z[\chi]; \mathbf{\Theta}) \int \mathrm{d}\chi' \, W^{(j)}(z[\chi']; \mathbf{\Theta}) \\ \int \mathrm{d}k \, k^{2} \,  P_{\delta, \mathrm{nl}}(k, z[\chi], z[\chi']; \mathbf{\Theta}) \, j_{\ell}(k \chi) \, j_{\ell}(k \chi')\,,
    \label{eq:method:fs:nonlimber:non_limber}
\end{multline}
\noindent where $\chi'\equiv z[\chi']$, $j_{\ell}(k\chi)$ are spherical Bessel functions of order $\ell$, and $\ell \in \mathbb{Z}^{0+}$. We can simplify Eq.~(\ref{eq:method:fs:nonlimber:non_limber}) by making the geometric approximation, that is taking the geometric mean of two equal-time power spectra \citep{castro2005weak, kitching2017unequal, kilbinger2017precision, bella2021unequal}, such that
\begin{equation}
    P_{\delta, \mathrm{nl}}(k, z[\chi], z[\chi']; \mathbf{\Theta}) = \sqrt{P_{\delta, \mathrm{nl}}(k, z[\chi]; \mathbf{\Theta}) \, P_{\delta, \mathrm{nl}}(k, z[\chi']; \mathbf{\Theta})}\,,
\end{equation}
\noindent which has been shown to be an accurate approximation both in the linear and non-linear regimes \citep{kitching2017unequal}.

\begin{figure}
    \centering
    \includegraphics[width=8.5cm]{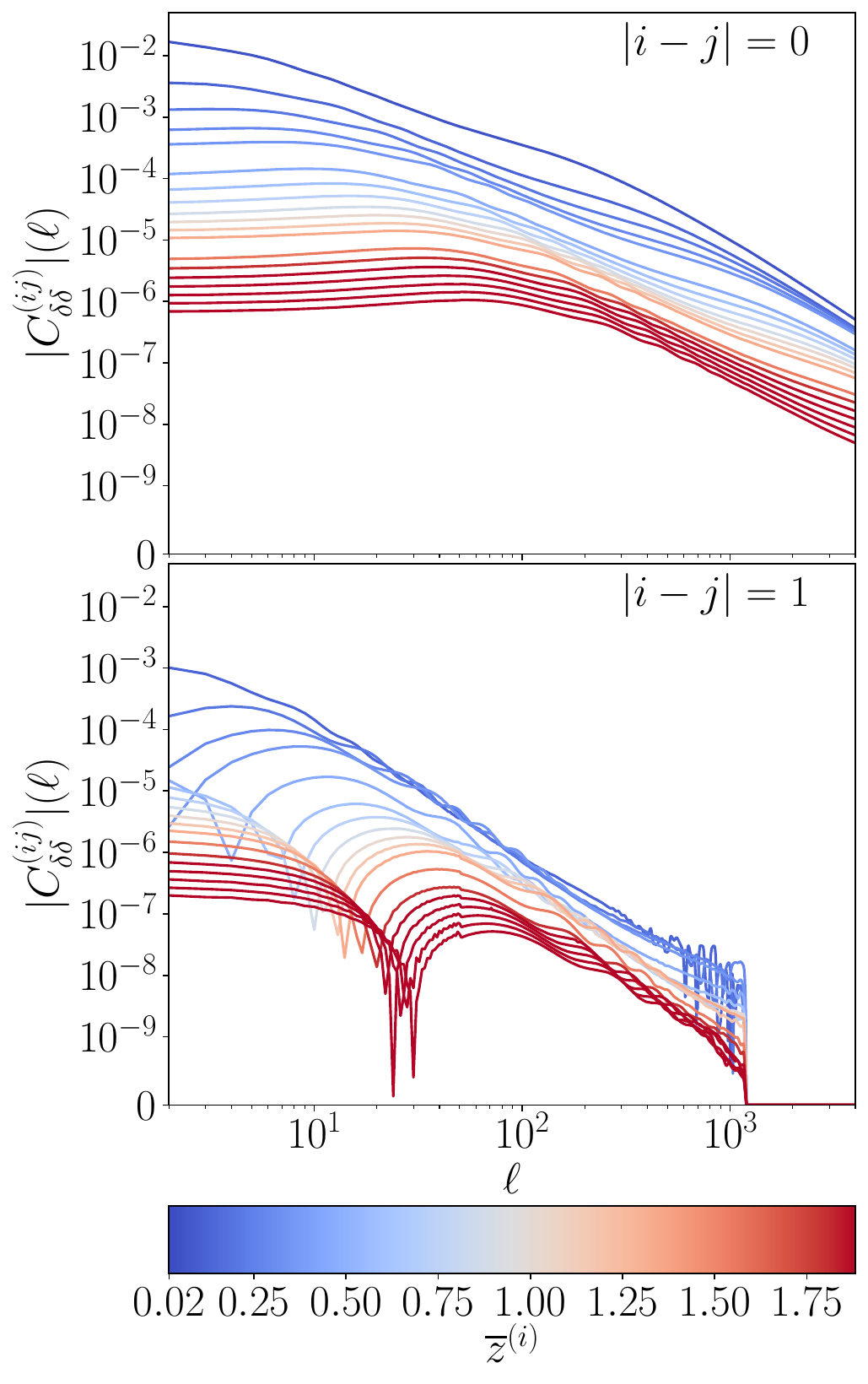}
    \caption{2D angular matter power spectra, $C_{\delta \delta}^{(i j)}(\ell)$, projected with \texttt{Levin} describing the correlations between the large-scale structure within a set of 19 concentric shells spaced along the line of sight as shown in Fig.~\ref{fig:kids_nz}. The top panel shows the autocorrelations of all shells, i.e. $|i-j| = 0$. The bottom panel shows the correlations of each shell with its nearest neighbour, i.e. $|i - j| = 1$ ($|i-j|>1$ are not shown as they do not have a large effect on matter fields within each shell, see \citealt{tessore2023glass}). The colour of each line is given by the mean redshift of the $i^\mathrm{th}$ bin, $\overline{z}^{(i)}$. The underlying linear 3D matter power spectrum is based on flat $\Lambda$CDM assuming $\Omega_{\mathrm{c}} = 0.05$, $\Omega_{\mathrm{b}} = 0.28$, $\sigma_{8} = 0.79$, $S_{8} = 0.84$ and $H_{0} = 67$ km s$^{-1}$ Mpc$^{-1}$ calculated using \texttt{CAMB} \citep{lewis2000efficient, lewis2002evolution, howlett2012cmb}, while the non-linear contribution is calculated using \texttt{HMcode-2016} \citep{mead2015an, mead2016accurate} assuming $A_{\mathrm{bary}} = 3.1$. The non-Limber projection done by \texttt{Levin} assumes $\ell_{\mathrm{max, nL}} = 1200$ which causes the $C_{\delta \delta}^{(ij)}$ with $|i-j| > 0$ go to a value of zero for $\ell > \ell_{\mathrm{max, nL}}$.}
    \label{fig:non_limber}
\end{figure}

Additionally, it is common to make the so-called Limber approximation \citep{limber1953analysis, kaiser1992galactic} which consists in approximating a spherical Bessel function as a single peak or delta function at $k\chi = \ell + 1/2$. This approximation is reasonably accurate in the auto-correlations for broad window functions and high angular modes, $C_{\delta \delta}^{(i i)}(\ell; \mathbf{\Theta})$,  but depending on the thickness of the shells, the Limber approximation can lead to substantial biases, especially, at large scales \citep{lemos2017effect}. However, when making the Limber approximation, all off-diagonal angular power spectra, $C_{\delta \delta}^{(i j)}$, where $i \neq j$, will be zero as long as their associated weight functions, $W^{(i)}(z[\chi]; \mathbf{\Theta})$, are non-overlapping, as is the case for the weights constructed in Eq.~(\ref{eq:method:fs:shells:weight}). This can be seen by looking at the full Taylor expansion of Eq.~(\ref{eq:method:fs:nonlimber:non_limber}) as given by \citep{loverde2008extended}
\begin{multline}
    C_{\delta \delta}^{(i j)}(\ell) = \int \frac{\mathrm{d}\chi}{\chi^2} \, W^{(i)}(z[\chi]) \, W^{(j)}(z[\chi]) \, P_{\delta, \mathrm{nl}} \Bigg(\frac{\ell +1/2}{\chi}, z[\chi] \Bigg) \\
    \Bigg\{ 1 - \frac{1}{(\ell + 1/2)^2} \Bigg[ \frac{\chi^2}{2} \Bigg( \frac{W^{(i)''}(z[\chi])}{W^{(i)}(z[\chi])} \, + \frac{W^{(j)''}(z[\chi])}{W^{(j)}(z[\chi])}\Bigg) \, \\ + 
    \frac{\chi^3}{6} \Bigg( \frac{W^{(i)'''}(z[\chi])}{W^{(i)}(z[\chi])}  + \frac{W^{(j)'''}(z[\chi])}{W^{(j)}(z[\chi])}\Bigg) \Bigg] + \mathcal{O}\Big([\ell + 1/2]^{-4} \Big) \Bigg\}\,,
    \label{eq:method:fs:nonlimber:extended_limber}
\end{multline}
\noindent where each apostrophe, $'$, denotes a partial derivative with respect to $\chi$, specifically $\partial/ \partial \chi$, while we also omit the explicit cosmological dependence of the power spectrum and the weight functions here for the sake of clarity. The non-Limber integral therefore also contributes to the autocorrelations, where $|i-j|=0$, particularly for small $\ell$ and the narrow weight functions considered here. If not considered this can induce a large bias in the angular power spectra of the shells which could significantly bias cosmological constraints from KiDS-1000. At the same time, we find that for the cross-correlations of adjacent shells, such that $|i-j|=1$, the angular power can be of the order of $10^{-1} C_{\delta \delta}^{(i i)}$ or less. In this case, the simulated $\delta^{(i)}(\bm{\theta}; \mathrm{\Theta})$ would have a percent-level bias while being void of any large-scale structure along the line of sight beyond the scales of individual shells. To avoid this, we chose to calculate the full non-Limber projection as given by Eq.~(\ref{eq:method:fs:nonlimber:non_limber}). The correlations between shell pairs which are not immediate neighbours, that is $|i-j|>1$, contribute less than $0.1\%$ to the overall angular power of a shell, so their effect is undetectable for KiDS.

Calculating the full non-Limber integral is a computationally expensive endeavour as it involves three numerical integrals over highly oscillatory Bessel functions. To address this, we present the \texttt{Levin} module\footnote{\texttt{Levin}; \url{https://github.com/rreischke/nonLimber_matter_shells}}. This is a novel Python module which implements the Levin integration method for oscillatory functions \citep{levin1996fast} in the context of non-Limber integrals for weak lensing and galaxy clustering which is discussed in more detail in Appendix~\ref{appendix:levin}.

The module allows us to divide the domain of the angular power spectra into three regions: a non-Limber domain, a second-order extended Limber domain and a Limber domain. In this work, we chose to split the calculation of $C_{\delta \delta}^{(i j)}(\ell)$ into two domains: non-Limber ($1 < \ell \leq \ell_{\mathrm{max, nL}}$) and second-order extended Limber given in Eq.~(\ref{eq:method:fs:nonlimber:extended_limber}) for $\ell_{\mathrm{max, nL}} < \ell \leq 30,000$. Over the former domain, \texttt{Levin} numerically integrates the expression shown in Eq.~(\ref{eq:method:fs:nonlimber:non_limber}) for the given matter power spectrum and the weights. Over the latter domain, we perform the numerical integration of Eq.~(\ref{eq:method:fs:nonlimber:extended_limber}) up to the second-order term in $(\ell+1/2)^{-1}$. This reduces the computational resources needed for the integration without any substantial loss of accuracy as long as $\ell_{\mathrm{max, nL}}$ is sufficiently large, since the residuals scale with $(\ell + 1/2)^{-4}$ \citep{loverde2008extended}. With the shells shown in Fig.~\ref{fig:kids_nz} weighted with Eq.~(\ref{eq:method:fs:shells:weight}), we obtain angular power spectra, $C_{\delta \delta}^{(i j)}(\ell; \mathbf{\Theta})$, shown in Fig.~\ref{fig:non_limber}. We note that, in Fig.~\ref{fig:non_limber}, the Limber extension to second order from Eq.~(\ref{eq:method:fs:nonlimber:extended_limber}) calculated for $\ell > 1200$ reduces roughly to the Limber approximation as the higher-order terms have a vanishing contribution to the overall angular power spectrum. In fact, in the cross-correlations between shells where $|i-j|>0$, the second order term in Eq.~(\ref{eq:method:fs:nonlimber:extended_limber}) becomes vanishingly small for these values of $\ell$. Consequently, we effectively apply a scale cut in the matter shell cross-correlations for $\ell>1200$. We find that this does not bias our analysis, as the relative contribution cross-correlations to the total $C^{(ij)}_{\delta \delta} (\ell)$ of a given shell is less than $0.1 \%$ for $\ell \geq 1200$ for all redshifts. This ensures sub-percent level accuracy of the input matter fields across all scales considered in this analysis.

With this configuration, we find that we can obtain computationally efficient and accurate (consistent with \texttt{CAMB} within $0.1\%$) angular power spectra which describe the correlations between the matter fields within each spherical shell of the simulation. This is in line with testing conducted on the \texttt{Levin} module in previous applications \citep{zieser2016the, mancini2018testing, mancini20183d, baleato2023the}. 

\subsection{Log-normal matter field simulations}\label{method:fs:glass}
With the input angular power spectra for a given cosmology computed as described in Appendix~\ref{appendix:levin}, the next step in the forward simulations involves sampling random matter fields consistent with these correlations. These fields are generated with \texttt{GLASS} (\citealt{tessore2023glass}). This framework allows us to create efficient and accurate random field simulations of large-scale structure matter fields as well as the associated weak gravitational lensing signals. Within the \texttt{GLASS} suite, many different choices can be made, while it also allows us to sample the angular positions, redshifts, shapes and shears of galaxies. Nevertheless, in this analysis, we only chose to use it to simulate the underlying matter overdensity fields, $\delta^{(i)}(\bm{\theta}; \mathbf{\Theta})$, as well as the associated convergence fields for each shell, $\kappa^{(i)}(\bm{\theta}; \mathbf{\Theta})$. These are then integrated along the line of sight to give the convergence field for a source within a given shell, $\kappa^{(i)}$.

It has been found that a good approximation to such fields is a log-normally distributed random field (which can be derived from a Gaussian random field). In fact, log-normal fields are even able to produce reasonable approximations to the three-point and four-point statistics measured from N-body simulations \citep{hall2022non, piras2023fast}. For this reason, log-normal random fields are a common approximation used for matter overdensity fields \citep{coles1991lognormal, bohm2017bayesian, abramo2016fourier, abramo2022fisher} as well as for convergence fields \citep{hilbert2011cosmic, clerkin2017testing, giocoli2017fast, gatti2020dark}. 

Log-normal fields within \texttt{GLASS} are based on Gaussian random fields. We can obtain these by taking the angular power spectra of the shells and assuming that they describe the correlations between log-normal fields, $\delta_{\mathrm{log}}^{(i)}(\bm{\theta}; \mathbf{\Theta})$, as follows
\begin{equation}
    \langle \delta_{\mathrm{log}}^{(i)}(\bm{\theta}; \mathbf{\Theta})  \, \delta_{\mathrm{log}}^{(j)*}(\bm{\theta}; \mathbf{\Theta})  \rangle =  C_{\delta \delta}^{(i j)}(\theta; \mathbf{\Theta})\,.
    \label{eq:method:fs:glass:lognormal}
\end{equation}
By defining the mean and the variance for $G_{\delta \delta}^{(i j)}(\theta; \mathbf{\Theta})$ as 
\begin{equation}
    \langle \delta_{\mathrm{G}}^{(i)}(\bm{\theta}; \mathbf{\Theta})  \, \delta_{\mathrm{G}}^{(j)*}(\bm{\theta}; \mathbf{\Theta})  \rangle =  G_{\delta \delta}^{(i j)}(\theta; \mathbf{\Theta})\,,
    \label{eq:method:fs:glass:gaussian}
\end{equation}
\noindent we can then sample $\delta_{\mathrm{G}}^{(i)}(\bm{\theta}; \mathbf{\Theta})$ for each shell. To achieve this with numerical methods, the 2D spherical shell of each layer is discretised using \texttt{HEALPix} pixels \citep{gorski2005healpix}. This means that $\bm{\theta}$ is discretised such that it can be mapped to a linearised variable given by $\theta_{m}$ where $m \in \{1, 2, 3, ..., 12 N_{\mathrm{side}}^{2} \}$ and $N_{\mathrm{side}}$ is the \texttt{HEALPix} resolution parameter. To obtain the final matter overdensity fields, we use the following expression \citep{coles1991lognormal, kayo2001probability, hilbert2011cosmic, xavier2016improving},
\begin{equation}
    \delta_{\mathrm{log}}^{(i)}(\theta_{m}; \mathbf{\Theta}) = e^{\delta_{\mathrm{G}}^{(i)}(\theta_{m}; \mathbf{\Theta})} - \lambda\,,
\end{equation}
\noindent where $\lambda$ is the shift of the log-normal distribution which is assumed to be $\lambda = 1$ for matter fields. For more details about this calculation, see \citealt{tessore2023glass}.

The layers of concentric log-normal matter overdensity fields within each shell give a full description of the large-scale structure. The resolution of this structure will of course be limited by the size of the shells along the line of sight, while being limited by the resolution of the discretised 2D pixels on the sphere. However, as discussed with regards to the resolution of shells in Sect.~\ref{method:fs:shells}, if we chose a sufficiently large $\ell_{\mathrm{max}}$ value up to which the input angular power spectra are calculated, we can accurately sample $\delta_{\mathrm{log}}^{(i)}(\theta_{m}; \mathbf{\Theta})$ up to an $N_{\mathrm{side}} \sim 0.5 \ell_{\mathrm{max}}$ \citep{leistedt2013estimating, alonso2019a}.

As  these simulations were performed prior to carrying out the weak gravitational lensing study, the next step in the forward simulations is to model how the image of galactic sources within a given shell will be distorted by weak gravitational lensing due to the matter overdensities the light encounters along the line of sight to the observer. Here, the geometry of the simulations aids us again, since the concentric volumes in comoving distance that each shell makes up intrinsically allow us to trace all possible light cones emanating from the observer.

We start from the definition of the convergence for a source located at $\bm{\theta}$ and $z$ along a continuous line of sight under the Born approximation \citep{bartelmann2001weak},
\begin{equation}
\begin{split}
    \kappa(\bm{\theta}, \mathit{z}; \mathbf{\Theta}) = \frac{3 \Omega_\mathrm{m}}{2} \int_{0}^{\mathit{z}}  \mathrm{d}\mathit{z}'& \, \frac{\mathit{f}_{\mathit{k}}(\mathit{z}'; \mathbf{\Theta}) \, [\mathit{f}_{\mathit{k}}(\mathit{z}'; \mathbf{\Theta})-\mathit{f}_{\mathit{k}}(\mathit{z}; \mathbf{\Theta})]}{\mathit{f}_{\mathit{k}}(\mathit{z}; \mathbf{\Theta})} \\ &\times \frac{1+\mathit{z}'}{E(\mathit{z}'; \mathbf{\Theta})} \,  \delta(\bm{\theta}, \mathit{z}'; \mathbf{\Theta})\,,
    \label{eq:method:fs:glass:continuous_kappa}
    \end{split}
\end{equation}
\noindent where $\Omega_\mathrm{m}$ is the matter density fraction at $z=0$, and $\Omega_\mathrm{m} \in \mathbf{\Theta}$. We then discretise the continuous 2D matter overdensity field along over $\bm{\theta}$ using \texttt{HEALPix} pixels, while also discretising along the line of sight using Eq.~(\ref{eq:method:fs:nonlimber:delta_theta}) and defining a mean redshift, $\overline{z}^{(i)}$, for a given weighted shell as follows
\begin{equation}
    \overline{\mathit{z}}^{(i)}(\mathbf{\Theta}) =  \frac{\int \mathrm{d}\mathit{z} \, \mathit{z} \, \mathit{W}^{(i)}(\mathit{z}; \mathbf{\Theta})}{\int \mathrm{d}\mathit{z}  \, \mathit{W}^{(i)}(\mathit{z}; \mathbf{\Theta})}\,.
\end{equation}

To accelerate the calculation of the discretised convergence fields for each shell, $\kappa^{(i)}$, within \texttt{GLASS}, it makes use of the fact that for all Robertson-Walker space-times, transverse comoving distances will scale in such a way that we can write down a recurrence relation from one interval to the next along the line of sight \citep{schneider2016generalized}. This gives a recurrence relation for the convergence field, $\kappa^{(i)}$, within shells where $i \geq 2$, which only depends on $\kappa^{(i-1)}$, $\kappa^{(i-2)}$ and $\delta^{(i-1)}$ (see \citealt{tessore2023glass} for details). Such a relation allows us to calculate the convergence fields for all shells, while only holding three fields in memory at a given time. This greatly reduces the amount of computational resources needed per realisation, so the large amount of simulations at different cosmologies, $\mathbf{\Theta}$, needed for this analysis can be obtained with fewer resources when compared to forward models in previous analysis (see Sect.~\ref{method:sbi:performance}).
Once we  have all the ingredients to sample galaxies and to shear their shapes, while simultaneously having the flexibility to add any relevant field-level systematics,  we can proceed to the next phase of the study.
\subsection{Intrinsic alignments: Non-linear alignment model}\label{method:fs:nla}
An important effect to consider when modelling weak gravitational lensing is intrinsic alignments (IAs). This refers to the fact that correlations between source galaxy shapes in different parts of the sky may not only be caused by weak gravitational lensing due to the foreground matter distribution. This happens due to two local processes which occur irrespective of weak gravitational lensing:  tidal alignments and tidal torquing. The latter is thought to arise when galaxy discs form perpendicular to the angular momentum axis of the tidal field, while tidal alignments occur when the gravitational tidal forces from neighbouring matter distributions cause the galaxies to align with the surrounding matter density field. This alignment leads to a coherent stretching or compression of the galaxy shapes as a function of the local matter overdensity field. Hence, the intrinsic ellipticities of source galaxies are systematically aligned with the underlying large-scale structure within which they form. Thus, the measured cosmic shear signal will be biased by these intrinsic alignments \citep{heavens2000intrinsic, king2002supressing, heymans2003weak, bridle2007dark}.

To model this effect, we followed the methodology set out in \citet{tessore2023glass}. In this prescription, for a given intrinsic alignment model, we define an associated effective convergence field, $\kappa^{(i)}_{\mathrm{IA}}$. This $\kappa^{(i)}_{\mathrm{IA}}$ is a useful construct which describes the contribution to the observed weak lensing signal from IA under the Born approximation. Alternatively, an IA model could also directly be implemented into the dark matter distribution of the matter fields which could be an interesting avenue for future work.

Based on the fact that for the IA model, we take into consideration $\kappa^{(i)}_{\mathrm{IA}} \propto \delta^{(i)}$, we can assume that, for a given shell, this field can be added linearly to the underlying convergence field due to weak lensing, as given by
\begin{equation}
    \kappa^{(i)}(\theta_{m}; \mathbf{\Theta}) \rightarrow \kappa^{(i)}(\theta_{m}; \mathbf{\Theta}) + \kappa_{\mathrm{IA}}^{(i)}(\theta_{m}; \mathbf{\Theta})\,.
\end{equation}
For simplicity and consistency with typical modelling assumptions in weak lensing surveys such as KiDS-1000 \citep{joachimi2021kids}, we chose to model $\kappa^{(i)}_{\mathrm{IA}}$ with the Non-Linear Alignment (NLA) model \citep{catelan2001intrinsic, hirata2004galaxy, bridle2007dark}, so we consider tidal alignments and do not consider tidal torquing. The NLA model assumes that the bias in the shear signal from IAs is linearly dependent on the projected local tidal field.  The 'non-linear' part of the NLA model then simply refers to the fact that it has been found that the modelling of IAs is more accurate when modelling the underlying large-scale structure using a non-linear matter power spectrum, rather than a linear one (\citealt{bridle2007dark}, similar to our approach described in Sect.~\ref{method:fs:cosm}). This means that it is also proportional to the local matter overdensity field, which is given by \citep{hirata2004galaxy},
\begin{equation}
    \kappa_{\mathrm{IA}}^{(i)}(\theta_{m}; \mathbf{\Theta}) = - A_{\mathrm{IA}} \frac{C_{1} \, \Omega_{\mathrm{m}} \, \overline{\rho}_{\mathrm{cr}}(\overline{z}^{(i)}; \mathbf{\Theta})}{D(\overline{z}^{(i)}; \mathbf{\Theta})} \, \delta^{(i)}(\theta_{m}; \mathbf{\Theta})\,,
    \label{eq:method:fs:nla:model}
\end{equation}
\noindent where $A_{\mathrm{IA}}$ is the intrinsic alignments amplitude (which we are treating as a nuisance parameter that is also sampled, see Sect.~\ref{method:sbi}), $C_{1}$ is a normalisation constant which we set to $C_{1} = 5 \times 10^{-14} h^{-2} M_{\odot}^{-1} \mathrm{Mpc}^{3}$ in agreement with the IA measurements at low redshifts by SuperCOSMOS \citep{brown2002measurement}, $\overline{\rho}_{\mathrm{cr}}(\overline{z}^{(i)}; \mathbf{\Theta})$ is the mean critical matter density as a function of redshift and $D(\overline{z}^{(i)}; \mathbf{\Theta})$ is the linear growth factor normalised to be unity at $z = 0$. We note that in Eq.~(\ref{eq:method:fs:nla:model}), we do not consider an explicit redshift dependence which is often expressed as a power-law term \citep{joachimi2011constraints}. We omit this term, as it is not considered in the fiducial KiDS-1000 analysis \citep{asgari2021kids, busch2022kids}, and it has been found that at least for $z < 1$, the IA signals do not vary largely with redshift \citep{fortuna2021kids}. 

Nevertheless, the NLA model still comes with some caveats. Mainly, it neglects non-linearities which may not be captured in the underlying matter power spectrum, such as source density weighing, expressly the next-to-leading order perturbations proportional to the source density considered in a full tidal alignment model \citep{blazek2015tidal}. This effect can in fact produce a signal in the shape measurements comparable to the correction due to non-linear structure growth \citep{krause2016impact}. Despite these limitations, the NLA model has been found to be accurate enough to model the IA signal seen by past and current weak lensing surveys \citep{joachimi2011constraints, blazek2011testing, heymans2013cfhtlens, krause2016impact, hilbert2017intrinsic}, including KiDS-1000 \citep{fortuna2021kids}.

\subsection{Galaxy positions and redshifts}\label{method:fs:salmo}
With the cosmological dependence of the underlying large-scale structure characterised and astrophysical effects such as intrinsic alignments and baryonic feedback included, the forward simulations still need to include observational biases from our instruments. We chose to apply these on the level of galaxies, namely, by biasing the sampled galaxy positions, redshifts, and shapes and shears in accordance with our models for the observational systematic effects.

When sampling the galaxies' positions on the sky, we consider the survey footprint, $\Omega_{\mathrm{survey}}$, the galaxy number density which the survey can observe, $n_{\mathrm{gal}}$, and its variability as a function of redshift and observational depth across the sky, namely, as a function of the local observing conditions. When sampling the galaxies' redshifts, we take into consideration the redshift distributions calibrated from photometry for a given survey, $n(z)$. Lastly, when determining the observed galaxy shapes, we consider the intrinsic ellipticity dispersion of the galaxies, $\sigma_{\mathrm{\epsilon}}$, the multiplicative and additive shear bias, while also including the effects on the measured galaxy shapes caused by the variation in the point spread function of the instrument as a function of the position in the sky.

Since we are interested in simulating a photometric survey, we must model the resolution along the line of sight realistically. This means that rather than modelling large-scale structure observables across $N_{\mathrm{shells}}$ concentric shells, they ought to be modelled across $N_{\mathrm{tomo}}$ tomographic bins. The tomographic bins of a given survey are defined by finite usually non-overlapping domains in photometric redshift, $z_{\mathrm{ph}}$, in which all observed galaxies are grouped, such that $z_{p} < z_{\mathrm{ph}} \leq z_{p+1}$, $p \in \{ 1, 2, 3, ..., N_{\mathrm{tomo}}\}$\footnote{Note: we use the indices $p$ and $q$ for tomographic bins and the indices $i$ and $j$ for underlying shells.}. The measured photometric redshift of a galaxy does not necessarily agree with its true/spectroscopic redshift, $z$. This means that galaxies which are grouped within a given photometric redshift bin, $p$, may not be located in this redshift range. To account for this, we can characterise a conditional redshift distribution for each tomographic bin $P(z|z_{\mathrm{ph}})$ by mapping the measured $z_{\mathrm{ph}}$ to known spectroscopic redshift measurements which have a comparatively low bias in the redshift estimate \citep{hildebrandt2021kids, busch2022kids}. Therefore, a given galaxy is in an associated pixel, $m$, shell, $i$, and tomographic bin, $p$. The probability of the galaxy being in tomographic bin, $p$, while it is sampled in shell, $i$, then depends on $P(z|z_{\mathrm{ph}})$.

In line with the KiDS-1000 analysis, we define the effective galaxy number density as follows
\begin{equation}
    n_{\mathrm{eff}}^{(p)} \equiv \frac{1}{|\Omega_{\mathrm{survey}}|}\frac{\Big[\sum_{i \in (p)} w_{i}\Big]^{2}}{\sum_{i \in (p)} w_{i}^2}\,,
    \label{eq:n_gal_kids}
\end{equation}
\noindent where $|\Omega_{\mathrm{survey}}|$ is the effective area covered by the survey footprint, and $w_{i}$ is the \texttt{lensfit} weight for a given galaxy, $i$. We note that the effective galaxy density within the KiDS-1000 gold sample is not the same as the galaxy density, $n_{\mathrm{gal}}$, when not correcting for the \texttt{lensfit} weights for each galaxy. However, as {KiDS-SBI} does not model galaxy weights, in the forward simulations $n_{\mathrm{eff}} = n_{\mathrm{gal}}$ since $w_{i} = 1 \forall i$.

Similarly, we define the galaxy intrinsic ellipticity dispersion, $\sigma_{\mathrm{\epsilon}}$, within a given tomographic bin as follows
\begin{equation}
    \sigma_{\mathrm{\epsilon}}^{(p)2} \equiv \frac{\sum_{i \in (p)} w_{i}^2 \, \Big( \epsilon_{\mathrm{obs}, i, 1}^{2} + \epsilon_{\mathrm{obs}, i, 2}^{2}\Big)}{\Big(1+\overline{M}^{(p)}\Big)^{2} \sum_{i \in (p)} w_{i}^2 },
    \label{eq:sigma_eps_kids}
\end{equation}
\noindent where $\epsilon_{\mathrm{obs}, i, 1} $ and $\epsilon_{\mathrm{obs}, i, 2}$ are the two components of the observed galaxy ellipticity, and $\overline{M}^{(p)}$ is the mean multiplicative shear bias in tomographic bin $p$ as calibrated in \citep{giblin2021kids}. Applying Eqs.~(\ref{eq:n_gal_kids}) and (\ref{eq:sigma_eps_kids}) to the KiDS-1000 gold sample, we obtain the values shown in Table~\ref{tab:dr4_values}. We note that due to the way $\sigma_{\epsilon}$ is defined in Eq.~(\ref{eq:sigma_eps_kids}), it incorporates both the noise due to intrinsic galaxy shapes as well as due to image noise in the survey.

\begin{table}
    \centering
    \caption{Parameters used to sample galaxies and their shapes in line with the expectations for the KiDS-1000 gold sample for each tomographic bin (from S1 to S5).}
    \begin{tabular}{ccccccc}
         \hline
         Bin & $\overline{n}_{\mathrm{eff}}$ &  $\overline{\sigma}_{\epsilon}/\sqrt{2}$ & $a_{n_{\mathrm{eff}}}$ & $b_{n_{\mathrm{eff}}}$ & $a_{\sigma_{\epsilon}}$ & $b_{\sigma_{\epsilon}}$ \\
         & [$\mathrm{arcmin}^{-2}$] & & & & $\times 10^{3}$ & \\
         \hline
         S1 & 0.62 & 0.27 & -0.035 & 0.72 & 1.81 & 0.267 \\
         S2 & 1.18 & 0.26 & -0.042 & 1.30 & 1.30 & 0.257 \\
         S3 & 1.85 & 0.28 & -0.243 & 2.58 & -0.86 & 0.280 \\
         S4 & 1.26 & 0.25 & -0.250 & 2.01 & -0.62 & 0.255 \\
         S5 & 1.31 & 0.27 & -0.416 & 2.56 & 2.39 & 0.261 \\
         \hline
    \end{tabular}
    \tablefoot{$\overline{n}_{\mathrm{eff}}$ is the mean galaxy number density for a given tomographic bin, $\overline{\sigma}_{\epsilon}/\sqrt{2}$ is the mean per-component shape dispersion, $a_{n_{\mathrm{eff}}}$ and $b_{n_{\mathrm{eff}}}$ are the slope and y-intercept, respectively, for the linear interpolation of the galaxy density as a function of the root-mean-square of the background noise in the KiDS catalogue, $\sigma_{\mathrm{rms}}$, according to Eq.~(\ref{eq:level_fitting_ngal}), while $a_{\sigma_{\epsilon}}$ and $b_{\sigma_{\epsilon}}$ are the parameters to linearly interpolate $\sigma_{\epsilon}$ from $\sigma_{\mathrm{rms}}$ according to Eq.~(\ref{eq:level_fitting_sigma_eps}).}
    \label{tab:dr4_values}
\end{table}

Although it does not have detectable effects on the cosmic shear signal, the {KiDS-SBI} forward simulations sample galaxy positions such that they trace the underlying matter fields, so spatially varying systematics may be accurately applied. Using the \texttt{SALMO}\footnote{Speedy Acquisition for Lensing and Matter Observables: \url{https://github.com/Linc-tw/salmo}} framework \citep{joachimi2021kids} for simulating galaxy catalogues, we Poisson-sample galaxies within each \texttt{HEALPix} pixel, $m$, on the sky using the following expectation for the galaxy counts
\begin{equation}
    \langle N_{m}^{(i)(p)}\rangle (\mathbf{\Theta}) = w_{m}(\Omega_{\mathrm{survey}}) \, \Big[1+b^{(i)}\delta^{(i)}(\theta_{m}; \mathbf{\Theta})\Big] \, P_{m}(p|i) \, n_{\mathrm{gal}, m}^{(p)} \, A_{\mathrm{pix}, m}\,,
    \label{eq:method:fs:salmo:position}
\end{equation}
\noindent where $w_{m}(\Omega_{\mathrm{survey}})$ is the weight for a given pixel $m$ which we take to be unity if a pixel is within the survey's mask/footprint, $\Omega_{\mathrm{survey}}$, and zero if it is not, $b^{(i)}$ is galaxy bias parameter within a given shell which we assume to be $b^{(i)} = 1 \forall i$ for the purposes of this cosmic shear analysis, $P_m(p|i)$ is the probability of a galaxy in pixel $m$ being detected within tomographic bin $p$ while being within shell $i$ (in other words, a discretised version of $P_m(z|z_{\mathrm{ph}})$), $n_{\mathrm{gal}, m}^{(p)}$ is the observed galaxy number per unit area for a given tomographic bin $p$ and a given pixel $m$ (here the pixel dependence accounts for the spatial variability of the detection limit of galaxies as a function of observing conditions, the number of exposures, etc.; see Sect.~\ref{method:fs:salmo_vd}), and $A_{\mathrm{pix}, m}$ is the area of pixel $m$. As $n_{\mathrm{gal}, m}^{(p)}$ changes due to variations in observational depth, it is more or less likely to observe high-redshift galaxies. Hence, a change in depth from one pixel to the next can change the observed redshift distribution (see Sect.~\ref{method:fs:salmo_vd}).

Taking $w_{m}(\Omega_{\mathrm{survey}})$, $P_{m}(p|i)$ and $n_{\mathrm{gal}, m}^{(p)}$ from the photometric survey's measurements, we can Poisson sample $\smash{N_{m}^{(i)(p)}(\mathbf{\Theta})}$ given the expectation value shown in Eq.~(\ref{eq:method:fs:salmo:position}) from which we can get the expected number of galaxy counts within each pixel of a tomographic bin $p$ by taking
\begin{equation}
    N_{m}^{(p)}(\mathbf{\Theta}) = \sum_{i=1}^{N_{\mathrm{shells}}} N_{m}^{(i)(p)}(\mathbf{\Theta})\,.
\end{equation}
With that, we then randomly sample each galaxy's right ascension and declination within each pixel $m$ assuming a uniform distribution. This means that below the scales of pixels, the simulations do not have any information on galaxy clustering.

To sample each galaxy's redshift, we can use the fact that we know exactly how many galaxies are in each shell $i$ for a given tomographic bin $p$ and pixel $m$ from $\smash{N_{m}^{(i)(p)}(\mathbf{\Theta})}$. We can therefore randomly assign an index $i$ and $p$ to each sampled galaxy in accordance with that number. We can then randomly sample a specific $z_{\mathrm{spec}}$ for each galaxy by assuming that the galaxies are uniformly distributed within a given shell. Again, this means that weak lensing and galaxy clustering measurements from the simulated galaxies do not contain any information about the large-scale structure along the line of sight below the scales of shells.

\begin{figure*}
    \centering
    \includegraphics[trim=0 0 0 60,clip,width=3.8cm]{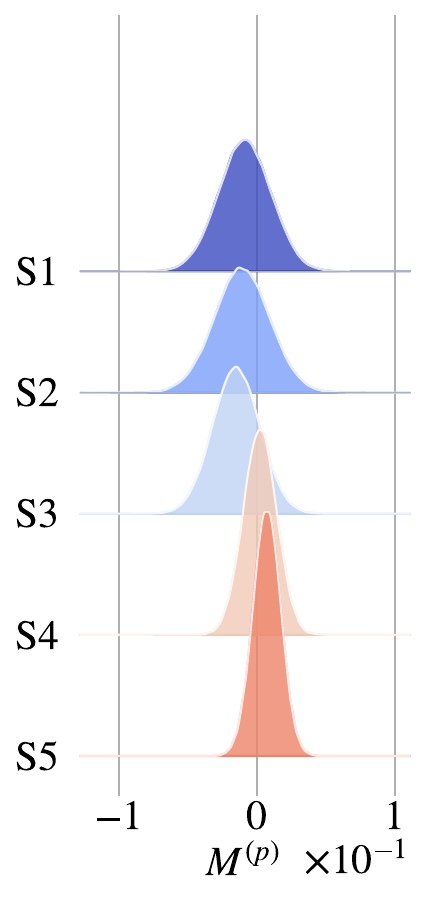}
    \includegraphics[trim=0 0 0 50,clip,width=3.7cm]{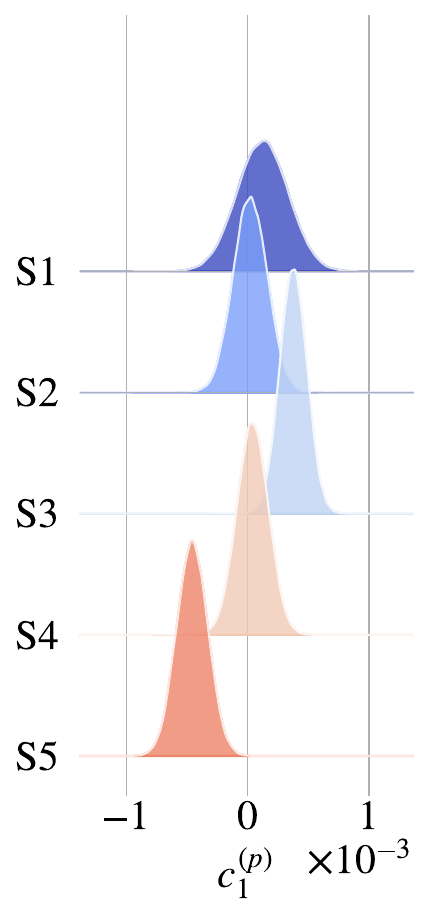}
    \includegraphics[trim=0 0 0 55,clip,width=3.4cm]{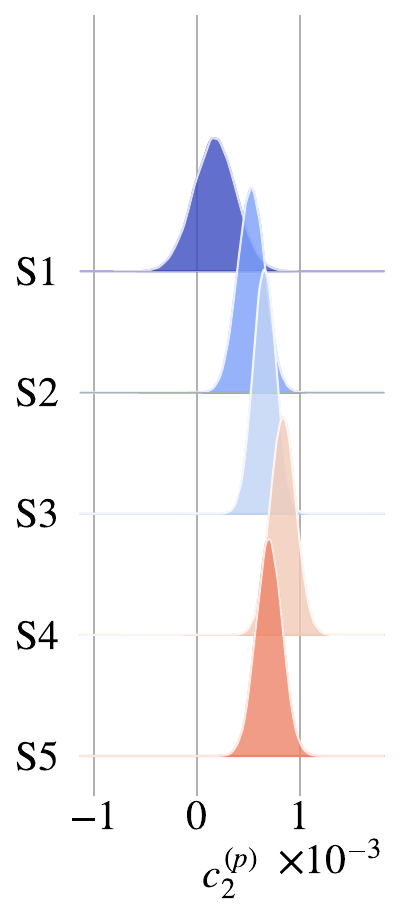}
    \includegraphics[trim=0 0 0 50,clip,width=3.45cm]{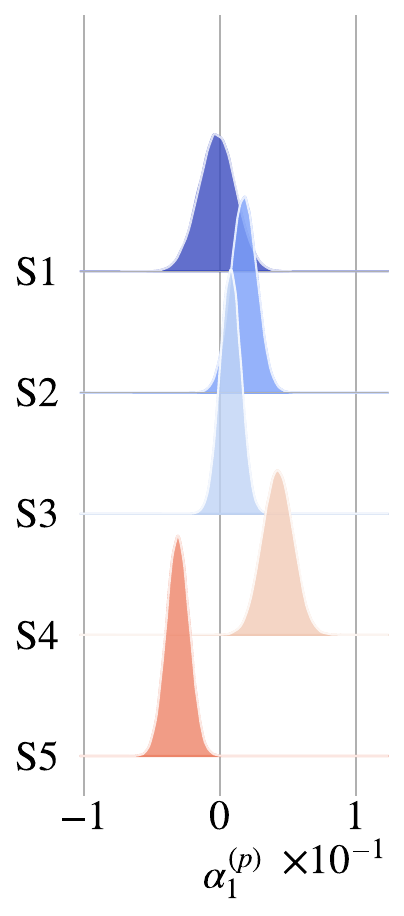}
    \includegraphics[trim=0 0 0 50,clip,width=3.5cm]{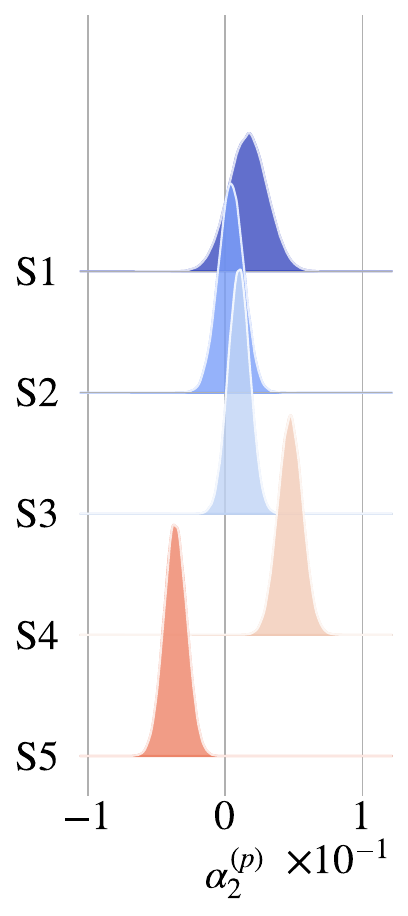}
    \caption{Distributions for each tomographic bin (S1 to S5) from which the shear bias parameters from KiDS-1000 \citep{giblin2021kids} shown in Eq.~(\ref{eq:method:fs:salmo_shears:bias}) are sampled. The first panel from the left shows the multiplicative shear bias, $M^{(p)}$. The second and third panel show the real and imaginary part of the additive shear bias, $c_{1}^{(p)}$ and $c_{2}^{(p)}$, respectively. The fourth and fifth panel show the real and imaginary part of the amplitude of the shear bias due to variations in the point-spread function, $\alpha_{1}^{(p)}$ and $\alpha_{2}^{(p)}$, respectively.}
    \label{fig:shear_bias}
\end{figure*}

\subsection{Galaxy shapes}\label{method:fs:salmo_shears}

To determine the galaxy shapes and shears, we begin by assigning intrinsic ellipticities. We define a given galaxy's shape through a complex ellipticity, $\epsilon$, and any galaxy ellipticity in this work as follows
\begin{equation}
    \epsilon = \epsilon_{1} + \mathrm{i} \, \epsilon_{2}\,,
\end{equation}
\noindent where $ \mathrm{i} \equiv \sqrt{-1}$. The intrinsic $\epsilon_{1}$ and $\epsilon_{2}$ of each galaxy are then sampled within \texttt{SALMO} as two independent normal random variates with zero mean and a variance equal to $\sigma_{\epsilon, m}^{(p)2}$. As with the galaxy density, $n_{\mathrm{gal}, m}^{(p)}$, the shape dispersion, $\sigma_{\epsilon, m}^{(p)}$, can also vary with tomographic bin and as a function of the position on the sky due to the effects of the spatial variability of observational depth. See Sect.~\ref{method:fs:salmo_vd} for a more detailed discussion. The ellipticity of each source galaxy is then altered through weak gravitational lensing by the matter along the line of sight. The exact shape distortion is given by \citep{seitz1996cluster, hu2000weak},
\begin{equation}
    \epsilon_{\mathrm{lensed}}(\mathbf{\Theta}) = \frac{\epsilon_{\mathrm{int}} + g(\mathbf{\Theta})}{1+ g^{*}(\mathbf{\Theta}) \epsilon_{\mathrm{int}}}\,,
    \label{eq:method:fs:salmo_shears:shear}
\end{equation}
\noindent where $\epsilon_{\mathrm{lensed}}(\mathbf{\Theta})$ is the lensed galaxy ellipticity, $\epsilon_{\mathrm{int}}$ is the intrinsic galaxy ellipticity and $g(\mathbf{\Theta})$ is the reduced shear ($g \in \mathbb{C}$),  given by
\begin{equation}
    g(\mathbf{\Theta}) = \frac{\gamma(\mathbf{\Theta})}{1-\kappa(\mathbf{\Theta})}\,,
    \label{eq:method:fs:salmo_shears:reduced}
\end{equation}
\noindent where $\gamma(\mathbf{\Theta})$ is the shear factor ($\gamma \in \mathbb{C}$). The use of Eq.~(\ref{eq:method:fs:salmo_shears:shear}) is an extension to the KiDS-1000 analysis which assumed the reduced shear approximation, where $\epsilon_{\mathrm{lensed}} \approx \epsilon_{\mathrm{int}} + g$. In any case, this should cause a negligible change in the measured shear considering the sensitivity of KiDS-1000 \citep{joachimi2021kids}. To calculate the shear field, it is useful to decompose the spin-0 convergence field into spherical harmonics as follows:
\begin{equation}
    \kappa_{\ell m}^{(i)}(\mathbf{\Theta}) = \int \mathbf{d}^{2}\bm{\theta} \,\kappa^{(i)}(\bm{\theta}; \mathbf{\Theta})\yellmarb{0}{*}{\ell m}(\bm{\theta})\,,
\end{equation}
\noindent where $\kappa_{\ell m}^{(i)}(\mathbf{\Theta})$ are the harmonic coefficients of the convergence field of a given shell $i$. We can then use the following relation to define the shear field's harmonic coefficients, $\gamma_{\ell m}^{(i)}$, given by \citep{peebles1973statistical},
\begin{equation}
    \gamma_{\ell m}^{(i)}(\mathbf{\Theta}) = - \Bigg(\frac{(\ell+2)(\ell-1)}{\ell(\ell+1)}\Bigg)^{1/2} \, \kappa_{\ell m}^{(i)}(\mathbf{\Theta})\,.
\end{equation}
The $\gamma_{\ell m}^{(i)}(\mathbf{\Theta})$ coefficients can then be used to calculate the spin-2 discrete shear field, $\gamma^{(i)}(\theta_{m}, \mathbf{\Theta})$. Having defined the effective $\kappa^{(i)}(\theta_{m}, \mathbf{\Theta})$ in Sect.~\ref{method:fs:glass} and~\ref{method:fs:nla}, while also defining $\gamma^{(i)}(\theta_{m}, \mathbf{\Theta})$, we can assign a value of $\kappa$ and $\gamma = \gamma_{1} +  \mathrm{i} \, \gamma_{2}$ to each galaxy sampled according to the procedure described in Sect.~\ref{method:fs:salmo}. This is done by taking the value of $\kappa^{(i)}(\theta_{m}, \mathbf{\Theta})$ and $\gamma^{(i)}(\theta_{m}, \mathbf{\Theta})$ within the shell $i$ and pixel $m$ in which a given galaxy is located. Between neighbouring pixels, the values of $\kappa^{(i)}(\theta_{m}, \mathbf{\Theta})$ and $\gamma^{(i)}(\theta_{m}, \mathbf{\Theta})$ are linearly interpolated. This choice may not be optimal \citep{tessore2023glass}, but we find that at the precision level of KiDS-1000, it does not lead to any significant biases (see Appendix~\ref{appendix:consistency}). We then use Eq.~(\ref{eq:method:fs:salmo_shears:reduced}) to calculate the associated reduced shear, $g(\mathbf{\Theta})$, as well as the lensed ellipticity, $\epsilon_{\mathrm{lensed}}(\mathbf{\Theta})$, by combining $g$ with the intrinsic ellipticity using Eq.~(\ref{eq:method:fs:salmo_shears:shear}).

\begin{figure*}
    \centering
    \includegraphics[trim=35 120 35 110,clip,width=18cm]{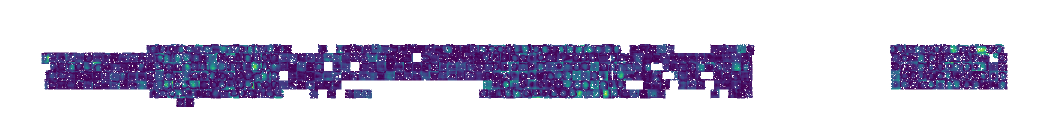}
    \includegraphics[trim=35 7 35 160,clip,width=18cm]{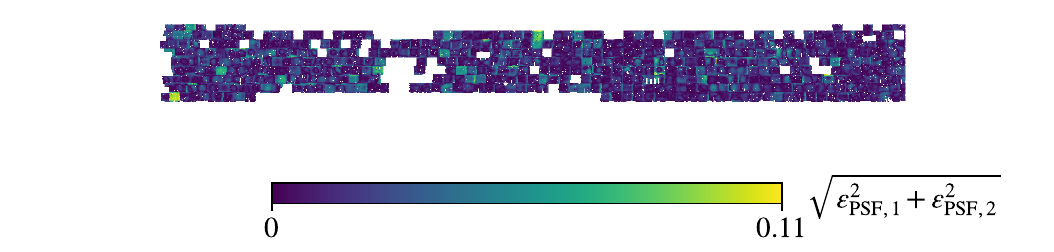}
    \caption{Cartesian spatial map ($N_{\mathrm{side}}=1024$) of the observed magnitude of the point-spread function ellipticities, $|\epsilon_{\mathrm{PSF}}| = \sqrt{\epsilon_{\mathrm{PSF},1}^2 + \epsilon_{\mathrm{PSF},2}^2}$, throughout the KiDS-1000 North field in the upper panel and the KiDS-1000 South field in the lower panel. $\epsilon_{\mathrm{PSF}}$ is added to the lensed galaxy shapes in the forward simulations within {KiDS-SBI} in accordance with Eq.~(\ref{eq:method:fs:salmo_shears:bias}).}
    \label{fig:psf_variation}
\end{figure*}

However, this lensed ellipticity, $\epsilon_{\mathrm{lensed}}(\mathbf{\Theta})$, is not the observed shape measurement in a weak gravitational lensing survey such as KiDS-1000. Additionally, the shape measurement may be distorted by instrumental effects or shape modelling inaccuracies. Some relevant effects are selection biases, noise biases from a low signal-to-noise ratio, biases in the galaxy weights, artefacts, non-linear CCD responses, asymmetries in the point spread function, etc. (see for example \citealt{mandelbaum2018weak} for a review). To account for this, it is common to parametrically estimate any residual systematics which may be affecting the shape measurements \citep{hildebrandt2017kids, hildebrandt2021kids, zuntz2018dark, giblin2021kids}. Within the forward simulations, we use the first-order parametric expansion of the observed galaxy shapes given by \citet{heymans2006the} as follows
\begin{equation}
    \epsilon_{\mathrm{obs}, i \in m}^{(p)}= \Big(1+M^{(p)}\Big)\epsilon_{\mathrm{lensed}, i} + \alpha^{(p)} \epsilon_{\mathrm{PSF}, m} + \beta^{(p)} \delta \epsilon_{\mathrm{PSF}, m} + c^{(p)}\,,
    \label{eq:method:fs:salmo_shears:bias}
\end{equation}
\noindent where $i$ is the index for a single galaxy within tomographic bin $p$ and pixel $m$, $M^{(p)}$ is the multiplicative shear bias as measured for tomographic bin, $p$, we have $\alpha^{(p)}$ as the fraction of the PSF ellipticity, which remains in the shear estimator, $\epsilon_{\mathrm{PSF}}(m)$ is the local PSF measured within pixel $m$ (see Fig.~\ref{fig:psf_variation} for a map of $\epsilon_{\mathrm{PSF}}$ in KiDS-1000), $\beta^{(p)}$ gives the amplitude of the shear bias due to residuals which are not taken into account by the PSF model, $\delta \epsilon_{\mathrm{PSF}}$ represents the residuals in question, and $c^{(p)}$ is the additive shear bias within a given tomographic bin, $p$. We note that spatially varying multiplicative shear biases, $M^{(p)}$, can also arise, but previous work has found that the mean bias across all positions is sufficient to accurately model the effect \citep{kitching2019propagating}. In accordance with previous findings \citep{giblin2021kids}, we take $\beta^{(p)} = 0 \, \forall p$. The amplitudes of the remaining systematics as calibrated by \citet{giblin2021kids} are shown in Fig.~\ref{fig:shear_bias}. It becomes apparent that these effects are relatively small which explains why only the multiplicative and additive bias is taken into consideration in the modelling on the fiducial KiDS-1000 analysis \citep{giblin2021kids, joachimi2021kids, asgari2021kids, heymans2021kids}. Additionally, the measured values of $M^{(p)}$, $\alpha^{(p)}$, and $c^{(p)}$ have associated uncertainties. To take this uncertainty into account and make sure it is represented in the effective likelihood, rather than using a fixed value, we randomly sample different values for  $M^{(p)}$, $\alpha^{(p)}$ and $c^{(p)}$ for each realisation of the forward simulations from a normal distribution as shown in Fig.~\ref{fig:shear_bias}. We note that we do not explicitly marginalise over these parameters, but as they are sampled randomly from predetermined probability density distributions, so that their associated uncertainty is propagated into the effective likelihood. This is similar to how the seed for each log-normal random field, and the galaxy positions which populate the associated matter fields are varied from one forward simulation to the next, such that cosmic variance enters the effective likelihood as well.

As a result, for given cosmological parameters, $\mathbf{\Theta}$, the {KiDS-SBI} forward simulations produce a galaxy catalogue containing the galaxies' position on the sky, their spectroscopic redshift, the tomographic bin in which they are detected and their observed ellipticities, while taking into account many relevant systematics which are considered in the modelling of weak gravitational lensing measurements.

\begin{figure*}
    \centering
    \includegraphics[trim=35 120 35 110,clip,width=18cm]{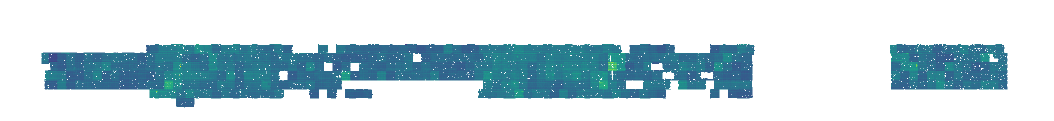}
    \includegraphics[trim=35 7 35 160,clip,width=18cm]{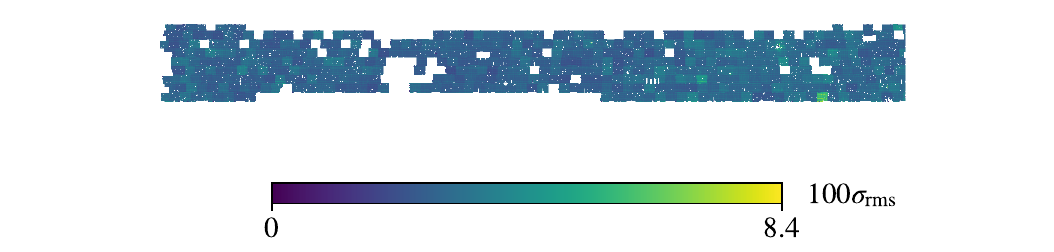}
    \caption{Cartesian spatial map ($N_{\mathrm{side}}=1024$) of root-mean-square of the observed background noise, $\sigma_{\mathrm{rms}}$, throughout the KiDS-1000 North field in the upper panel and the KiDS-1000 South field in the lower panel.}
    \label{fig:level_map}
\end{figure*}

\begin{figure}
    \centering
    \includegraphics[trim=0 20 0 20,clip,width=8cm]{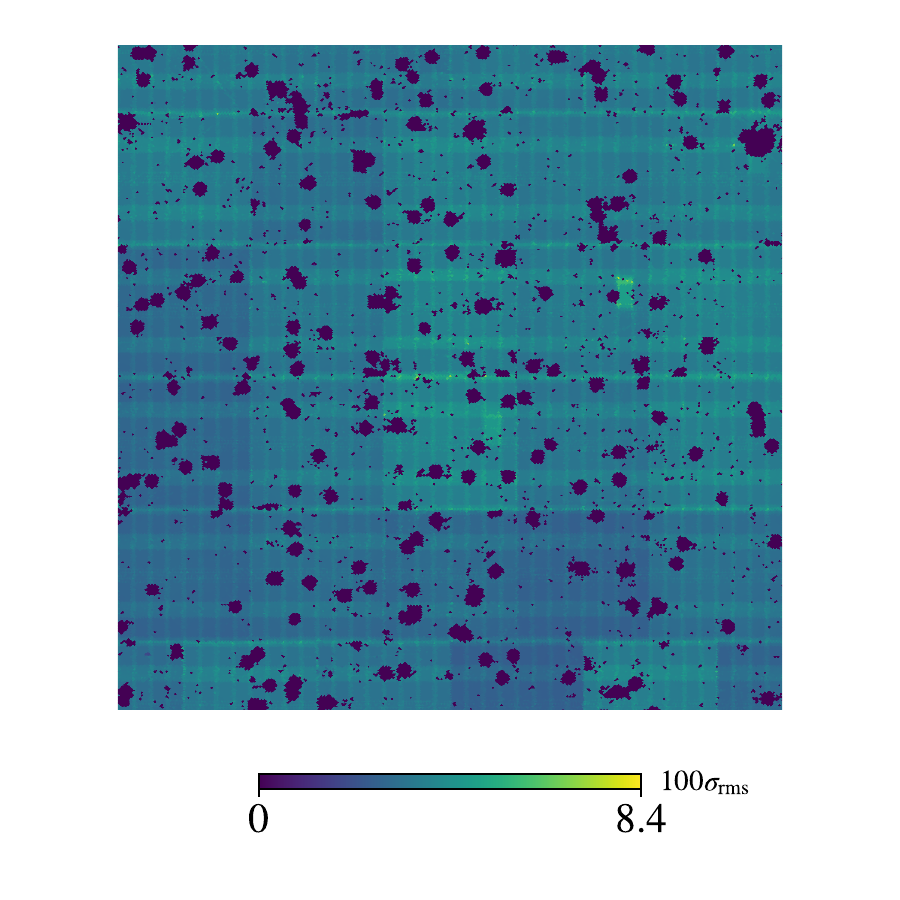}
    \caption{Spatial map ($N_{\mathrm{side}}=4096$) in a Cartesian projection of root-mean-square of the observed background noise, $\sigma_{\mathrm{rms}}$, for a $5^{\circ} \times 5^{\circ}$ patch of the KiDS-1000 North field at a right ascension (RA) of 180$^{\circ}$ and a declination (DEC) of 0$^{\circ}$.}
    \label{fig:level_map_zoom}
\end{figure}

\subsection{Variable depth}\label{method:fs:salmo_vd}
As alluded to in Sect.~\ref{method:fs:salmo} and~\ref{method:fs:salmo_shears}, when randomly sampling galaxies and their intrinsic ellipticities, the parameters which charaterise the probability distributions depend on the location on the sky, namely, the pixel $m$ in question. Specifically, the number of galaxies which is sampled (in other words, 'observed') within a given pixel depends on the average galaxy density, $n_{\mathrm{gal}, m}^{(p)}$, measured within that pixel. The galaxy density varies from pixel to pixel as it depends on the observational depth of the survey in different parts of the sky, and on anisotropies and time variations in the atmospheric seeing (particularly, for surveys such as KiDS which only visit each pointing in the sky once at a given time). Since many parts of the sky are observed at different times, as is the case for ground-based telescopes such as the VLT Survey Telescope and the Visible and the Infrared Survey Telescope for Astronomy (VISTA) used for KiDS-1000, the atmospheric conditions and background light will be different for each pointing for a given exposure time. Thus, the depth of observations may change as seeing, the point spread functions and/or the background flux varies (for example, due to moonlight, zodiacal light, or galactic absorption). 

In addition, surveys such as KiDS often have overlapping pointings in order to ensure that the footprint is observed without gaps. However, this implies that galactic sources which happen to be located near the edge of a pointing will be observed more often in different overlapping fields than a source located in the centre of a pointing. Consequently, near the edge of a pointing, there will tend to be a higher S/N which allows for deeper observations. Since these systematic effects in the observational depth can happen at fixed scales and/or have certain periodicities, they can induce significant systematic effects into galaxy clustering and weak gravitational lensing signals. For cosmic shear in KiDS-1000, the bias can be near $1\%$ in the signal and an average of ${\sim} 10 \%$ in the standard deviation of cosmic shear observables \citep{heydenreich2020the, joachimi2021kids}.

As the systematics modify the local observational depth, the galaxies which can be observed within a given pixel are selected according to their magnitude. This will systematically bias the dispersion of intrinsic ellipticities observed, $\sigma_{\epsilon, m}^{(p)}$, and further exacerbate the effect of observational depth variability on the cosmic shear signal.

\begin{figure*}
    \centering
    \includegraphics[width=16cm, trim={0cm 0 0cm 0},clip]{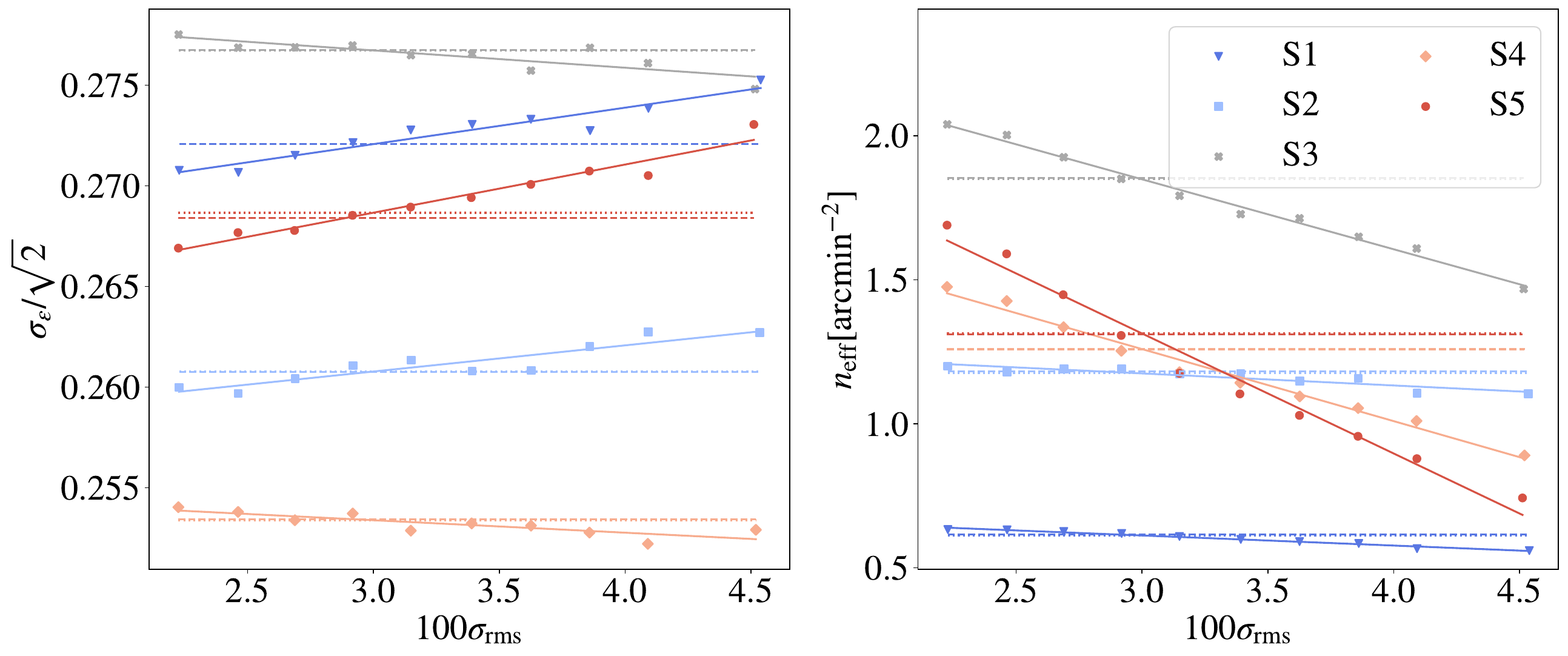}
    \caption{The dependence of the per-component Gaussian shape dispersion, $\sigma_{\epsilon}/\sqrt{2}$, (left panel) and the galaxy density, $n_{\mathrm{eff}}$, (right panel) on the root-mean square of the background noise, $\sigma_{\mathrm{rms}}$ in the KiDS-1000 DR4 data. For both panels, the data points represent the mean $\sigma_{\epsilon}$ or $n_{\mathrm{eff}}$ of ten equi-populated bins in $\sigma_{\mathrm{rms}}$ with their boundaries in $\{ 1.70, \, 2.33, \, 2.57, \, 2.80,  \, 3.04, \, 3.28, \, 3.52, \, 3.76, \, 4.00, \, 4.23, \, 12.96 \}$. The solid line shows the linear fit to the aforementioned data points of their respective tomographic bin according to Eqs.~(\ref{eq:level_fitting_ngal}) and (\ref{eq:level_fitting_sigma_eps}). The parameters obtained from this fit are given in Table~\ref{tab:dr4_values}. The dotted horizontal lines show the mean values of $\sigma_{\epsilon}$ and $n_{\mathrm{eff}}$ calculated from the galaxy samples with variable depth per tomographic bin, while the dashed horizontal lines show the values of $\sigma_{\epsilon}$ and $n_{\mathrm{eff}}$ for the respective galaxy samples without any spatial variations in the observational depth. Both of these lines agree exceptionally well by construction, so that for some source bins there is not any observable difference between them.}
    \label{fig:level_var}
\end{figure*}

To take all these effects into consideration within the forward simulations, we calibrate the spatial variability of $n_{\mathrm{eff}, m}^{(p)}$ and $\sigma_{\epsilon, m}^{(p)}$ from the KiDS-1000 measurements directly. Rather than defining maps over the entire survey footprint of both quantities for each tomographic bin, we use a map of a direct estimator of observational depth which also correlates with $n_{\mathrm{eff}, m}^{(p)}$ and $\sigma_{\epsilon, m}^{(p)}$: the root-mean square of the background noise, $\sigma_{\mathrm{rms}}$. This quantity gives the detection threshold of counts above the background, and is the dimensionless analogue to the detection threshold typically expressed in mag arcsec$^{-2}$. Since it directly characterises the detection probability of an object as a function of its angular position, $\sigma_{\mathrm{rms}}$ incorporates variations in the magnitude limit or the seeing \citep{johnston2021organised}. We found this quantity to be a good indicator of variable depth as it correlates well with the measured $n_{\mathrm{eff}, m}^{(p)}$ and $\sigma_{\epsilon, m}^{(p)}$, while being uncorrelated with the measured photometric redshifts and magnitudes in KiDS-1000.

A similar approach to modelling variable depth has been taken in \citet{joachimi2021kids}. However, in that analysis, the variable depth along the KiDS-1000 footprint was modelled with a direct estimate of the magnitude limit in the $r$-band (the band in which galaxy shape measurements are made in KiDS). We have found that the magnitude limit associated with a given galaxy in the KiDS-1000 gold sample is correlated with the galaxy's observed $r$-band magnitude as well as the photometric redshift estimate for the galaxy. This means that a selection according to the magnitude limit in the $r$-band can bias cosmological estimates, as the selection is not independent of the cosmic shear signal. In contrast, $\sigma_{\mathrm{rms}}$ is uncorrelated with the $r$-band magnitude measurements and the photometric redshifts in the KiDS-1000 gold sample, while still being highly correlated with the local magnitude limit in the $r$-band. Thus, we conclude it is a more direct parametrisation of the variations across the survey footprint of the galaxy selection.

As can be seen in Fig.~\ref{fig:level_map} and ~\ref{fig:level_map_zoom}, a map of $\sigma_{\mathrm{rms}}$ across the KiDS North and South fields clearly shows variations over different pointings as well as with areas of increased overlap. In addition, Fig.~\ref{fig:level_var} shows that $n_{\mathrm{eff}}^{(p)}$ and $\sigma_{\epsilon}^{(p)}$ vary linearly with $\sigma_{\mathrm{rms}}$ for KiDS-1000 data. Thanks to this fact, we do not need to model the variation of $n_{\mathrm{eff}, m}^{(p)}$ and $\sigma_{\epsilon, m}^{(p)}$ with the \texttt{HEALPix} pixel $m$ with individual maps for each tomographic bin $p$. Instead, using the linear relations shown in Fig.~\ref{fig:level_var}, we may assign a value for $n_{\mathrm{eff}, m}^{(p)}$ and $\sigma_{\epsilon, m}^{(p)}$ for each $\sigma_{\mathrm{rms}, m}$ (specifically the mean value of level within each pixel $m$ contained within the footprint observed by KiDS-1000) given that
\begin{align}
    \label{eq:level_fitting_ngal}
    n_{\mathrm{eff}, m}^{(p)} = & \; a_{n_{\mathrm{eff}}}^{(p)} \sigma_{\mathrm{rms}, m} + b_{n_{\mathrm{eff}}}^{(p)}\,,\\
    \label{eq:level_fitting_sigma_eps}
    \sigma_{\epsilon, m}^{(p)} = & \;a_{\sigma_{\epsilon}}^{(p)} \sigma_{\mathrm{rms}, m} + b_{\sigma_{\epsilon}}^{(p)}\,,  
\end{align}
\noindent where $a_{n_{\mathrm{eff}}}^{(p)}$ and $a_{\sigma_{\epsilon}}^{(p)}$ are the slopes of the linear fits shown in Fig.~\ref{fig:level_var} of galaxy density and galaxy shape dispersion for a given tomographic bin, respectively, and $b_{n_{\mathrm{eff}}}^{(p)}$ and $b_{\sigma_{\epsilon}}^{(p)}$ are the associated y-intercepts. The values of these parameters are shown in Table~\ref{tab:dr4_values}. This drastically helps the performance of the forward simulations, as this allows us to just use a single map of $\sigma_{\mathrm{rms}}$ to model the variable depth rather than requiring $2N_{\mathrm{tomo}}$ maps for it. Such a linear relation may not exist in other weak lensing surveys, but it should still be possible to numerically calibrate the relations dependence of $n_{\mathrm{eff}}^{(p)}$ and $\sigma_{\epsilon}^{(p)}$ on $\sigma_{\mathrm{rms}}$ to define an interpolation.

\begin{figure*}
    \centering
    \includegraphics[width=16cm]{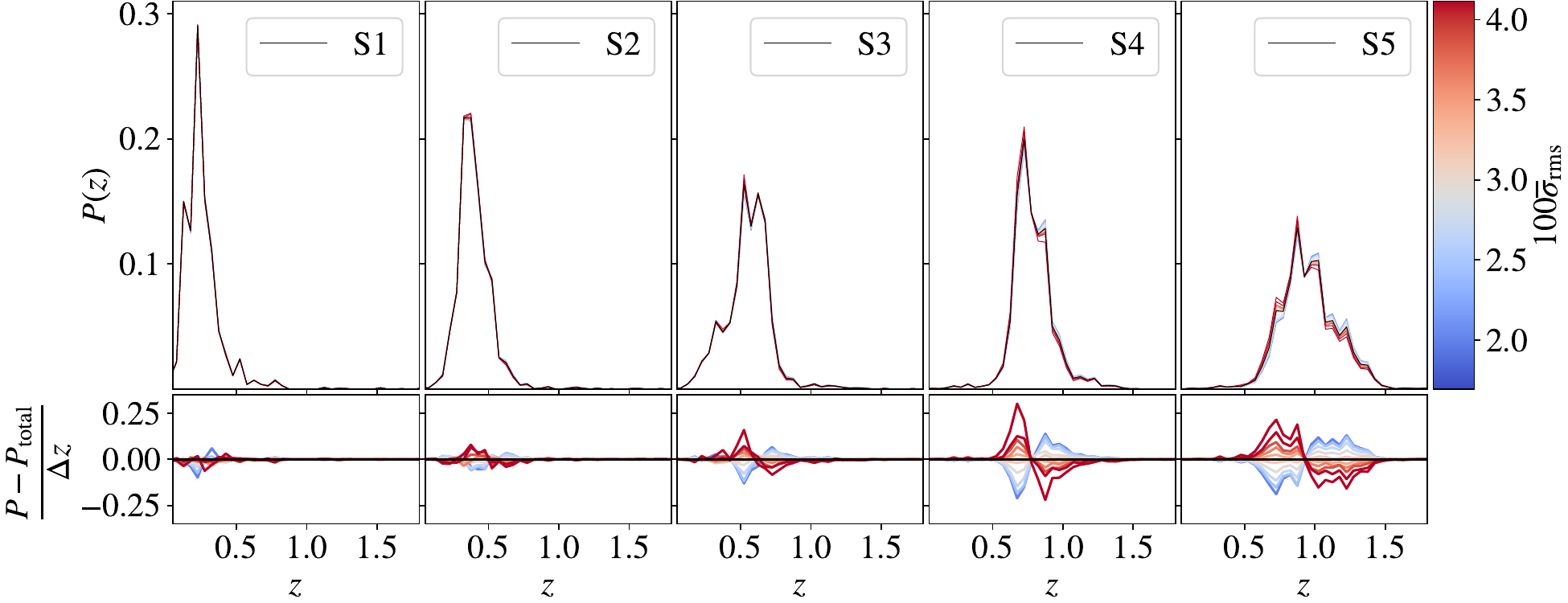}
    \caption{In the upper panels, the normalised redshift distributions, $P(z)$, for each tomographic bin (S1 to S5). The redshift distribution from the entire KiDS-1000 DR4 galaxy sample, $P_{\mathrm{total}}(z)$, is shown in black, while the other ten redshift distributions are derived from 10 equi-populated subsamples of DR4 based on their observational depth (i.e. the mean value of the root-mean-square of the background noise, $\overline{\sigma}_{\mathrm{rms}}$) which is shown with its respective colour. The lower panels show the associated residual change in the redshift distributions with respect to $P_{\mathrm{total}}(z)$ per unit redshift. It is apparent that variable depth mostly affects the source distributions at high redshifts, while the effect tends to decrease the mean of the redshift distribution with increasing $\sigma_{\mathrm{rms}}$.}
    \label{fig:nofz_variation}
\end{figure*}

These inhomogeneities in the galaxy selection across the survey footprint also change the redshift distribution of galaxies. As $\sigma_{\mathrm{rms}}$ decreases, the local magnitude limit increases and fainter galaxies can be detected by the survey. As more distant galaxies with high redshifts also tend to be fainter, variable depth can shift the redshift distributions of the local galaxy population as can be seen in Fig.~\ref{fig:nofz_variation}. The redshift distributions shown in Fig.~\ref{fig:nofz_variation} have been calculated using the same approach as in KiDS-1000 where the photometric redshifts are calibrated from spectroscopic samples using self-organising maps \citep{wright2019kids}. The self-organising map has been applied to ten equi-populated subsamples of each tomographic bin of the KiDS-1000 gold sample which are binned in $100\sigma_{\mathrm{rms}}$ along the following boundaries: $\{ 1.7, \, 2.33, \, 2.57, \, 2.8,  \, 3.04, \, 3.28, \, 3.52, \, 3.76, \, 4, \, 4.23, \, 12.96 \}$. To account for variable depth in our forward model, depending on the value of $\sigma_{\mathrm{rms}}$ in a given pixel, $m$, on the survey footprint, we sample galaxy redshifts in that pixel from the associated redshift distribution from Fig.~\ref{fig:nofz_variation}. This effect is important to model as an unaccounted shift in the redshift distribution can lead to significant biases in the cosmic shear signal \citep{heydenreich2020the, baleato2023the}.

To summarise, in {KiDS-SBI}, we account for the effects of spatial variations in the observational depth on galaxy number density, galaxy shape dispersion and redshift distributions. Moreover, we also consider anisotropies in the shear bias by including distortions to the lensed galaxy ellipticities calibrated from the spatial variation in the point-spread function ellipticity in the KiDS-1000 gold sample (see Sect.~\ref{method:fs:salmo_shears}).

We note that the galaxy selection function considered here does not account for systematic effects which have been previously corrected in KiDS data, such as source blending. Thanks to the selection and calibration of the data discussed in Sect.~\ref{data}, the KiDS-1000 source catalogue as well as the simulations thereof can be treated as a sample of non-overlapping sources. Any residual effects from source blending which remain after the measurement and calibration stages of the data are included within the shear biases and weights. Consequently, when quantifying the variable depth of the survey, any residual effect on blending is compensated by the inclusion of the galaxy weights and the multiplicative shear bias in the definitions of the effective galaxy density and the intrinsic ellipticity dispersion of galaxies, in Eqs.~(\ref{eq:n_gal_kids}) and~(\ref{eq:sigma_eps_kids}).

\section{Measurements}\label{measurements}
After sampling the galaxy shapes and shear, then applying all relevant instrumental systematics within {KiDS-SBI}, we can go on to measure the galaxy shapes, along with the cosmic shear summary statistics in the same manner as the measurements done for the KiDS-1000 gold sample's weak lensing catalogue. The only difference between the measurements of the forward-modelled catalogues and the KiDS-1000 gold sample is the values of the galaxy weights, $w_{i}$. In the forward simulations, the galaxy weights are set to unity for all sampled galaxies, $w_{i} = 1 \forall i$, while the weights used in the measurements from the KiDS-1000 gold sample are given by the \texttt{lensfit} weights (see details in Sect.~\ref{data}). To avoid biasing the statistics of the galaxy populations in the forward simulations with respect to the real data, we sample the galaxy positions and shapes from the galaxy densities, $n_{\mathrm{gal}}$, and intrinsic shape dispersions, $\sigma_{\epsilon}$, where the corrections for the \texttt{lensfit} weight and the multiplicative shear bias have been applied, as defined in Eqs.~(\ref{eq:n_gal_kids}) and (\ref{eq:sigma_eps_kids}).

\subsection{Shape measurements}\label{method:fs:shapes}
From the KiDS-1000 gold sample data as well as the forward simulated catalogues, we measure the observed galaxy ellipticity, $\epsilon_{\mathrm{obs}}$, for each galaxy per tomographic bin. To this end, we apply the same corrections to the shear measurements as described in \citet{giblin2021kids}. The observed shape measurements are corrected as follows
\begin{equation}
    \epsilon_{\mathrm{obs}, i}^{\mathrm{corr} \, (p)}(\mathbf{\Theta}) = \frac{1}{1 + \overline{M}^{(p)}} \frac{\sum_{i \in p}{w_i (\epsilon_{\mathrm{obs}, i}(\mathbf{\Theta}) - \langle \epsilon_{\mathrm{obs}}(\mathbf{\Theta}) \rangle )}}{\sum_{i \in p}{w_i}}\,,
\end{equation}
\noindent where $w_i$ is the galaxy \texttt{lensfit} weight for galaxy $i$ in the KiDS-1000 gold sample (for the case of the simulations, $w_i = 1 \forall i$), $\langle \epsilon_{\mathrm{obs}}(\mathbf{\Theta}) \rangle$ compensates for any additive biases \citep{asgari2019consistent}, that is the $c^{(p)}$ term and the mean of $\alpha^{(p)}$ in Eq.~(\ref{eq:method:fs:salmo_shears:bias}), and it is defined as the mean observed shape of all galaxies within tomographic bin $p$, and $\overline{M}^{(p)}$ is the mean multiplicative shear bias measured for each tomographic bin $p$ as shown in Fig.~\ref{fig:shear_bias}. We note that $\overline{M}^{(p)}$ is not necessarily the same value as the value of $M^{(p)}$ applied in Eq.~(\ref{eq:method:fs:salmo_shears:bias}). For each instance of the simulations, a different $M^{(p)}$ is drawn from a Gaussian probability distribution with mean $\overline{M}^{(p)}$ and the standard deviations shown in Fig.~\ref{fig:shear_bias}. Regardless of the value of $M^{(p)}$ drawn, the shear measurements of each simulation are corrected by the same mean $\overline{M}^{(p)}$. Any discrepancy between these values is going to introduce noise in the simulations which accounts for the uncertainty on the shear bias measurements.

\subsection{Observed angular power spectra}\label{method:fs:pcl}

With a full simulated cosmic shear catalogue akin to the KiDS-1000 gold sample, we can map any set of cosmological and astrophysical parameters, $\mathbf{\Theta}$, to a corresponding set of cosmic shear measurements from the KiDS survey as required for simulation-based inference. Although theoretically possible, it is computationally impractical to conduct cosmological inference at the level of catalogues. For this reason, it is useful to compress the catalogues down to useful statistics which still retain most of the relevant cosmological information about the underlying large-scale structure.

We chose to compress the catalogues down to two-point statistics. The main reason for this is that the forward simulation pipeline is designed to be only percent-level accurate in two-point statistics. Higher-order correlations of the shear field are captured to a certain degree within the simulations, but their accuracy is limited \citep{hall2022non, piras2023fast}. In addition, the use of two-point statistics assures that the only distinguishing feature of our analysis when compared to the fiducial cosmic shear KiDS-1000 analyses \citep{loureiro2021kids, asgari2021kids} is the dropping of the assumption of a Gaussian likelihood, so we can carry out a direct comparison between the analyses. This way we can ensure that any non-Gaussianities in our likelihood are either attributed to inherent non-Gaussianities in the likelihood of two-point statistics or non-Gaussianities induced by systematic effects.

Due to its computational efficiency when compared to spatial two-point correlation functions, we chose pseudo-Cls as our two-point statistic. We decompose the observed shear field, $ \epsilon_{\mathrm{obs}}^{\mathrm{corr} \, (p)}(\theta_{m}; \mathbf{\Theta})$, into its curl-free E-modes and its divergence-free B-modes as follows
\begin{equation}
    \epsilon_{\mathrm{obs}}^{\mathrm{corr} \, (p)}(\theta; \mathbf{\Theta}, \Omega_{\mathrm{survey}}) = \sum_{\ell=0}^{\ell_{\mathrm{max}}} \sum_{m=-\ell}^{\ell} (\tilde{E}_{\ell m}^{(p)} (\mathbf{\Theta}) +  \mathrm{i} \tilde{B}_{\ell m}^{(p)}(\mathbf{\Theta})) \yellmarb{2}{}{\ell m}(\theta) \,,
    \label{eq:method:fs:pcl:epsilon_obs}
\end{equation}
\noindent where $\yellmarb{\pm 2}{\mathstrut}{\ell m}(\theta)$ are the spin-2 spherical harmonic functions which define an orthonormal basis for the observed shear field, such that 
\begin{equation}
    \int \mathrm{d}^{2}\theta \yellmarb{\pm 2}{\mathstrut}{\ell m}(\bm{\theta}) \yellmarb{\pm 2}{*}{\ell' m'}(\bm{\theta}) = \delta^{\mathrm{K}}_{\ell \ell'} \delta^{\mathrm{K}}_{m m'}\,,
\end{equation}
\noindent while $\tilde{E}_{\ell m}^{(p)} (\mathbf{\Theta})$ and $\tilde{B}_{\ell m}^{(p)} (\mathbf{\Theta})$ are spherical harmonic coefficients of the curl-free and the divergence-free observed shear fields within each tomographic bin, respectively, and $\delta^{\mathrm{K}}$ is the Kronecker delta. These coefficients are defined as follows:
\begin{align}
    \tilde{E}_{\ell m}^{(p)} (\mathbf{\Theta})\;= \;\frac{1}{2}\; \int \mathrm{d}^2 & \theta \, [\epsilon_{\mathrm{obs}}^{\mathrm{corr} \, (\mathit{p})}(\bm{\theta}; \mathbf{\Theta})\yellmarb{2}{*}{\ell m}(\bm{\theta})\nonumber \\
    & + \epsilon_{\mathrm{obs}}^{\mathrm{corr} \, (\mathit{p})*}(\bm{\theta}; \mathbf{\Theta})\yellmarb{-2}{*}{\ell m}(\bm{\theta})]\,,\\
    \tilde{B}_{\ell m}^{(p)} (\mathbf{\Theta}) = \frac{-\mathrm{i}}{2} \int \mathrm{d}^2{\theta} & \,[\epsilon_{\mathrm{obs}}^{\mathrm{corr} \, (\mathit{p})}(\bm{\theta}; \mathbf{\Theta})\yellmarb{2}{*}{\ell m}(\bm{\theta}) \nonumber \\
   & - \epsilon_{\mathrm{obs}}^{\mathrm{corr} \, (\mathit{p})*}(\theta; \mathbf{\Theta})\yellmarb{-2}{*}{\ell m}(\bm{\theta})]\,.
\end{align}

From these coefficients, we then calculate the pseudo-Cls, $\tilde{C}^{(pq)}_{\epsilon \epsilon}(\ell; \mathbf{\Theta})$, as follows

\begin{equation}
    \tilde{C}^{(pq)}_{\epsilon \epsilon, \mu}(\ell; \mathbf{\Theta}) = \begin{pmatrix}  \tilde{C}_{\epsilon \epsilon}^{EE (pq)}(\ell; \mathbf{\Theta}) \\  \tilde{C}_{\epsilon \epsilon}^{EB (pq)}(\ell; \mathbf{\Theta}) \\  \tilde{C}_{\epsilon \epsilon}^{BB (pq)}(\ell; \mathbf{\Theta}) \end{pmatrix} = \frac{1}{2\ell + 1} \sum_{m = -\ell}^{\ell}  \begin{pmatrix}  \tilde{E}^{(p)}_{\ell m}\tilde{E}^{(q)*}_{\ell m}  (\mathbf{\Theta}) \\  \tilde{E}^{(p)}_{\ell m}\tilde{B}^{(q)*}_{\ell m}  (\mathbf{\Theta})\\  \tilde{B}^{(p)}_{\ell m}\tilde{B}^{(q)*}_{\ell m}  (\mathbf{\Theta}) \end{pmatrix}\,,
    \label{eq:method:fs:pcl:pcl}
\end{equation}
\noindent where $\mu \in \{1, 2, 3\}$ and $1$ stands for the EE component, $2$ for the EB component, and $3$ for the BB component. We note that $\tilde{C}_{\epsilon \epsilon}^{EB}(\ell; \mathbf{\Theta}) = \tilde{C}_{\epsilon \epsilon}^{BE}(\ell; \mathbf{\Theta})$, since $ \tilde{E}_{\ell m}\tilde{B}_{\ell m}^{*}  (\mathbf{\Theta})=  \tilde{B}_{\ell m}\tilde{E}_{\ell m}^{*}  (\mathbf{\Theta})$. 

Going forward, similar to other previous analyses \citep{hikage2019cosmology, loureiro2021kids, troester2022joint}, we only take into consideration $\tilde{C}^{EE(pq)}_{\epsilon \epsilon}(\ell; \mathbf{\Theta})$, so any cosmological signal which may have ended up in  $\tilde{C}^{EB(pq)}_{\epsilon \epsilon}(\ell; \mathbf{\Theta})$ or  $\tilde{C}^{BB(pq)}_{\epsilon \epsilon}(\ell; \mathbf{\Theta})$ because of EE to EB, or EE to BB mode mixing will be lost. $\tilde{C}^{EE(pq)}_{\epsilon \epsilon}(\ell; \mathbf{\Theta})$ is expected to still contain all of the cosmological cosmic shear signal because the shear field resulting from a scalar gravitational field in $\Lambda$CDM is predicted to be a curl-free field \citep{bartelmann2001weak, hikage2011shear, kilbinger2015cosmology}. 

To account for the EE signal from the correlations between the intrinsic galaxy shapes, namely the shape noise bias, we subtract the mean shape noise power spectrum as follows
\begin{equation}
    \tilde{C}^{(pq)}_{\epsilon \epsilon}(\ell; \mathbf{\Theta}) = \tilde{C}^{EE(pq)}_{\epsilon \epsilon}(\ell; \mathbf{\Theta}) - \delta^{\mathrm{K}}_{pq} \langle \tilde{C}^{EE(pq)}_{\mathrm{noise}}(\ell) \rangle\,,
\end{equation}
\noindent where $\tilde{C}^{(pq)}_{\epsilon \epsilon}(\ell; \mathbf{\Theta})$ is the E-mode pseudo angular power spectrum for cosmic shear with the shape noise bias subtracted, while $\langle \tilde{C}^{EE(pq)}_{\mathrm{noise}}(\ell) \rangle$ is the mean of the curl-free angular power spectrum of the shape noise. The latter is estimated as follows \citep{becker2016cosmic, hikage2019cosmology, nicola2021cosmic, loureiro2021kids},
\begin{equation}
    \tilde{C}^{EE(pp)}_{\mathrm{noise}}(\ell) = \frac{1}{2\ell + 1} \sum_{m = -\ell}^{\ell} \tilde{E}^{\mathrm{rand} \, (p)}_{\ell m} (\mathbf{\Theta}) \tilde{E}^{\mathrm{rand} \, (p)*}_{\ell m}  (\mathbf{\Theta})\,,
\end{equation}
\noindent where $\tilde{E}^{\mathrm{rand} \, (p)}_{\ell m}$ is the curl-free spherical harmonic coefficient of the randomly rotated shear values, $\epsilon^{\mathrm{rand} \, (p)}_{\mathrm{obs}}(\theta_{m}; \mathbf{\Theta})$. In turn, we define this field from the following galaxy shear values as follows:
\begin{align}
   & \epsilon^{\mathrm{rand}}_{\mathrm{obs}, i}(\mathbf{\Theta}) = \;\epsilon^{\mathrm{rand} \, }_{\mathrm{obs}, i, 1}(\mathbf{\Theta}) +  \mathrm{i} \epsilon^{\mathrm{rand} \, }_{\mathrm{obs}, i, 2}(\mathbf{\Theta})\,,\\
  &  \epsilon^{\mathrm{rand} \, }_{\mathrm{obs}, i, 1} = \;\epsilon^{\mathrm{corr} \, }_{\mathrm{obs}, i, 1} \, \mathrm{cos}(\theta_{\mathrm{rand}, i}) - \epsilon^{\mathrm{corr} \, }_{\mathrm{obs}, i, 2} \, \mathrm{sin}(\theta_{\mathrm{rand}, i})\,,\\
  &  \epsilon^{\mathrm{rand} \, }_{\mathrm{obs}, i, 2} = \; \epsilon^{\mathrm{corr} \, }_{\mathrm{obs}, i, 2} \, \mathrm{cos}(\theta_{\mathrm{rand}, i}) + \epsilon^{\mathrm{corr} \, }_{\mathrm{obs}, i, 1} \, \mathrm{sin}(\theta_{\mathrm{rand}, i})\,,
\end{align}
\noindent where $\theta_{\mathrm{rand}, i}$ is a randomly drawn angle for each galaxy $i$ from a uniform distribution where $\theta_{\mathrm{rand}, i} \in [0, 2\pi)$. To calculate the mean of the curl-free angular power spectrum of the shape noise, $\langle \tilde{C}^{EE(pq)}_{\mathrm{noise}}(\ell) \rangle$, we take mean of all modes in $\tilde{C}^{EE(pq)}_{\mathrm{noise}}(\ell)$ as follows
\begin{equation}
    \langle \tilde{C}^{EE(pq)}_{\mathrm{noise}}(\ell) \rangle =  \frac{\sum_{\ell = \ell_{\mathrm{min}}}^{\ell_{\mathrm{max}}}(2 \ell + 1) \, \tilde{C}^{EE(pq)}_{\mathrm{noise}}(\ell)}{\sum_{\ell = \ell_{\mathrm{min}}}^{\ell_{\mathrm{max}}} (2 \ell + 1)}\,,
    \label{eq:pseudo_cl_noise_weighted_mean}
\end{equation}
\noindent where the mean is weighted by a factor of $2 \ell + 1$ to account for the fact that each angular scale, $\ell$, contains $2 \ell + 1$ modes, $m$. We find setting $\ell_{\mathrm{min}} {\sim} 200$ with  $\ell_{\mathrm{max}} = 1500$ yields optimal results, as high angular scales are typically shape noise-dominated, and thus put better constraints on the average shape noise bias. In fact, we find that the measured average shape noise bias, $\langle \tilde{C}^{EE(pq)}_{\mathrm{noise}}(\ell) \rangle$, agrees within $<0.5\%$ with the underlying shape noise signal computed directly from the known intrinsic galaxy shapes. Additionally, we find that $\langle \tilde{C}^{EE(pq)}_{\mathrm{noise}}(\ell) \rangle$ is self-consistent with $\langle \tilde{C}^{BB(pq)}_{\mathrm{noise}}(\ell) \rangle$ within $<0.01\%$. The introduction of a weighted mean given by Eq.~(\ref{eq:pseudo_cl_noise_weighted_mean}) allows us to achieve this level of precision with only a single random rotation of the galaxies' shapes. Instead, previous approaches relied on creating many different instances of random rotations and then computed $\langle \tilde{C}^{EE(pq)}_{\mathrm{noise}}(\ell) \rangle$ by taking the average over all the instances while weighting each angular scale, $\ell$, equally (see for instance \citealt{loureiro2021kids}). When weighting each $\ell$ equally, the low-$\ell$ scales have an equally strong pull on the estimator as the high-$\ell$ scales which increases the statistical uncertainty on $\langle \tilde{C}^{EE(pq)}_{\mathrm{noise}}(\ell) \rangle$ as the large scales are mostly dominated by the large-scale structure signal. For this reason, it is common to take the mean over many realisations of the shape noise to reduce the variance of the mean. Although such approaches achieve similar levels of precision when estimating the shape noise bias as the approach shown in this work, computing the angular power spectra for hundreds of realisations of randomly rotated galaxy shape catalogues can become computationally expensive and time-consuming. Hence, it is not feasible for an SBI analysis which requires ${\sim}$$10^4$ realisations and we opt to use the weighted mean of a single realisation of a random rotation to estimate the shape noise bias.

Furthermore, we bin the noise-free observed angular power spectrum, $\tilde{C}^{(pq)}_{\epsilon \epsilon}(\ell; \mathbf{\Theta})$, into 8 log-spaced bins between $\ell = 76$ and $\ell = 1500$ in line with \citet{loureiro2021kids}. For this, we chose the pseudo-Cl binning scheme described in \citet{brown2005cosmic} given by
\begin{equation}
    \tilde{C}^{(pq)}_{\epsilon \epsilon, L}(\mathbf{\Theta}) = \frac{1}{2 \pi} \sum_{\ell = \ell_{L}}^{\ell_{L+1}} \frac{\ell(\ell+1)}{(\ell_{L+1} - \ell_{L})} \tilde{C}^{(pq)}_{\epsilon \epsilon}(\ell; \mathbf{\Theta}),
\end{equation}
\noindent where $\ell_{L}$ and $\ell_{L+1}$ are the lower and upper limits of the $L^{\mathrm{th}}$ bin, respectively.

We note that we did not choose to deconvolve the pseudo-Cls to estimate the full-sky E-mode angular power spectra for cosmic shear. We find that with the sky-coverage of KiDS-1000 and the complexity of the geometry of the KiDS-1000 footprint, the mixing matrix (see Appendix~\ref{appendix:signal}) is not necessarily invertible, and the deconvolution is not single-valued.

We therefore obtain, for a given forward simulation, a cosmic shear measurement like the one shown in Fig.~\ref{fig:pcl_measurement} for a single run of the forward simulations. The calculation of the data vector calculated from the theory prediction shown in Fig.~\ref{fig:pcl_measurement} is described in Appendix~\ref{appendix:signal}. As can be seen from Fig.~\ref{fig:pcl_measurement}, the measured cosmic shear signal is consistent with theory predictions which we find is the case at all cosmologies throughout the prior volume (see Appendix~\ref{appendix:consistency}). This shows that the log-normal random fields accurately recover the two-point statistics of the galaxy populations throughout parameter space, as expected.

As seen in Fig.~\ref{fig:pcl_measurement}, for the five tomographic bins in KiDS-1000, we obtain $8 \times N_{\mathrm{tomo}}(N_{\mathrm{tomo}} + 1)/2$ data points in $\tilde{C}^{(pq)}_{\epsilon \epsilon, L}(\mathbf{\Theta})$, namely, a 120D data vector. The analysis choices described within Sect~\ref{fs} and \ref{measurements} in conjunction constitute the forward model assumed within this SBI analysis, and it is henceforth labelled as the anisotropic systematics model, in reference to the fact that the model includes novel features such as variable depth, PSF variation, and dropping of the reduced shear approximation.

To summarise, each data vector depends on 12 parameters in $\mathbf{\Theta}$: 5 cosmological parameters ($\sigma_{8}$, the root-mean-square matter fluctuation over 8 Mpc$/h_{0}$; $\omega_b$, baryonic matter density; $\omega_c$, cold dark matter density; $n_s$, the scalar spectral index of the primordial density fluctuation power spectrum, $P_{\mathcal{R}}$; and $h_0$\footnote{Note: throughout we deviate from standard notation and denote the dimensionless Hubble constant as  $h_0$.}, the normalised Hubble constant), and 7 astrophysical and nuisance parameters related to systematics: $A_{\mathrm{bary}}$, the baryonic feedback amplitude within the non-linear 3D matter power spectrum; $A_{\mathrm{IA}}$, the galaxy intrinsic alignment amplitude of the NLA model; and five correlated $\delta_{z}$ parameters which define the shift in the mean of the source redshift distribution, $P(z|z_{\mathrm{ph}})$, of each tomographic bin. All other parameters on which the simulation depends are fixed from run to run (such as $\omega_k = 0$, the equation of state of dark energy, $w = - 1$, $\sum m_{\nu} = 0.06$ eV, etc.). The exceptions to this are the seed used to sample the matter overdensities and the galaxies, and the amplitudes of the shear biases as described in Sect.~\ref{method:fs:salmo_shears}. These values are varied from run to run in order to simulate cosmic variance, shape noise as well as the uncertainty in the shear bias, respectively.

\section{Simulation-based inference (SBI)}\label{method:sbi}

The final data vector as defined by the forward simulations is used to determine the effective likelihood of the data using sequential neural likelihood estimation based on the density estimation likelihood-free inference package (\texttt{pyDELFI}\footnote{\url{https://github.com/justinalsing/pydelfi}}, \citealt{alsing2019fast}, or \texttt{DELFI} henceforth to denote the method and software package) using the same analysis pipeline outlined in \citetalias{lin2022a}.

\begin{figure}
    \centering
    \begin{tikzpicture}[node distance=1.25cm]
        \node (start) [startstop, text width=5cm, yshift=0cm] {Cosmological parameters, $\bm{\Theta}$};
         \node (sbi) [process, below of=start, text width=4cm, yshift=0cm] {Forward simulations};
        \node (pcl) [startstop, below of=sbi, yshift=0cm, text width=5cm] {Mock data: Pseudo-Cl, $\Tilde{C}_{\epsilon \epsilon}^{(p q)}(\ell)$};
        \node (comp) [process, below of=pcl, yshift=0cm, text width=4cm] {Score compression};
         \node (delfi) [process, below of=comp, yshift=0cm, text width=5cm] {Train Neural Density Estimators};
         \node (likelihood) [startstop, below of=delfi, yshift=-0.2cm, text width=1.5cm] {Effective likelihood};
         \node (data) [startstop, below of=delfi, xshift=3cm, text width=1.5cm, yshift=-0.2cm] {Observed data};
         \node (priors) [startstop, below of=likelihood, xshift=-2cm, yshift=0cm, text width=2cm] {Priors};
         \node (posterior) [startstop, below of=likelihood, xshift=+2cm, yshift=0cm, text width=2cm] {Posterior};
        \draw [arrow] (start) -- (sbi);
        \draw [arrow] (sbi) -- (pcl);
        \draw [arrow] (pcl) -- (comp);
        \draw [arrow] (comp) -- (delfi);
        \draw [arrow] (delfi) -- (likelihood);
        \draw [arrow] (data) -- (likelihood);
        \draw (likelihood.south) -| ($(likelihood.south) + (0,0mm)$) |- (priors.east);
        \draw [arrow] (priors)--(posterior);

        \draw [arrow] (delfi.east) -| ($(delfi.east) + (+15mm,0mm)$) |- (start.east) node[midway, right, yshift=-3cm, xshift=1cm, rotate=-90,anchor=north]  {Active learning};
    \end{tikzpicture}
\caption{Flowchart describing the structure of the simulation-based inference pipeline. The dark blue rounded boxes represent the inputs and outputs which are given to the simulation-based inference pipeline. The grey rectangular boxes show steps in the inference pipeline.}
\label{fig:sbi_diagram}
\end{figure}
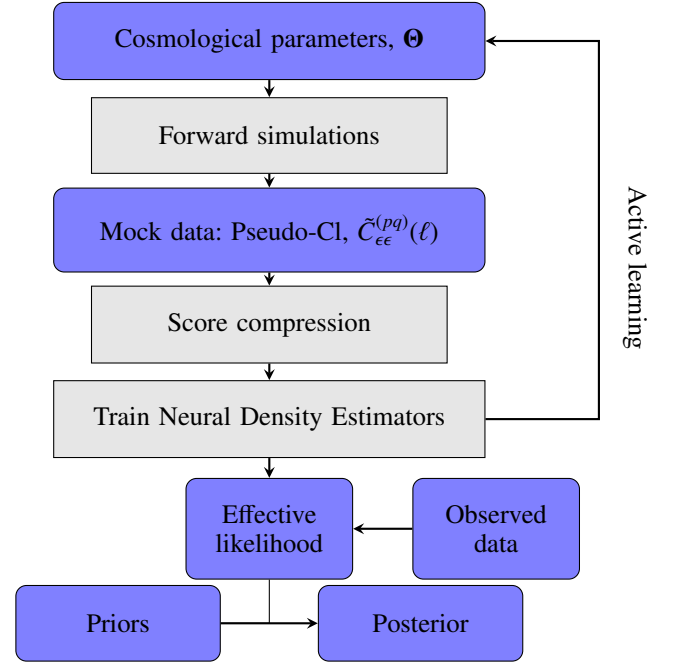

To implement this, we create a bespoke interface that allows \texttt{DELFI} to be the sampler for \texttt{CosmoSIS} \citep{zuntz2015cosmosis}. Initially, the sampler chooses a fixed number of points within the considered parameter space based on the hypercube-generating algorithm from PyDOE\footnote{\url{https://github.com/tisimst/pyDOE}} with some minor modifications (see \citetalias{lin2022a} for more details). Subsequently, as is outlined in Fig.~\ref{fig:sbi_diagram}, the measured pseudo-Cls for each evaluation of the parameters on the hypercube are compressed further using score compression (see Sect.~\ref{method:sbi:score}). This reduces the dimensionality of the data vector to the size of the parameter vector (in this case, seven dimensions). The compressed data is then used to train neural density estimators through \texttt{DELFI} (see Sect.~\ref{method:sbi:delfi}). 

As the ensemble of neural density estimators evaluates the sampling distribution throughout the hypercube, through the use of active learning, \texttt{DELFI} finds areas in parameter space where the likelihood is undersampled to select new parameter vectors (see \citealt{alsing2019fast} for details). As found in \citetalias{lin2022a}, this can halve the number of simulations needed to accurately train \texttt{DELFI} by drawing more simulations in the parameter region of interest when compared to a Latin hypercube alone. Once the ensemble of neural density estimators has reached the stopping criterion, controlled for over-fitting by comparison to a hold-out validation data set, we can calculate an effective likelihood from the learnt sampling distribution conditioned on the observed data vector (which can be a mock or measured from real data). We can then obtain a posterior distribution by sampling the product of the likelihood and the prior. This approach of learning the sampling distribution conditioned on the observed compressed data to yield an effective likelihood has the advantage that if the prior or data vector were to be changed, there is nothing new that must be re-trained within \texttt{DELFI}.

We refer to the model choices that we make within {KiDS-SBI} as the anisotropic systematics model. The model follows the choices outlined in Sects.~\ref{fs} and~\ref{measurements}. It is designed to be consistent with previous analyses of KiDS-1000, specifically, with \citet{loureiro2021kids}, while adding additional realism. Both \citet{loureiro2021kids} and this work assume a flat $\Lambda$CDM cosmology to model the cosmic shear signal, both use pseudo-Cls as their data vector of choice, and both consider systematics such as multiplicative shear bias, variable depth in the uncertainty and intrinsic alignments in the signal. The anisotropic systematics model differs from \citet{loureiro2021kids} in that it also includes variable depth in the signal modelling (as described in Sect.~\ref{method:fs:salmo_vd}), it considers the effect of intrinsic alignments on the likelihood, and it takes into account the variance in the additive and PSF shear biases (as described in Sect.~\ref{method:fs:salmo_shears}). This is a consequence of the fact that in {KiDS-SBI} the effects which are modelled in the signal are intrinsically considered in the uncertainty model. In \citet{loureiro2021kids}, these effects are only considered separately as they pertain to the signal or the uncertainty, not both at once.

\subsection{Pipeline performance}\label{method:sbi:performance}

A single such evaluation of the forward simulations from cosmological parameters to the shear pseudo-Cls as described in Sect.~\ref{fs} runs within ${\sim}$20 minutes on a single core (with $N_{\mathrm{shells}} = 19$, $N_{\mathrm{tomo}} = 5$ and $N_{\mathrm{side}} = 1024$). We chose this spatial resolution $N_{\mathrm{side}}$, because it implies that the measured two-point statistics should be representative of the input angular power spectra up to an $\ell_{\mathrm{max}} {\sim} 2 \times N_{\mathrm{side}}$ \citep{leistedt2013estimating, alonso2019a}. In addition, for the KiDS-1000's galaxy number density, the pixel size at an $N_{\mathrm{side}} = 1024$ is sufficient for almost all pixels to contain at least one galaxy. This suppresses mode-mixing in the pseudo-Cls due to random masking of empty pixels which were observed, as at this resolution all pixels are on average populated by at least one galaxy (see Appendix~\ref{appendix:signal}). 

\begin{figure}
    \centering
    \includegraphics[width=9cm]{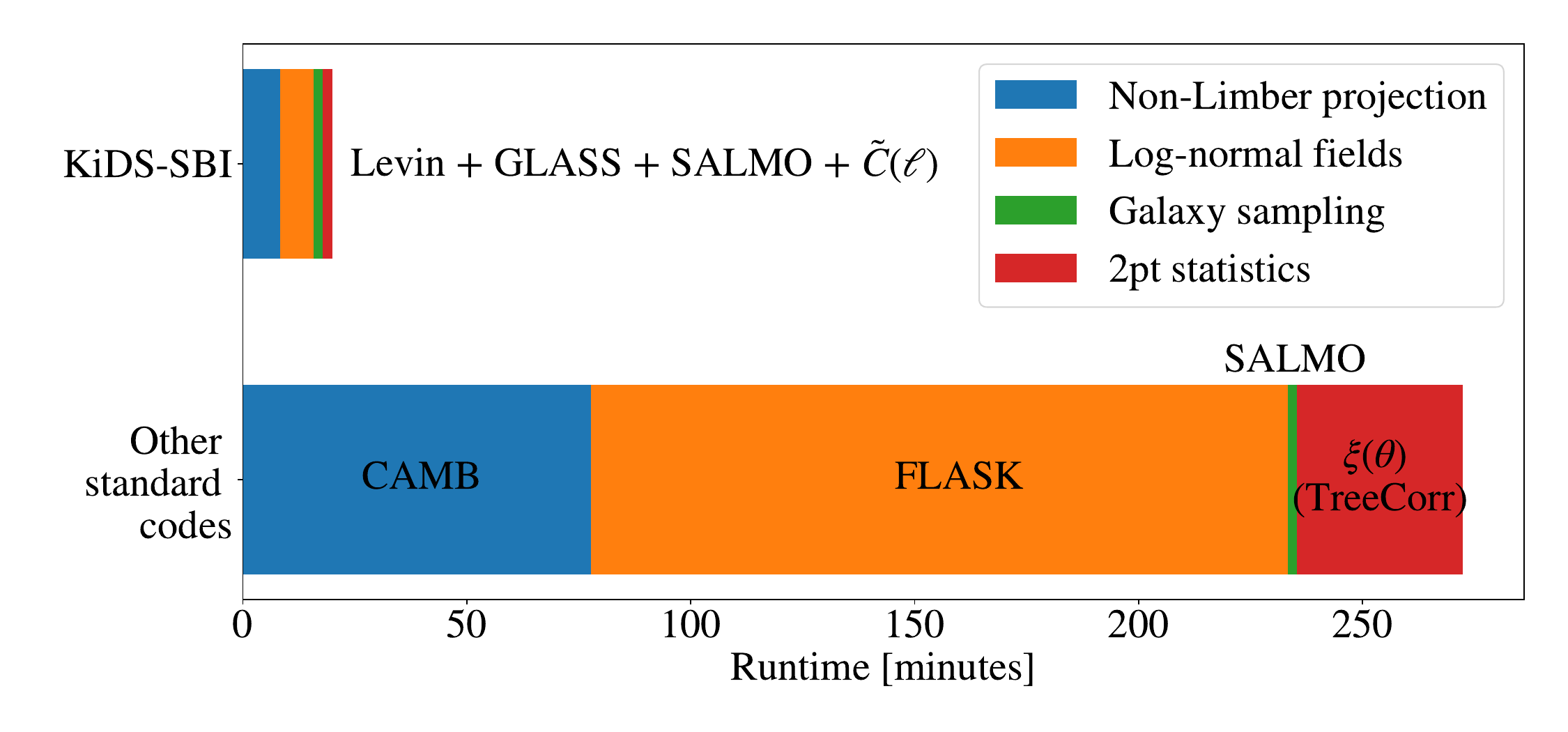}
    \caption{Bar chart comparing the run-time of a single evaluation of {KiDS-SBI} (above) versus a single evaluation of a simulation based on \citet{joachimi2021kids} (below), both on a single core ($N_{\mathrm{shells}} = 19$, $N_{\mathrm{tomo}} = 5$ and $N_{\mathrm{side}} = 1024$). Both suites of simulations use \texttt{CAMB} \citep{lewis2000efficient, lewis2002evolution, howlett2012cmb} to compute the 3D matter power spectrum. For the reference simulations, we use the non-Limber projection built into \texttt{CAMB} with \textit{limber\_phi\_lmin} $= 1200$ rather than \texttt{Levin} with $\ell_{\mathrm{max, nL}} = 1200$. We run \texttt{FLASK} \citep{xavier2016improving} rather than \texttt{GLASS} \citep{tessore2023glass} to compute the underlying matter and convergence fields of each of the 19 shells. Subsequently, we sample galaxies using \texttt{SALMO} in both cases, and then calculate the spatial two-point correlation functions, $\xi_{\pm}(\theta)$, rather than calculating the angular power spectra, $\tilde{C}(\ell)$. To calculate $\xi_{\pm}(\theta)$ in the reference simulations, we use \texttt{TreeCorr} \citep{jarvis2004the}.}
    \label{fig:runtime}
\end{figure}

In comparison (as can be seen in the bar chart in Fig.~\ref{fig:runtime}) log-normal random field simulations similar to the ones presented in \citet{joachimi2021kids}, take ${\sim}$280 minutes to compute a single forward simulation when run with the same accuracy and precision settings on a single core. In practice, the simulations described in \citet{joachimi2021kids} are used exclusively for the characterisation of a numerical covariance matrix. Hence, only a single evaluation of the underlying power spectrum from \texttt{CAMB} is necessary for every simulation. At the same time, the \citet{joachimi2021kids} simulations have a lower resolution of shells along the line of sight, so the runtime of \texttt{FLASK} \citep{xavier2016improving} in the original setting is lower than the runtime indicated in Fig.~\ref{fig:runtime}. We also note that the runtime of \texttt{TreeCorr} \citep{jarvis2004the} could be drastically reduced from the single-core runtime shown in Fig.~\ref{fig:runtime} by increasing the number of cores and/or by pixelising the galaxy shears prior to calculating the spatial two-point correlation functions, $\xi(\theta)$.

These gains in speed are driven by four main factors: the fast non-Limber integration using the Levin method, the use of recurrence relations to calculate the convergence within \texttt{GLASS}, the choice of using a quick-to-compute summary statistic like angular power spectra, $\tilde{C}(\ell)$, and the efficient subtraction of the shape noise bias from a single random rotation of the galaxy shapes. This is a substantial improvement with respect to other typical codes used for such simulations, while also improving accuracy with respect to previous models by including systematics, improving the resolution along the line of sight, and dropping common approximations like the Limber and reduced shear approximation. Thanks to this, it becomes feasible to increase the precision of our simulations as necessary and add the realism discussed in Sect.~\ref{method:fs:salmo}, \ref{method:fs:salmo_shears}, and \ref{method:fs:salmo_vd}, while still being able to compute the ${\sim}$$10^4$ forward simulations needed to adequately characterise the effective likelihood \citetalias{lin2022a}.

\subsection{Parameters and priors}

The parameters which are varied in the simulation-based inference in this work are shown in Table~\ref{tab:sbi_priors}. Their priors match those of the previous KiDS-1000 cosmic shear analyses \citep{asgari2021kids, heymans2021kids, loureiro2021kids, busch2022kids, troester2022joint}.

To avoid overinformative priors, most are flat top-hat functions. For the same reason, the top-hat priors in $S_{8}$, $h_{0}$, $\omega_{\mathrm{b}}$, $A_{\mathrm{IA}}$, and $A_{\mathrm{bary}}$ are chosen to be wide. The prior on $n_{\mathrm{s}}$ spans a smaller range around the theoretical value of unity for scale-invariant primordial fluctuations. This avoids artefacts within the prior volume as this parameter is not well constrained by weak gravitational lensing. The prior on $\omega_{\mathrm{c}}$ is defined to be consistent with a range in $\Omega_{\mathrm{m}} \in [0.188, 0.408]$, where the limits are given by the $\pm 5 \sigma$ intervals of the marginal constraints from independent measurements of luminosity distance to Type Ia Supernovae in \citet{scolnic2018the}. See \citet{joachimi2021kids} for more details on the motivations for the chosen priors. The priors on the nuisance parameters capturing any shifts in the mean of each of the five tomographic bins, $\bm{\delta}_{z}$, are given by a multivariate Gaussian, since the $\delta_{z}$ of a given tomographic bin is not independent of the shifts in the other bins. This is quantified by the covariance, $\bm{C}_{z}$, as estimated in \citet{hildebrandt2021kids}.

\begin{table*}
    \centering
    \caption{Parameters that are varied within the simulation-based inference pipeline and their priors.}
    \begin{tabular}{cccccc}
        \hline
        Parameter & Symbol & Prior type & Prior range & Initial fiducial & Final fiducial\\
        \hline
        Density fluctuation amp. & $S_{8}$ & Flat & [0.1, 1.3] & 0.760 & 0.739 \\
        Hubble constant & $h_0$ & Flat & [0.64, 0.82] & 0.767 & 0.653\\
        Cold dark matter density & $\omega_{\mathrm{c}}$ & Flat & [0.051, 0.255] & 0.118 & 0.135 \\
        Baryonic matter density & $\omega_{\mathrm{b}}$ & Flat & [0.019, 0.026] & 0.026 & 0.023 \\
        Scalar spectral index & $n_{\mathrm{s}}$ & Flat & [0.84, 1.1] & 0.901 & 0.972\\
        \hline
        Intrinsic alignment amp. & $A_{\mathrm{IA}}$ & Flat & [-6, 6] & 0.264 & 0.603\\
        Baryon feedback amp. & $A_{\mathrm{bary}}$ & Flat & [2, 3.13] & 3.10 & 2.66\\
        Redshift displacement & $\bm{\delta}_{z}$ & Gaussian & $\mathcal{N}(\bm{0}, \bm{C}_{z})$ & $\mathbf{0}$ & $\mathbf{0}$\\
        \hline
        Multiplicative shear bias & $M^{(p)}$ & Gaussian & $\mathcal{N}(\overline{M}^{(p)}, \sigma^{(p)}_{M})$ & $\overline{M}^{(p)}$  & $\overline{M}^{(p)}$ \\
        Additive shear bias & $c_{1,2}^{(p)}$ & Gaussian & $\mathcal{N}(\overline{c}_{1,2}^{(p)}, \sigma^{(p)}_{c_{1,2}})$ & $\overline{c}_{1,2}^{(p)}$  & $\overline{c}_{1,2}^{(p)}$\\
        PSF variation shear bias & $\alpha_{1,2}^{(p)}$ & Gaussian & $\mathcal{N}(\overline{\alpha}_{1,2}^{(p)}, \sigma^{(p)}_{\alpha_{1,2}})$ & $\overline{\alpha}_{1,2}^{(p)}$ & $\overline{\alpha}_{1,2}^{(p)}$\\
        \hline
    \end{tabular}
    \tablefoot{The prior ranges are selected to be exactly in line with previous KiDS-1000 analyses \citep{asgari2021kids, heymans2021kids, loureiro2021kids, busch2022kids, troester2022joint}. The upper five rows show the cosmological parameters of interest, while the three rows below show the nuisance parameters which quantify systematic biases. The last three rows show shear bias parameters for each tomographic bin, $p$, as given in Fig.~\ref{fig:shear_bias} which are sampled implicitly within each simulation by drawing from the prior distribution, so any posterior is pre-marginalised over these parameters. For flat priors, the lower and upper limits of the normalised rectangular function defines the prior. For the Gaussian prior on $\bm{\delta}_{z}$, we use a 5D multivariate Gaussian with its mean at the zero vector and the covariance, $\bm{C}_{z}$, defined by the one estimated in \citep{hildebrandt2021kids}. We note that for simplicity, the $\bm{\delta}_{z}$ are implicitly marginalised throughout this analysis. The fiducial values given here constitute the fiducial parameter choice for the score compression described in Sect.~\ref{method:sbi:score}.}
    \label{tab:sbi_priors}
\end{table*}

\subsection{Score compression}\label{method:sbi:score}
To improve the computational efficiency of the analysis and to facilitate the use of \texttt{DELFI}, we compress each of the 120-dimensional data vectors measured from each forward simulation using score compression as described in \citet{alsing2019fast}.

If the likelihood is known a priori, we can compress a given data vector to a summary of the same dimensionality as the degrees of freedom in the assumed model, such that Fisher information is conserved \citep{zablocki2016extreme, alsing2018generalized, alsing2018massive, alsing2019fast}. This allows us to compress the data vector, $\bm{d}$, down to a vector, $\bm{t}$, with the dimensionality of $|\bm{\Theta}|$ as follows
\begin{equation}
    \bm{t} = \bm{\nabla} \mathcal{L}(\bm{d} | \bm{\Theta}_{*})\,,
    \label{eq:score_compression}
\end{equation}
\noindent where $\mathcal{L}$ is the log-likelihood distribution evaluated at the fiducial cosmology, $\bm{\Theta}_{*}$. 

In this work, the exact form of the likelihood is not known a priori, as the main motivation is to characterise the form of the effective likelihood. However, it is still possible to perform score compression on the data by assuming an analytical form for the likelihood. To this end, we assume a Gaussian likelihood with a fixed covariance matrix from 1,600 forward simulations at the fiducial cosmology given in Table~\ref{tab:sbi_priors} and evaluate its gradient through a five-point stencil near the fiducial cosmology. In this form with a Gaussian likelihood, this compression is equivalent to \texttt{MOPED} \citep{heavens2000massive} or a linear compression based on Karhunen-Loéve eigenvalue decomposition \citep{tegmark1997karhunen}.

This compression is optimal if the true likelihood is Gaussian, which it is asymptotically near the peak of the likelihood and the chosen fiducial set of parameters is equal to the true parameter values. The downside of this is that the compression can lose information if this is not the case. Nevertheless, it was found in \citetalias{lin2022a} that, in {KiDS-SBI}, the score compression is robust to suboptimal choices of fiducial parameter values and data covariance.

Additionally, we find that the 12 parameters shown in Table~\ref{tab:sbi_priors} are not necessary to characterise the degrees of freedom of the 120-dimensional cosmic shear pseudo-Cls. As all $\delta_{z}$ are broadly consistent with zero, we may compress all data vectors to a 7D summary which is still capable of capturing all the complexity in the data. In any case, the $\bm{\delta}_{z}$ parameters are still explicitly varied within the simulations to propagate any uncertainties on the mean of the tomographic bins. For simplicity, henceforth all posteriors are implicitly marginalised over the five $\delta_{z}$ parameters.

\subsection{Density estimation likelihood-free inference (\texttt{DELFI})} \label{method:sbi:delfi}
To estimate the effective likelihood for the anisotropic systematics model, we train neural density estimators (NDEs) using \texttt{DELFI} \citep{alsing2019fast} such that the NDEs learn the sampling distribution between the compressed data and the input cosmological parameters. To this end, we initialise an ensemble of six independent conditional Masked Autoregressive Flows (MAFs;  \citealt{papamakarios2017masked}). Each MAF is made up of between three and eight Masked Autoencoders for Density Estimation (MADEs; \citealt{germain2015made}) each with two hidden layers of 50 neurons. For a detailed definition and description of these network architectures, see \citet{alsing2019fast}.

As the first step in the exploration of the prior space, the ensemble of these neural density estimators (NDEs) initially learns the sampling distribution from a set 2000 points on a Latin hypercube. Based on this, we then make use of the active learning feature within \texttt{DELFI} to sample additional points in parameter space which efficiently contribute to learning the effective likelihood within prior space. To ensure that the sampling distribution density, $P(\bm{t}| \bm{\Theta}, \bm{w})$ is well learnt, recursive sampling continues until the validation loss stops decreasing for 20 training epochs, where the loss function is defined as

\begin{equation}
    - \mathrm{ln} [U(\bm{w})] = - \sum_{i=1}^{N_{\mathrm{sims}}} \mathrm{ln}[P(\bm{t}_{i}|\bm{\Theta}_{i}, \bm{w})],
    \label{eq:training_loss}
\end{equation}
\noindent where $N_{\mathrm{sims}}$ is the number of forward simulations and the loss function, $ - \mathrm{ln} [U(\bm{w})]$, is defined such that it is a Monte Carlo estimate of the Kullback-Leibler divergence between the learnt effective likelihood and the true effective likelihood (see \citealt{alsing2019fast} for a more detailed explanation). This process is repeated over an ensemble of many independent neural density estimator networks, where each network learns the conditional likelihood independently. The final estimate of the effective likelihood is then given by the weighted sum of the learnt distributions from each neural density estimator.

We find that for the anisotropic systematics model, this occurs already after 5,000 realisations. We note that the NDEs require fewer realistic {KiDS-SBI} forward simulations to minimise the loss function as is the case in \citetalias{lin2022a} where the forward simulations were idealised random samples from a Gaussian distribution. In any case, to ensure that the prior space is densely sampled, we chose to train the NDEs on 18,000 realisations. This is facilitated by the computational efficiency of the forward simulations making additional realisations relatively inexpensive.

Upon training each of the six MAF NDEs using forward simulations independently, the NDEs are stacked while weighting each according to their relative validation losses. From the stacked NDE, we obtain the final posterior distributions by sampling with \texttt{nautilus} \citep{lange2023nautilus} and check for consistency with \texttt{emcee} \citep{emcee}.

As advised in \citetalias{lin2022a}, to minimise the loss of information through from the compression step described in Sect.~\ref{method:sbi:score}, once the simulation-based inference has been conducted the first time on the KiDS-1000 data for a given forward model, we repeat the inference again a second time while taking the fiducial model parameters assumed in the score compression to be given by the maximum a posteriori (MAP) from the initial inference (see Tab.~\ref{tab:sbi_fiducial_results}). We note that in the second inference, we do not change the forward model in any way, and keep the exact same realisations of the model as in the first analysis. Only the fiducial data vector used to characterise the likelihood in Eq.~(\ref{eq:score_compression}) are updated to re-compress all simulated data vectors. Considering that the initial fiducial parameters based on previous KiDS-1000 analyses are mostly consistent with the MAP which we infer in the initial analysis, while score compression is also robust against the biased fiducial parameters \citep{lin2022a}, we find that this step only marginally alters the inferred posterior.

\begin{figure}
    \centering
    \includegraphics[width=8cm]{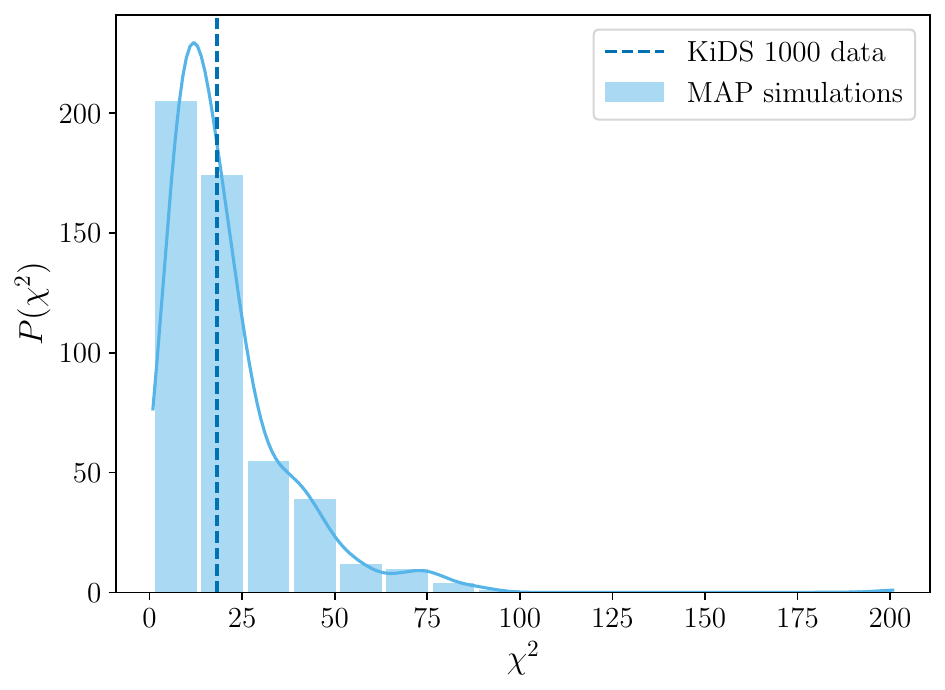}
    \caption{Goodness-of-fit test implemented following \cite{gelman1996posterior}. A distribution of $\chi^2$ values for 500 forward simulations at the MAP cosmology is evaluated from a $\chi^2$ function characterised by a Gaussian likelihood defined by a numerical covariance. The vertical dashed line indicates the $\chi^2$ for the MAP when analysing the real KiDS-1000 data vector. The solid blue line shows the histogram upon the application of a smoothing kernel. Since the data's $\chi^2$ falls near the mean and the mode of the $\chi^2$ distribution from random realisations of the anisotropic systematics model, we conclude that the model evaluated at the MAP cosmology provides a good fit to the data, i.e. the data is a feasible realisation of the model.}
    \label{fig:sbi_goodness_of_fit}
\end{figure}

\section{Goodness-of-fit in SBI} \label{method:gof}
As the inference is performed using SBI, we cannot perform a goodness of fit test the traditional way when we have access to an analytical likelihood. However, we still wish to keep this metric and so create a novel implementation of the $\chi^2$ goodness-of-fit test. We follow the work of \cite{gelman1996posterior} and implement a SBI goodness-of-fit test from simulations in the \texttt{ECP-GF}\footnote{\url{https://github.com/Kiyam/ecp_gf_tests}\label{note:gf}} module.

The methodology is as follows, initially, we find the maximum a posteriori (MAP) values from the posterior sampled in the SBI using the Nelder-Mead \citep{nelder1965simplex} optimiser. As the parameters of $n_s$ and $A_{\mathrm{bary}}$ are unconstrained by our analysis, and the MAP can therefore vary freely across the prior, instead of making use of the MAP cosmology, we run simulations at the cosmology chosen to perform score compression for those parameters, which is already a pseudo maximum likelihood estimation (see \cite{lin2022a} for more details).

\begin{figure}
    \centering
    \includegraphics[width=8cm, clip]{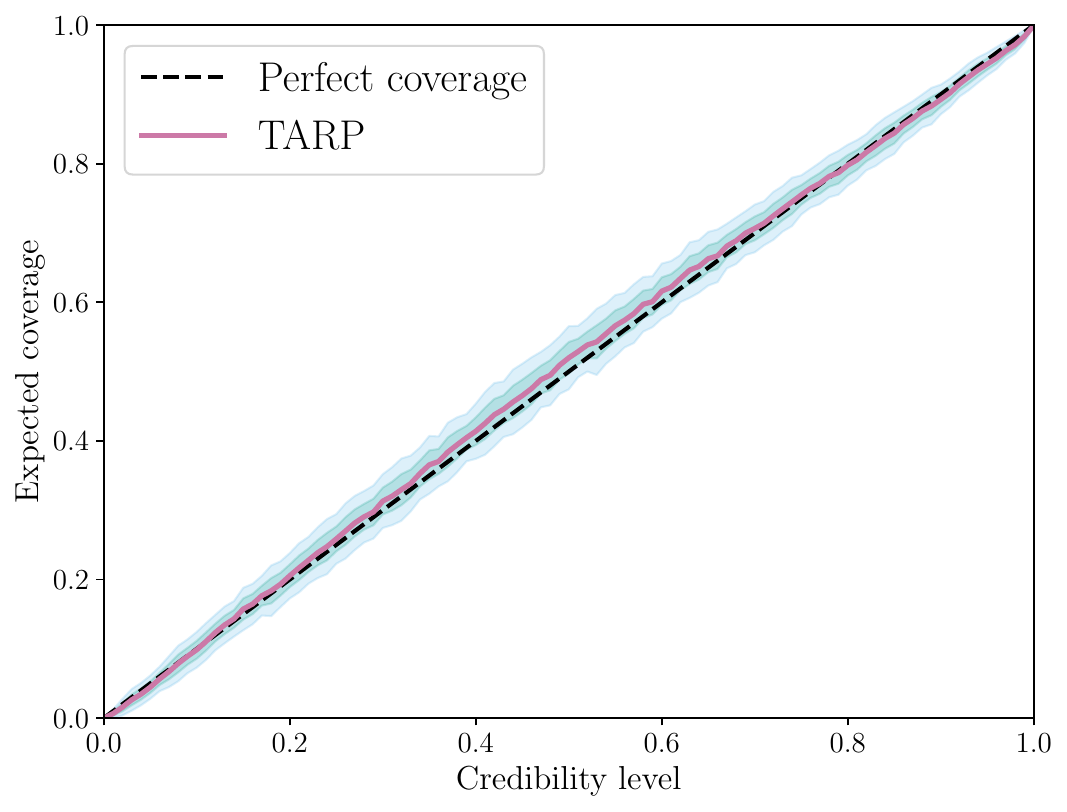}
    \caption{The expected coverage probability versus the credibility level as defined in the Tests of Accuracy with Random Points (TARP) described in \citet{lemos2023sampling} for the posterior shown in Fig.~\ref{fig:sbi_validation_posterior} assuming the anisotropic systematics model. The dark blue region indicates the 1$\sigma$ fluctuations in the TARP obtained from bootstrapping over 100 realisations, while the light blue region indicates the $2\sigma$ region. The dashed line is a reference line for a perfectly linear relation. The credibility level gives the fraction of the total probability density of the learnt posterior being considered, while the expected coverage probability measures the fraction of posterior samples which have a posterior probability smaller than the best estimate at a given credibility level. We note the relation, in this case, indicates slight underconfidence, such that the posteriors in Figs.~\ref{fig:sbi_validation_posterior} and~\ref{fig:sbi_results} can be considered as conservative. The ideal case is completely within the $1\sigma$ confidence interval of the TARP test.}
    \label{fig:sbi_validation_coverage}
\end{figure}

At the MAP cosmological parameter values, we simulate the sampling distribution by running a further 500 forward simulations while randomly varying noise realisations. We can thus define a $\chi^2$ discrepancy measure as follows
\begin{equation}
    \chi^2(\boldsymbol{t}|\boldsymbol{\Theta}) \coloneqq(\boldsymbol{t}_i - \mathrm{E}[\boldsymbol{t}_*|\boldsymbol{\Theta}_*])^\mathrm{T}{(\mathrm{Cov}(\boldsymbol{t}_*|\boldsymbol{\Theta}_*)})^{-1}(\boldsymbol{t}_i - \mathrm{E}[\boldsymbol{t}_*|\boldsymbol{\Theta}_*])\,,
\end{equation}
\noindent where $\bm{t}_*$ is the compressed data vector at the fiducial cosmology, $\bm{\Theta}_*$ the fiducial cosmology and $\bm{t}_i$ one of $n$ randomly varied noise realisations of the compressed data vector at the MAP parameter values. A fully Bayesian implementation would simply require forward simulations to be realised from posterior parameter draws as opposed to only at the inferred MAP parameter value. We have stuck to making use of simulations at the MAP to be analogous to the traditional KiDS-1000 analysis. Furthermore, as we are making use of linear score compression, we have linearised our model, and thus our goodness of fit is analogous to the classic $\chi^2$ test. The main difference is that the degrees of freedom of the maximum $\chi^2$ does not trivially reflect the degrees of freedom of our model due to the effects of optimal score compression in performing data dimensionality reduction.

For our setup, we evaluate a mean model data vector and a covariance from a set of 1,600 simulations run at the fiducial cosmology. It should be noted that these simulations are the same as the ones required for the score compression step described in Sect.~\ref{method:sbi:score}, so this comes without any additional computational costs. Using this newly obtained model data vector and covariance, and the $\chi^2$ discrepancy measure defined above, we calculate the discrepancy measure for the 100 separate forward simulations from the anisotropic systematics model, and the KiDS-1000 gold sample data, which are depicted in the left panel of Fig.~\ref{fig:sbi_goodness_of_fit}. As we obtain a probability to exceed (PTE) of 0.42, we find a good fit to the data.
We note that in this test, the degrees of freedom are not recovered directly due to the effects of normalisation and optimal score compression in doing dimensionality reduction.

%%%%%%%%%%%%%%%%%%%%%%%%%%%%%%%%%%%%%%%%%%%%%%%%%%
\section{Validation} \label{validation}
The SBI pipeline presented in this work has previously been extensively tested in an idealised case. From the findings in \citetalias{lin2022a}, we conclude that the SBI pipeline based on \texttt{DELFI} implemented within {KiDS-SBI} robustly and accurately recovers the posterior distribution of $\Lambda$CDM cosmological parameters from KiDS-1000 cosmic shear data, even if the data compression loses information. However, \citetalias{lin2022a} only considered simulated data whose likelihood is by construction Gaussian. For this reason, it becomes necessary to test the robustness of the SBI pipeline again for the anisotropic systematics forward model. We find in Appendix~\ref{validation:paramter_recovery} that accurately and robustly recovers the underlying posterior distribution of $\Lambda$CDM parameters.

\subsection{Coverage}

\begin{figure*}
    \centering
    Noiseless mock data \hskip 6cm Noisy mock data\par\medskip
    \includegraphics[width=8.5cm]{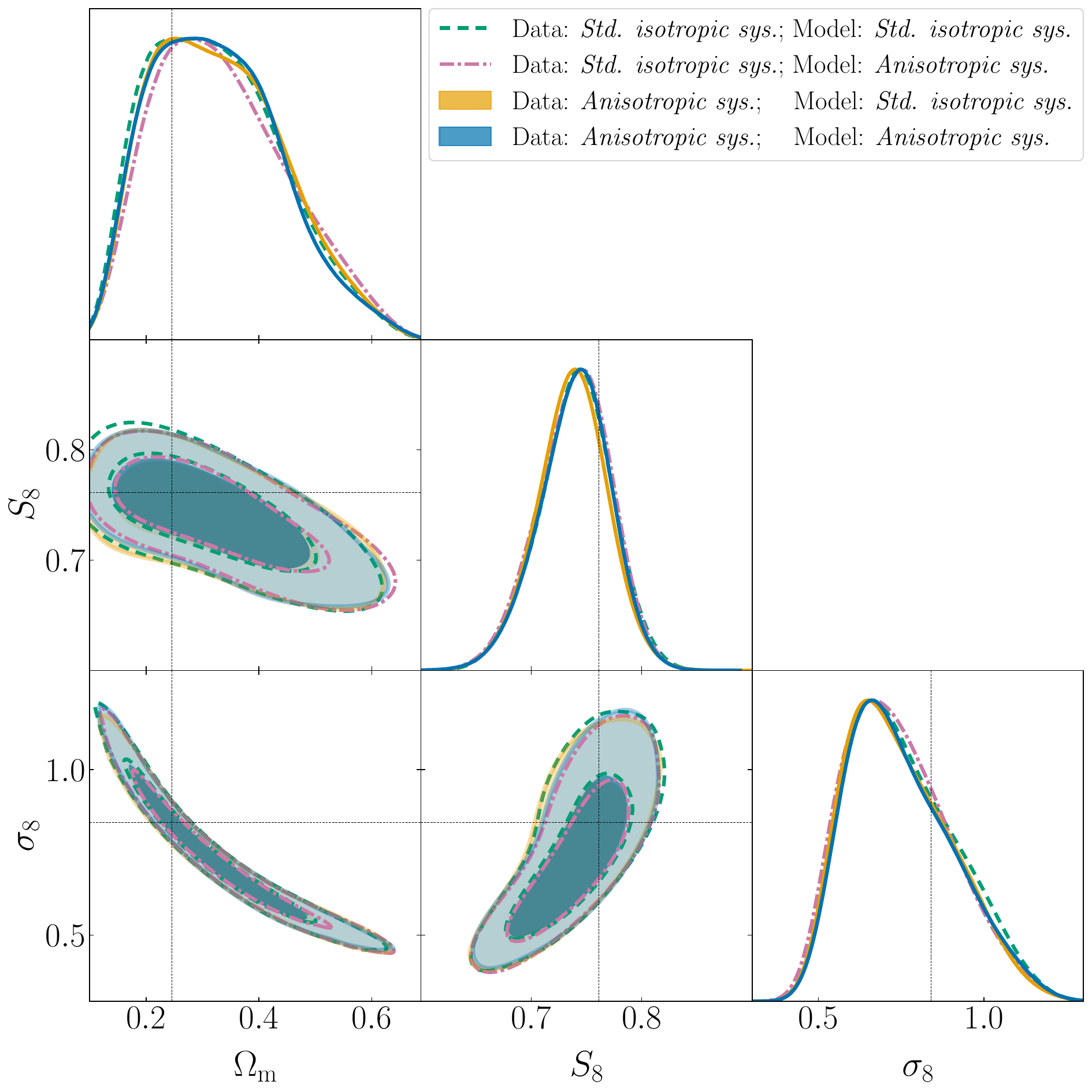}
    \includegraphics[width=8.5cm]{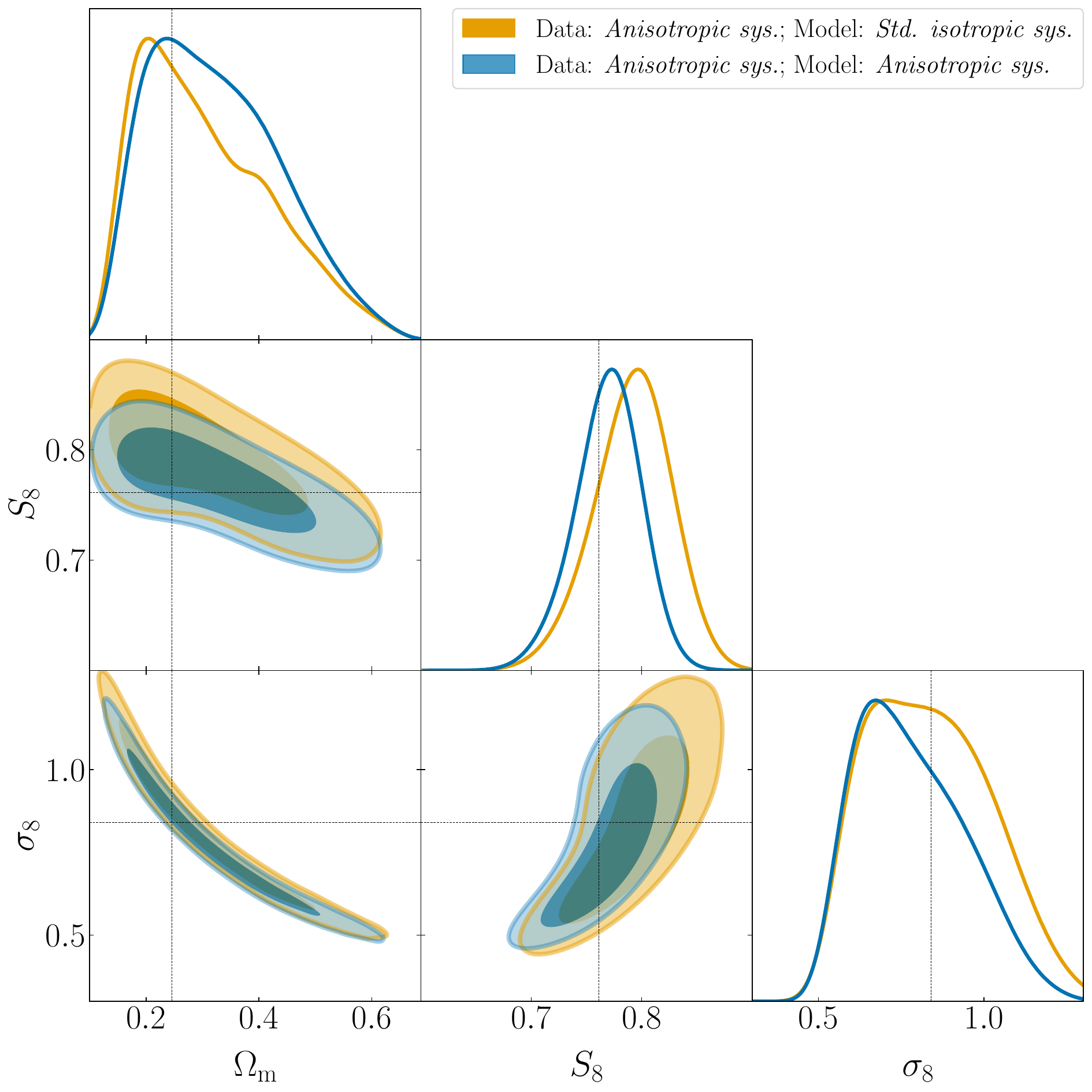}
    \caption{Contour plots comparing the effects of model misspecification for different model cases and noise levels in the data. \textbf{Left panel}: Comparison of marginalised posterior contours from noise-free data assuming the anisotropic systematics model (which includes variable depth and PSF variations) in blue and green, and the standard isotropic systematics model (which is in line with the analysis assumptions in \citealt{asgari2021kids}, i.e. it does not model variable depth or PSF variations) in orange and pink. The orange and the blue posteriors are derived from the same data vector: a noise-free realisation from the anisotropic systematics model at the fiducial cosmology shown in Table~\ref{tab:sbi_priors}; while the pink and the green posteriors are derived from a noise-free realisation from the standard isotropic systematics model at the same cosmology. \textbf{Right panel}: Comparison of marginalised posterior contours from noisy data simulated with anisotropic systematics model with the same seed as in the noise-free case, and subsequently analysed with the SBI based on the anisotropic systematics model in blue, and standard isotropic systematics model in orange. In both panels, the black solid line indicates the true cosmology assumed for the mock data vector (see Table~\ref{tab:sbi_validation_truth}).}
    \label{fig:sbi_validation_vd_impact}
\end{figure*}

To determine whether the learnt posterior distribution from the anisotropic systematics model is representative of the true underlying posterior and is unbiased, we conduct the Tests of Accuracy with Random Points (TARP) shown in Fig.~\ref{fig:sbi_validation_coverage}. TARP \citep{lemos2023sampling} measures the expected coverage probability of random posterior samples within a given credibility level of the learnt posterior empirically. Whilst the original authors have their implementation of the coverage test, we made use of our implementation (\texttt{ECP-GF}\footref{note:gf}) to interface with the outputs from our simulations and DELFI in a more straightforward manner.  

We did not perform the coverage test over our entire parameter space, as there are many cosmological parameters that are entirely unconstrained by our inference. Instead, the coverage test was only performed on $S_8$ and $\Omega_\mathrm{m}$ as those are the parameters of greatest interest for cosmic shear.

To perform this coverage test, we first ran 500 new simulations at the fiducial cosmology for $A_\mathrm{IA}$, $n_\mathrm{s}$, $h_0$, $A_\mathrm{bary}$, freely varying the correlated $\delta_z [5]$ values, but varied $S_8 \in [0.5, 0.9]$, $\omega_\mathrm{b} \in [0.019, 0.026]$, $\omega_\mathrm{c} \in [0.051, 0.255]$ and $h_0 \in [0.64, 0.82]$, with a latin hypercube. A slightly narrower range was chosen for $S_8$ than the prior for the full inference so that the testing points were positioned closer to the final parameter posterior range of interest. For each of these simulations, the noise realisation was allowed to freely vary. We also performed a coverage test with $A_\mathrm{IA}$ also varied, and obtained similar results to what is shown in Fig.~\ref{fig:sbi_validation_coverage}.

We then used our trained \texttt{DELFI} model to perform inference on the new suite of 500 simulations, each acting as a mock data vector and thus obtaining 1000 posterior points for all 500 simulations. From the posterior samples, the coverage probability was calculated empirically for a given credibility level (see \citealt{lemos2023sampling} for a detailed discussion of the algorithm). Due to the inherently stochastic nature of the coverage probability test, the test was bootstrapped 100 times for each credibility level. Figure~\ref{fig:sbi_validation_coverage} depicts both the mean and two standard deviations of the spread of the bootstrapped TARP coverage test.

For a well-estimated posterior that is not biased and representative of the true underlying posterior, we expect the coverage probability to be directly proportional to the credibility level. Any biases in the learnt posterior would lead to deviations from linearity in the TARP.

As can be seen from Fig.~\ref{fig:sbi_validation_coverage}, the learnt posterior is very close to being ideal. As the mean coverage is seen to be slightly lower than the credibility level but always within one standard deviation of the ideal case. This means that the learnt posterior is slightly underconfident. This is acceptable as it implies that the {KiDS-SBI} posterior estimates are conservative but not biased. We repeat this test while also varying $A_\mathrm{IA}$, and find similar results.

\subsection{Impact of variable depth and shear bias}\label{validation:impact_vd}
To assess the impact of the additional observational systematics considered in the anisotropic systematics model, we re-train a set of NDE ensembles on another 18,000 simulations of a different model which does not consider variable depth or additional shear biases while following the same procedure as outlined in Sect.~\ref{method:sbi}. We refer to this model as the standard isotropic systematics model as it is more in line with previous 'standard' KiDS-1000 analyses such as \citet{asgari2021kids}, \citet{busch2022kids}, \citet{li2023kids}, or \citet{deskids2023}. This model considers the cosmological signal, intrinsic alignments, and baryonic feedback in the same way as the anisotropic systematics model, but it has two main differences: it does not consider variable depth, and it does not consider the additive or PSF shear biases. Firstly, when sampling galaxies as described in Eq.~(\ref{eq:method:fs:salmo:position}) the galaxy density, $n_{\mathrm{gal}}$, and the photometric redshift distribution, $P(p|i)$, are assumed to be isotropic for a given tomographic bin. Secondly, when biasing the observed galaxy shapes according to the shear biases measured in KiDS-1000, the standard isotropic systematics model only considers the multiplicative shear bias; that is, only $M^{(p)} \neq 0$ while $\alpha^{(p)} = 0, \, c^{(p)} = 0 \, \forall p$.

This simplification of the forward model impacts the measured cosmic shear signal in a few ways. Firstly, from just comparing the measured two-point statistics, as can be seen in more detail in Appendix~\ref{method:consistency:var_depth}, we find that variable depth can contribute up to $1\%$ to the measured two-point shear signal in KiDS-1000, which is consistent with the predictions from \citet{heydenreich2020the}. At the same time, variable depth can significantly alter the uncertainty of the two-point statistic (by $\lesssim 20\%$; \citealt{joachimi2021kids}). As the intrinsic galaxy shape dispersion and the galaxy density vary anisotropically, the shape noise of the sample is no longer determined by the mean value per tomographic bin. Instead, the shape noise becomes an anisotropic distribution too which in the case of KiDS-1000 is skewed below the mean (see Appendix~\ref{method:consistency:var_depth}). In addition, as can be seen by eye from Fig.~\ref{fig:level_map} and~\ref{fig:level_map_zoom}, variable depth can add angular correlations to the data at the scales of pixels, pointing overlaps, whole pointings or even over the whole footprint, while also changing the signal along the line of sight anisotropically as shown in Fig.~\ref{fig:nofz_variation}. These additional correlations can significantly alter the covariance of the two-point statistic or may even lead to non-Gaussian noise.

In fact, these non-cosmological correlations are significant enough for us to find that the overall uncertainty on the two-point statistic of a model including variable depth is larger than the uncertainty in a model which does not consider the effect. Hence, variable depth contributes approximately $1\%$ (${\sim} 0.2 \sigma$) additional non-cosmological two-point signal.

\begin{figure}
    \centering
    \includegraphics[trim=0 0 0 0, width=9cm]{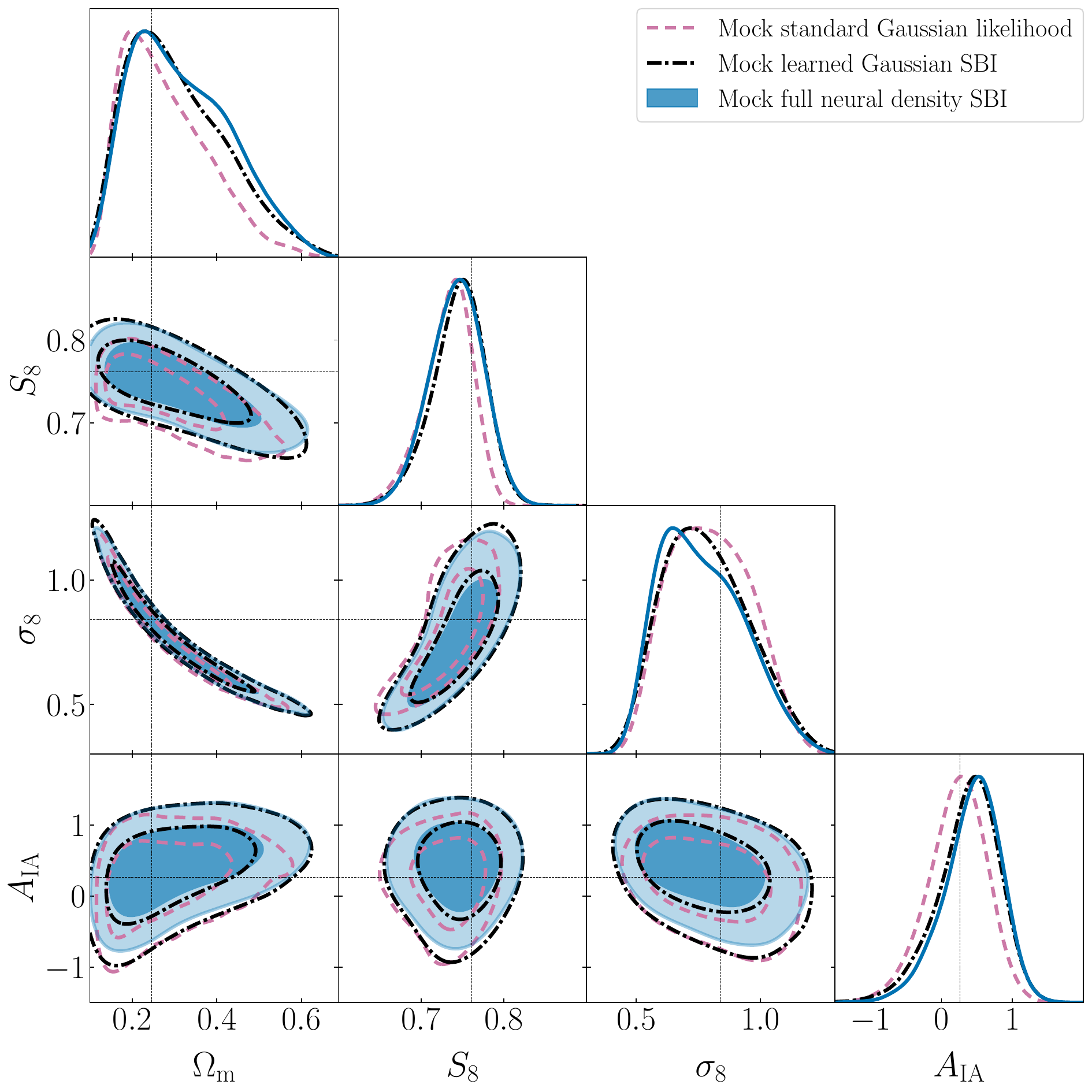}
    \caption{Comparison of likelihood models with a noise-free mock data vector from the anisotropic systematics forward model. Shown are marginalised posterior contours: in blue, assuming the anisotropic systematics model within the standard setup of {KiDS-SBI} as described in Sect.~\ref{method:sbi:delfi}; in black, assuming the same anisotropic systematics forward simulations to train a Mixture Density Network (MDN; \citealt{bishop1994mixture}) which is made up of a single multivariate Gaussian; and in pink, applying a standard analysis which assumes a Gaussian likelihood characterised by a numerical sample covariance derived from anisotropic systematics forward simulations and a model given by the analytic pseudo-Cl model described in Appendix~\ref{appendix:signal}. All posteriors are evaluated assuming the same noise realisation and the same model parameters (see Table~\ref{tab:sbi_validation_truth}).}
    \label{fig:sbi_mcmc_v_sbi}
\end{figure}

The other distinguishing systematic between the anisotropic systematics model and the standard isotropic systematics model is the shear bias. When including additive and PSF shear biases in the anisotropic systematics model, which can also add a per-cent level (up to ${\sim} 0.3 \sigma$, but less than $0.1 \sigma$ at most angular scales) non-cosmological signal to the pseudo-Cls, mostly driven by the PSF variation shear bias (see Appendix~\ref{method:consistency:var_depth} for details). This is consistent with the negligible impact on the measurement found in previous studies \citep{giblin2021kids}. At the same time, the inclusion of the $\alpha_{1}^{(p)}$, $\alpha_{2}^{(p)}$,  $c_{1}^{(p)}$ and $c_{2}^{(p)}$ parameters for each tomographic bin and their associated uncertainties adds additional variance to the anisotropic systematics model when we pre-marginalise over these 20 parameters. Therefore, the additive and PSF variation shear biases are yet another source of additional non-cosmological signal and variance.

These effects clearly propagate to the posteriors as can be seen when comparing the full simulation-based inference based on a likelihood learnt from 18,000 realisations of the standard isotropic systematics model to a likelihood based on 18,000 realisations of the anisotropic systematics model in Fig.~\ref{fig:sbi_validation_vd_impact}.

To assess the effect of model misspecification and whether the inclusion or exclusion of variable depth and shear biases can bias the SBI of cosmological parameters, we repeat the inference of a mock data vector, but this time we generate a random realisation from the standard isotropic systematics model and analyse it using the likelihood learnt from the forward simulation which assumes the anisotropic systematics model, and vice versa. This allows us to determine whether the inclusion of the effects in the forward model would bias our inference if they were not actually present in the data that is analysed. We also show these posteriors in Fig.~\ref{fig:sbi_validation_vd_impact} where we see that when the data model and the likelihood model are mismatched, the constraints are consistent within a given model. This confirms that, given the learned likelihood, the bias in the two-point shear signal from both variable depth and PSF shear bias is negligible in KiDS-1000. These findings are consistent with previous work which found this to be the case for KiDS-1000 for variable depth \citep{heydenreich2020the, joachimi2021kids} and for the PSF shear bias \citep{giblin2021kids}.

Nonetheless, systematics such as variable depth and the PSF shear bias do not only affect the measured signal, but also its uncertainty. As shown in the right panel of Fig.~\ref{fig:sbi_validation_vd_impact}, when analysing the same noisy mock data vector simulated with the anisotropic systematics model with the SBI trained on both models (anisotropic systematics and standard isotropic systematics), the marginal in $S_{8}$ shifts considerably by about $0.7\sigma$. At the same time, the constraints based on the anisotropic systematics model give ${\sim}5\%$ smaller $1\sigma$ confidence intervals in the $S_{8}$ marginals than standard isotropic systematics model. As each model characterises a different effective likelihood, when analysing the same data vector with both models, each data point is weighted differently depending on the model assumed during the inference. Depending on the noise realisation of the mock data and the effective likelihood of a given model, the posterior may vary considerably. This shows that when assessing the impact of a systematic a measurement, it is not sufficient only checking its contribution to the signal with respect to the noise. Simultaneously, it is important to assess the impact of the systematic on the likelihood itself.

Of course, the noise realisation depicted in the right panel of Fig.~\ref{fig:sbi_validation_vd_impact} is only one possible realisation of many. We find that for other noise realisations from mocks, the difference in the cosmological constraints between anisotropic systematics and standard isotropic systematics models varies as each posterior scatters as shown in Fig.~\ref{fig:sbi_validation_seed_comparison}. Consequently, neglecting variable depth and the PSF shear bias in the forward modelling, while it is present in the data that is analysed, can bias the inferred value of $S_8$ by up to ${\sim}1\sigma$. As discussed in Sect.~\ref{results:real_data}, this is consistent with what we find for the KiDS-1000 data.

\begin{figure*}
    \centering
    \includegraphics[width=12cm]{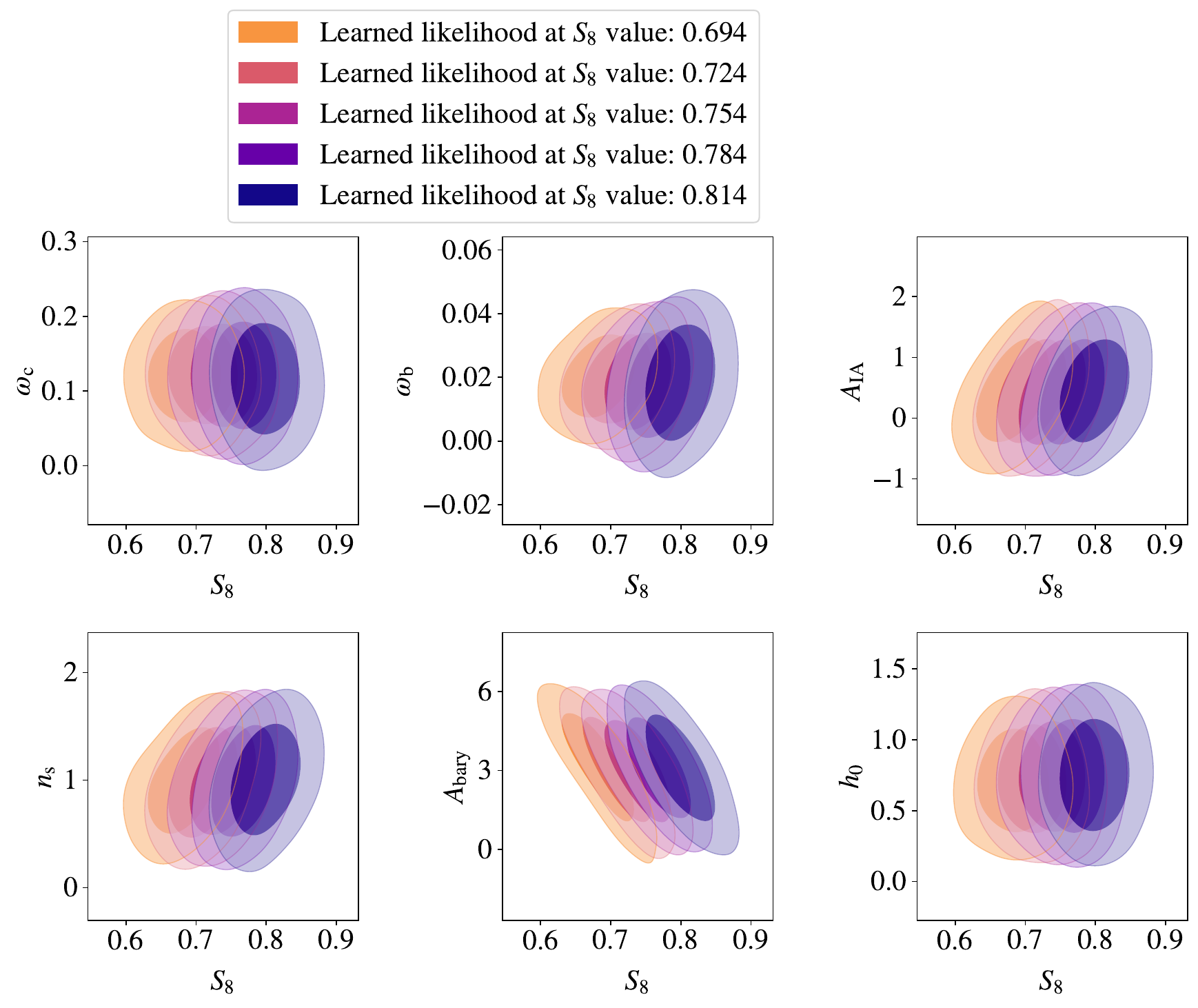}
    \caption{Likelihood marginals in the compressed data space for five different sets of cosmological parameters given the anisotropic systematics model within {KiDS-SBI} over the prior space shown in Table~\ref{tab:sbi_priors}. The compressed data values are labelled according to the cosmological parameter with which they are most correlated (see Sect.~\ref{method:sbi:score} for details). For the orange contours, the input data vector is set to $S_8=0.694$, $S_8=0.724$ for the pink contours, $S_8=0.754$ for the purple contours, $S_8=0.784$ for the purple contours, and $S_8=0.814$ for the blue contours. All other cosmological parameters are taken to be the same as in Table~\ref{tab:sbi_validation_truth}.}
    \label{fig:cosmic_var_likelihood}
\end{figure*}

\section{Impact of the Gaussian likelihood assumption}\label{gaussian_likelihood}
One of the main aims of this work is to determine whether the standard assumption of the likelihood of cosmic shear two-point statistics being Gaussian holds under realistic conditions while including systematics. To test this assumption, we performed the same analysis of a mock data vector using SBI (as described in Sect.~\ref{method:sbi} and Appendix~\ref{validation:paramter_recovery}) again, but with different inference methods.

As shown in Fig.~\ref{fig:sbi_mcmc_v_sbi}, we compare the posterior from the SBI analysis learnt from forward simulations based on the anisotropic systematics model to a standard Gaussian likelihood analysis and an SBI analysis which forces the learnt likelihood to be Gaussian. The standard analysis assumes a Gaussian likelihood characterised by the same numerical covariance from 1,600 realisations of the anisotropic systematics model fixed at the fiducial cosmology in Table~\ref{tab:sbi_priors}. The model evaluations are calculated from analytical theory for the cosmic shear signal as described in Appendix~\ref{appendix:signal}, and the sampling is done with the \texttt{nautilus} sampler \citep{lange2023nautilus} over the same 12D prior space defined in Table~\ref{tab:sbi_priors}. In contrast, the Gaussian SBI analysis makes the exact same modelling choices as the main SBI analysis described Sect.~\ref{method:sbi} and Appendix~\ref{validation:paramter_recovery} with one change: instead of using ensembles of MAFs made up of MADEs as the architecture of the neural density estimators when learning the effective likelihood, we use an ensemble of three single-component Mixture Density Networks (MDNs; \citealt{bishop1994mixture}) to learn the likelihood from the 18,000 realisation of the anisotropic systematics model. As a consequence, the ensemble of MDNs forces the learnt likelihood to be described by a single multi-variate Gaussian distribution. Notably, this is different from a standard Gaussian likelihood analysis with a Gaussian likelihood in that the Gaussian SBI analysis allows the covariance of the likelihood to vary with cosmology.

From Fig.~\ref{fig:sbi_mcmc_v_sbi}, we can conclude that the full non-Gaussian SBI is in exceptionally good agreement with the SBI analysis which forces the likelihood to be Gaussian but allows for cosmology dependence in the uncertainty (particularly, when the mock data vector includes realistic noise). Thus, the uncertainty as modelled by the anisotropic systematics model can be considered to be approximately Gaussian at a given cosmology, since a Gaussian NDE network is fully capable of learning an unbiased likelihood. We conclude that, for a fixed cosmology, the assumption of a Gaussian likelihood for the pseudo-Cl cosmic shear signal for $\ell>76$ holds well. This is in agreement with previous studies that found this to be the case in angular power spectra for $\ell>50$ \citep{hamimeche2008likelihood, schneider2009constrained, sellentin2018the, lin2020non}.

At the same time, we find that the standard Gaussian likelihood analysis has a noticeably different posterior to both SBI analyses despite the underlying data vector being identical, and the covariance being sampled from the same anisotropic systematics model. In particular, the $S_8$ marginal is appreciably narrower in the Gaussian likelihood analysis than in the SBI analyses. In the noiseless case shown in Fig.~\ref{fig:sbi_mcmc_v_sbi}, we find that the SBI gives $S_{8} = 0.743^{+0.034}_{-0.031}   $, while the standard Gaussian likelihood analysis gives $S_{8} =0.733^{+0.034}_{-0.024}   $, specifically the $1\sigma$ confidence interval in $S_{8}$ is ${\sim}11\%$ narrower than in the SBI analysis. When adding realistic noise to the data vector from the same seed, we find that the SBI recovers $S_{8} = 0.762^{+0.032}_{-0.027}   $, while the Gaussian likelihood analysis recovers $S_{8} = 0.744^{+0.031}_{-0.022}$, specifically the $1\sigma$ confidence interval in $S_{8}$ is also ${\sim}10\%$ narrower than in the SBI analysis. Additionally, when noise is introduced, the Gaussian likelihood analysis appears more biased with respect to the truth ($S_{8} = 0.756$) than the SBI.

The good agreement between the Gaussian likelihood SBI based on MDNs and the non-Gaussian likelihood SBI based on MAFs indicates that both analyses appear to be sensitive to some form of cosmology dependence in the uncertainty model, as that is the only effect which is not accounted for in the standard Gaussian likelihood analysis.

To investigate the origin of this effect further, we sample the learnt likelihood from the full SBI analysis directly at different values of $S_{8}$ in Fig.~\ref{fig:cosmic_var_likelihood}. In this plot, we show the 7D score-compressed data space and label each element of the data vector according to the model parameter with which they are most correlated. Then, we vary the value of $S_{8}$ while keeping all other parameters fixed at the fiducial cosmology, and plot the likelihood of the data given the chosen model parameters. We find that as $S_{8}$ increases the marginal likelihoods of the elements of the score-compressed data vector increase linearly. In particular, the data elements most correlated with $\omega_{c}$, $\omega_{b}$, $n_{s}$, and $h_{0}$ show the strongest linear correlation with $S_{8}$. The implication of this is that the anisotropic systematics model entails that data  at higher assumed $S_{8}$ values has larger uncertainties than data modelled for low $S_{8}$ values, and therefore the learnt likelihood is different from a Gaussian likelihood with the covariance evaluated at a fixed cosmology. This effect appears to drive the increase in the uncertainty on $S_{8}$ of the SBI-based posteriors with respect to the Gaussian likelihood analysis in Fig.~\ref{fig:sbi_mcmc_v_sbi}.

In Appendix~\ref{appendix:cosmology dependence}, we conduct a detailed investigation into the physical origin of this observed cosmology dependence in the learnt likelihood of cosmic shear two-point statistics. We find that, as has been noted in previous work \citep{eifler2009dependence, reischke2017variations}, the cosmic variance of the cosmic shear two-point signal (at least, for $76 \leq \ell \leq 1500$) is measurably cosmology-dependent, and mostly $S_8$-dependent. We also determine that this effect is expected to be important enough to become detectable in most tomographic bin combinations of the KiDS-1000 gold sample in the regime of $\ell \sim 10^2$. Since the cosmic variance is imprinted into the forward simulations at the step of sampling log-normal random matter fields (see Sect.~\ref{method:fs:glass}), the cosmology dependence of the cosmic variance is present in any forward model presented in this work irrespective of the observational systematic effects considered (namely both the anisotropic systematics and the standard isotropic systematics model are affected by this).

With this being the case, the noise of the measured cosmic shear pseudo-Cls may contain some information which may impact the inferred cosmology from the SBI. Hence, it is important to validate whether the variance realised by log-normal random simulations does not bias the cosmological inference. Firstly, it has been found in previous studies that log-normal simulations can incorporate accurate higher-order statistics into the sampled matter, galaxy and shear fields when compared to N-body dark matter simulations \citep{hall2022non, piras2023fast}, so log-normal fields can incorporate three- or four-point statistics to model the variance of two-point statistics. Additionally, we find in our own testing that a sample covariance of cosmic shear two-point statistics in KiDS-1000 as generated from log-normal simulations agrees within $10\%$ with analytical covariance matrices computed from analytical considerations \citep{joachimi2021kids, reischke2024kids}, so the variance of the cosmic shear signal as modelled within KiDS-SBI is consistent with analytical theory.

\section{Cosmological inference from KiDS-1000 data}\label{results:real_data}
\subsection{Blinding}
Throughout this analysis, we followed a procedure to avoid any form of unconscious bias when making choices within the {KiDS-SBI} models. Although the KiDS-1000 two-point shear statistics have been measured before, we fixed the forward model used in the analysis before ever evaluating learnt likelihood function for the KiDS data.

All authors did not conduct any measurements on the KiDS-1000 data nor was it analysed with the SBI pipeline until the forward model as well as all validation tests were finalised and fixed. Finally, upon evaluating the learnt likelihood for the KiDS data, we initially inferred an anomalously low $S_{8}$ with the SBI and the standard MCMC caused by an inconsistency between the measurement code for the forward simulations and the real data. Upon re-measuring the KiDS data in a consistent fashion, we obtain the pseudo-Cls from the KiDS-1000 gold sample as shown in Fig.~\ref{fig:pcl_measurement} and ran it through the entire {KiDS-SBI} pipeline without making any adjustments to the methodology described in this paper.

\subsection{Cosmological parameter constraints}
With the learnt likelihood found to be unbiased, and with an understanding of the sensitivity of the likelihood to systematics and cosmology, we proceed to analyse the cosmic shear pseudo-Cls measured from the KiDS-1000 gold sample described in Sect.~\ref{data} (see Fig.~\ref{fig:pcl_measurement} for the full measurement).

\begin{figure*}
    \centering
    \includegraphics[width=13cm]{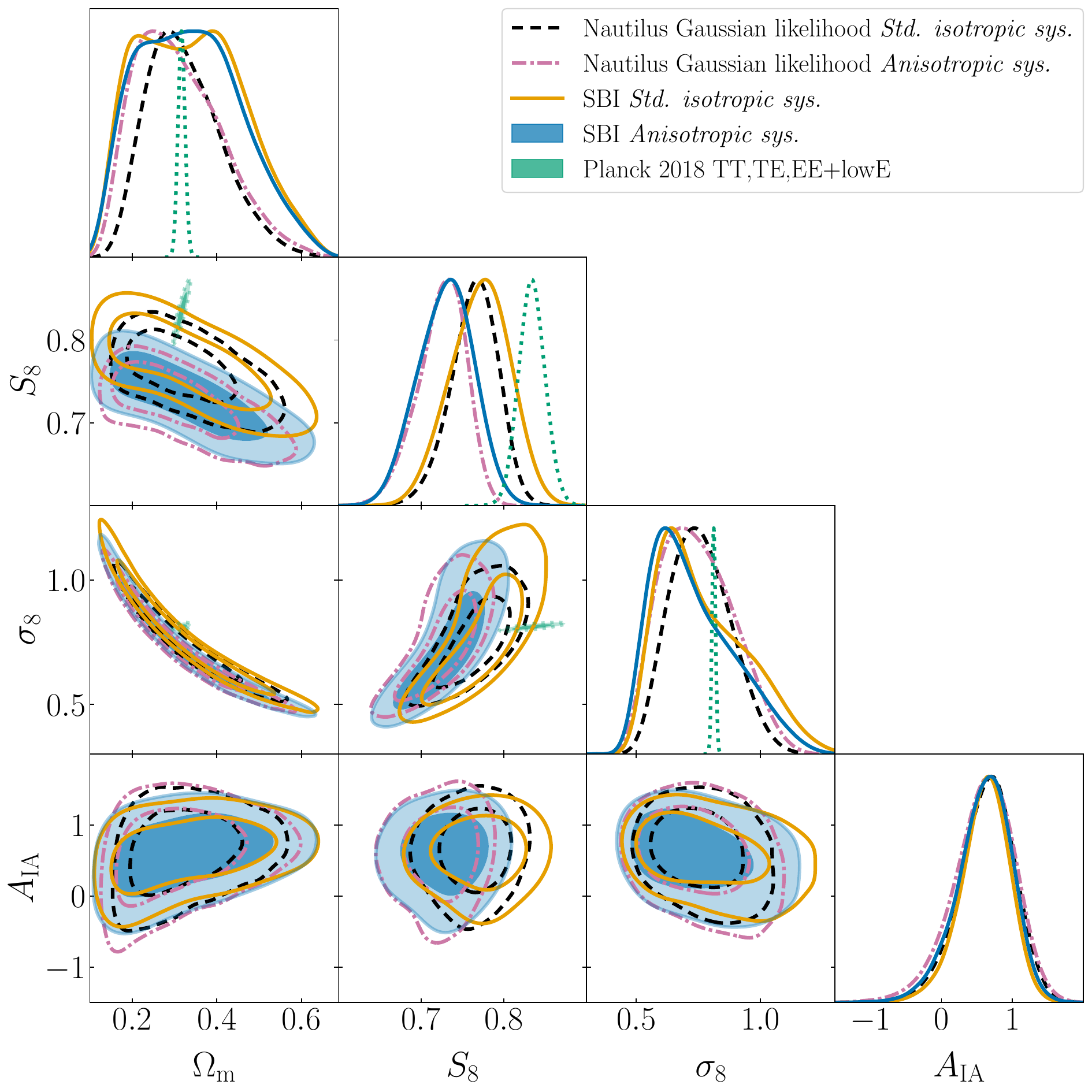}
    \caption{Posterior contours of the main constrained cosmological parameters from the {KiDS-SBI} analysis of the KiDS-1000 cosmic shear data assuming the anisotropic systematics model (in blue), which incorporates additional systematics such as variable depth and shear biases, compared against posterior contours from other analyses. In pink, we show the contours for the equivalent analysis assuming a Gaussian likelihood sampled with \texttt{nautilus} \citep{lange2023nautilus}. In orange, we show the posterior from the same data while assuming the standard isotropic systematics model, which considers the systematic effects which are typically modelled in standard cosmic shear analyses. In black, we show the posterior sampled with \texttt{nautilus} based on a Gaussian likelihood defined by the standard isotropic systematics model. The green contour shows the posterior from the cosmic microwave background constraints from the TT,TE,EE+lowE modes \citep{planck2020planck}. We note that the Planck contours do not have any marginals in $A_{\mathrm{IA}}$ as the CMB is not sensitive to the intrinsic alingments of galaxies.}
    \label{fig:sbi_results}
\end{figure*}

\begin{table}
    \centering
    \caption{Main inferred cosmological parameters varied within the anisotropic systematics model from the KiDS-1000 gold sample.}
    \begin{tabular}{ccc}
        \hline
        Parameter             & Marginal $\pm 1 \sigma$      & MAP $\pm$ PJ-HPD          \\ \hline \\[-0.25cm]
        $S_{8}$                  & $0.731\pm 0.033$    & $0.743^{+0.015}_{-0.051}$ \\[0.15cm]
        $\sigma_{8}$              & $0.73^{+0.10}_{-0.21}$     & $0.72^{+0.09}_{-0.20}$ \\[0.15cm]
        $\Omega_{\mathrm{m}}$     & $0.337^{+0.097}_{-0.150}$       & $0.323^{+0.220}_{-0.060}$ \\[0.1cm]
        \hline
    \end{tabular}
    \tablefoot{The second column shows the marginal as well as the upper and lower 68$\%$ confidence intervals, i.e. $1\sigma$, of the marginals. The third column shows the multivariate maximum a posteriori (MAP), and the uncertainties are defined as the upper and lower 68$\%$ confidence intervals, i.e. $1\sigma$, given by the projected joint highest posterior density, PJ-HPD \citep{robert2007bayesian, joachimi2021kids}.}
    \label{tab:sbi_fiducial_results}
\end{table}

Our main results are based on the likelihood learnt from the anisotropic systematics model in {KiDS-SBI} which incorporates variable depth and additional shear biases when compared to previous KiDS analyses (for example \citealt{asgari2021kids}). The marginalised posterior and the constraints on the main parameters are shown in Fig.~\ref{fig:sbi_results} and Table~\ref{tab:sbi_fiducial_results}, respectively (see Appendix~\ref{appendix:full_posteriors} for the full posteriors and all parameter best estimates). We report the maximum a posteriori (MAP) $\pm$ the projected joint highest posterior density (PJ-HPD; \citealt{robert2007bayesian, joachimi2021kids}) of $0.743^{+0.015}_{-0.051}$, and a mean marginal with 68$\%$ confidence intervals of $0.731\pm 0.033$. We note that Fig.~\ref{fig:sbi_results} also shows that the SBI significantly broadens the posterior in $\Omega_{\mathrm{m}}$ with respect to a Gaussian likelihood analysis, while the mean value of $\Omega_{\mathrm{m}}$ from SBI is highly consistent with the value measured from the early Universe with Planck 2018 TT,TE,EE+lowE \citep{planck2020planck}.

As shown in Fig.~\ref{fig:sbi_goodness_of_fit}, we find that the MAP best-fit cosmology given the anisotropic systematics model provides a good fit to the data with a probability-to-exceed (PTE) of 0.42. Additionally, we plot the line of best fit at the MAP cosmology in Fig.~\ref{fig:pcl_measurement} and find that it is in good agreement with the measured data vector and the analytical theory at the MAP cosmology.

In addition to our main results based on the anisotropic systematics model, we re-analyse the KiDS-1000 gold sample data with the SBI pipeline based on the standard isotropic systematics model which makes similar assumptions to previous KiDS-1000 analyses and does not consider the effects of variable depth and some shear biases (see Sect.~\ref{validation:impact_vd} for details). When considering this model instead, we find a MAP$\pm$PJ-HPD of $S_{8} = 0.780^{+0.020}_{-0.048}$, and a mean marginal with 68$\%$ confidence intervals of $0.772^{+0.038}_{-0.032}$, while the fit to the data at the MAP is similarly good with a $\chi^2$ that gives a PTE of $0.52$. With the standard isotropic systematics model, the marginal in $S_8$ is shifted by $0.9 \sigma$ upwards. This is broadly consistent with the expectation from the validation discussed in Sect.~\ref{validation:impact_vd}, specifically that the extra systematics in the anisotropic systematics model can bias the inference when the data is noisy despite the bias in the signal itself being negligible at the precision of KiDS-1000. This is driven by the uncertainty modelling being considerably different in both models (see Sect.~\ref{validation:impact_vd}): the shape noise model is different, variable depth changes the correlation functions due to the anisotropic selection, and the inclusion of additional shear biases adds statistical noise from 20 more parameters. Consequently, both models respond differently to the noise measured in the KiDS-1000 data.

We also conduct a Bayesian model comparison between the two models by evaluating the Bayesian evidence, $Z$, for each posterior using \texttt{nautilus} \citep{lange2023nautilus}. We find that for the anisotropic systematics model, $\mathrm{ln}(Z) = -14.18$, while for the standard isotropic systematics model, $\mathrm{ln}(Z) = -15.07$. Hence, assuming equal prior probability of both models, the log of the Bayes factor, $|\mathrm{ln}(B)| = 1.11$ which implies no significant preference for either model when compared to the other.

As also shown in Fig.~\ref{fig:sbi_results}, we find that when re-analysing the KiDS-1000 data assuming a Gaussian likelihood instead, we find results consistent with the testing conducted on mocks (see Fig.~\ref{fig:sbi_mcmc_v_sbi}): the Gaussian likelihood assumption causes $S_{8}$ to be overconstrained by ${\sim}10\%$ with respect to the SBI irrespective of how the observational systematics are modelled as it neglects to consider the cosmology dependence of cosmic variance. Additionally, the posterior mean marginal in $S_{8}$ assuming the Gaussian likelihood shifts from $0.764^{+0.031}_{-0.025} $ when deriving the sample covariance from the standard isotropic systematics model to a value of $0.725^{+0.034}_{-0.023}$ when assuming the anisotropic systematics model in the covariance instead. We note that both assume the same modelling in the signal described in Appendix~\ref{appendix:signal}.

We highlight that the Gaussian likelihood analysis based on the standard isotropic systematics model agrees  well with the constraints from \citet{asgari2021kids} where $S_{8} = 0.758^{+0.017}_{-0.026}$. We note that the constraints in our work are broader despite making consistent modelling assumptions. This is caused by the use of a different MCMC sampler. Here we have employed the \texttt{nautilus} sampler \citep{lange2023nautilus}, while in \citet{asgari2021kids} the \texttt{multinest} sampler \citep{feroz2009multinest} was used. It has been found that \texttt{multinest} can be overconfident by approximately ${\sim}10\%$ \citep{lemos2023robust} which we also find in our analysis and show in Appendix~\ref{appendix:sampler_comparison}. When accounting for the difference in sampler used, we find that the Gaussian likelihood constraints from the standard isotropic systematics model agree exceptionally well with the results from \citet{asgari2021kids}.

Likewise the inferred $S_{8}$ value when assuming a Gaussian likelihood derived from the anisotropic systematics model is consistent with the result from \citet{loureiro2021kids} where the mean marginal is measured to be $0.742^{+0.034}_{-0.023}$. Despite both analyses being based on pseudo-Cls, the one presented in this work finds a slightly lower value of $S_{8}$ (${\sim}0.27\sigma$) which can be driven by the additional signal and uncertainty induced by the PSF variations into the anisotropic systematics model (see Appendix~\ref{method:consistency:var_depth}) which was not considered in the modelling of \citet{loureiro2021kids}.

\begin{figure}
    \centering
    \includegraphics[width=8cm]{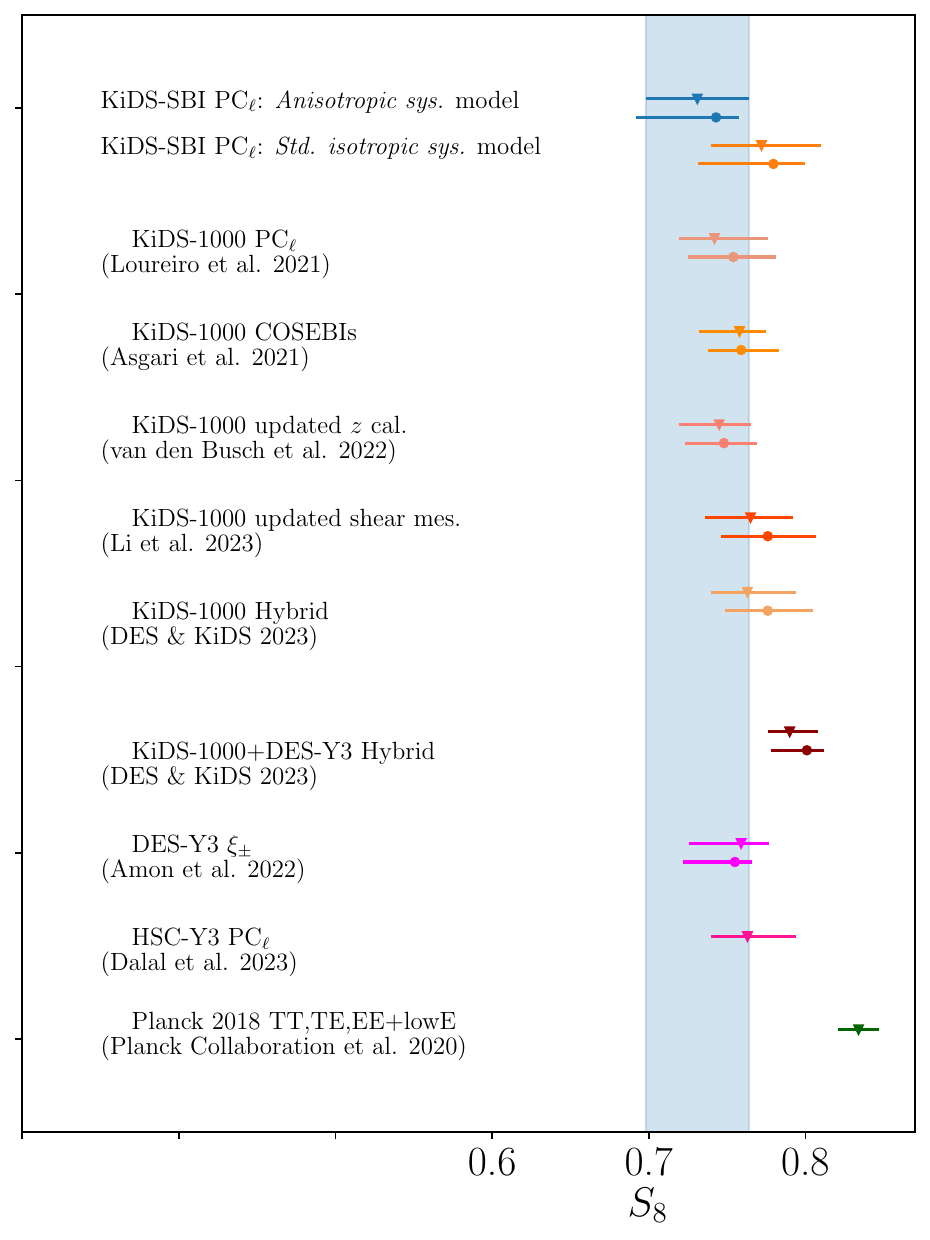}
    \caption{Comparison between different constraints of the growth of structure parameter, $S_{8}$, from cosmic shear measurements and from the cosmic microwave background. The circular points indicate MAP+PJ-HPD constraints and the triangular points reflect estimates from the marginal posterior mean and 68$\%$ confidence intervals. The blue dashed regions mark the $1\sigma$ region from the {KiDS-SBI} constraint based on the anisotropic systematics model shown in this work. }
    \label{fig:whisker}
\end{figure}

When comparing the constraints from the anisotropic systematics model SBI to previous cosmic shear analyses in Fig.~\ref{fig:whisker}, we find that our results are in agreement within $1\sigma$ of $S_8$ best-fits from all previous KiDS-1000 analyses \citep{asgari2021kids, loureiro2021kids, busch2022kids, li2023kids}, the HSC-Y3 pseudo-Cl analysis \citep{dalal2023hyper}, and the DES-Y3 two-point analysis \citep{amon2022dark, secco2022dark}. The one exception is the anisotropic systematics DES-Y3 analysis \citep{deskids2023} whose marginal from the main hybrid analysis is different by $1.65\sigma$ from our marginal estimate of $S_8$ based on the anisotropic systematics model. However, in the same work, all re-analyses of the KiDS-1000 data alone (with KiDS-like, DES-like, and hybrid analysis) are consistent within less than $1\sigma$ with the best-fit for $S_8$ presented in this work. These discrepancies may be partially attributed to cosmic variance, different modelling choices for certain systematics, and the impact of the cosmology dependence in the learnt likelihood which is included in our SBI analysis, but it is not considered in any other previous analyses.

In fact, as discussed in detail in Sect.~\ref{gaussian_likelihood}, since the uncertainty is cosmology-dependent and scales with $S_8$ in {KiDS-SBI}, the SBI infers a larger uncertainty in $S_{8}$ on top of the added uncertainty from variable depth and shear biases. Additionally, variable depth and the PSF shear bias can impact the likelihood to significantly re-weight the data vector despite their impact on the signal being negligible. This may explain why the best estimate for $S_{8}$ from the anisotropic systematics model is about $0.5\sigma$ lower than previous standard analyses of KiDS-1000.

When comparing the results from the SBI with the anisotropic systematics model to early-Universe probes of large-scale structure, such as the results from the Planck collaboration shown in Fig.~\ref{fig:sbi_results}, we find that the main {KiDS-SBI} $S_{8}$ marginal is in $2.9 \sigma$ tension with the best estimate of $S_{8} = 0.834 \pm 0.016$ from Planck 2018 TT,TE,EE+lowE \citep{planck2020planck}.

\section{Conclusions}\label{conclusions}
We performed a novel simulation-based inference (SBI) analysis, \textit{KiDS-SBI}, of the cosmic shear two-point statistics as measured by the Kilo-Degree Survey's fourth data release (KiDS-1000; \citealt{kuijken2019the}).

We present a new suite of stochastic forward simulations of cosmic shear observables within {KiDS-SBI,} which take into consideration all the systematic effects which are typically modelled in a weak lensing analysis plus additional systematics that have the potential to significantly bias the signal. At the same time, {KiDS-SBI} introduces considerable improvements in accuracy and performance with respect to similar simulation environments in previous works \citep{joachimi2021kids}: non-Limber projections of angular power spectra (with the new \texttt{Levin} module), higher resolution along the line of sight, inclusion of field-level intrinsic alignments, consideration of variable depth and anisotropic shear biases, and significant computational performance improvements.

On the basis of these forward simulations, we implemented the first simulation-based inference analysis of large-scale structure observables as measured by KiDS-1000 with the same level of complexity as other standard analyses in the field. This has enabled us to fully propagate all uncertainties from the data vector all the way to the inferred cosmological parameters in a Bayesian way. The analysis is based on neural density estimators which learn the effective likelihood through the mapping between the model parameters and the forward-simulated data vectors. We conducted an inference of the cosmological parameters in flat $\Lambda$CDM from cosmic shear two-point statistics, specifically, pseudo-Cls. We assumed a forward model consistent with previous KiDS-1000 cosmic shear analyses \citep{asgari2021kids, loureiro2021kids}, which additionally includes variable depth and anisotropic shear biases: the anisotropic systematics model. Variable depth accounts for the anisotropy in the galaxy selection function in KiDS-1000 and how it impacts the measured galaxy densities, shapes, and redshifts. The inclusion of anisotropic shear biases models the angular variation of the residual bias in the galaxy shapes due to variations in the point-spread function of the telescope from the mean model used in the shape calibration.

To assess the quality of any best fit obtained from {KiDS-SBI}, we also introduced a Bayesian goodness-of-fit measurement based on \citet{gelman1996posterior}. From a small number of realisations of the forward model, we characterised a distribution of $\chi^2$ values for the given model. By determining where the $\chi^2$ of the best fit on the KiDS-1000 data lies, we evaluated the quality of the fit and whether the data stand as a feasible realisation of the forward model. On this basis, we were able to conclude that the KiDS-1000 data is fitted well by the anisotropic systematics model at the maximum a posteriori cosmology (with a probability-to-exceed of 0.42).

Through the testing conducted in this work and in a previous paper \citep{lin2022a}, we determine that {KiDS-SBI} recovers accurate and robust posteriors of the 12D space of cosmological and astrophysical parameters. At the same time, we find that the likelihood learnt from {KiDS-SBI} forward simulations is well approximated by a Gaussian distribution at a fixed cosmology. However, we determined that the uncertainty on the cosmic shear signal measurably varies with the assumed cosmological parameters, which are driven by the cosmic variance scaling with the underlying value of $S_{8}$,  as  noted in previous works \citep{eifler2009dependence}. Consequently, we find that the uncertainty of the two-point statistics in KiDS-1000 is measurably cosmology-dependent which increases the nominal uncertainty of $S_{8}$ by ${\sim}10\%$. Hence, the standard assumption of the covariance of a Gaussian likelihood being fixed for all parameter points in the prior space does not necessarily hold. Therefore, such an assumption may not be valid for current Stage-III galaxy surveys and this issue may be also be relevant for upcoming stage-IV galaxy surveys, such as \emph{Euclid} \citep{laureijs2011euclid} or Rubin \citep{lsst2009lsst}.  Addressing this assumption further will  improve the precision of cosmic shear measurements.

Given the anisotropic systematics model in {KiDS-SBI}, we obtain a mean marginal with 68$\%$ confidence intervals of $0.731\pm 0.033$. These constraints are in agreement within $1\sigma$ with previous cosmic shear analyses of KiDS-1000 \citep{asgari2021kids, loureiro2021kids, busch2022kids, li2023kids}, HSC-Y3 \citep{li2023hyper, dalal2023hyper}, and DES-Y3 \citep{amon2022dark, secco2022dark}. The tension in $S_{8}$ from the KiDS-1000 analysis, in this work and measurements from the cosmic microwave background from \emph{Planck} 2018 TT,TE,EE+lowE \citep{planck2020planck}, is at $2.9\sigma$, which is similar to the $3.0\sigma$ tension found in an equivalent Gaussian likelihood analysis of KiDS-1000 \citep{asgari2021kids}. Alternatively, this is also close to the $2.8\sigma$ tension found the analysis of KiDS-1000, which incorporated variable depth in the Gaussian likelihood \citep{loureiro2021kids}, and the $3.1\sigma$ tension found in the KiDS-1000 beyond $\Lambda$CDM \citep{troester2021kids}. This establishes that the $\sigma_{8}$-tension between late-Universe and early-Universe probes of large-scale structure still persists even when we drops the assumption of a Gaussian likelihood for the cosmic shear signal and considers additional contaminant systematics.

To evaluate and isolate the impact of the inclusion of variable depth and shear biases in the anisotropic systematics model, we reproduce the {KiDS-SBI} analysis with a second model which makes the same assumptions, but does not consider these specific systematic effects: the standard isotropic systematics model (which is more in line with previous KiDS-1000 analyses, such as \citealt{asgari2021kids}). We find that the inclusion of both systematics in the forward model increases the $S_{8}$ marginal by $0.9\sigma$ or $5\%$ and decreases the marginal $1\sigma$ confidence intervals by $6\%$. We find that this is not driven by these systematics' impact on the signal but, rather, on their impact on the uncertainty model. The variable depth alone appears to impact $S_{8}$ by approx. $0.38 \sigma$ when comparing the Gaussian likelihood results from \citet{loureiro2021kids} with the constraints from the standard isotropic systematics model. The variations in the PSF modelling residuals appear to account for approx. $0.36 \sigma$ in the shift in $S_{8}$ when comparing the Gaussian likelihood results from from the anisotropic systematics model with the constraints from \citet{loureiro2021kids}. However, we note that the PSF variations and the variable depth are known to be correlated as both are influenced by some of the same observational systematics, such as atmospheric seeing or thermal effects in the telescope optics; thus, it may not be physically sound to consider these effects separately.

In any case, this reflects the fact that observational anisotropies, such as variable depth and PSF shear variation bias, can significantly alter the uncertainty modelling despite having a negligible impact on the shear signal at the precision levels of KiDS-1000. The KiDS-1000 data appears to be consistent with the presence of these effects since the anisotropic systematics model provides a good fit to the data, but when examining the Bayesian evidence, we do not find a clear preference between the two models by the data. Nevertheless, both systematic effects have been independently measured in the KiDS-1000 weak lensing data \citep{joachimi2021kids, giblin2021kids}.

This work stresses the importance of testing the common assumption of Gaussian likelihoods in realistic conditions. At the same time, we also highlight the potential of simulation-based inference to rigorously perform Bayesian uncertainty propagation even when the models of the uncertainty of the signal or the systematics are analytically intractable. The approach using SBI with likelihood-based neural density estimation comes with the added advantage that it does not require the expensive re-evaluation of the posterior distribution if the input measurements or priors are updated, as would be the case for a standard MCMC analysis. Since the likelihood for a given forward model is learnt independently of the input data and the priors, SBI can significantly improve the efficiency of inference analysis where the forward model is not changed between measurements. All these advantages showcase the potential of simulation-based inference becoming the method of choice in future parameter inference and model testing analyses in observational cosmology.

We conclude that SBI is a powerful tool that may be able to tackle many of the physical and statistical challenges faced by future galaxy surveys, as their observations become more precise, less limited by statistical noise, and more limited by the accurate modelling of systematics. To address this need, future works should expand upon the forward model in {KiDS-SBI} to also include other probes of large-scale structure such as galaxy clustering and galaxy-galaxy lensing, while incorporating the relevant modelling for the relevant systematic effects, such as magnification bias, non-linear galaxy bias, and field-level baryonic feedback.

\begin{acknowledgements}
We thank Stephen Feeney, Justin Alsing, Hendrik Hildebrandt, Marika Asgari and Alan Heavens for helpful discussions. We are grateful to the referee for the constructive comments on the manuscript. MvWK thanks the Science and Technology Facilities Council for support. KL thanks the UCL Cosmoparticle Initiative. NT and BJ are supported by the ERC-selected UKRI Frontier Research Grant EP/Y03015X/1 and the UK Space Agency grants ST/W002574/1 and ST/X00208X/1. BJ is also supported by STFC Consolidated Grant ST/V000780/1. AL acknowledges support from the research project grant `Understanding the Dynamic Universe' funded by the Knut and Alice Wallenberg Foundation under Dnr KAW 2018.0067. RR is supported by the European Research Council (Grant No. 770935). AHW is supported by the Deutsches Zentrum für Luft- und Raumfahrt (DLR), made possible by the Bundesministerium für Wirtschaft und Klimaschutz, and acknowledges funding from the German Science Foundation DFG, via the Collaborative Research Center SFB1491 'Cosmic Interacting Matters - From Source to Signal'. This work used facilities provided by the UCL Cosmoparticle Initiative. We thank Dr Edd Edmonson for technical support at the UCL HPC facilities. The results in this paper are based on observations made with ESO Telescopes at the La Silla Paranal Observatory under programme IDs: 088.D-4013, 092.A-0176, 092.D-0370, 094.D-0417, 177.A-3016, 177.A-3017, 177.A-3018 and 179.A-2004, and on data products produced by the KiDS consortium. The KiDS production team acknowledges support from: Deutsche Forschungsgemeinschaft, ERC, NOVA and NWO-M grants; Target; the University of Padova, and the University Federico II (Naples). Contributions to the data processing for VIKING were made by the VISTA Data Flow System at CASU, Cambridge and WFAU, Edinburgh. The work presented in this paper made use of the following software packages: \texttt{Numpy} \citep{harris2020array}, \texttt{SciPy} \citep{virtanen2020scipy}, \texttt{AstroPy} \citep{astropy2018astropy}, \texttt{HealPy} \citep{zonca2019healpy}, \texttt{GetDist} \citep{antony2019getdist}, \texttt{Jupyter Lab} \citep{kluyver2016jupyter}, \texttt{numba} \citep{lam2015numba}, and \texttt{Matplotlib} \citep{hunter2007matplotlib}.
\end{acknowledgements}

\bibliographystyle{aa}
\bibliography{bibliography}

\begin{thebibliography}{191}
\expandafter\ifx\csname natexlab\endcsname\relax\def\natexlab#1{#1}\fi

\bibitem[{{Abbott} {et~al.}(2022){Abbott}, {Aguena}, {Alarcon}, {Allam},
  {Alves}, {Amon}, {Andrade-Oliveira}, {Annis}, {Avila}, {Bacon}, {Baxter},
  {Bechtol}, {Becker}, {Bernstein}, {Bhargava}, {Birrer}, {Blazek},
  {Brandao-Souza}, {Bridle}, {Brooks}, {Buckley-Geer}, {Burke}, {Camacho},
  {Campos}, {Carnero Rosell}, {Carrasco Kind}, {Carretero}, {Castander},
  {Cawthon}, {Chang}, {Chen}, {Chen}, {Choi}, {Conselice}, {Cordero},
  {Costanzi}, {Crocce}, {da Costa}, {da Silva Pereira}, {Davis}, {Davis}, {De
  Vicente}, {DeRose}, {Desai}, {Di Valentino}, {Diehl}, {Dietrich}, {Dodelson},
  {Doel}, {Doux}, {Drlica-Wagner}, {Eckert}, {Eifler}, {Elsner}, {Elvin-Poole},
  {Everett}, {Evrard}, {Fang}, {Farahi}, {Fernandez}, {Ferrero}, {Fert{\'e}},
  {Fosalba}, {Friedrich}, {Frieman}, {Garc{\'\i}a-Bellido}, {Gatti},
  {Gaztanaga}, {Gerdes}, {Giannantonio}, {Giannini}, {Gruen}, {Gruendl},
  {Gschwend}, {Gutierrez}, {Harrison}, {Hartley}, {Herner}, {Hinton},
  {Hollowood}, {Honscheid}, {Hoyle}, {Huff}, {Huterer}, {Jain}, {James},
  {Jarvis}, {Jeffrey}, {Jeltema}, {Kovacs}, {Krause}, {Kron}, {Kuehn},
  {Kuropatkin}, {Lahav}, {Leget}, {Lemos}, {Liddle}, {Lidman}, {Lima}, {Lin},
  {MacCrann}, {Maia}, {Marshall}, {Martini}, {McCullough}, {Melchior},
  {Mena-Fern{\'a}ndez}, {Menanteau}, {Miquel}, {Mohr}, {Morgan}, {Muir},
  {Myles}, {Nadathur}, {Navarro-Alsina}, {Nichol}, {Ogando}, {Omori},
  {Palmese}, {Pandey}, {Park}, {Paz-Chinch{\'o}n}, {Petravick}, {Pieres},
  {Plazas Malag{\'o}n}, {Porredon}, {Prat}, {Raveri}, {Rodriguez-Monroy},
  {Rollins}, {Romer}, {Roodman}, {Rosenfeld}, {Ross}, {Rykoff}, {Samuroff},
  {S{\'a}nchez}, {Sanchez}, {Sanchez}, {Sanchez Cid}, {Scarpine}, {Schubnell},
  {Scolnic}, {Secco}, {Serrano}, {Sevilla-Noarbe}, {Sheldon}, {Shin}, {Smith},
  {Soares-Santos}, {Suchyta}, {Swanson}, {Tabbutt}, {Tarle}, {Thomas}, {To},
  {Troja}, {Troxel}, {Tucker}, {Tutusaus}, {Varga}, {Walker}, {Weaverdyck},
  {Wechsler}, {Weller}, {Yanny}, {Yin}, {Zhang}, {Zuntz}, \& {DES
  Collaboration}}]{abbott2022dark}
{Abbott}, T.~M.~C., {Aguena}, M., {Alarcon}, A., {et~al.} 2022, \prd, 105,
  023520

\bibitem[{{Abramo} {et~al.}(2022){Abramo}, {Dinarte Ferri}, {Tashiro}, \&
  {Loureiro}}]{abramo2022fisher}
{Abramo}, L.~R., {Dinarte Ferri}, J.~V., {Tashiro}, I.~L., \& {Loureiro}, A.
  2022, \jcap, 2022, 073

\bibitem[{{Abramo} {et~al.}(2016){Abramo}, {Secco}, \&
  {Loureiro}}]{abramo2016fourier}
{Abramo}, L.~R., {Secco}, L.~F., \& {Loureiro}, A. 2016, \mnras, 455, 3871

\bibitem[{{Aihara} {et~al.}(2022){Aihara}, {AlSayyad}, {Ando}, {Armstrong},
  {Bosch}, {Egami}, {Furusawa}, {Furusawa}, {Harasawa}, {Harikane}, {Hsieh},
  {Ikeda}, {Ito}, {Iwata}, {Kodama}, {Koike}, {Kokubo}, {Komiyama}, {Li},
  {Liang}, {Lin}, {Lupton}, {Lust}, {MacArthur}, {Mawatari}, {Mineo},
  {Miyatake}, {Miyazaki}, {More}, {Morishima}, {Murayama}, {Nakajima},
  {Nakata}, {Nishizawa}, {Oguri}, {Okabe}, {Okura}, {Ono}, {Osato}, {Ouchi},
  {Pan}, {Plazas Malag{\'o}n}, {Price}, {Reed}, {Rykoff}, {Shibuya},
  {Simunovic}, {Strauss}, {Sugimori}, {Suto}, {Suzuki}, {Takada}, {Takagi},
  {Takata}, {Takita}, {Tanaka}, {Tang}, {Taranu}, {Terai}, {Toba}, {Turner},
  {Uchiyama}, {Vijarnwannaluk}, {Waters}, {Yamada}, {Yamamoto}, \&
  {Yamashita}}]{aihara2022third}
{Aihara}, H., {AlSayyad}, Y., {Ando}, M., {et~al.} 2022, \pasj, 74, 247

\bibitem[{{Alonso} {et~al.}(2019){Alonso}, {Sanchez}, {Slosar}, \& {LSST Dark
  Energy Science Collaboration}}]{alonso2019a}
{Alonso}, D., {Sanchez}, J., {Slosar}, A., \& {LSST Dark Energy Science
  Collaboration}. 2019, \mnras, 484, 4127

\bibitem[{Alsing {et~al.}(2019)Alsing, Charnock, Feeney, \&
  Wandelt}]{alsing2019fast}
Alsing, J., Charnock, T., Feeney, S., \& Wandelt, B. 2019, Monthly Notices of
  the Royal Astronomical Society, 488, 4440

\bibitem[{{Alsing} \& {Wandelt}(2018)}]{alsing2018generalized}
{Alsing}, J. \& {Wandelt}, B. 2018, \mnras, 476, L60

\bibitem[{Alsing {et~al.}(2018)Alsing, Wandelt, \& Feeney}]{alsing2018massive}
Alsing, J., Wandelt, B., \& Feeney, S. 2018, Monthly Notices of the Royal
  Astronomical Society, 477, 2874

\bibitem[{{Amon} \& {Efstathiou}(2022)}]{amon2022a}
{Amon}, A. \& {Efstathiou}, G. 2022, \mnras, 516, 5355

\bibitem[{{Amon} {et~al.}(2022){Amon}, {Gruen}, {Troxel}, {MacCrann},
  {Dodelson}, {Choi}, {Doux}, {Secco}, {Samuroff}, {Krause}, {Cordero},
  {Myles}, {DeRose}, {Wechsler}, {Gatti}, {Navarro-Alsina}, {Bernstein},
  {Jain}, {Blazek}, {Alarcon}, {Fert{\'e}}, {Lemos}, {Raveri}, {Campos},
  {Prat}, {S{\'a}nchez}, {Jarvis}, {Alves}, {Andrade-Oliveira}, {Baxter},
  {Bechtol}, {Becker}, {Bridle}, {Camacho}, {Carnero Rosell}, {Carrasco Kind},
  {Cawthon}, {Chang}, {Chen}, {Chintalapati}, {Crocce}, {Davis}, {Diehl},
  {Drlica-Wagner}, {Eckert}, {Eifler}, {Elvin-Poole}, {Everett}, {Fang},
  {Fosalba}, {Friedrich}, {Gaztanaga}, {Giannini}, {Gruendl}, {Harrison},
  {Hartley}, {Herner}, {Huang}, {Huff}, {Huterer}, {Kuropatkin}, {Leget},
  {Liddle}, {McCullough}, {Muir}, {Pandey}, {Park}, {Porredon}, {Refregier},
  {Rollins}, {Roodman}, {Rosenfeld}, {Ross}, {Rykoff}, {Sanchez},
  {Sevilla-Noarbe}, {Sheldon}, {Shin}, {Troja}, {Tutusaus}, {Tutusaus},
  {Varga}, {Weaverdyck}, {Yanny}, {Yin}, {Zhang}, {Zuntz}, {Aguena}, {Allam},
  {Annis}, {Bacon}, {Bertin}, {Bhargava}, {Brooks}, {Buckley-Geer}, {Burke},
  {Carretero}, {Costanzi}, {da Costa}, {Pereira}, {De Vicente}, {Desai},
  {Dietrich}, {Doel}, {Ferrero}, {Flaugher}, {Frieman}, {Garc{\'\i}a-Bellido},
  {Gaztanaga}, {Gerdes}, {Giannantonio}, {Gschwend}, {Gutierrez}, {Hinton},
  {Hollowood}, {Honscheid}, {Hoyle}, {James}, {Kron}, {Kuehn}, {Lahav}, {Lima},
  {Lin}, {Maia}, {Marshall}, {Martini}, {Melchior}, {Menanteau}, {Miquel},
  {Mohr}, {Morgan}, {Ogando}, {Palmese}, {Paz-Chinch{\'o}n}, {Petravick},
  {Pieres}, {Romer}, {Sanchez}, {Scarpine}, {Schubnell}, {Serrano}, {Smith},
  {Soares-Santos}, {Tarle}, {Thomas}, {To}, {Weller}, \& {DES
  Collaboration}}]{amon2022dark}
{Amon}, A., {Gruen}, D., {Troxel}, M.~A., {et~al.} 2022, \prd, 105, 023514

\bibitem[{{Asgari} {et~al.}(2019){Asgari}, {Heymans}, {Hildebrandt}, {Miller},
  {Schneider}, {Amon}, {Choi}, {Erben}, {Georgiou}, {Harnois-Deraps}, \&
  {Kuijken}}]{asgari2019consistent}
{Asgari}, M., {Heymans}, C., {Hildebrandt}, H., {et~al.} 2019, \aap, 624, A134

\bibitem[{{Asgari} {et~al.}(2021){Asgari}, {Lin}, {Joachimi}, {Giblin},
  {Heymans}, {Hildebrandt}, {Kannawadi}, {St{\"o}lzner}, {Tr{\"o}ster}, {van
  den Busch}, {Wright}, {Bilicki}, {Blake}, {de Jong}, {Dvornik}, {Erben},
  {Getman}, {Hoekstra}, {K{\"o}hlinger}, {Kuijken}, {Miller}, {Radovich},
  {Schneider}, {Shan}, \& {Valentijn}}]{asgari2021kids}
{Asgari}, M., {Lin}, C.-A., {Joachimi}, B., {et~al.} 2021, \aap, 645, A104

\bibitem[{{Astropy Collaboration} {et~al.}(2018){Astropy Collaboration},
  {Price-Whelan}, {Sip{\H{o}}cz}, {G{\"u}nther}, {Lim}, {Crawford}, {Conseil},
  {Shupe}, {Craig}, {Dencheva}, {Ginsburg}, {VanderPlas}, {Bradley},
  {P{\'e}rez-Su{\'a}rez}, {de Val-Borro}, {Aldcroft}, {Cruz}, {Robitaille},
  {Tollerud}, {Ardelean}, {Babej}, {Bach}, {Bachetti}, {Bakanov}, {Bamford},
  {Barentsen}, {Barmby}, {Baumbach}, {Berry}, {Biscani}, {Boquien}, {Bostroem},
  {Bouma}, {Brammer}, {Bray}, {Breytenbach}, {Buddelmeijer}, {Burke},
  {Calderone}, {Cano Rodr{\'\i}guez}, {Cara}, {Cardoso}, {Cheedella}, {Copin},
  {Corrales}, {Crichton}, {D'Avella}, {Deil}, {Depagne}, {Dietrich}, {Donath},
  {Droettboom}, {Earl}, {Erben}, {Fabbro}, {Ferreira}, {Finethy}, {Fox},
  {Garrison}, {Gibbons}, {Goldstein}, {Gommers}, {Greco}, {Greenfield},
  {Groener}, {Grollier}, {Hagen}, {Hirst}, {Homeier}, {Horton}, {Hosseinzadeh},
  {Hu}, {Hunkeler}, {Ivezi{\'c}}, {Jain}, {Jenness}, {Kanarek}, {Kendrew},
  {Kern}, {Kerzendorf}, {Khvalko}, {King}, {Kirkby}, {Kulkarni}, {Kumar},
  {Lee}, {Lenz}, {Littlefair}, {Ma}, {Macleod}, {Mastropietro}, {McCully},
  {Montagnac}, {Morris}, {Mueller}, {Mumford}, {Muna}, {Murphy}, {Nelson},
  {Nguyen}, {Ninan}, {N{\"o}the}, {Ogaz}, {Oh}, {Parejko}, {Parley}, {Pascual},
  {Patil}, {Patil}, {Plunkett}, {Prochaska}, {Rastogi}, {Reddy Janga},
  {Sabater}, {Sakurikar}, {Seifert}, {Sherbert}, {Sherwood-Taylor}, {Shih},
  {Sick}, {Silbiger}, {Singanamalla}, {Singer}, {Sladen}, {Sooley},
  {Sornarajah}, {Streicher}, {Teuben}, {Thomas}, {Tremblay}, {Turner},
  {Terr{\'o}n}, {van Kerkwijk}, {de la Vega}, {Watkins}, {Weaver}, {Whitmore},
  {Woillez}, {Zabalza}, \& {Astropy Contributors}}]{astropy2018astropy}
{Astropy Collaboration}, {Price-Whelan}, A.~M., {Sip{\H{o}}cz}, B.~M., {et~al.}
  2018, \aj, 156, 123

\bibitem[{{Baleato Lizancos} \& {White}(2023)}]{baleato2023the}
{Baleato Lizancos}, A. \& {White}, M. 2023, \jcap, 2023, 044

\bibitem[{{Bartelmann} \& {Schneider}(2001)}]{bartelmann2001weak}
{Bartelmann}, M. \& {Schneider}, P. 2001, \physrep, 340, 291

\bibitem[{{Beaumont}(2019)}]{beaumont2019approximate}
{Beaumont}, M.~A. 2019, Annual Review of Statistics and Its Application, 6, 379

\bibitem[{{Becker} {et~al.}(2016){Becker}, {Troxel}, {MacCrann}, {Krause},
  {Eifler}, {Friedrich}, {Nicola}, {Refregier}, {Amara}, {Bacon}, {Bernstein},
  {Bonnett}, {Bridle}, {Busha}, {Chang}, {Dodelson}, {Erickson}, {Evrard},
  {Frieman}, {Gaztanaga}, {Gruen}, {Hartley}, {Jain}, {Jarvis}, {Kacprzak},
  {Kirk}, {Kravtsov}, {Leistedt}, {Peiris}, {Rykoff}, {Sabiu}, {S{\'a}nchez},
  {Seo}, {Sheldon}, {Wechsler}, {Zuntz}, {Abbott}, {Abdalla}, {Allam},
  {Armstrong}, {Banerji}, {Bauer}, {Benoit-L{\'e}vy}, {Bertin}, {Brooks},
  {Buckley-Geer}, {Burke}, {Capozzi}, {Carnero Rosell}, {Carrasco Kind},
  {Carretero}, {Castander}, {Crocce}, {Cunha}, {D'Andrea}, {da Costa}, {DePoy},
  {Desai}, {Diehl}, {Dietrich}, {Doel}, {Fausti Neto}, {Fernandez}, {Finley},
  {Flaugher}, {Fosalba}, {Gerdes}, {Gruendl}, {Gutierrez}, {Honscheid},
  {James}, {Kuehn}, {Kuropatkin}, {Lahav}, {Li}, {Lima}, {Maia}, {March},
  {Martini}, {Melchior}, {Miller}, {Miquel}, {Mohr}, {Nichol}, {Nord},
  {Ogando}, {Plazas}, {Reil}, {Romer}, {Roodman}, {Sako}, {Sanchez},
  {Scarpine}, {Schubnell}, {Sevilla-Noarbe}, {Smith}, {Soares-Santos},
  {Sobreira}, {Suchyta}, {Swanson}, {Tarle}, {Thaler}, {Thomas}, {Vikram},
  {Walker}, \& {Dark Energy Survey Collaboration}}]{becker2016cosmic}
{Becker}, M.~R., {Troxel}, M.~A., {MacCrann}, N., {et~al.} 2016, \prd, 94,
  022002

\bibitem[{{Ben{\'\i}tez}(2000)}]{benitez2000bayesian}
{Ben{\'\i}tez}, N. 2000, \apj, 536, 571

\bibitem[{{Bennett} {et~al.}(2003){Bennett}, {Halpern}, {Hinshaw}, {Jarosik},
  {Kogut}, {Limon}, {Meyer}, {Page}, {Spergel}, {Tucker}, {Wollack}, {Wright},
  {Barnes}, {Greason}, {Hill}, {Komatsu}, {Nolta}, {Odegard}, {Peiris},
  {Verde}, \& {Weiland}}]{bennett2003first}
{Bennett}, C.~L., {Halpern}, M., {Hinshaw}, G., {et~al.} 2003, \apjs, 148, 1

\bibitem[{{Bertin} \& {Arnouts}(1996)}]{bertin1996}
{Bertin}, E. \& {Arnouts}, S. 1996, \aaps, 117, 393

\bibitem[{Bishop(1994)}]{bishop1994mixture}
Bishop, C.~M. 1994, Mixture density networks (Aston University)

\bibitem[{{Blazek} {et~al.}(2011){Blazek}, {McQuinn}, \&
  {Seljak}}]{blazek2011testing}
{Blazek}, J., {McQuinn}, M., \& {Seljak}, U. 2011, \jcap, 2011, 010

\bibitem[{{Blazek} {et~al.}(2015){Blazek}, {Vlah}, \&
  {Seljak}}]{blazek2015tidal}
{Blazek}, J., {Vlah}, Z., \& {Seljak}, U. 2015, \jcap, 2015, 015

\bibitem[{{Bocquet} {et~al.}(2024){Bocquet}, {Grandis}, {Bleem}, {Klein},
  {Mohr}, {Schrabback}, {Abbott}, {Ade}, {Aguena}, {Alarcon}, {Allam}, {Allen},
  {Alves}, {Amon}, {Anderson}, {Annis}, {Ansarinejad}, {Austermann}, {Avila},
  {Bacon}, {Bayliss}, {Beall}, {Bechtol}, {Becker}, {Bender}, {Benson},
  {Bernstein}, {Bhargava}, {Bianchini}, {Brodwin}, {Brooks}, {Bryant},
  {Campos}, {Canning}, {Carlstrom}, {Carnero Rosell}, {Carrasco Kind},
  {Carretero}, {Castander}, {Cawthon}, {Chang}, {Chang}, {Chaubal}, {Chen},
  {Chiang}, {Choi}, {Chou}, {Citron}, {Corbett Moran}, {Cordero}, {Costanzi},
  {Crawford}, {Crites}, {da Costa}, {Pereira}, {Davis}, {Davis}, {DeRose},
  {Desai}, {de Haan}, {Diehl}, {Dobbs}, {Dodelson}, {Doux}, {Drlica-Wagner},
  {Eckert}, {Elvin-Poole}, {Everett}, {Everett}, {Ferrero}, {Fert{\'e}},
  {Flores}, {Frieman}, {Gallicchio}, {Garc{\'\i}a-Bellido}, {Gatti}, {George},
  {Giannini}, {Gladders}, {Gruen}, {Gruendl}, {Gupta}, {Gutierrez},
  {Halverson}, {Harrison}, {Hartley}, {Herner}, {Hinton}, {Holder},
  {Hollowood}, {Holzapfel}, {Honscheid}, {Hrubes}, {Huang}, {Hubmayr}, {Huff},
  {Huterer}, {Irwin}, {James}, {Jarvis}, {Khullar}, {Kim}, {Knox}, {Kraft},
  {Krause}, {Kuehn}, {Kuropatkin}, {K{\'e}ruzor{\'e}}, {Lahav}, {Lee}, {Leget},
  {Li}, {Lin}, {Lowitz}, {MacCrann}, {Mahler}, {Mantz}, {Marshall},
  {McCullough}, {McDonald}, {McMahon}, {Mena-Fern{\'a}ndez}, {Menanteau},
  {Meyer}, {Miquel}, {Montgomery}, {Myles}, {Natoli}, {Navarro-Alsina},
  {Nibarger}, {Noble}, {Novosad}, {Ogando}, {Omori}, {Padin}, {Pandey},
  {Paschos}, {Patil}, {Pieres}, {Plazas Malag{\'o}n}, {Porredon}, {Prat},
  {Pryke}, {Raveri}, {Reichardt}, {Roberson}, {Rollins}, {Romero}, {Roodman},
  {Ruhl}, {Rykoff}, {Saliwanchik}, {Salvati}, {S{\'a}nchez}, {Sanchez},
  {Sanchez Cid}, {Saro}, {Schaffer}, {Secco}, {Sevilla-Noarbe}, {Sharon},
  {Sheldon}, {Shin}, {Sievers}, {Smecher}, {Smith}, {Somboonpanyakul},
  {Sommer}, {Stalder}, {Stark}, {Stephen}, {Strazzullo}, {Suchyta}, {Tarle},
  {To}, {Troxel}, {Tucker}, {Tutusaus}, {Varga}, {Veach}, {Vieira},
  {Vikhlinin}, {von der Linden}, {Wang}, {Weaverdyck}, {Weller}, {Whitehorn},
  {Wu}, {Yanny}, {Yefremenko}, {Yin}, {Young}, {Zebrowski}, {Zhang}, {Zohren},
  {Zuntz}, {(SPT}, \& {DES Collaborations)}}]{bocquet2024spt}
{Bocquet}, S., {Grandis}, S., {Bleem}, L.~E., {et~al.} 2024, \prd, 110, 083510

\bibitem[{{B{\"o}hm} {et~al.}(2017){B{\"o}hm}, {Hilbert}, {Greiner}, \&
  {En{\ss}lin}}]{bohm2017bayesian}
{B{\"o}hm}, V., {Hilbert}, S., {Greiner}, M., \& {En{\ss}lin}, T.~A. 2017,
  \prd, 96, 123510

\bibitem[{Bridle \& King(2007)}]{bridle2007dark}
Bridle, S. \& King, L. 2007, New Journal of Physics, 9, 444

\bibitem[{{Brown} {et~al.}(2005){Brown}, {Castro}, \&
  {Taylor}}]{brown2005cosmic}
{Brown}, M.~L., {Castro}, P.~G., \& {Taylor}, A.~N. 2005, \mnras, 360, 1262

\bibitem[{{Brown} {et~al.}(2002){Brown}, {Taylor}, {Hambly}, \&
  {Dye}}]{brown2002measurement}
{Brown}, M.~L., {Taylor}, A.~N., {Hambly}, N.~C., \& {Dye}, S. 2002, \mnras,
  333, 501

\bibitem[{{Castro} {et~al.}(2005){Castro}, {Heavens}, \&
  {Kitching}}]{castro2005weak}
{Castro}, P.~G., {Heavens}, A.~F., \& {Kitching}, T.~D. 2005, \prd, 72, 023516

\bibitem[{{Catelan} {et~al.}(2001){Catelan}, {Kamionkowski}, \&
  {Blandford}}]{catelan2001intrinsic}
{Catelan}, P., {Kamionkowski}, M., \& {Blandford}, R.~D. 2001, \mnras, 320, L7

\bibitem[{{Chen} {et~al.}(2023){Chen}, {Harness}, \&
  {Melchior}}]{chen2022lightweight}
{Chen}, A., {Harness}, A., \& {Melchior}, P. 2023, Journal of Astronomical
  Telescopes, Instruments, and Systems, 9, 025002

\bibitem[{{Chiu} {et~al.}(2023){Chiu}, {Klein}, {Mohr}, \&
  {Bocquet}}]{chiu2023cosmological}
{Chiu}, I.~N., {Klein}, M., {Mohr}, J., \& {Bocquet}, S. 2023, \mnras, 522,
  1601

\bibitem[{{Clerkin} {et~al.}(2017){Clerkin}, {Kirk}, {Manera}, {Lahav},
  {Abdalla}, {Amara}, {Bacon}, {Chang}, {Gazta{\~n}aga}, {Hawken}, {Jain},
  {Joachimi}, {Vikram}, {Abbott}, {Allam}, {Armstrong}, {Benoit-L{\'e}vy},
  {Bernstein}, {Bernstein}, {Bertin}, {Brooks}, {Burke}, {Rosell}, {Carrasco
  Kind}, {Crocce}, {Cunha}, {D'Andrea}, {da Costa}, {Desai}, {Diehl},
  {Dietrich}, {Eifler}, {Evrard}, {Flaugher}, {Fosalba}, {Frieman}, {Gerdes},
  {Gruen}, {Gruendl}, {Gutierrez}, {Honscheid}, {James}, {Kent}, {Kuehn},
  {Kuropatkin}, {Lima}, {Melchior}, {Miquel}, {Nord}, {Plazas}, {Romer},
  {Roodman}, {Sanchez}, {Schubnell}, {Sevilla-Noarbe}, {Smith},
  {Soares-Santos}, {Sobreira}, {Suchyta}, {Swanson}, {Tarle}, \&
  {Walker}}]{clerkin2017testing}
{Clerkin}, L., {Kirk}, D., {Manera}, M., {et~al.} 2017, \mnras, 466, 1444

\bibitem[{{Coles} \& {Jones}(1991)}]{coles1991lognormal}
{Coles}, P. \& {Jones}, B. 1991, \mnras, 248, 1

\bibitem[{{Dalal} {et~al.}(2023){Dalal}, {Li}, {Nicola}, {Zuntz}, {Strauss},
  {Sugiyama}, {Zhang}, {Rau}, {Mandelbaum}, {Takada}, {More}, {Miyatake},
  {Kannawadi}, {Shirasaki}, {Taniguchi}, {Takahashi}, {Osato}, {Hamana},
  {Oguri}, {Nishizawa}, {Malag{\'o}n}, {Sunayama}, {Alonso}, {Slosar}, {Luo},
  {Armstrong}, {Bosch}, {Hsieh}, {Komiyama}, {Lupton}, {Lust}, {MacArthur},
  {Miyazaki}, {Murayama}, {Nishimichi}, {Okura}, {Price}, {Tait}, {Tanaka}, \&
  {Wang}}]{dalal2023hyper}
{Dalal}, R., {Li}, X., {Nicola}, A., {et~al.} 2023, \prd, 108, 123519

\bibitem[{{Dark Energy Survey and Kilo-Degree Survey Collaboration}
  {et~al.}(2023){Dark Energy Survey and Kilo-Degree Survey Collaboration},
  {Abbott}, {Aguena}, {Alarcon}, {Alves}, {Amon}, {Andrade-Oliveira}, {Asgari},
  {Avila}, {Bacon}, {Bechtol}, {Becker}, {Bernstein}, {Bertin}, {Bilicki},
  {Blazek}, {Bocquet}, {Brooks}, {Burger}, {Burke}, {Camacho}, {Campos},
  {Carnero Rosell}, {Carrasco Kind}, {Carretero}, {Castander}, {Cawthon},
  {Chang}, {Chen}, {Choi}, {Conselice}, {Cordero}, {Crocce}, {da Costa}, {da
  Silva Pereira}, {Dalal}, {Davis}, {de Jong}, {DeRose}, {Desai}, {Diehl},
  {Dodelson}, {Doel}, {Doux}, {Drlica-Wagner}, {Dvornik}, {Eckert}, {Eifler},
  {Elvin-Poole}, {Everett}, {Fang}, {Ferrero}, {Fert{\'e}}, {Flaugher},
  {Friedrich}, {Frieman}, {Garc{\'\i}a-Bellido}, {Gatti}, {Giannini}, {Giblin},
  {Gruen}, {Gruendl}, {Gutierrez}, {Harrison}, {Hartley}, {Herner}, {Heymans},
  {Hildebrandt}, {Hinton}, {Hoekstra}, {Hollowood}, {Honscheid}, {Huang},
  {Huff}, {Huterer}, {James}, {Jarvis}, {Jeffrey}, {Jeltema}, {Joachimi},
  {Joudaki}, {Kannawadi}, {Krause}, {Kuehn}, {Kuijken}, {Kuropatkin}, {Lahav},
  {Leget}, {Lemos}, {Li}, {Li}, {Liddle}, {Lima}, {Lin}, {Lin}, {MacCrann},
  {Mahony}, {Marshall}, {McCullough}, {Mena-Fern{\'a}ndez}, {Menanteau},
  {Miquel}, {Mohr}, {Muir}, {Myles}, {Napolitano}, {Navarro-Alsina}, {Ogando},
  {Palmese}, {Pandey}, {Park}, {Paterno}, {Peacock}, {Petravick}, {Pieres},
  {Plazas Malag{\'o}n}, {Porredon}, {Prat}, {Radovich}, {Raveri}, {Reischke},
  {Robertson}, {Rollins}, {Romer}, {Roodman}, {Rykoff}, {Samuroff},
  {S{\'a}nchez}, {Sanchez}, {Sanchez}, {Schneider}, {Secco}, {Sevilla-Noarbe},
  {Shan}, {Sheldon}, {Shin}, {Sif{\'o}n}, {Smith}, {Soares-Santos},
  {St{\"o}lzner}, {Suchyta}, {Swanson}, {Tarle}, {Thomas}, {To}, {Troxel},
  {Tr{\"o}ster}, {Tutusaus}, {van den Busch}, {Varga}, {Walker}, {Weaverdyck},
  {Wechsler}, {Weller}, {Wiseman}, {Wright}, {Yanny}, {Yin}, {Yoon}, {Zhang},
  \& {Zuntz}}]{deskids2023}
{Dark Energy Survey and Kilo-Degree Survey Collaboration}, {Abbott}, T.~M.~C.,
  {Aguena}, M., {et~al.} 2023, The Open Journal of Astrophysics, 6, 36

\bibitem[{{de Jong} {et~al.}(2017){de Jong}, {Verdoes Kleijn}, {Erben},
  {Hildebrandt}, {Kuijken}, {Sikkema}, {Brescia}, {Bilicki}, {Napolitano},
  {Amaro}, {Begeman}, {Boxhoorn}, {Buddelmeijer}, {Cavuoti}, {Getman}, {Grado},
  {Helmich}, {Huang}, {Irisarri}, {La Barbera}, {Longo}, {McFarland},
  {Nakajima}, {Paolillo}, {Puddu}, {Radovich}, {Rifatto}, {Tortora},
  {Valentijn}, {Vellucci}, {Vriend}, {Amon}, {Blake}, {Choi}, {Conti}, {Gwyn},
  {Herbonnet}, {Heymans}, {Hoekstra}, {Klaes}, {Merten}, {Miller}, {Schneider},
  \& {Viola}}]{dejong2017third}
{de Jong}, J. T.~A., {Verdoes Kleijn}, G.~A., {Erben}, T., {et~al.} 2017, \aap,
  604, A134

\bibitem[{{de la Bella} {et~al.}(2021){de la Bella}, {Tessore}, \&
  {Bridle}}]{bella2021unequal}
{de la Bella}, L.~F., {Tessore}, N., \& {Bridle}, S. 2021, \jcap, 2021, 001

\bibitem[{{Edge} {et~al.}(2013){Edge}, {Sutherland}, {Kuijken}, {Driver},
  {McMahon}, {Eales}, \& {Emerson}}]{edge2013the}
{Edge}, A., {Sutherland}, W., {Kuijken}, K., {et~al.} 2013, The Messenger, 154,
  32

\bibitem[{{Eifler} {et~al.}(2009){Eifler}, {Schneider}, \&
  {Hartlap}}]{eifler2009dependence}
{Eifler}, T., {Schneider}, P., \& {Hartlap}, J. 2009, \aap, 502, 721

\bibitem[{{Erben} {et~al.}(2005){Erben}, {Schirmer}, {Dietrich}, {Cordes},
  {Haberzettl}, {Hetterscheidt}, {Hildebrandt}, {Schmithuesen}, {Schneider},
  {Simon}, {Deul}, {Hook}, {Kaiser}, {Radovich}, {Benoist}, {Nonino}, {Olsen},
  {Prandoni}, {Wichmann}, {Zaggia}, {Bomans}, {Dettmar}, \&
  {Miralles}}]{erben2005gabods}
{Erben}, T., {Schirmer}, M., {Dietrich}, J.~P., {et~al.} 2005, Astronomische
  Nachrichten, 326, 432

\bibitem[{{Euclid Collaboration} {et~al.}(2023){Euclid Collaboration}, {Ajani},
  {Baldi}, {Barthelemy}, {Boyle}, {Burger}, {Cardone}, {Cheng}, {Codis},
  {Giocoli}, {Harnois-D{\'e}raps}, {Heydenreich}, {Kansal}, {Kilbinger},
  {Linke}, {Llinares}, {Martinet}, {Parroni}, {Peel}, {Pires}, {Porth},
  {Tereno}, {Uhlemann}, {Vicinanza}, {Vinciguerra}, {Aghanim}, {Auricchio},
  {Bonino}, {Branchini}, {Brescia}, {Brinchmann}, {Camera}, {Capobianco},
  {Carbone}, {Carretero}, {Castander}, {Castellano}, {Cavuoti}, {Cimatti},
  {Cledassou}, {Congedo}, {Conselice}, {Conversi}, {Corcione}, {Courbin},
  {Cropper}, {Da Silva}, {Degaudenzi}, {Di Giorgio}, {Dinis}, {Douspis},
  {Dubath}, {Dupac}, {Farrens}, {Ferriol}, {Fosalba}, {Frailis}, {Franceschi},
  {Galeotta}, {Garilli}, {Gillis}, {Grazian}, {Grupp}, {Hoekstra}, {Holmes},
  {Hornstrup}, {Hudelot}, {Jahnke}, {Jhabvala}, {K{\"u}mmel}, {Kitching},
  {Kunz}, {Kurki-Suonio}, {Lilje}, {Lloro}, {Maiorano}, {Mansutti}, {Marggraf},
  {Markovic}, {Marulli}, {Massey}, {Mei}, {Mellier}, {Meneghetti}, {Moresco},
  {Moscardini}, {Niemi}, {Nightingale}, {Nutma}, {Padilla}, {Paltani},
  {Pedersen}, {Pettorino}, {Polenta}, {Poncet}, {Popa}, {Raison}, {Renzi},
  {Rhodes}, {Riccio}, {Romelli}, {Roncarelli}, {Rossetti}, {Saglia}, {Sapone},
  {Sartoris}, {Schneider}, {Schrabback}, {Secroun}, {Seidel}, {Serrano},
  {Sirignano}, {Stanco}, {Starck}, {Tallada-Cresp{\'\i}}, {Taylor},
  {Toledo-Moreo}, {Torradeflot}, {Tutusaus}, {Valentijn}, {Valenziano},
  {Vassallo}, {Wang}, {Weller}, {Zamorani}, {Zoubian}, {Andreon}, {Bardelli},
  {Boucaud}, {Bozzo}, {Colodro-Conde}, {Di Ferdinando}, {Fabbian}, {Farina},
  {Graci{\'a}-Carpio}, {Keih{\"a}nen}, {Lindholm}, {Maino}, {Mauri},
  {Neissner}, {Schirmer}, {Scottez}, {Zucca}, {Akrami}, {Baccigalupi},
  {Balaguera-Antol{\'\i}nez}, {Ballardini}, {Bernardeau}, {Biviano},
  {Blanchard}, {Borgani}, {Borlaff}, {Burigana}, {Cabanac}, {Cappi},
  {Carvalho}, {Casas}, {Castignani}, {Castro}, {Chambers}, {Cooray}, {Coupon},
  {Courtois}, {Davini}, {de la Torre}, {De Lucia}, {Desprez}, {Dole},
  {Escartin}, {Escoffier}, {Ferrero}, {Finelli}, {Ganga}, {Garcia-Bellido},
  {George}, {Giacomini}, {Gozaliasl}, {Hildebrandt}, {Jimenez Mu{\~n}oz},
  {Joachimi}, {Kajava}, {Kirkpatrick}, {Legrand}, {Loureiro}, {Magliocchetti},
  {Maoli}, {Marcin}, {Martinelli}, {Martins}, {Matthew}, {Maurin}, {Metcalf},
  {Monaco}, {Morgante}, {Nadathur}, {Nucita}, {Popa}, {Potter}, {Pourtsidou},
  {P{\"o}ntinen}, {Reimberg}, {S{\'a}nchez}, {Sakr}, {Schneider}, {Sefusatti},
  {Sereno}, {Shulevski}, {Spurio Mancini}, {Steinwagner}, {Teyssier},
  {Valiviita}, {Veropalumbo}, {Viel}, \& {Zinchenko}}]{euclid2023euclid}
{Euclid Collaboration}, {Ajani}, V., {Baldi}, M., {et~al.} 2023, \aap, 675,
  A120

\bibitem[{{Fenech Conti} {et~al.}(2017){Fenech Conti}, {Herbonnet}, {Hoekstra},
  {Merten}, {Miller}, \& {Viola}}]{fenech2017calibration}
{Fenech Conti}, I., {Herbonnet}, R., {Hoekstra}, H., {et~al.} 2017, \mnras,
  467, 1627

\bibitem[{{Feroz} {et~al.}(2009){Feroz}, {Hobson}, \&
  {Bridges}}]{feroz2009multinest}
{Feroz}, F., {Hobson}, M.~P., \& {Bridges}, M. 2009, \mnras, 398, 1601

\bibitem[{{Fluri} {et~al.}(2022){Fluri}, {Kacprzak}, {Lucchi}, {Schneider},
  {Refregier}, \& {Hofmann}}]{fluri2022full}
{Fluri}, J., {Kacprzak}, T., {Lucchi}, A., {et~al.} 2022, \prd, 105, 083518

\bibitem[{{Foreman-Mackey} {et~al.}(2013){Foreman-Mackey}, {Hogg}, {Lang}, \&
  {Goodman}}]{emcee}
{Foreman-Mackey}, D., {Hogg}, D.~W., {Lang}, D., \& {Goodman}, J. 2013, PASP,
  125, 306

\bibitem[{{Fortuna} {et~al.}(2021){Fortuna}, {Hoekstra}, {Johnston}, {Vakili},
  {Kannawadi}, {Georgiou}, {Joachimi}, {Wright}, {Asgari}, {Bilicki},
  {Heymans}, {Hildebrandt}, {Kuijken}, \& {von
  Wietersheim-Kramsta}}]{fortuna2021kids}
{Fortuna}, M.~C., {Hoekstra}, H., {Johnston}, H., {et~al.} 2021, \aap, 654, A76

\bibitem[{{Franco-Abell{\'a}n} {et~al.}(2024){Franco-Abell{\'a}n},
  {Ca{\~n}as-Herrera}, {Martinelli}, {Savchenko}, {Sciotti}, \&
  {Weniger}}]{abellan2024fast}
{Franco-Abell{\'a}n}, G., {Ca{\~n}as-Herrera}, G., {Martinelli}, M., {et~al.}
  2024, \jcap, 2024, 057

\bibitem[{{Garrel} {et~al.}(2022){Garrel}, {Pierre}, {Valageas}, {Eckert},
  {Marulli}, {Veropalumbo}, {Pacaud}, {Clerc}, {Sereno}, {Umetsu},
  {Moscardini}, {Bhargava}, {Adami}, {Chiappetti}, {Gastaldello},
  {Koulouridis}, {Le Fevre}, \& {Plionis}}]{garrel2022xxl}
{Garrel}, C., {Pierre}, M., {Valageas}, P., {et~al.} 2022, \aap, 663, A3

\bibitem[{{Gatti} {et~al.}(2020){Gatti}, {Chang}, {Friedrich}, {Jain}, {Bacon},
  {Crocce}, {DeRose}, {Ferrero}, {Fosalba}, {Gaztanaga}, {Gruen}, {Harrison},
  {Jeffrey}, {MacCrann}, {McClintock}, {Secco}, {Whiteway}, {Abbott}, {Allam},
  {Annis}, {Avila}, {Brooks}, {Buckley-Geer}, {Burke}, {Carnero Rosell},
  {Carrasco Kind}, {Carretero}, {Cawthon}, {da Costa}, {De Vicente}, {Desai},
  {Diehl}, {Doel}, {Eifler}, {Estrada}, {Everett}, {Evrard}, {Frieman},
  {Garc{\'\i}a-Bellido}, {Gerdes}, {Gruendl}, {Gschwend}, {Gutierrez}, {James},
  {Johnson}, {Krause}, {Kuehn}, {Lima}, {Maia}, {March}, {Marshall},
  {Melchior}, {Menanteau}, {Miquel}, {Palmese}, {Paz-Chinch{\'o}n}, {Plazas},
  {S{\'a}nchez}, {Sanchez}, {Scarpine}, {Schubnell}, {Santiago},
  {Sevilla-Noarbe}, {Smith}, {Soares-Santos}, {Suchyta}, {Swanson}, {Tarle},
  {Thomas}, {Troxel}, {Zuntz}, {Zuntz}, \& {DES Collaboration}}]{gatti2020dark}
{Gatti}, M., {Chang}, C., {Friedrich}, O., {et~al.} 2020, \mnras, 498, 4060

\bibitem[{{Gatti} {et~al.}(2024){Gatti}, {Jeffrey}, {Whiteway}, {Williamson},
  {Jain}, {Ajani}, {Anbajagane}, {Giannini}, {Zhou}, {Porredon}, {Prat},
  {Yamamoto}, {Blazek}, {Kacprzak}, {Samuroff}, {Alarcon}, {Amon}, {Bechtol},
  {Becker}, {Bernstein}, {Campos}, {Chang}, {Chen}, {Choi}, {Davis}, {Derose},
  {Diehl}, {Dodelson}, {Doux}, {Eckert}, {Elvin-Poole}, {Everett}, {Ferte},
  {Gruen}, {Gruendl}, {Harrison}, {Hartley}, {Herner}, {Huff}, {Jarvis},
  {Kuropatkin}, {Leget}, {MacCrann}, {McCullough}, {Myles}, {Navarro-Alsina},
  {Pandey}, {Raveri}, {Rollins}, {Roodman}, {Sanchez}, {Secco},
  {Sevilla-Noarbe}, {Sheldon}, {Shin}, {Troxel}, {Tutusaus}, {Varga}, {Yanny},
  {Yin}, {Zhang}, {Zuntz}, {Aguena}, {Alves}, {Annis}, {Brooks}, {Carretero},
  {Castander}, {Cawthon}, {Costanzi}, {da Costa}, {Pereira}, {Evrard},
  {Flaugher}, {Fosalba}, {Frieman}, {Garc{\'\i}a-Bellido}, {Gerdes}, {Gruen},
  {Gruendl}, {Gschwend}, {Gutierrez}, {Hollowood}, {Honscheid}, {James},
  {Kuehn}, {Lahav}, {Lee}, {Marshall}, {Mena-Fern{\'a}ndez}, {Menanteau},
  {Miquel}, {Ogando}, {Pereira}, {Pieres}, {Plazas Malag{\'o}n}, {Sanchez},
  {Smith}, {Suchyta}, {Swanson}, {Tarle}, {Weaverdyck}, {Weller}, {Wiseman}, \&
  {DES Collaboration}}]{gatti2023dark}
{Gatti}, M., {Jeffrey}, N., {Whiteway}, L., {et~al.} 2024, \prd, 109, 063534

\bibitem[{{Gatti} {et~al.}(2021){Gatti}, {Sheldon}, {Amon}, {Becker}, {Troxel},
  {Choi}, {Doux}, {MacCrann}, {Navarro-Alsina}, {Harrison}, {Gruen},
  {Bernstein}, {Jarvis}, {Secco}, {Fert{\'e}}, {Shin}, {McCullough}, {Rollins},
  {Chen}, {Chang}, {Pandey}, {Tutusaus}, {Prat}, {Elvin-Poole}, {Sanchez},
  {Plazas}, {Roodman}, {Zuntz}, {Abbott}, {Aguena}, {Allam}, {Annis}, {Avila},
  {Bacon}, {Bertin}, {Bhargava}, {Brooks}, {Burke}, {Carnero Rosell}, {Carrasco
  Kind}, {Carretero}, {Castander}, {Conselice}, {Costanzi}, {Crocce}, {da
  Costa}, {Davis}, {De Vicente}, {Desai}, {Diehl}, {Dietrich}, {Doel},
  {Drlica-Wagner}, {Eckert}, {Everett}, {Ferrero}, {Frieman},
  {Garc{\'\i}a-Bellido}, {Gerdes}, {Giannantonio}, {Gruendl}, {Gschwend},
  {Gutierrez}, {Hartley}, {Hinton}, {Hollowood}, {Honscheid}, {Hoyle}, {Huff},
  {Huterer}, {Jain}, {James}, {Jeltema}, {Krause}, {Kron}, {Kuropatkin},
  {Lima}, {Maia}, {Marshall}, {Miquel}, {Morgan}, {Myles}, {Palmese},
  {Paz-Chinch{\'o}n}, {Rykoff}, {Samuroff}, {Sanchez}, {Scarpine}, {Schubnell},
  {Serrano}, {Sevilla-Noarbe}, {Smith}, {Soares-Santos}, {Suchyta}, {Swanson},
  {Tarle}, {Thomas}, {To}, {Tucker}, {Varga}, {Wechsler}, {Weller}, {Wester},
  \& {Wilkinson}}]{gatti2021dark}
{Gatti}, M., {Sheldon}, E., {Amon}, A., {et~al.} 2021, \mnras, 504, 4312

\bibitem[{Gelman {et~al.}(1996)Gelman, Meng, \& Stern}]{gelman1996posterior}
Gelman, A., Meng, X.-L., \& Stern, H. 1996, Statistica sinica, 733

\bibitem[{{Gerardi} {et~al.}(2021){Gerardi}, {Feeney}, \&
  {Alsing}}]{gerardi2021unbiased}
{Gerardi}, F., {Feeney}, S.~M., \& {Alsing}, J. 2021, \prd, 104, 083531

\bibitem[{Germain {et~al.}(2015)Germain, Gregor, Murray, \&
  Larochelle}]{germain2015made}
Germain, M., Gregor, K., Murray, I., \& Larochelle, H. 2015, in Proceedings of
  Machine Learning Research, Vol.~37, Proceedings of the 32nd International
  Conference on Machine Learning, ed. F.~Bach \& D.~Blei (Lille, France: PMLR),
  881--889

\bibitem[{{Ghirardini} {et~al.}(2024){Ghirardini}, {Bulbul}, {Artis}, {Clerc},
  {Garrel}, {Grandis}, {Kluge}, {Liu}, {Bahar}, {Balzer}, {Chiu}, {Comparat},
  {Gruen}, {Kleinebreil}, {Krippendorf}, {Merloni}, {Nandra}, {Okabe},
  {Pacaud}, {Predehl}, {Ramos-Ceja}, {Reiprich}, {Sanders}, {Schrabback},
  {Seppi}, {Zelmer}, {Zhang}, {Bornemann}, {Brunner}, {Burwitz}, {Coutinho},
  {Dennerl}, {Freyberg}, {Friedrich}, {Gaida}, {Gueguen}, {Haberl}, {Kink},
  {Lamer}, {Li}, {Liu}, {Maitra}, {Meidinger}, {Mueller}, {Miyatake},
  {Miyazaki}, {Robrade}, {Schwope}, \& {Stewart}}]{ghirardini2024the}
{Ghirardini}, V., {Bulbul}, E., {Artis}, E., {et~al.} 2024, \aap, 689, A298

\bibitem[{{Giblin} {et~al.}(2021){Giblin}, {Heymans}, {Asgari}, {Hildebrandt},
  {Hoekstra}, {Joachimi}, {Kannawadi}, {Kuijken}, {Lin}, {Miller},
  {Tr{\"o}ster}, {van den Busch}, {Wright}, {Bilicki}, {Blake}, {de Jong},
  {Dvornik}, {Erben}, {Getman}, {Napolitano}, {Schneider}, {Shan}, \&
  {Valentijn}}]{giblin2021kids}
{Giblin}, B., {Heymans}, C., {Asgari}, M., {et~al.} 2021, \aap, 645, A105

\bibitem[{{Giocoli} {et~al.}(2017){Giocoli}, {Di Meo}, {Meneghetti}, {Jullo},
  {de la Torre}, {Moscardini}, {Baldi}, {Mazzotta}, \&
  {Metcalf}}]{giocoli2017fast}
{Giocoli}, C., {Di Meo}, S., {Meneghetti}, M., {et~al.} 2017, \mnras, 470, 3574

\bibitem[{{G{\'o}rski} {et~al.}(2005){G{\'o}rski}, {Hivon}, {Banday},
  {Wandelt}, {Hansen}, {Reinecke}, \& {Bartelmann}}]{gorski2005healpix}
{G{\'o}rski}, K.~M., {Hivon}, E., {Banday}, A.~J., {et~al.} 2005, \apj, 622,
  759

\bibitem[{{Guzik} \& {Bernstein}(2005)}]{guzik2005inhomogeneous}
{Guzik}, J. \& {Bernstein}, G. 2005, \prd, 72, 043503

\bibitem[{{Hall} \& {Taylor}(2022)}]{hall2022non}
{Hall}, A. \& {Taylor}, A. 2022, \prd, 105, 123527

\bibitem[{{Hamimeche} \& {Lewis}(2008)}]{hamimeche2008likelihood}
{Hamimeche}, S. \& {Lewis}, A. 2008, \prd, 77, 103013

\bibitem[{{Handley} \& {Lemos}(2021)}]{handley2021quantifying}
{Handley}, W. \& {Lemos}, P. 2021, \prd, 103, 063529

\bibitem[{{Harris} {et~al.}(2020){Harris}, {Millman}, {van der Walt},
  {Gommers}, {Virtanen}, {Cournapeau}, {Wieser}, {Taylor}, {Berg}, {Smith},
  {Kern}, {Picus}, {Hoyer}, {van Kerkwijk}, {Brett}, {Haldane}, {del R{\'\i}o},
  {Wiebe}, {Peterson}, {G{\'e}rard-Marchant}, {Sheppard}, {Reddy}, {Weckesser},
  {Abbasi}, {Gohlke}, \& {Oliphant}}]{harris2020array}
{Harris}, C.~R., {Millman}, K.~J., {van der Walt}, S.~J., {et~al.} 2020, \nat,
  585, 357

\bibitem[{{Heavens} {et~al.}(2000{\natexlab{a}}){Heavens}, {Refregier}, \&
  {Heymans}}]{heavens2000intrinsic}
{Heavens}, A., {Refregier}, A., \& {Heymans}, C. 2000{\natexlab{a}}, \mnras,
  319, 649

\bibitem[{{Heavens} {et~al.}(2000{\natexlab{b}}){Heavens}, {Jimenez}, \&
  {Lahav}}]{heavens2000massive}
{Heavens}, A.~F., {Jimenez}, R., \& {Lahav}, O. 2000{\natexlab{b}}, \mnras,
  317, 965

\bibitem[{{Heydenreich} {et~al.}(2020){Heydenreich}, {Schneider},
  {Hildebrandt}, {Asgari}, {Heymans}, {Joachimi}, {Kuijken}, {Lin},
  {Tr{\"o}ster}, \& {van den Busch}}]{heydenreich2020the}
{Heydenreich}, S., {Schneider}, P., {Hildebrandt}, H., {et~al.} 2020, \aap,
  634, A104

\bibitem[{{Heymans} {et~al.}(2013){Heymans}, {Grocutt}, {Heavens}, {Kilbinger},
  {Kitching}, {Simpson}, {Benjamin}, {Erben}, {Hildebrandt}, {Hoekstra},
  {Mellier}, {Miller}, {Van Waerbeke}, {Brown}, {Coupon}, {Fu},
  {Harnois-D{\'e}raps}, {Hudson}, {Kuijken}, {Rowe}, {Schrabback}, {Semboloni},
  {Vafaei}, \& {Velander}}]{heymans2013cfhtlens}
{Heymans}, C., {Grocutt}, E., {Heavens}, A., {et~al.} 2013, \mnras, 432, 2433

\bibitem[{{Heymans} \& {Heavens}(2003)}]{heymans2003weak}
{Heymans}, C. \& {Heavens}, A. 2003, \mnras, 339, 711

\bibitem[{{Heymans} {et~al.}(2021){Heymans}, {Tr{\"o}ster}, {Asgari}, {Blake},
  {Hildebrandt}, {Joachimi}, {Kuijken}, {Lin}, {S{\'a}nchez}, {van den Busch},
  {Wright}, {Amon}, {Bilicki}, {de Jong}, {Crocce}, {Dvornik}, {Erben},
  {Fortuna}, {Getman}, {Giblin}, {Glazebrook}, {Hoekstra}, {Joudaki},
  {Kannawadi}, {K{\"o}hlinger}, {Lidman}, {Miller}, {Napolitano}, {Parkinson},
  {Schneider}, {Shan}, {Valentijn}, {Verdoes Kleijn}, \&
  {Wolf}}]{heymans2021kids}
{Heymans}, C., {Tr{\"o}ster}, T., {Asgari}, M., {et~al.} 2021, \aap, 646, A140

\bibitem[{{Heymans} {et~al.}(2006){Heymans}, {Van Waerbeke}, {Bacon}, {Berge},
  {Bernstein}, {Bertin}, {Bridle}, {Brown}, {Clowe}, {Dahle}, {Erben}, {Gray},
  {Hetterscheidt}, {Hoekstra}, {Hudelot}, {Jarvis}, {Kuijken}, {Margoniner},
  {Massey}, {Mellier}, {Nakajima}, {Refregier}, {Rhodes}, {Schrabback}, \&
  {Wittman}}]{heymans2006the}
{Heymans}, C., {Van Waerbeke}, L., {Bacon}, D., {et~al.} 2006, \mnras, 368,
  1323

\bibitem[{{Hikage} {et~al.}(2019){Hikage}, {Oguri}, {Hamana}, {More},
  {Mandelbaum}, {Takada}, {K{\"o}hlinger}, {Miyatake}, {Nishizawa}, {Aihara},
  {Armstrong}, {Bosch}, {Coupon}, {Ducout}, {Ho}, {Hsieh}, {Komiyama},
  {Lanusse}, {Leauthaud}, {Lupton}, {Medezinski}, {Mineo}, {Miyama},
  {Miyazaki}, {Murata}, {Murayama}, {Shirasaki}, {Sif{\'o}n}, {Simet},
  {Speagle}, {Spergel}, {Strauss}, {Sugiyama}, {Tanaka}, {Utsumi}, {Wang}, \&
  {Yamada}}]{hikage2019cosmology}
{Hikage}, C., {Oguri}, M., {Hamana}, T., {et~al.} 2019, \pasj, 71, 43

\bibitem[{{Hikage} {et~al.}(2011){Hikage}, {Takada}, {Hamana}, \&
  {Spergel}}]{hikage2011shear}
{Hikage}, C., {Takada}, M., {Hamana}, T., \& {Spergel}, D. 2011, \mnras, 412,
  65

\bibitem[{{Hilbert} {et~al.}(2011){Hilbert}, {Hartlap}, \&
  {Schneider}}]{hilbert2011cosmic}
{Hilbert}, S., {Hartlap}, J., \& {Schneider}, P. 2011, \aap, 536, A85

\bibitem[{{Hilbert} {et~al.}(2017){Hilbert}, {Xu}, {Schneider}, {Springel},
  {Vogelsberger}, \& {Hernquist}}]{hilbert2017intrinsic}
{Hilbert}, S., {Xu}, D., {Schneider}, P., {et~al.} 2017, \mnras, 468, 790

\bibitem[{{Hildebrandt} {et~al.}(2021){Hildebrandt}, {van den Busch}, {Wright},
  {Blake}, {Joachimi}, {Kuijken}, {Tr{\"o}ster}, {Asgari}, {Bilicki}, {de
  Jong}, {Dvornik}, {Erben}, {Getman}, {Giblin}, {Heymans}, {Kannawadi}, {Lin},
  \& {Shan}}]{hildebrandt2021kids}
{Hildebrandt}, H., {van den Busch}, J.~L., {Wright}, A.~H., {et~al.} 2021,
  \aap, 647, A124

\bibitem[{{Hildebrandt} {et~al.}(2017){Hildebrandt}, {Viola}, {Heymans},
  {Joudaki}, {Kuijken}, {Blake}, {Erben}, {Joachimi}, {Klaes}, {Miller},
  {Morrison}, {Nakajima}, {Verdoes Kleijn}, {Amon}, {Choi}, {Covone}, {de
  Jong}, {Dvornik}, {Fenech Conti}, {Grado}, {Harnois-D{\'e}raps}, {Herbonnet},
  {Hoekstra}, {K{\"o}hlinger}, {McFarland}, {Mead}, {Merten}, {Napolitano},
  {Peacock}, {Radovich}, {Schneider}, {Simon}, {Valentijn}, {van den Busch},
  {van Uitert}, \& {Van Waerbeke}}]{hildebrandt2017kids}
{Hildebrandt}, H., {Viola}, M., {Heymans}, C., {et~al.} 2017, \mnras, 465, 1454

\bibitem[{{Hirata} {et~al.}(2004){Hirata}, {Mandelbaum}, {Seljak}, {Guzik},
  {Padmanabhan}, {Blake}, {Brinkmann}, {Bud{\'a}vari}, {Connolly}, {Csabai},
  {Scranton}, \& {Szalay}}]{hirata2004galaxy}
{Hirata}, C.~M., {Mandelbaum}, R., {Seljak}, U., {et~al.} 2004, \mnras, 353,
  529

\bibitem[{{Hivon} {et~al.}(2002){Hivon}, {G{\'o}rski}, {Netterfield}, {Crill},
  {Prunet}, \& {Hansen}}]{hivon2002master}
{Hivon}, E., {G{\'o}rski}, K.~M., {Netterfield}, C.~B., {et~al.} 2002, \apj,
  567, 2

\bibitem[{{Howlett} {et~al.}(2012){Howlett}, {Lewis}, {Hall}, \&
  {Challinor}}]{howlett2012cmb}
{Howlett}, C., {Lewis}, A., {Hall}, A., \& {Challinor}, A. 2012, \jcap, 2012,
  027

\bibitem[{{Hu} {et~al.}(2022){Hu}, {Khaire}, {Hennawi}, {Walther}, {Hiss},
  {Alsing}, {O{\~n}orbe}, {Lukic}, \& {Davies}}]{hu2022measuring}
{Hu}, T., {Khaire}, V., {Hennawi}, J.~F., {et~al.} 2022, \mnras, 515, 2188

\bibitem[{{Hu}(2000)}]{hu2000weak}
{Hu}, W. 2000, \prd, 62, 043007

\bibitem[{{Hunter}(2007)}]{hunter2007matplotlib}
{Hunter}, J.~D. 2007, Computing in Science and Engineering, 9, 90

\bibitem[{{Jarvis} {et~al.}(2004){Jarvis}, {Bernstein}, \&
  {Jain}}]{jarvis2004the}
{Jarvis}, M., {Bernstein}, G., \& {Jain}, B. 2004, \mnras, 352, 338

\bibitem[{{Jeffrey} {et~al.}(2021){Jeffrey}, {Alsing}, \&
  {Lanusse}}]{jeffrey2021}
{Jeffrey}, N., {Alsing}, J., \& {Lanusse}, F. 2021, \mnras, 501, 954

\bibitem[{{Jeffrey} {et~al.}(2024){Jeffrey}, {Whiteway}, {Gatti}, {Williamson},
  {Alsing}, {Porredon}, {Prat}, {Doux}, {Jain}, {Chang}, {Cheng}, {Kacprzak},
  {Lemos}, {Alarcon}, {Amon}, {Bechtol}, {Becker}, {Bernstein}, {Campos},
  {Rosell}, {Chen}, {Choi}, {DeRose}, {Drlica-Wagner}, {Eckert}, {Everett},
  {Fert{\'e}}, {Gruen}, {Gruendl}, {Herner}, {Jarvis}, {McCullough}, {Myles},
  {Navarro-Alsina}, {Pandey}, {Raveri}, {Rollins}, {Rykoff}, {S{\'a}nchez},
  {Secco}, {Sevilla-Noarbe}, {Sheldon}, {Shin}, {Troxel}, {Tutusaus}, {Varga},
  {Yanny}, {Yin}, {Zuntz}, {Aguena}, {Allam}, {Alves}, {Bacon}, {Bocquet},
  {Brooks}, {da Costa}, {Davis}, {De Vicente}, {Desai}, {Diehl}, {Ferrero},
  {Frieman}, {Garc{\'\i}a-Bellido}, {Gaztanaga}, {Giannini}, {Gutierrez},
  {Hinton}, {Hollowood}, {Honscheid}, {Huterer}, {James}, {Lahav}, {Lee},
  {Marshall}, {Mena-Fern{\'a}ndez}, {Miquel}, {Pieres}, {Malag{\'o}n},
  {Roodman}, {Sako}, {Sanchez}, {Sanchez Cid}, {Smith}, {Suchyta}, {Swanson},
  {Tarle}, {Tucker}, {Weaverdyck}, {Weller}, {Wiseman}, \&
  {Yamamoto}}]{jeffrey2024des}
{Jeffrey}, N., {Whiteway}, L., {Gatti}, M., {et~al.} 2024, \mnras
  [\eprint[arXiv]{2403.02314}]

\bibitem[{{Joachimi} {et~al.}(2021){Joachimi}, {Lin}, {Asgari}, {Tr{\"o}ster},
  {Heymans}, {Hildebrandt}, {K{\"o}hlinger}, {S{\'a}nchez}, {Wright},
  {Bilicki}, {Blake}, {van den Busch}, {Crocce}, {Dvornik}, {Erben}, {Getman},
  {Giblin}, {Hoekstra}, {Kannawadi}, {Kuijken}, {Napolitano}, {Schneider},
  {Scoccimarro}, {Sellentin}, {Shan}, {von Wietersheim-Kramsta}, \&
  {Zuntz}}]{joachimi2021kids}
{Joachimi}, B., {Lin}, C.~A., {Asgari}, M., {et~al.} 2021, \aap, 646, A129

\bibitem[{{Joachimi} {et~al.}(2011){Joachimi}, {Mandelbaum}, {Abdalla}, \&
  {Bridle}}]{joachimi2011constraints}
{Joachimi}, B., {Mandelbaum}, R., {Abdalla}, F.~B., \& {Bridle}, S.~L. 2011,
  \aap, 527, A26

\bibitem[{{Johnston} {et~al.}(2021){Johnston}, {Wright}, {Joachimi}, {Bilicki},
  {Elisa Chisari}, {Dvornik}, {Erben}, {Giblin}, {Heymans}, {Hildebrandt},
  {Hoekstra}, {Joudaki}, \& {Vakili}}]{johnston2021organised}
{Johnston}, H., {Wright}, A.~H., {Joachimi}, B., {et~al.} 2021, \aap, 648, A98

\bibitem[{{Joudaki} {et~al.}(2018){Joudaki}, {Blake}, {Johnson}, {Amon},
  {Asgari}, {Choi}, {Erben}, {Glazebrook}, {Harnois-D{\'e}raps}, {Heymans},
  {Hildebrandt}, {Hoekstra}, {Klaes}, {Kuijken}, {Lidman}, {Mead}, {Miller},
  {Parkinson}, {Poole}, {Schneider}, {Viola}, \& {Wolf}}]{joudaki2018kids}
{Joudaki}, S., {Blake}, C., {Johnson}, A., {et~al.} 2018, \mnras, 474, 4894

\bibitem[{{Kaiser}(1992)}]{kaiser1992galactic}
{Kaiser}, N. 1992, \apj, 388, 272

\bibitem[{{Kaiser}(1998)}]{kaiser1998weak}
{Kaiser}, N. 1998, \apj, 498, 26

\bibitem[{{Kannawadi} {et~al.}(2019){Kannawadi}, {Hoekstra}, {Miller}, {Viola},
  {Fenech Conti}, {Herbonnet}, {Erben}, {Heymans}, {Hildebrandt}, {Kuijken},
  {Vakili}, \& {Wright}}]{kannawadi2019towards}
{Kannawadi}, A., {Hoekstra}, H., {Miller}, L., {et~al.} 2019, \aap, 624, A92

\bibitem[{{Kayo} {et~al.}(2001){Kayo}, {Taruya}, \&
  {Suto}}]{kayo2001probability}
{Kayo}, I., {Taruya}, A., \& {Suto}, Y. 2001, \apj, 561, 22

\bibitem[{{Kilbinger}(2015)}]{kilbinger2015cosmology}
{Kilbinger}, M. 2015, Reports on Progress in Physics, 78, 086901

\bibitem[{{Kilbinger} {et~al.}(2017){Kilbinger}, {Heymans}, {Asgari},
  {Joudaki}, {Schneider}, {Simon}, {Van Waerbeke}, {Harnois-D{\'e}raps},
  {Hildebrandt}, {K{\"o}hlinger}, {Kuijken}, \&
  {Viola}}]{kilbinger2017precision}
{Kilbinger}, M., {Heymans}, C., {Asgari}, M., {et~al.} 2017, \mnras, 472, 2126

\bibitem[{{King} \& {Schneider}(2002)}]{king2002supressing}
{King}, L. \& {Schneider}, P. 2002, \aap, 396, 411

\bibitem[{{Kitching} \& {Heavens}(2017)}]{kitching2017unequal}
{Kitching}, T.~D. \& {Heavens}, A.~F. 2017, \prd, 95, 063522

\bibitem[{{Kitching} {et~al.}(2019){Kitching}, {Paykari}, {Hoekstra}, \&
  {Cropper}}]{kitching2019propagating}
{Kitching}, T.~D., {Paykari}, P., {Hoekstra}, H., \& {Cropper}, M. 2019, The
  Open Journal of Astrophysics, 2, 5

\bibitem[{{Kluyver} {et~al.}(2016){Kluyver}, {Ragan-Kelley}, {P{\'e}rez},
  {Granger}, {Bussonnier}, {Frederic}, {Kelley}, {Hamrick}, {Grout}, {Corlay},
  {Ivanov}, {Avila}, {Abdalla}, {Willing}, \& {Jupyter Development
  Team}}]{kluyver2016jupyter}
{Kluyver}, T., {Ragan-Kelley}, B., {P{\'e}rez}, F., {et~al.} 2016, in IOS Press
  (IOS Press), 87--90

\bibitem[{{Krause} {et~al.}(2016){Krause}, {Eifler}, \&
  {Blazek}}]{krause2016impact}
{Krause}, E., {Eifler}, T., \& {Blazek}, J. 2016, \mnras, 456, 207

\bibitem[{{Kuijken}(2011)}]{kuijken2011omegacam}
{Kuijken}, K. 2011, The Messenger, 146, 8

\bibitem[{{Kuijken} {et~al.}(2019){Kuijken}, {Heymans}, {Dvornik},
  {Hildebrandt}, {de Jong}, {Wright}, {Erben}, {Bilicki}, {Giblin}, {Shan},
  {Getman}, {Grado}, {Hoekstra}, {Miller}, {Napolitano}, {Paolilo}, {Radovich},
  {Schneider}, {Sutherland}, {Tewes}, {Tortora}, {Valentijn}, \& {Verdoes
  Kleijn}}]{kuijken2019the}
{Kuijken}, K., {Heymans}, C., {Dvornik}, A., {et~al.} 2019, \aap, 625, A2

\bibitem[{{Kuijken} {et~al.}(2015){Kuijken}, {Heymans}, {Hildebrandt},
  {Nakajima}, {Erben}, {de Jong}, {Viola}, {Choi}, {Hoekstra}, {Miller}, {van
  Uitert}, {Amon}, {Blake}, {Brouwer}, {Buddendiek}, {Conti}, {Eriksen},
  {Grado}, {Harnois-D{\'e}raps}, {Helmich}, {Herbonnet}, {Irisarri},
  {Kitching}, {Klaes}, {La Barbera}, {Napolitano}, {Radovich}, {Schneider},
  {Sif{\'o}n}, {Sikkema}, {Simon}, {Tudorica}, {Valentijn}, {Verdoes Kleijn},
  \& {van Waerbeke}}]{kuijiken2015gravitational}
{Kuijken}, K., {Heymans}, C., {Hildebrandt}, H., {et~al.} 2015, \mnras, 454,
  3500

\bibitem[{{Lam} {et~al.}(2015){Lam}, {Pitrou}, \& {Seibert}}]{lam2015numba}
{Lam}, S.~K., {Pitrou}, A., \& {Seibert}, S. 2015, in Proc. Second Workshop on
  the LLVM Compiler Infrastructure in HPC, 1--6

\bibitem[{{Lange}(2023)}]{lange2023nautilus}
{Lange}, J.~U. 2023, \mnras, 525, 3181

\bibitem[{{Laureijs} {et~al.}(2011){Laureijs}, {Amiaux}, {Arduini},
  {Augu{\`e}res}, {Brinchmann}, {Cole}, {Cropper}, {Dabin}, {Duvet}, {Ealet},
  {Garilli}, {Gondoin}, {Guzzo}, {Hoar}, {Hoekstra}, {Holmes}, {Kitching},
  {Maciaszek}, {Mellier}, {Pasian}, {Percival}, {Rhodes}, {Saavedra Criado},
  {Sauvage}, {Scaramella}, {Valenziano}, {Warren}, {Bender}, {Castander},
  {Cimatti}, {Le F{\`e}vre}, {Kurki-Suonio}, {Levi}, {Lilje}, {Meylan},
  {Nichol}, {Pedersen}, {Popa}, {Rebolo Lopez}, {Rix}, {Rottgering},
  {Zeilinger}, {Grupp}, {Hudelot}, {Massey}, {Meneghetti}, {Miller}, {Paltani},
  {Paulin-Henriksson}, {Pires}, {Saxton}, {Schrabback}, {Seidel}, {Walsh},
  {Aghanim}, {Amendola}, {Bartlett}, {Baccigalupi}, {Beaulieu}, {Benabed},
  {Cuby}, {Elbaz}, {Fosalba}, {Gavazzi}, {Helmi}, {Hook}, {Irwin}, {Kneib},
  {Kunz}, {Mannucci}, {Moscardini}, {Tao}, {Teyssier}, {Weller}, {Zamorani},
  {Zapatero Osorio}, {Boulade}, {Foumond}, {Di Giorgio}, {Guttridge}, {James},
  {Kemp}, {Martignac}, {Spencer}, {Walton}, {Bl{\"u}mchen}, {Bonoli},
  {Bortoletto}, {Cerna}, {Corcione}, {Fabron}, {Jahnke}, {Ligori}, {Madrid},
  {Martin}, {Morgante}, {Pamplona}, {Prieto}, {Riva}, {Toledo}, {Trifoglio},
  {Zerbi}, {Abdalla}, {Douspis}, {Grenet}, {Borgani}, {Bouwens}, {Courbin},
  {Delouis}, {Dubath}, {Fontana}, {Frailis}, {Grazian}, {Koppenh{\"o}fer},
  {Mansutti}, {Melchior}, {Mignoli}, {Mohr}, {Neissner}, {Noddle}, {Poncet},
  {Scodeggio}, {Serrano}, {Shane}, {Starck}, {Surace}, {Taylor},
  {Verdoes-Kleijn}, {Vuerli}, {Williams}, {Zacchei}, {Altieri}, {Escudero
  Sanz}, {Kohley}, {Oosterbroek}, {Astier}, {Bacon}, {Bardelli}, {Baugh},
  {Bellagamba}, {Benoist}, {Bianchi}, {Biviano}, {Branchini}, {Carbone},
  {Cardone}, {Clements}, {Colombi}, {Conselice}, {Cresci}, {Deacon}, {Dunlop},
  {Fedeli}, {Fontanot}, {Franzetti}, {Giocoli}, {Garcia-Bellido}, {Gow},
  {Heavens}, {Hewett}, {Heymans}, {Holland}, {Huang}, {Ilbert}, {Joachimi},
  {Jennins}, {Kerins}, {Kiessling}, {Kirk}, {Kotak}, {Krause}, {Lahav}, {van
  Leeuwen}, {Lesgourgues}, {Lombardi}, {Magliocchetti}, {Maguire}, {Majerotto},
  {Maoli}, {Marulli}, {Maurogordato}, {McCracken}, {McLure}, {Melchiorri},
  {Merson}, {Moresco}, {Nonino}, {Norberg}, {Peacock}, {Pello}, {Penny},
  {Pettorino}, {Di Porto}, {Pozzetti}, {Quercellini}, {Radovich}, {Rassat},
  {Roche}, {Ronayette}, {Rossetti}, {Sartoris}, {Schneider}, {Semboloni},
  {Serjeant}, {Simpson}, {Skordis}, {Smadja}, {Smartt}, {Spano}, {Spiro},
  {Sullivan}, {Tilquin}, {Trotta}, {Verde}, {Wang}, {Williger}, {Zhao},
  {Zoubian}, \& {Zucca}}]{laureijs2011euclid}
{Laureijs}, R., {Amiaux}, J., {Arduini}, S., {et~al.} 2011, arXiv e-prints,
  arXiv:1110.3193

\bibitem[{{Leclercq}(2018)}]{leclercq2018bayesian}
{Leclercq}, F. 2018, \prd, 98, 063511

\bibitem[{{Legin} {et~al.}(2021){Legin}, {Hezaveh}, {Perreault Levasseur}, \&
  {Wandelt}}]{legin2021simulation}
{Legin}, R., {Hezaveh}, Y., {Perreault Levasseur}, L., \& {Wandelt}, B. 2021,
  arXiv e-prints, arXiv:2112.05278

\bibitem[{{Leistedt} {et~al.}(2013){Leistedt}, {Peiris}, {Mortlock},
  {Benoit-L{\'e}vy}, \& {Pontzen}}]{leistedt2013estimating}
{Leistedt}, B., {Peiris}, H.~V., {Mortlock}, D.~J., {Benoit-L{\'e}vy}, A., \&
  {Pontzen}, A. 2013, \mnras, 435, 1857

\bibitem[{{Lemos} {et~al.}(2017){Lemos}, {Challinor}, \&
  {Efstathiou}}]{lemos2017effect}
{Lemos}, P., {Challinor}, A., \& {Efstathiou}, G. 2017, \jcap, 2017, 014

\bibitem[{{Lemos} {et~al.}(2023{\natexlab{a}}){Lemos}, {Coogan}, {Hezaveh}, \&
  {Perreault-Levasseur}}]{lemos2023sampling}
{Lemos}, P., {Coogan}, A., {Hezaveh}, Y., \& {Perreault-Levasseur}, L.
  2023{\natexlab{a}}, 40th International Conference on Machine Learning, 202,
  19256

\bibitem[{{Lemos} {et~al.}(2023{\natexlab{b}}){Lemos}, {Cranmer}, {Abidi},
  {Hahn}, {Eickenberg}, {Massara}, {Yallup}, \& {Ho}}]{lemos2023robust}
{Lemos}, P., {Cranmer}, M., {Abidi}, M., {et~al.} 2023{\natexlab{b}}, Machine
  Learning: Science and Technology, 4, 01LT01

\bibitem[{{Lemos} {et~al.}(2021){Lemos}, {Jeffrey}, {Whiteway}, {Lahav},
  {Libeskind}, \& {Hoffman}}]{lemos2021sum}
{Lemos}, P., {Jeffrey}, N., {Whiteway}, L., {et~al.} 2021, \prd, 103, 023009

\bibitem[{{Lemos} {et~al.}(2023{\natexlab{c}}){Lemos}, {Parker}, {Hahn}, {Ho},
  {Eickenberg}, {Hou}, {Massara}, {Modi}, {Moradinezhad Dizgah},
  {R{\'e}galdo-Saint Blancard}, \& {Spergel}}]{lemos2023simbig}
{Lemos}, P., {Parker}, L.~H., {Hahn}, C., {et~al.} 2023{\natexlab{c}}, in
  Machine Learning for Astrophysics, 18

\bibitem[{{Leonard} {et~al.}(2023){Leonard}, {Ferreira}, {Fang}, {Reischke},
  {Schoeneberg}, {Tr{\"o}ster}, {Alonso}, {Campagne}, {Lanusse}, {Slosar}, \&
  {Ishak}}]{leonard2023the}
{Leonard}, C.~D., {Ferreira}, T., {Fang}, X., {et~al.} 2023, The Open Journal
  of Astrophysics, 6, 8

\bibitem[{Levin(1996)}]{levin1996fast}
Levin, D. 1996, Journal of Computational and Applied Mathematics, 67, 95

\bibitem[{{Lewis}(2019)}]{antony2019getdist}
{Lewis}, A. 2019, arXiv e-prints, arXiv:1910.13970

\bibitem[{{Lewis} \& {Challinor}(2002)}]{lewis2002evolution}
{Lewis}, A. \& {Challinor}, A. 2002, \prd, 66, 023531

\bibitem[{{Lewis} {et~al.}(2000){Lewis}, {Challinor}, \&
  {Lasenby}}]{lewis2000efficient}
{Lewis}, A., {Challinor}, A., \& {Lasenby}, A. 2000, \apj, 538, 473

\bibitem[{{Li} {et~al.}(2023{\natexlab{a}}){Li}, {Hoekstra}, {Kuijken},
  {Asgari}, {Bilicki}, {Giblin}, {Heymans}, {Hildebrandt}, {Joachimi},
  {Miller}, {van den Busch}, {Wright}, {Kannawadi}, {Reischke}, \&
  {Shan}}]{li2023kids}
{Li}, S.-S., {Hoekstra}, H., {Kuijken}, K., {et~al.} 2023{\natexlab{a}}, \aap,
  679, A133

\bibitem[{{Li} {et~al.}(2023{\natexlab{b}}){Li}, {Zhang}, {Sugiyama}, {Dalal},
  {Terasawa}, {Rau}, {Mandelbaum}, {Takada}, {More}, {Strauss}, {Miyatake},
  {Shirasaki}, {Hamana}, {Oguri}, {Luo}, {Nishizawa}, {Takahashi}, {Nicola},
  {Osato}, {Kannawadi}, {Sunayama}, {Armstrong}, {Bosch}, {Komiyama}, {Lupton},
  {Lust}, {MacArthur}, {Miyazaki}, {Murayama}, {Nishimichi}, {Okura}, {Price},
  {Tait}, {Tanaka}, \& {Wang}}]{li2023hyper}
{Li}, X., {Zhang}, T., {Sugiyama}, S., {et~al.} 2023{\natexlab{b}}, \prd, 108,
  123518

\bibitem[{{Limber}(1953)}]{limber1953analysis}
{Limber}, D.~N. 1953, \apj, 117, 134

\bibitem[{{Lin} \& {Kilbinger}(2015)}]{lin2015a}
{Lin}, C.-A. \& {Kilbinger}, M. 2015, \aap, 583, A70

\bibitem[{{Lin} {et~al.}(2016){Lin}, {Kilbinger}, \& {Pires}}]{lin2016a}
{Lin}, C.-A., {Kilbinger}, M., \& {Pires}, S. 2016, \aap, 593, A88

\bibitem[{{Lin} {et~al.}(2020){Lin}, {Harnois-D{\'e}raps}, {Eifler},
  {Pospisil}, {Mandelbaum}, {Lee}, {Singh}, \& {LSST Dark Energy Science
  Collaboration}}]{lin2020non}
{Lin}, C.-H., {Harnois-D{\'e}raps}, J., {Eifler}, T., {et~al.} 2020, \mnras,
  499, 2977

\bibitem[{{Lin} {et~al.}(2023){Lin}, {von Wietersheim-Kramsta}, {Joachimi}, \&
  {Feeney}}]{lin2022a}
{Lin}, K., {von Wietersheim-Kramsta}, M., {Joachimi}, B., \& {Feeney}, S. 2023,
  \mnras, 524, 6167

\bibitem[{{Loureiro} {et~al.}(2023){Loureiro}, {Whiteway}, {Sellentin}, {Silva
  Lafaurie}, {Jaffe}, \& {Heavens}}]{loureiro2023almanac}
{Loureiro}, A., {Whiteway}, L., {Sellentin}, E., {et~al.} 2023, The Open
  Journal of Astrophysics, 6, 6

\bibitem[{{Loureiro} {et~al.}(2022){Loureiro}, {Whittaker}, {Spurio Mancini},
  {Joachimi}, {Cuceu}, {Asgari}, {St{\"o}lzner}, {Tr{\"o}ster}, {Wright},
  {Bilicki}, {Dvornik}, {Giblin}, {Heymans}, {Hildebrandt}, {Shan}, {Amara},
  {Auricchio}, {Bodendorf}, {Bonino}, {Branchini}, {Brescia}, {Capobianco},
  {Carbone}, {Carretero}, {Castellano}, {Cavuoti}, {Cimatti}, {Cledassou},
  {Congedo}, {Conversi}, {Copin}, {Corcione}, {Cropper}, {Da Silva}, {Douspis},
  {Dubath}, {Duncan}, {Dupac}, {Dusini}, {Farrens}, {Ferriol}, {Fosalba},
  {Frailis}, {Franceschi}, {Fumana}, {Garilli}, {Gillis}, {Giocoli}, {Grazian},
  {Grupp}, {Haugan}, {Holmes}, {Hormuth}, {Jahnke}, {K{\"u}mmel}, {Kermiche},
  {Kiessling}, {Kilbinger}, {Kitching}, {Kuijken}, {Kunz}, {Kurki-Suonio},
  {Ligori}, {Lilje}, {Lloro}, {Mansutti}, {Marggraf}, {Markovic}, {Marulli},
  {Massey}, {Meneghetti}, {Meylan}, {Moresco}, {Morin}, {Moscardini}, {Munari},
  {Niemi}, {Padilla}, {Paltani}, {Pasian}, {Pedersen}, {Pettorino}, {Pires},
  {Poncet}, {Popa}, {Raison}, {Rhodes}, {Rix}, {Roncarelli}, {Saglia},
  {Schneider}, {Secroun}, {Serrano}, {Sirignano}, {Sirri}, {Stanco}, {Starck},
  {Tallada-Cresp{\'\i}}, {Taylor}, {Tereno}, {Toledo-Moreo}, {Torradeflot},
  {Valentijn}, {Wang}, {Welikala}, {Weller}, {Zamorani}, {Zoubian}, {Andreon},
  {Baldi}, {Camera}, {Farinelli}, {Polenta}, \& {Tessore}}]{loureiro2021kids}
{Loureiro}, A., {Whittaker}, L., {Spurio Mancini}, A., {et~al.} 2022, \aap,
  665, A56

\bibitem[{{LoVerde} \& {Afshordi}(2008)}]{loverde2008extended}
{LoVerde}, M. \& {Afshordi}, N. 2008, \prd, 78, 123506

\bibitem[{{LSST Science Collaboration} {et~al.}(2009){LSST Science
  Collaboration}, {Abell}, {Allison}, {Anderson}, {Andrew}, {Angel}, {Armus},
  {Arnett}, {Asztalos}, {Axelrod}, {Bailey}, {Ballantyne}, {Bankert},
  {Barkhouse}, {Barr}, {Barrientos}, {Barth}, {Bartlett}, {Becker}, {Becla},
  {Beers}, {Bernstein}, {Biswas}, {Blanton}, {Bloom}, {Bochanski}, {Boeshaar},
  {Borne}, {Bradac}, {Brandt}, {Bridge}, {Brown}, {Brunner}, {Bullock},
  {Burgasser}, {Burge}, {Burke}, {Cargile}, {Chandrasekharan}, {Chartas},
  {Chesley}, {Chu}, {Cinabro}, {Claire}, {Claver}, {Clowe}, {Connolly}, {Cook},
  {Cooke}, {Cooray}, {Covey}, {Culliton}, {de Jong}, {de Vries}, {Debattista},
  {Delgado}, {Dell'Antonio}, {Dhital}, {Di Stefano}, {Dickinson}, {Dilday},
  {Djorgovski}, {Dobler}, {Donalek}, {Dubois-Felsmann}, {Durech},
  {Eliasdottir}, {Eracleous}, {Eyer}, {Falco}, {Fan}, {Fassnacht}, {Ferguson},
  {Fernandez}, {Fields}, {Finkbeiner}, {Figueroa}, {Fox}, {Francke}, {Frank},
  {Frieman}, {Fromenteau}, {Furqan}, {Galaz}, {Gal-Yam}, {Garnavich},
  {Gawiser}, {Geary}, {Gee}, {Gibson}, {Gilmore}, {Grace}, {Green}, {Gressler},
  {Grillmair}, {Habib}, {Haggerty}, {Hamuy}, {Harris}, {Hawley}, {Heavens},
  {Hebb}, {Henry}, {Hileman}, {Hilton}, {Hoadley}, {Holberg}, {Holman},
  {Howell}, {Infante}, {Ivezic}, {Jacoby}, {Jain}, {R}, {Jedicke}, {Jee},
  {Garrett Jernigan}, {Jha}, {Johnston}, {Jones}, {Juric}, {Kaasalainen},
  {Styliani}, {Kafka}, {Kahn}, {Kaib}, {Kalirai}, {Kantor}, {Kasliwal},
  {Keeton}, {Kessler}, {Knezevic}, {Kowalski}, {Krabbendam}, {Krughoff},
  {Kulkarni}, {Kuhlman}, {Lacy}, {Lepine}, {Liang}, {Lien}, {Lira}, {Long},
  {Lorenz}, {Lotz}, {Lupton}, {Lutz}, {Macri}, {Mahabal}, {Mandelbaum},
  {Marshall}, {May}, {McGehee}, {Meadows}, {Meert}, {Milani}, {Miller},
  {Miller}, {Mills}, {Minniti}, {Monet}, {Mukadam}, {Nakar}, {Neill}, {Newman},
  {Nikolaev}, {Nordby}, {O'Connor}, {Oguri}, {Oliver}, {Olivier}, {Olsen},
  {Olsen}, {Olszewski}, {Oluseyi}, {Padilla}, {Parker}, {Pepper}, {Peterson},
  {Petry}, {Pinto}, {Pizagno}, {Popescu}, {Prsa}, {Radcka}, {Raddick},
  {Rasmussen}, {Rau}, {Rho}, {Rhoads}, {Richards}, {Ridgway}, {Robertson},
  {Roskar}, {Saha}, {Sarajedini}, {Scannapieco}, {Schalk}, {Schindler},
  {Schmidt}, {Schmidt}, {Schneider}, {Schumacher}, {Scranton}, {Sebag},
  {Seppala}, {Shemmer}, {Simon}, {Sivertz}, {Smith}, {Allyn Smith}, {Smith},
  {Spitz}, {Stanford}, {Stassun}, {Strader}, {Strauss}, {Stubbs}, {Sweeney},
  {Szalay}, {Szkody}, {Takada}, {Thorman}, {Trilling}, {Trimble}, {Tyson}, {Van
  Berg}, {Vanden Berk}, {VanderPlas}, {Verde}, {Vrsnak}, {Walkowicz},
  {Wandelt}, {Wang}, {Wang}, {Warner}, {Wechsler}, {West}, {Wiecha},
  {Williams}, {Willman}, {Wittman}, {Wolff}, {Wood-Vasey}, {Wozniak}, {Young},
  {Zentner}, \& {Zhan}}]{lsst2009lsst}
{LSST Science Collaboration}, {Abell}, P.~A., {Allison}, J., {et~al.} 2009,
  arXiv e-prints, arXiv:0912.0201

\bibitem[{{Lu} {et~al.}(2023){Lu}, {Haiman}, \& {Li}}]{lu2023cosmological}
{Lu}, T., {Haiman}, Z., \& {Li}, X. 2023, \mnras, 521, 2050

\bibitem[{{Mandelbaum}(2018)}]{mandelbaum2018weak}
{Mandelbaum}, R. 2018, \araa, 56, 393

\bibitem[{{Mantz} {et~al.}(2015){Mantz}, {von der Linden}, {Allen},
  {Applegate}, {Kelly}, {Morris}, {Rapetti}, {Schmidt}, {Adhikari}, {Allen},
  {Burchat}, {Burke}, {Cataneo}, {Donovan}, {Ebeling}, {Shandera}, \&
  {Wright}}]{mantz2015weighing}
{Mantz}, A.~B., {von der Linden}, A., {Allen}, S.~W., {et~al.} 2015, \mnras,
  446, 2205

\bibitem[{{McFarland} {et~al.}(2013){McFarland}, {Verdoes-Kleijn}, {Sikkema},
  {Helmich}, {Boxhoorn}, \& {Valentijn}}]{mcfarland2013the}
{McFarland}, J.~P., {Verdoes-Kleijn}, G., {Sikkema}, G., {et~al.} 2013,
  Experimental Astronomy, 35, 45

\bibitem[{{Mead} {et~al.}(2021){Mead}, {Brieden}, {Tr{\"o}ster}, \&
  {Heymans}}]{mead2021hmcode}
{Mead}, A.~J., {Brieden}, S., {Tr{\"o}ster}, T., \& {Heymans}, C. 2021, \mnras,
  502, 1401

\bibitem[{{Mead} {et~al.}(2016){Mead}, {Heymans}, {Lombriser}, {Peacock},
  {Steele}, \& {Winther}}]{mead2016accurate}
{Mead}, A.~J., {Heymans}, C., {Lombriser}, L., {et~al.} 2016, \mnras, 459, 1468

\bibitem[{{Mead} {et~al.}(2015){Mead}, {Peacock}, {Heymans}, {Joudaki}, \&
  {Heavens}}]{mead2015an}
{Mead}, A.~J., {Peacock}, J.~A., {Heymans}, C., {Joudaki}, S., \& {Heavens},
  A.~F. 2015, \mnras, 454, 1958

\bibitem[{{Miller} {et~al.}(2013){Miller}, {Heymans}, {Kitching}, {van
  Waerbeke}, {Erben}, {Hildebrandt}, {Hoekstra}, {Mellier}, {Rowe}, {Coupon},
  {Dietrich}, {Fu}, {Harnois-D{\'e}raps}, {Hudson}, {Kilbinger}, {Kuijken},
  {Schrabback}, {Semboloni}, {Vafaei}, \& {Velander}}]{miller2013bayesian}
{Miller}, L., {Heymans}, C., {Kitching}, T.~D., {et~al.} 2013, \mnras, 429,
  2858

\bibitem[{{Miller} {et~al.}(2007){Miller}, {Kitching}, {Heymans}, {Heavens}, \&
  {van Waerbeke}}]{miller2007bayesian}
{Miller}, L., {Kitching}, T.~D., {Heymans}, C., {Heavens}, A.~F., \& {van
  Waerbeke}, L. 2007, \mnras, 382, 315

\bibitem[{{Mishra-Sharma} \& {Cranmer}(2022)}]{mishra2022neural}
{Mishra-Sharma}, S. \& {Cranmer}, K. 2022, \prd, 105, 063017

\bibitem[{{Miyatake} {et~al.}(2023){Miyatake}, {Sugiyama}, {Takada},
  {Nishimichi}, {Li}, {Shirasaki}, {More}, {Kobayashi}, {Nishizawa}, {Rau},
  {Zhang}, {Takahashi}, {Dalal}, {Mandelbaum}, {Strauss}, {Hamana}, {Oguri},
  {Osato}, {Luo}, {Kannawadi}, {Hsieh}, {Armstrong}, {Bosch}, {Komiyama},
  {Lupton}, {Lust}, {MacArthur}, {Miyazaki}, {Murayama}, {Okura}, {Price},
  {Sunayama}, {Tait}, {Tanaka}, \& {Wang}}]{miyatake2023hyper}
{Miyatake}, H., {Sugiyama}, S., {Takada}, M., {et~al.} 2023, \prd, 108, 123517

\bibitem[{{Moser} {et~al.}(2024){Moser}, {Kacprzak}, {Fischbacher},
  {Refregier}, {Grimm}, \& {Tortorelli}}]{moser2024simulation}
{Moser}, B., {Kacprzak}, T., {Fischbacher}, S., {et~al.} 2024, \jcap, 2024, 049

\bibitem[{Nelder \& Mead(1965)}]{nelder1965simplex}
Nelder, J.~A. \& Mead, R. 1965, The computer journal, 7, 308

\bibitem[{{Nicola} {et~al.}(2021){Nicola}, {Garc{\'\i}a-Garc{\'\i}a}, {Alonso},
  {Dunkley}, {Ferreira}, {Slosar}, \& {Spergel}}]{nicola2021cosmic}
{Nicola}, A., {Garc{\'\i}a-Garc{\'\i}a}, C., {Alonso}, D., {et~al.} 2021,
  \jcap, 2021, 067

\bibitem[{{Papamakarios} \& {Murray}(2016)}]{papamakarios2016fast}
{Papamakarios}, G. \& {Murray}, I. 2016, arXiv e-prints, arXiv:1605.06376

\bibitem[{Papamakarios {et~al.}(2017)Papamakarios, Pavlakou, \&
  Murray}]{papamakarios2017masked}
Papamakarios, G., Pavlakou, T., \& Murray, I. 2017, in Advances in Neural
  Information Processing Systems, ed. I.~Guyon, U.~V. Luxburg, S.~Bengio,
  H.~Wallach, R.~Fergus, S.~Vishwanathan, \& R.~Garnett, Vol.~30 (Curran
  Associates, Inc.)

\bibitem[{{Peebles}(1973)}]{peebles1973statistical}
{Peebles}, P.~J.~E. 1973, \apj, 185, 413

\bibitem[{{Piras} {et~al.}(2023){Piras}, {Joachimi}, \&
  {Villaescusa-Navarro}}]{piras2023fast}
{Piras}, D., {Joachimi}, B., \& {Villaescusa-Navarro}, F. 2023, \mnras, 520,
  668

\bibitem[{{Planck Collaboration} {et~al.}(2020){Planck Collaboration},
  {Aghanim}, {Akrami}, {Ashdown}, {Aumont}, {Baccigalupi}, {Ballardini},
  {Banday}, {Barreiro}, {Bartolo}, {Basak}, {Battye}, {Benabed}, {Bernard},
  {Bersanelli}, {Bielewicz}, {Bock}, {Bond}, {Borrill}, {Bouchet}, {Boulanger},
  {Bucher}, {Burigana}, {Butler}, {Calabrese}, {Cardoso}, {Carron},
  {Challinor}, {Chiang}, {Chluba}, {Colombo}, {Combet}, {Contreras}, {Crill},
  {Cuttaia}, {de Bernardis}, {de Zotti}, {Delabrouille}, {Delouis}, {Di
  Valentino}, {Diego}, {Dor{\'e}}, {Douspis}, {Ducout}, {Dupac}, {Dusini},
  {Efstathiou}, {Elsner}, {En{\ss}lin}, {Eriksen}, {Fantaye}, {Farhang},
  {Fergusson}, {Fernandez-Cobos}, {Finelli}, {Forastieri}, {Frailis},
  {Fraisse}, {Franceschi}, {Frolov}, {Galeotta}, {Galli}, {Ganga},
  {G{\'e}nova-Santos}, {Gerbino}, {Ghosh}, {Gonz{\'a}lez-Nuevo}, {G{\'o}rski},
  {Gratton}, {Gruppuso}, {Gudmundsson}, {Hamann}, {Handley}, {Hansen},
  {Herranz}, {Hildebrandt}, {Hivon}, {Huang}, {Jaffe}, {Jones}, {Karakci},
  {Keih{\"a}nen}, {Keskitalo}, {Kiiveri}, {Kim}, {Kisner}, {Knox},
  {Krachmalnicoff}, {Kunz}, {Kurki-Suonio}, {Lagache}, {Lamarre}, {Lasenby},
  {Lattanzi}, {Lawrence}, {Le Jeune}, {Lemos}, {Lesgourgues}, {Levrier},
  {Lewis}, {Liguori}, {Lilje}, {Lilley}, {Lindholm}, {L{\'o}pez-Caniego},
  {Lubin}, {Ma}, {Mac{\'\i}as-P{\'e}rez}, {Maggio}, {Maino}, {Mandolesi},
  {Mangilli}, {Marcos-Caballero}, {Maris}, {Martin}, {Martinelli},
  {Mart{\'\i}nez-Gonz{\'a}lez}, {Matarrese}, {Mauri}, {McEwen}, {Meinhold},
  {Melchiorri}, {Mennella}, {Migliaccio}, {Millea}, {Mitra},
  {Miville-Desch{\^e}nes}, {Molinari}, {Montier}, {Morgante}, {Moss}, {Natoli},
  {N{\o}rgaard-Nielsen}, {Pagano}, {Paoletti}, {Partridge}, {Patanchon},
  {Peiris}, {Perrotta}, {Pettorino}, {Piacentini}, {Polastri}, {Polenta},
  {Puget}, {Rachen}, {Reinecke}, {Remazeilles}, {Renzi}, {Rocha}, {Rosset},
  {Roudier}, {Rubi{\~n}o-Mart{\'\i}n}, {Ruiz-Granados}, {Salvati}, {Sandri},
  {Savelainen}, {Scott}, {Shellard}, {Sirignano}, {Sirri}, {Spencer},
  {Sunyaev}, {Suur-Uski}, {Tauber}, {Tavagnacco}, {Tenti}, {Toffolatti},
  {Tomasi}, {Trombetti}, {Valenziano}, {Valiviita}, {Van Tent}, {Vibert},
  {Vielva}, {Villa}, {Vittorio}, {Wandelt}, {Wehus}, {White}, {White},
  {Zacchei}, \& {Zonca}}]{planck2020planck}
{Planck Collaboration}, {Aghanim}, N., {Akrami}, Y., {et~al.} 2020, \aap, 641,
  A6

\bibitem[{{Porqueres} {et~al.}(2022){Porqueres}, {Heavens}, {Mortlock}, \&
  {Lavaux}}]{porqueres2022lifting}
{Porqueres}, N., {Heavens}, A., {Mortlock}, D., \& {Lavaux}, G. 2022, \mnras,
  509, 3194

\bibitem[{{Porqueres} {et~al.}(2023){Porqueres}, {Heavens}, {Mortlock},
  {Lavaux}, \& {Makinen}}]{porqueres2023field}
{Porqueres}, N., {Heavens}, A., {Mortlock}, D., {Lavaux}, G., \& {Makinen},
  T.~L. 2023, arXiv e-prints, arXiv:2304.04785

\bibitem[{Pritchard {et~al.}(1999)Pritchard, Seielstad, Perez-Lezaun, \&
  Feldman}]{pritchard1999population}
Pritchard, J.~K., Seielstad, M.~T., Perez-Lezaun, A., \& Feldman, M.~W. 1999,
  Molecular biology and evolution, 16, 1791

\bibitem[{{Reinecke}(2011)}]{reinecke2011libpsht}
{Reinecke}, M. 2011, \aap, 526, A108

\bibitem[{{Reischke} {et~al.}(2017){Reischke}, {Kiessling}, \&
  {Sch{\"a}fer}}]{reischke2017variations}
{Reischke}, R., {Kiessling}, A., \& {Sch{\"a}fer}, B.~M. 2017, \mnras, 465,
  4016

\bibitem[{{Reischke} {et~al.}(2024){Reischke}, {Unruh}, {Asgari}, {Dvornik},
  {Hildebrandt}, {Joachimi}, {Porth}, {von Wietersheim-Kramsta}, {van den
  Busch}, {St{\"o}lzner}, {Wright}, {Yan}, {Bilicki}, {Burger},
  {Harnois-Deraps}, {Georgiou}, {Heymans}, {Jalan}, {Joudaki}, {Kuijken}, {Li},
  {Linke}, {Mahony}, {Sciotti}, \& {Tr{\"o}ster}}]{reischke2024kids}
{Reischke}, R., {Unruh}, S., {Asgari}, M., {et~al.} 2024, arXiv e-prints,
  arXiv:2410.06962

\bibitem[{Robert {et~al.}(2007)}]{robert2007bayesian}
Robert, C.~P. {et~al.} 2007, The Bayesian choice: from decision-theoretic
  foundations to computational implementation, Vol.~2 (Springer)

\bibitem[{Rubin(1984)}]{rubin1984bayesianly}
Rubin, D.~B. 1984, The Annals of Statistics, 1151

\bibitem[{{Schneider} {et~al.}(2022){Schneider}, {Giri}, {Amodeo}, \&
  {Refregier}}]{schneider2022constraining}
{Schneider}, A., {Giri}, S.~K., {Amodeo}, S., \& {Refregier}, A. 2022, \mnras,
  514, 3802

\bibitem[{{Schneider} \& {Teyssier}(2015)}]{schneider2015a}
{Schneider}, A. \& {Teyssier}, R. 2015, \jcap, 2015, 049

\bibitem[{{Schneider} {et~al.}(2019){Schneider}, {Teyssier}, {Stadel},
  {Chisari}, {Le Brun}, {Amara}, \& {Refregier}}]{schneider2019quantifying}
{Schneider}, A., {Teyssier}, R., {Stadel}, J., {et~al.} 2019, \jcap, 2019, 020

\bibitem[{{Schneider}(2016)}]{schneider2016generalized}
{Schneider}, P. 2016, \aap, 592, L6

\bibitem[{{Schneider} \& {Hartlap}(2009)}]{schneider2009constrained}
{Schneider}, P. \& {Hartlap}, J. 2009, \aap, 504, 705

\bibitem[{{Scolnic} {et~al.}(2018){Scolnic}, {Jones}, {Rest}, {Pan},
  {Chornock}, {Foley}, {Huber}, {Kessler}, {Narayan}, {Riess}, {Rodney},
  {Berger}, {Brout}, {Challis}, {Drout}, {Finkbeiner}, {Lunnan}, {Kirshner},
  {Sanders}, {Schlafly}, {Smartt}, {Stubbs}, {Tonry}, {Wood-Vasey}, {Foley},
  {Hand}, {Johnson}, {Burgett}, {Chambers}, {Draper}, {Hodapp}, {Kaiser},
  {Kudritzki}, {Magnier}, {Metcalfe}, {Bresolin}, {Gall}, {Kotak}, {McCrum}, \&
  {Smith}}]{scolnic2018the}
{Scolnic}, D.~M., {Jones}, D.~O., {Rest}, A., {et~al.} 2018, \apj, 859, 101

\bibitem[{{Secco} {et~al.}(2022){Secco}, {Samuroff}, {Krause}, {Jain},
  {Blazek}, {Raveri}, {Campos}, {Amon}, {Chen}, {Doux}, {Choi}, {Gruen},
  {Bernstein}, {Chang}, {DeRose}, {Myles}, {Fert{\'e}}, {Lemos}, {Huterer},
  {Prat}, {Troxel}, {MacCrann}, {Liddle}, {Kacprzak}, {Fang}, {S{\'a}nchez},
  {Pandey}, {Dodelson}, {Chintalapati}, {Hoffmann}, {Alarcon}, {Alves},
  {Andrade-Oliveira}, {Baxter}, {Bechtol}, {Becker}, {Brandao-Souza},
  {Camacho}, {Carnero Rosell}, {Carrasco Kind}, {Cawthon}, {Cordero}, {Crocce},
  {Davis}, {Di Valentino}, {Drlica-Wagner}, {Eckert}, {Eifler}, {Elidaiana},
  {Elsner}, {Elvin-Poole}, {Everett}, {Fosalba}, {Friedrich}, {Gatti},
  {Giannini}, {Gruendl}, {Harrison}, {Hartley}, {Herner}, {Huang}, {Huff},
  {Jarvis}, {Jeffrey}, {Kuropatkin}, {Leget}, {Muir}, {Mccullough}, {Navarro
  Alsina}, {Omori}, {Park}, {Porredon}, {Rollins}, {Roodman}, {Rosenfeld},
  {Ross}, {Rykoff}, {Sanchez}, {Sevilla-Noarbe}, {Sheldon}, {Shin}, {Troja},
  {Tutusaus}, {Varga}, {Weaverdyck}, {Wechsler}, {Yanny}, {Yin}, {Zhang},
  {Zuntz}, {Abbott}, {Aguena}, {Allam}, {Annis}, {Bacon}, {Bertin}, {Bhargava},
  {Bridle}, {Brooks}, {Buckley-Geer}, {Burke}, {Carretero}, {Costanzi}, {da
  Costa}, {De Vicente}, {Diehl}, {Dietrich}, {Doel}, {Ferrero}, {Flaugher},
  {Frieman}, {Garc{\'\i}a-Bellido}, {Gaztanaga}, {Gerdes}, {Giannantonio},
  {Gschwend}, {Gutierrez}, {Hinton}, {Hollowood}, {Honscheid}, {Hoyle},
  {James}, {Jeltema}, {Kuehn}, {Lahav}, {Lima}, {Lin}, {Maia}, {Marshall},
  {Martini}, {Melchior}, {Menanteau}, {Miquel}, {Mohr}, {Morgan}, {Ogando},
  {Palmese}, {Paz-Chinch{\'o}n}, {Petravick}, {Pieres}, {Plazas Malag{\'o}n},
  {Rodriguez-Monroy}, {Romer}, {Sanchez}, {Scarpine}, {Schubnell}, {Scolnic},
  {Serrano}, {Smith}, {Soares-Santos}, {Suchyta}, {Swanson}, {Tarle}, {Thomas},
  {To}, \& {DES Collaboration}}]{secco2022dark}
{Secco}, L.~F., {Samuroff}, S., {Krause}, E., {et~al.} 2022, \prd, 105, 023515

\bibitem[{{Seitz} \& {Schneider}(1996)}]{seitz1996cluster}
{Seitz}, S. \& {Schneider}, P. 1996, \aap, 305, 383

\bibitem[{{Sellentin} \& {Heavens}(2018)}]{sellentin2018on}
{Sellentin}, E. \& {Heavens}, A.~F. 2018, \mnras, 473, 2355

\bibitem[{{Sellentin} {et~al.}(2018){Sellentin}, {Heymans}, \&
  {Harnois-D{\'e}raps}}]{sellentin2018the}
{Sellentin}, E., {Heymans}, C., \& {Harnois-D{\'e}raps}, J. 2018, \mnras, 477,
  4879

\bibitem[{{Shirasaki} {et~al.}(2019){Shirasaki}, {Hamana}, {Takada},
  {Takahashi}, \& {Miyatake}}]{shirasaki2019mock}
{Shirasaki}, M., {Hamana}, T., {Takada}, M., {Takahashi}, R., \& {Miyatake}, H.
  2019, \mnras, 486, 52

\bibitem[{{Spergel} {et~al.}(2015){Spergel}, {Gehrels}, {Baltay}, {Bennett},
  {Breckinridge}, {Donahue}, {Dressler}, {Gaudi}, {Greene}, {Guyon}, {Hirata},
  {Kalirai}, {Kasdin}, {Macintosh}, {Moos}, {Perlmutter}, {Postman},
  {Rauscher}, {Rhodes}, {Wang}, {Weinberg}, {Benford}, {Hudson}, {Jeong},
  {Mellier}, {Traub}, {Yamada}, {Capak}, {Colbert}, {Masters}, {Penny},
  {Savransky}, {Stern}, {Zimmerman}, {Barry}, {Bartusek}, {Carpenter}, {Cheng},
  {Content}, {Dekens}, {Demers}, {Grady}, {Jackson}, {Kuan}, {Kruk}, {Melton},
  {Nemati}, {Parvin}, {Poberezhskiy}, {Peddie}, {Ruffa}, {Wallace}, {Whipple},
  {Wollack}, \& {Zhao}}]{spergel2015wide}
{Spergel}, D., {Gehrels}, N., {Baltay}, C., {et~al.} 2015, arXiv e-prints,
  arXiv:1503.03757

\bibitem[{{Spurio Mancini} {et~al.}(2023){Spurio Mancini}, {Docherty}, {Price},
  \& {McEwen}}]{mancini2022bayesian}
{Spurio Mancini}, A., {Docherty}, M.~M., {Price}, M.~A., \& {McEwen}, J.~D.
  2023, RAS Techniques and Instruments, 2, 710

\bibitem[{{Spurio Mancini} {et~al.}(2018{\natexlab{a}}){Spurio Mancini},
  {Reischke}, {Pettorino}, {Sch{\"a}fer}, \&
  {Zumalac{\'a}rregui}}]{mancini2018testing}
{Spurio Mancini}, A., {Reischke}, R., {Pettorino}, V., {Sch{\"a}fer}, B.~M., \&
  {Zumalac{\'a}rregui}, M. 2018{\natexlab{a}}, \mnras, 480, 3725

\bibitem[{{Spurio Mancini} {et~al.}(2018{\natexlab{b}}){Spurio Mancini},
  {Taylor}, {Reischke}, {Kitching}, {Pettorino}, {Sch{\"a}fer}, {Zieser}, \&
  {Merkel}}]{mancini20183d}
{Spurio Mancini}, A., {Taylor}, P.~L., {Reischke}, R., {et~al.}
  2018{\natexlab{b}}, \prd, 98, 103507

\bibitem[{{Sugiyama} {et~al.}(2022){Sugiyama}, {Takada}, {Miyatake},
  {Nishimichi}, {Shirasaki}, {Kobayashi}, {Mandelbaum}, {More}, {Takahashi},
  {Osato}, {Oguri}, {Coupon}, {Hikage}, {Hsieh}, {Komiyama}, {Leauthaud}, {Li},
  {Luo}, {Lupton}, {Murayama}, {Nishizawa}, {Park}, {Price}, {Simet},
  {Speagle}, {Strauss}, \& {Tanaka}}]{sugiyama2022hsc}
{Sugiyama}, S., {Takada}, M., {Miyatake}, H., {et~al.} 2022, \prd, 105, 123537

\bibitem[{{Taylor} {et~al.}(2019){Taylor}, {Kitching}, {Alsing}, {Wandelt},
  {Feeney}, \& {McEwen}}]{taylor2019cosmic}
{Taylor}, P.~L., {Kitching}, T.~D., {Alsing}, J., {et~al.} 2019, \prd, 100,
  023519

\bibitem[{{Tegmark} {et~al.}(1997){Tegmark}, {Taylor}, \&
  {Heavens}}]{tegmark1997karhunen}
{Tegmark}, M., {Taylor}, A.~N., \& {Heavens}, A.~F. 1997, \apj, 480, 22

\bibitem[{{Tessore} {et~al.}(2023){Tessore}, {Loureiro}, {Joachimi}, {von
  Wietersheim-Kramsta}, \& {Jeffrey}}]{tessore2023glass}
{Tessore}, N., {Loureiro}, A., {Joachimi}, B., {von Wietersheim-Kramsta}, M.,
  \& {Jeffrey}, N. 2023, The Open Journal of Astrophysics, 6, 11

\bibitem[{{Tr{\"o}ster} {et~al.}(2021){Tr{\"o}ster}, {Asgari}, {Blake},
  {Cataneo}, {Heymans}, {Hildebrandt}, {Joachimi}, {Lin}, {S{\'a}nchez},
  {Wright}, {Bilicki}, {Bose}, {Crocce}, {Dvornik}, {Erben}, {Giblin},
  {Glazebrook}, {Hoekstra}, {Joudaki}, {Kannawadi}, {K{\"o}hlinger}, {Kuijken},
  {Lidman}, {Lombriser}, {Mead}, {Parkinson}, {Shan}, {Wolf}, \&
  {Xia}}]{troester2021kids}
{Tr{\"o}ster}, T., {Asgari}, M., {Blake}, C., {et~al.} 2021, \aap, 649, A88

\bibitem[{{Tr{\"o}ster} {et~al.}(2022){Tr{\"o}ster}, {Mead}, {Heymans}, {Yan},
  {Alonso}, {Asgari}, {Bilicki}, {Dvornik}, {Hildebrandt}, {Joachimi},
  {Kannawadi}, {Kuijken}, {Schneider}, {Shan}, {van Waerbeke}, \&
  {Wright}}]{troester2022joint}
{Tr{\"o}ster}, T., {Mead}, A.~J., {Heymans}, C., {et~al.} 2022, \aap, 660, A27

\bibitem[{{Upham} {et~al.}(2021){Upham}, {Brown}, \&
  {Whittaker}}]{upham2021sufficiency}
{Upham}, R.~E., {Brown}, M.~L., \& {Whittaker}, L. 2021, \mnras, 503, 1999

\bibitem[{{van den Busch} {et~al.}(2022){van den Busch}, {Wright},
  {Hildebrandt}, {Bilicki}, {Asgari}, {Joudaki}, {Blake}, {Heymans},
  {Kannawadi}, {Shan}, \& {Tr{\"o}ster}}]{busch2022kids}
{van den Busch}, J.~L., {Wright}, A.~H., {Hildebrandt}, H., {et~al.} 2022,
  \aap, 664, A170

\bibitem[{{Virtanen} {et~al.}(2020){Virtanen}, {Gommers}, {Oliphant},
  {Haberland}, {Reddy}, {Cournapeau}, {Burovski}, {Peterson}, {Weckesser},
  {Bright}, {van der Walt}, {Brett}, {Wilson}, {Millman}, {Mayorov}, {Nelson},
  {Jones}, {Kern}, {Larson}, {Carey}, {Polat}, {Feng}, {Moore}, {VanderPlas},
  {Laxalde}, {Perktold}, {Cimrman}, {Henriksen}, {Quintero}, {Harris},
  {Archibald}, {Ribeiro}, {Pedregosa}, {van Mulbregt}, \& {SciPy 1. 0
  Contributors}}]{virtanen2020scipy}
{Virtanen}, P., {Gommers}, R., {Oliphant}, T.~E., {et~al.} 2020, Nature
  Methods, 17, 261

\bibitem[{{Wright} {et~al.}(2019){Wright}, {Hildebrandt}, {Kuijken}, {Erben},
  {Blake}, {Buddelmeijer}, {Choi}, {Cross}, {de Jong}, {Edge},
  {Gonzalez-Fernandez}, {Gonz{\'a}lez Solares}, {Grado}, {Heymans}, {Irwin},
  {Kupcu Yoldas}, {Lewis}, {Mann}, {Napolitano}, {Radovich}, {Schneider},
  {Sif{\'o}n}, {Sutherland}, {Sutorius}, \& {Verdoes Kleijn}}]{wright2019kids}
{Wright}, A.~H., {Hildebrandt}, H., {Kuijken}, K., {et~al.} 2019, \aap, 632,
  A34

\bibitem[{{Wright} {et~al.}(2020){Wright}, {Hildebrandt}, {van den Busch}, \&
  {Heymans}}]{wright2020photometric}
{Wright}, A.~H., {Hildebrandt}, H., {van den Busch}, J.~L., \& {Heymans}, C.
  2020, \aap, 637, A100

\bibitem[{{Xavier} {et~al.}(2016){Xavier}, {Abdalla}, \&
  {Joachimi}}]{xavier2016improving}
{Xavier}, H.~S., {Abdalla}, F.~B., \& {Joachimi}, B. 2016, \mnras, 459, 3693

\bibitem[{{Zablocki} \& {Dodelson}(2016)}]{zablocki2016extreme}
{Zablocki}, A. \& {Dodelson}, S. 2016, \prd, 93, 083525

\bibitem[{{Zaldarriaga} \& {Seljak}(1997)}]{zaldarriaga1997all}
{Zaldarriaga}, M. \& {Seljak}, U. 1997, \prd, 55, 1830

\bibitem[{{Zieser} \& {Merkel}(2016)}]{zieser2016the}
{Zieser}, B. \& {Merkel}, P.~M. 2016, \mnras, 459, 1586

\bibitem[{{Zonca} {et~al.}(2019){Zonca}, {Singer}, {Lenz}, {Reinecke},
  {Rosset}, {Hivon}, \& {Gorski}}]{zonca2019healpy}
{Zonca}, A., {Singer}, L., {Lenz}, D., {et~al.} 2019, The Journal of Open
  Source Software, 4, 1298

\bibitem[{{Zuntz} {et~al.}(2015){Zuntz}, {Paterno}, {Jennings}, {Rudd},
  {Manzotti}, {Dodelson}, {Bridle}, {Sehrish}, \&
  {Kowalkowski}}]{zuntz2015cosmosis}
{Zuntz}, J., {Paterno}, M., {Jennings}, E., {et~al.} 2015, Astronomy and
  Computing, 12, 45

\bibitem[{{Zuntz} {et~al.}(2018){Zuntz}, {Sheldon}, {Samuroff}, {Troxel},
  {Jarvis}, {MacCrann}, {Gruen}, {Prat}, {S{\'a}nchez}, {Choi}, {Bridle},
  {Bernstein}, {Dodelson}, {Drlica-Wagner}, {Fang}, {Gruendl}, {Hoyle}, {Huff},
  {Jain}, {Kirk}, {Kacprzak}, {Krawiec}, {Plazas}, {Rollins}, {Rykoff},
  {Sevilla-Noarbe}, {Soergel}, {Varga}, {Abbott}, {Abdalla}, {Allam}, {Annis},
  {Bechtol}, {Benoit-L{\'e}vy}, {Bertin}, {Buckley-Geer}, {Burke}, {Carnero
  Rosell}, {Carrasco Kind}, {Carretero}, {Castander}, {Crocce}, {Cunha},
  {D'Andrea}, {da Costa}, {Davis}, {Desai}, {Diehl}, {Dietrich}, {Doel},
  {Eifler}, {Estrada}, {Evrard}, {Fausti Neto}, {Fernandez}, {Flaugher},
  {Fosalba}, {Frieman}, {Garc{\'\i}a-Bellido}, {Gaztanaga}, {Gerdes},
  {Giannantonio}, {Gschwend}, {Gutierrez}, {Hartley}, {Honscheid}, {James},
  {Jeltema}, {Johnson}, {Johnson}, {Kuehn}, {Kuhlmann}, {Kuropatkin}, {Lahav},
  {Li}, {Lima}, {Maia}, {March}, {Martini}, {Melchior}, {Menanteau}, {Miller},
  {Miquel}, {Mohr}, {Neilsen}, {Nichol}, {Ogando}, {Roe}, {Romer}, {Roodman},
  {Sanchez}, {Scarpine}, {Schindler}, {Schubnell}, {Smith}, {Smith},
  {Soares-Santos}, {Sobreira}, {Suchyta}, {Swanson}, {Tarle}, {Thomas},
  {Tucker}, {Vikram}, {Walker}, {Wechsler}, {Zhang}, \& {DES
  Collaboration}}]{zuntz2018dark}
{Zuntz}, J., {Sheldon}, E., {Samuroff}, S., {et~al.} 2018, \mnras, 481, 1149

\end{thebibliography}

\begin{appendix}
\section{Numerical non-Limber integration} \label{appendix:levin}
Levin's method \citep{levin1996fast} casts a quadrature problem of an oscillatory integral into the solution of a system of ordinary and linear differential equations. Assuming a set basis for the solution the solution to the set of differential equations can be cast into a simple linear algebra problem. It has been used in a variety of cosmic shear applications \citep{zieser2016the, mancini2018testing, mancini20183d, baleato2023the}. The method relies on integrals of the type:
\begin{align}\label{eq:i1}
I [h(x,k)] = \int_{x_1}^{x_2} \mathrm{d} x \,  \langle \boldsymbol{f} \left( x, k \right), \boldsymbol{w}(k,x)\rangle \;,
\end{align}
where $\langle\cdot,\cdot\rangle$ denotes a scalar product, which, in this case, is taken between the vector of non-oscillatory functions $\boldsymbol{f}$ and the oscillatory part $\boldsymbol{w}$, which has to satisfy:
\begin{equation}
    \frac{\mathrm{d}\boldsymbol{w}(x)}{\mathrm{d}x} = \boldsymbol{A} \boldsymbol{w}(x)\,,   
\end{equation}
with a matrix $\boldsymbol{A}$. Therefore any oscillatory function with recursion relations is very well suited for this formalism.

The integral in Eq.~\ref{eq:i1} is approximated by finding a vector $\boldsymbol{p}$ such that:
\begin{align}
\left\langle {\bf p}, {\bf w} \right\rangle' = \left\langle {\bf p}' + {\sf A}^T{\bf p}, {\bf w} \right\rangle \approx \left\langle {\bf F}, {\bf w} \right\rangle\,.
\end{align}
Thus
$\left\langle {\bf p}, {\bf w} \right\rangle' = \left\langle {\bf F}, {\bf w}\right\rangle$, at $n$ collocation points $x_j , j = 1, 2, . . . , n$, is subject to the following set of equations:
\begin{align}
  \left\langle {\bf p}' + {\sf A}^T {\bf p} - {\bf F}, {\bf w} \right\rangle (x_j) = 0, \quad j = 1, ..., n\,.
\end{align}
Here the trivial solution is the null vector:
\begin{align}\label{zeros}
{\bf p}'(x_j) + {\sf A}^T (x_j) {\bf p}(x_j) = {\bf F}(x_j)\,.
\end{align}
In a last step a solution is constructed for ${\bf p}$ using $n$ differentiable basis functions $u_m(x)$ that ${\bf p}$ can be expanded in:
\begin{align}
p_i(x) = \sum_m c_i^{(m)} u_m(x), \quad i= 1,...,d; \quad m = 1,...,n\,.
\end{align}
Implying:
\begin{align}\label{740}
  \sum_m c_i^{(m)}u_m'(x_j) + \sum_{m,q}A_{q i} c_q^{(m)} u_m(x_j) = F_i(x_j)\,,
\end{align}
where $i,q = 1, ...,d; j,m = 1, ...,n$\,. 
Any basis is suitable, here we chose equidistant collocation points $x_j$
and the $n$ lowest-order polynomials as basis functions. Finally, the vector ${\bf w}$ depends on the specific type of integral considered. Equation~\ref{eq:method:fs:nonlimber:non_limber} allows for two possibilities: $(i)$ first  integrate over the co-moving distance $\chi$: $(ii)$ first integrate over the wave-vector $k$. Both approaches have advantages and disadvantages: due to the separation of the $\chi$ and $\chi^\prime$ dependence, integrating over the co-moving distance first will be quicker. However, the residual integral over $k$ can still be very oscillatory and must be sampled accordingly, especially with narrow window functions $W(\chi)$. The second option is to integrate over $k$ first, this smooths out the oscillations from the Bessel function. However, there are two remaining integrals, thus slowing down the computation, especially for many radial bins. If, however, the radial bins do not overlap, the resulting spectrum for bins further apart than a single bin index will be very small and can be neglected. A more detailed discussion on this is provided in \citet{leonard2023the}.
For integrals of the type
\begin{equation}
\begin{split}
    I_{\ell,k} = &\; \int \mathrm{d}x f_k(x) j_{\ell}(kx)\;, \\
    I_{\ell,k_1,k_2} = &\; \int \mathrm{d}x f_{k_1,k_2}(x) j_{\ell}(k_1x)j_{\ell}(k_2x)\,,
    \end{split}
\end{equation}
we chose
\begin{equation}
\begin{split}
        w (x) = & \;\begin{pmatrix}
        j_{\ell}(xk) \\
        j_{\ell + 1}(xk)
    \end{pmatrix} \;, \; \; \boldsymbol{A}= 
    \begin{pmatrix}
        \frac{\ell}{x} & -k \\
        k & -\frac{\ell + 2}{x}
    \end{pmatrix} \,; \\
    w (x) =&\;\begin{pmatrix}
        j_{\ell}(xk_1)j_{\ell}(xk_2) \\
        j_{\ell+1}(xk_1)j_{\ell}(xk_2) \\
        j_{\ell}(xk_1)j_{\ell+1}(xk_2) \\
        j_{\ell+1}(xk_1)j_{\ell+1}(xk_2) 
    \end{pmatrix} \;,\; \; \boldsymbol{A}= 
    \begin{pmatrix}
        \frac{2\ell}{x} & -k_1 & -k_2 & 0  \\
        k_1 & -\frac{2}{x} & 0 &  - k_2 \\
        k_2 & 0 & -\frac{2}{x} & -k_1 \\
        0 &  k_2 & k_1 & -\frac{2\ell +2}{x}
    \end{pmatrix}
    \end{split}\,,
\end{equation}
respectively.

\section{Accuracy of the cosmic shear signal}\label{appendix:consistency}
\begin{figure}
    \centering
    \includegraphics[width=9cm]{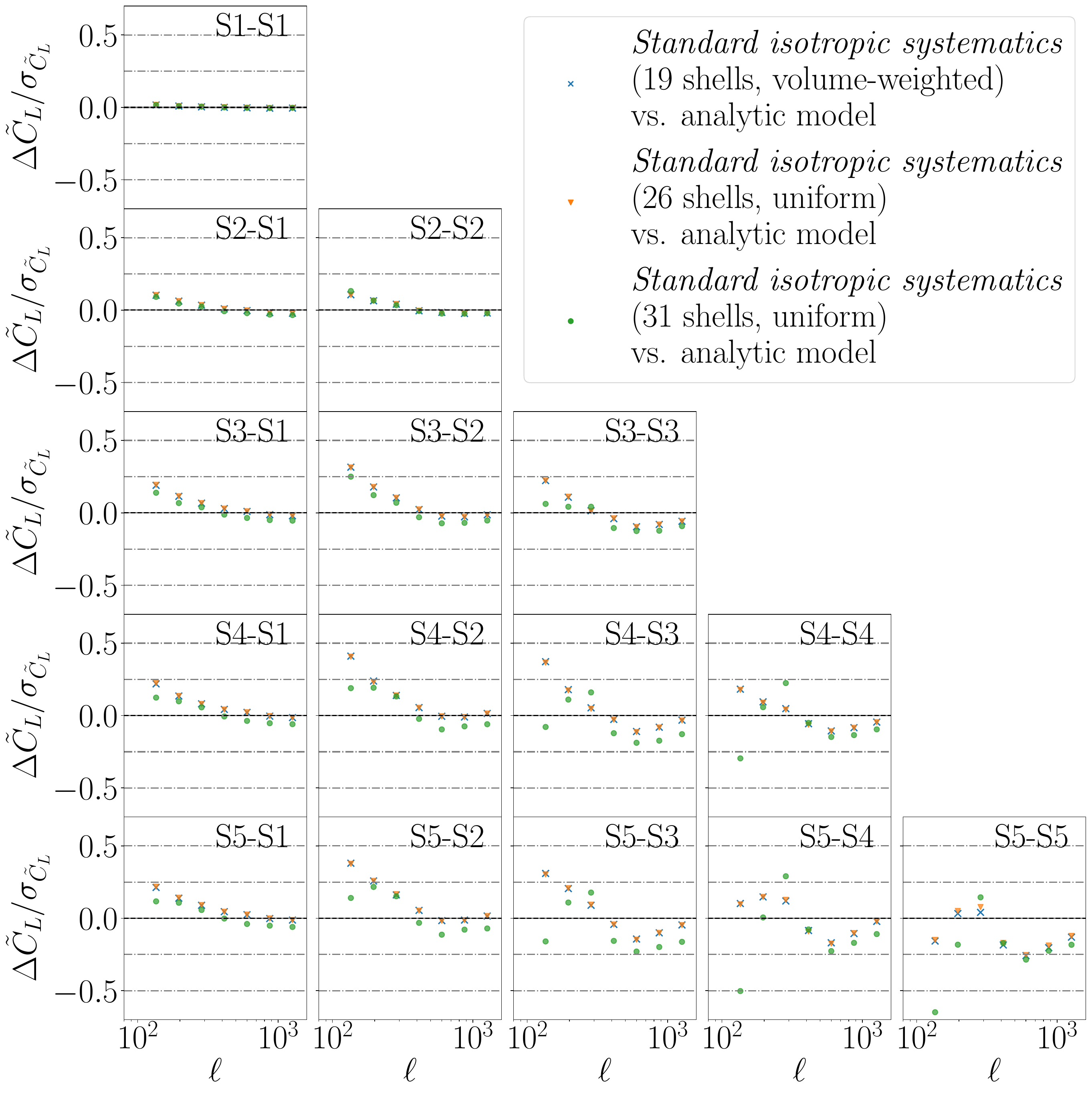}
    \caption{Relative difference between the mean pseudo-Cls measured from the standard isotropic systematics model in {KiDS-SBI} with different line-of-sight resolutions and the expectation from analytical theory as described in Appendix~\ref{appendix:signal}. The blue crosses show the standard isotropic systematics model with the resolution assumed in the analysis discussed in this work: 19 shells with a comoving-volume weighted kernel. The orange triangles and the green circles show difference in the signal of standard isotropic systematics model with a uniform kernel and at higher resolutions: 26 shells and 31 shells, respectively. All are evaluated at the MAP cosmology inferred from the KiDS-1000 data using the standard isotropic systematics in {KiDS-SBI} (see Table~\ref{tab:sbi_full_results}).}
    \label{fig:robustness}
\end{figure}

To test the robustness of the cosmic shear signal as predicted by the log-normal simulations generated in {KiDS-SBI} in addition to the self-consistency tests already conducted within Sect.~\ref{fs}, we perform a comparison of the noise-free pseudo-Cls as measured in {KiDS-SBI} assuming the standard isotropic systematics model (see Sects.~\ref{method:fs:pcl} and~\ref{validation:impact_vd}) to the pseudo-Cls predicted from analytical theory (see Appendix~\ref{appendix:signal}). This comparison is shown for the \textit{KiDS-SBI} MAP cosmology in Fig.~\ref{fig:robustness}. When assuming 19 shells and a comoving-volume weighted kernel as discussed in Sect.~\ref{method:fs:shells}, we find per-cent or sub-percent level agreement for all tomographic bin combinations and scales (within $0.5 \sigma$ or less). The agreement between the models is also shown to be good in Fig.~\ref{fig:sbi_mcmc_v_sbi} where the posterior from the SBI analysis based on {KiDS-SBI} is unbiased with respect to the posterior sampled with an MCMC while assuming analytical theory. Overall, we find that for small $\ell$ the forward simulations predict additional power with respect to the pseudo-Cls predicted from theory. This can be explained by discretisation effects along the line of sight due to the division of the observable Universe into concentric shells weighted by a window function (see Sect.~\ref{method:fs:shells}), and by the linear interpolation of convergence and shear across \texttt{HEALPix} pixels which are built into \texttt{SALMO}. Both effects have been shown to produce a similar minor excess in angular power in testing conducted in \citet{tessore2023glass}.

When increasing the resolution along the line of sight by increasing the number of shells, and when changing the kernel to a uniform window function as suggested in \citet{tessore2023glass}, we find that for 26 shells the agreement with the signal predicted by analytical theory is marginally better, but not enough to impact the SBI. When increasing the number of shells further to 31 shells, such that each shell has a width of ${\sim}100$ Mpc $h^{-1}$, the agreement with analytical theory is poorer for large $\ell$ and for low $\ell$ in the autocorrelations, as discretisation effects become important. Consequently, we conclude that the choice of 19 shells allows us to carry out an accurate forward modelling of the cosmic shear signal in \texttt{KiDS-SBI}, while still allowing for computational efficiency (see Sect.~\ref{method:sbi:performance}). 

These effects are not significant enough to bias the cosmological inference at the precision of KiDS-1000, particularly, since they are more prevalent in tomographic bin combinations with a lower signal-to-noise ratio. However, future simulation-based inference analyses of galaxy surveys with smaller uncertainties (for instance stage-IV surveys) based on similar simulations should aim to account for such effects.

Moreover, Fig.~\ref{fig:sbi_mcmc_v_sbi} shows that in the bin combinations at higher redshifts (with the highest signal-to-noise ratio) the angular power in the high $\ell$ modes tends to be slightly underestimated by the standard isotropic systematics model when compared to the analytical theory. We find that this is driven by more cosmic shear signal leaking from the EE modes to the BB modes than predicted by the mixing matrix discussed in Appendix~\ref{appendix:signal}. This should not bias the SBI analysis, since the KiDS-1000 data and the forward simulations share the same measurement procedure, so mode mixing effects should be intrinsically ingrained in both the real and the forward-modelled data vectors.

\section{Effects of variable depth and shear biases}\label{method:consistency:var_depth}
\begin{figure}
    \centering
    \includegraphics[width=7cm]{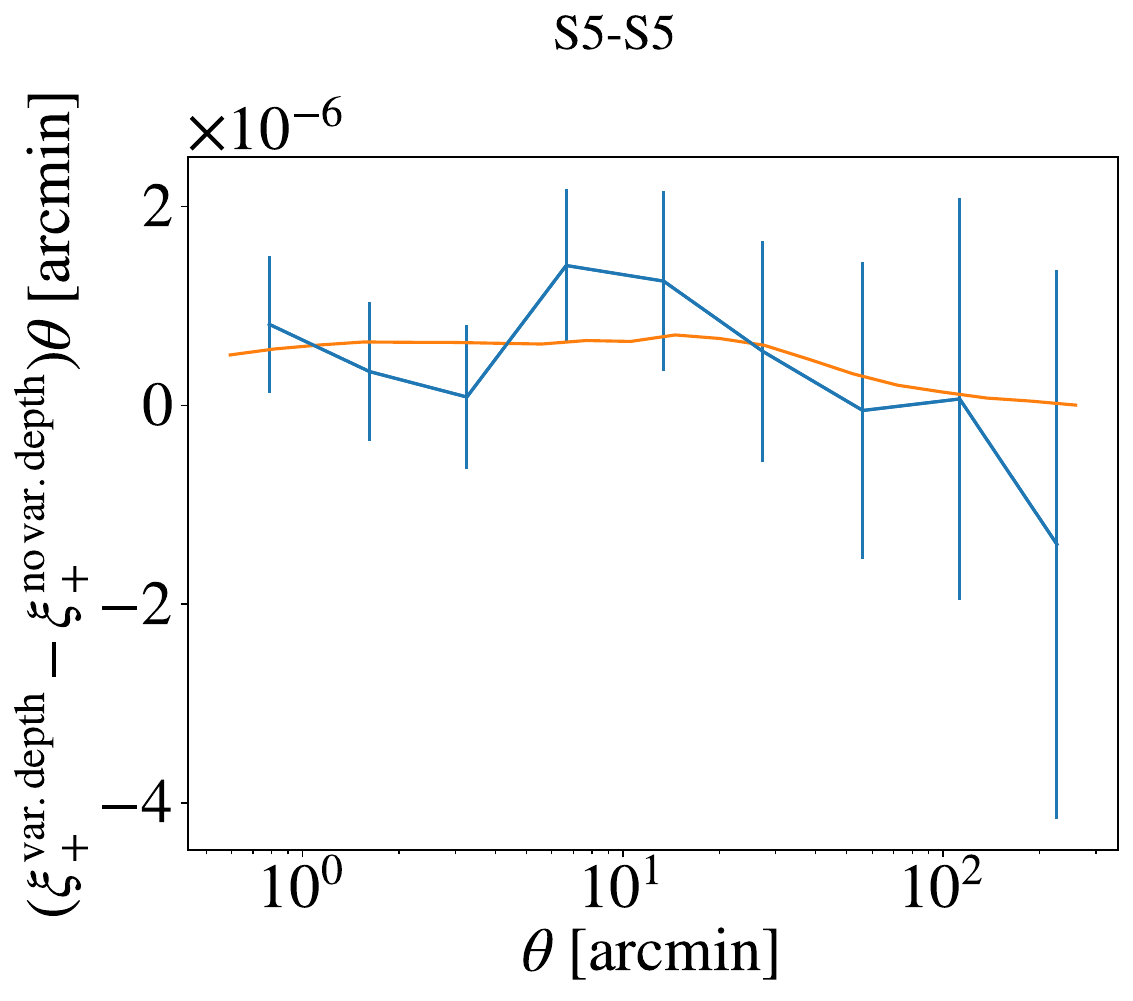}
    \caption{The impact of variable depth in KiDS-1000 on the cosmic shear real-space two-point correlation functions, $\xi(\theta)$ in the autocorrelation of the S5 bin. The blue line shows the difference in the mean signal from 5,000 realisations of the anisotropic systematics model with variable depth versus the mean signal from the same number of realisations when assuming that the observational depth is isotropic. The orange line shows the same difference in signal as predicted for KiDS-1000 by the analytical model in \citet{heydenreich2020the}. We note that variable depth constitutes approximately $1\%$ of the measured cosmic shear signal in KiDS-1000.}
    \label{fig:vd_theory_vs_vd_fs}
\end{figure}

To quantify the physical effect of the inclusion of variable depth and shear biases into the anisotropic systematics model as discussed in Sect.~\ref{method:fs:salmo_vd} and~\ref{method:fs:salmo_shears}, we isolate the bias each systematic causes in the observed two-point cosmic shear signal. First, we forward model variable depth in KiDS-1000 as described in Sect.~\ref{method:fs:salmo_vd}, and compare it to a set of forward simulations where the galaxy density, intrinsic galaxy shape dispersion and the photometric redshift distributions are isotropic for each tomographic bin, as is the case in the standard isotropic systematics model. When comparing the cosmic shear signal through the spatial two-point correlations functions (2PCF) measured from \texttt{TreeCorr} \citep{jarvis2004the} in Fig.~\ref{fig:vd_theory_vs_vd_fs}, we find that variable depth adds additional two-point correlation signal at all scales and particularly at scales where $\theta < 100 \, \mathrm{arcmin}$. In this case, we measure the cosmic shear signal with 2PCF instead of pseudo-Cls as it simplifies the comparison of the results with previous work. This additional signal is equivalent to about $1\%$ (${\sim}0.2\sigma$) additional non-cosmological shear signal which is in agreement with predictions made for KiDS-1000 from \citet{heydenreich2020the}. The effect is similar, but slightly smaller in the lower redshift tomographic bin combinations of KiDS-1000 where the shear signal is weaker.

\begin{figure}
    \centering
    \includegraphics[width=8cm]{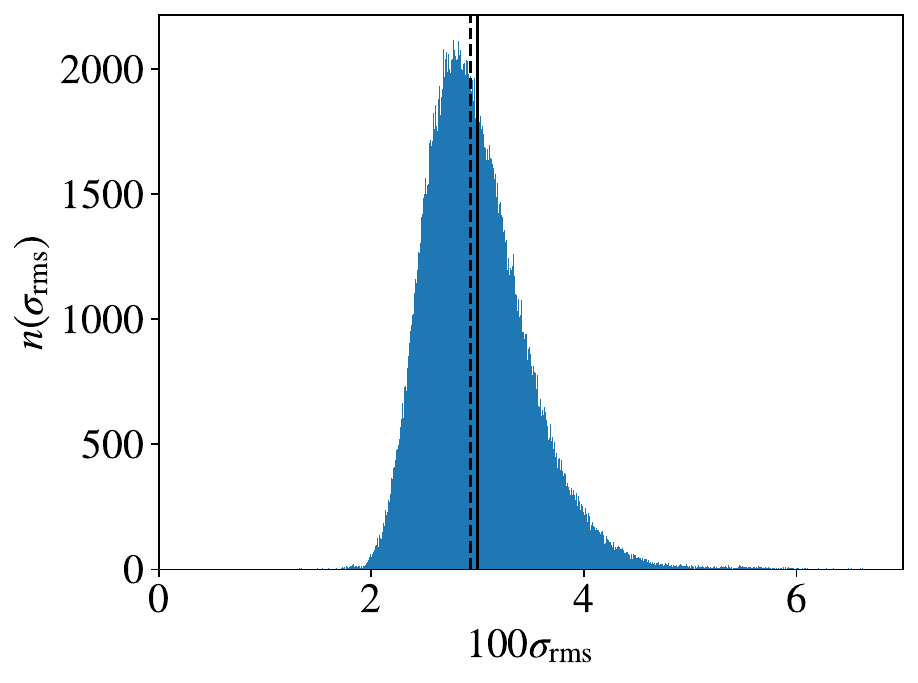}
    \caption{Histogram showing the distribution of the pixelised root-mean-square of the background noise, $\sigma_{\mathrm{rms}}$, within the KiDS-1000 footprint shown in Fig.~\ref{fig:level_map}. The solid black line shows the mean value of $\sigma_{\mathrm{rms}}$, while the dashed black line indicates the median of the distribution. This shows that the majority of the pixels in the KiDS-1000 footprint have lower $\sigma_{\mathrm{rms}}$ than the mean which is realised in the case of isotropic depth, i.e. in the standard isotropic systematics model.}
    \label{fig:level_hist}
\end{figure}

 Additionally, when modelling variable depth, the shape noise becomes anisotropic across the sky as a function of the variations in the effective galaxy density and the intrinsic galaxy shape dispersion. When considering a Gaussian likelihood, this implies that the shape noise scales with $\sigma_{\epsilon}^2/n_{\mathrm{eff}}$ \citep{kaiser1992galactic, kaiser1998weak}. As discussed in Sect.~\ref{method:fs:salmo_vd} both quantities scale linearly with the variable depth measure, $\sigma_{\mathrm{rms}}$, which implies that the shape noise in a given tomographic bin either scales with $\sigma_{\mathrm{rms}}^{3}$ in S1, S2 and S5, or with $\sigma_{\mathrm{rms}}^{-1}$ in S3 and S5. Since in KiDS-1000 the distribution of $\sigma_{\mathrm{rms}}$ is skewed below the mean (see Fig.~\ref{fig:level_hist}), variable depth in fact reduces the overall contribution of the shape noise to the uncertainty of the cosmic shear signal. Nevertheless, as shown in  Figs.~\ref{fig:level_map} and~\ref{fig:level_map_zoom}, variable depth also causes additional non-cosmological angular correlations in the data at the scales of pixels, pointing overlaps, whole pointings or even over the whole footprint, while also changing the signal along the line of sight anisotropically as shown in Fig.~\ref{fig:nofz_variation}. These additional correlations can significantly alter the covariance of the two-point statistic or may even lead to non-Gaussian noise. Such correlations can exacerbate the uncertainty contributions from cosmic variance and super-sample covariance, which causes the noise predicted in KiDS-1000 by a model including variable depth to ultimately exceed the noise predicted by a model which does not consider the systematic \citep{reischke2024kids}).

\begin{figure}
    \centering
    \includegraphics[width=9cm]{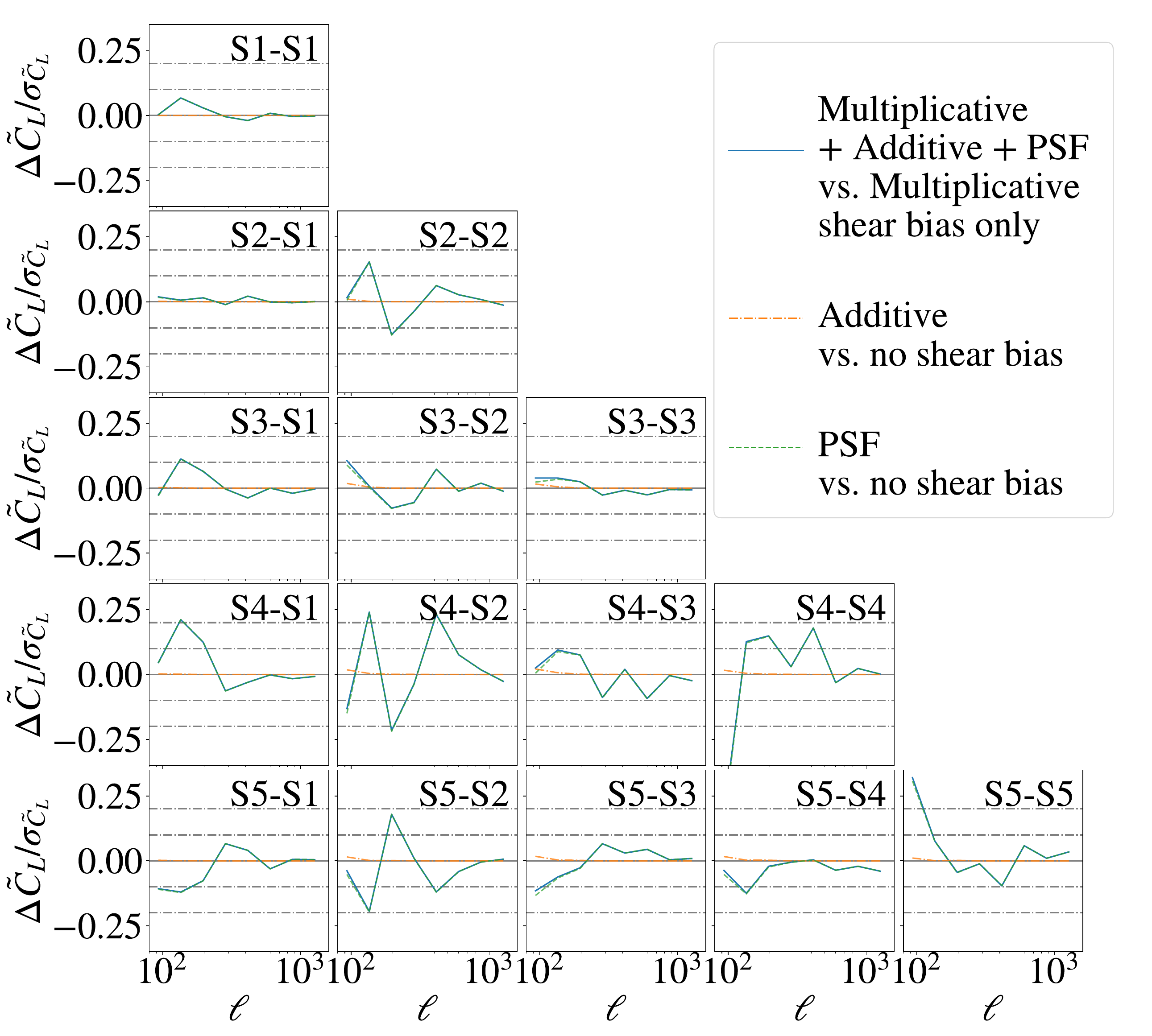}
    \caption{Relative difference between the mean pseudo-Cls measured in {KiDS-SBI} for different combinations of shear biases in KiDS-1000 as described in Sect.~\ref{method:fs:salmo_shears}. The blue solid line compares the pseudo-Cls of a model that adds multiplicative, additive and PSF shear biases (as is the case in the anisotropic systematics model) to the signal measured from a model which only considers multiplicative shear bias (i.e. the standard isotropic systematics model). The orange dot-dashed line shows the effect of the additive shear bias alone, while the green dashed line shows the effect of the PSF shear bias alone. We note that, in particular, the PSF shear bias causes an up to $0.3 \sigma$ bias in the pseudo-Cls, especially, at large scales.}
    \label{fig:shear_biases}
\end{figure}

When isolating the effect of additive and PSF variation shear bias on the observed cosmic shear signal as described in Sect.~\ref{method:fs:salmo_shears} in Fig.~\ref{fig:shear_biases}, we find that, while the additive shear bias has a negligible effect on the cosmic shear signal, the shear bias originating from PSF variations can bias the pseudo-Cls by up to $0.3\sigma$ particularly at large scales. Hence, the inclusion of the PSF shear bias can contribute a percentage level non-cosmological signal on certain scales.

\section{Theoretical signal modelling} \label{appendix:signal}
To provide consistency checks, we also theoretically model the expected signal for a given mock. We model the shape-shape angular power spectrum, $C_{\epsilon \epsilon}^{(pq)}(\ell) $, as follows

\begin{equation}
    C_{\epsilon \epsilon}^{(pq)}(\ell) = C_{\mathrm{g g}}^{(pq)}(\ell) + C_{\mathrm{gI}}^{(pq)}(\ell) + C_{\mathrm{Ig}}^{(pq)}(\ell) + C_{\mathrm{II}}^{(pq)}(\ell)\,,
\end{equation}
\noindent where $\epsilon$ indicates correlations with a galaxy shape field, $\mathrm{g}$ indicates correlations with a cosmic shear field, and $\mathrm{I}$ indicates correlations between galaxy shapes due to intrinsic alignments. To model each of these terms, we calculate the matter power spectrum using \texttt{CAMB} and proceed to make a Limber projection to obtain the following expression
\begin{equation}
    C_{a b}^{(p q)}(\ell) = \int_{0}^{\infty}  \frac{d\chi}{f_{k}^2(\chi)} \, W_{a}^{(p)}(\chi) \, W_{b}^{(q)}(\chi) \, P_{\delta} \bigg(\frac{\ell +1/2}{f_{k}(\chi)}, \chi \bigg)\,,
\end{equation}
\noindent where $W_{a}$ is the kernel for a given field, such that $a,b \in \{\mathrm{g}, \mathrm{I}\}$. For cosmic shear, the kernel is given by 
\begin{equation}
     W_{\mathrm{g}}^{(p)}(\chi) = 
        \frac{3 H_{\mathrm{0}}^{2} \Omega_{\mathrm{m}}}{2}
        \frac{f_{{k}}(\chi)}{a(\chi)}
        \int^{\chi_{\mathrm{hor}}}_{\chi} \mathrm{d}\chi'
        n_{\mathrm{S}}^{(p)}(\chi') \,
        \frac{\mathit{f}_{{k}}(\chi' - \chi)}{\mathit{f}_{{k}}(\chi')}\,,
\end{equation}
\noindent where $n_{\mathrm{S}}^{(p)}$ is the redshift distribution of the objects in $p^{th}$ tomographic bin as shown in Fig.~\ref{fig:kids_nz}. For consistency with the forward modelling described in Sect.~\ref{method:fs:nla}, we chose an intrinsic alignment kernel in accordance with the non-linear alignment model (NLA; \citealt{catelan2001intrinsic, hirata2004galaxy, bridle2007dark}) given by
    \begin{equation}
        W_{\mathrm{I}}^{(p)}(\chi) =
        - A_{\mathrm{IA}}
        \frac{C_{1} \overline{\rho}_{\mathrm{cr}}(\chi) \, \Omega_{\mathrm{m}}}{D(\chi)}
        \, n_{\mathrm{S}}^{(p)}(\chi)\,.
    \end{equation}

Knowing the full-sky angular power spectra, $C^{(pq)}_{\epsilon \epsilon, \nu}(\ell'; \mathbf{\Theta})$, we can define the pseudo-Cl, $\tilde{C}^{(pq)}_{\epsilon \epsilon, \mu}(\ell; \mathbf{\Theta})$, as follows
\begin{multline}
    \tilde{C}^{(pq)}_{\epsilon \epsilon, \mu}(\ell; \mathbf{\Theta}) = \sum_{\ell' = 0} ^{\ell'_{\mathrm{max}}} \sum_{\ell'' = 0} ^{\ell''_{\mathrm{max}}} \sum_{\nu = 1} ^{3} \sum_{\nu' = 1} ^{3} M_{\mu \nu', \ell \ell''}
    M'^{(pq)}_{\nu' \nu, \ell'' \ell'} C^{(pq)}_{\epsilon \epsilon, \nu}(\ell'; \mathbf{\Theta})\,,
    \label{eq:method:fs:pcl:mixing}
\end{multline}

\noindent where $\mu, \nu, \nu' \in \{1, 2, 3\}$ such that $1$ stands for the EE component, $2$ for the BB component and $3$ for the EB component, $M_{\ell \ell''}$ is the mixing matrix of the survey mask, $\mathrm{W}(\theta)$, which does not vary between tomographic bins, and $M'^{(pq)}_{\ell'' \ell'}$ is the mixing matrix of the effective mask imposed by the average random variations in the observed galaxies in a given tomographic bin over many realisations.

\begin{figure*}
    \centering\includegraphics[trim=3cm 30cm 1.5cm 30cm,clip,width=15.5cm]{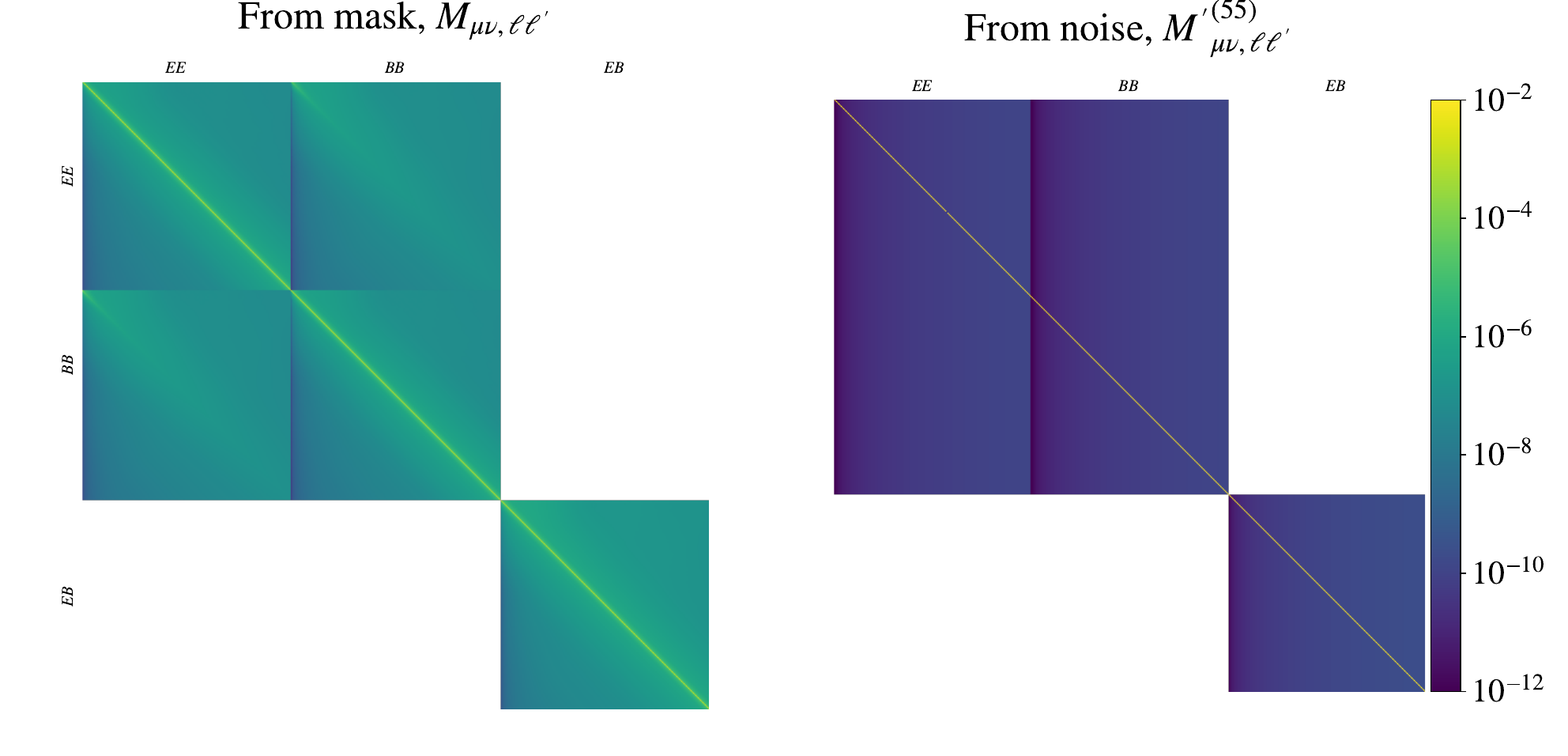}
    \caption{Bitmap of the mixing matrices to model the theoretical signal for the pseudo-Cls as seen by KiDS-1000. On the right panel is the mixing matrix derived from the KiDS-1000 mask, $M_{\mu \nu', \ell \ell''}$, (the input mask is at a resolution of $N_{\mathrm{side}} = 1024$). On the left panel is the mixing matrix caused by the selection due to randomly sampling galaxies for the auto-correlation of the fifth tomographic bin, $M'^{(55)}_{\nu' \nu, \ell'' \ell'}$. Both matrices are decomposed into block matrices separating the $EE \rightarrow EE$, $EE \rightarrow BB$ and $EB \rightarrow EB$ mixing. Each block matrix has dimensions of $8193 \times 8193$ with $\ell \in \{\ell \in \mathbb{Z}^{0+}| \ell \leq 8192\}$. }
    \label{fig:mixmat}
\end{figure*}

The first mixing matrix from the left in Eq.~(\ref{eq:method:fs:pcl:mixing}), $M_{\ell \ell''}$, is the standard block mixing matrix to account for the partial sky coverage in a survey \citep{hivon2002master, bennett2003first, brown2005cosmic}. The second mixing matrix, $M'^{(pq)}_{\ell'' \ell'}$, accounts for an additional effect: the mode mixing induced by a mask of uncorrelated noise originating from the random sampling described in Sect.~\ref{method:fs:salmo}. In other words, when randomly Poisson sampling galaxies within a discrete pixel $k$ given a probability of $P^{(m)}(N)$, there will be a non-zero probability of observing zero counts in that given pixel. In essence, all the pixels which are sampled to have zero galaxy counts impose an additional masking on our sample. This can become significant at the resolutions and galaxy densities which we probe in this work. For example, for an \texttt{HealPix} $N_{\mathrm{side}}= 2048$ and a galaxy density of $n=1 \,\mathrm{arcmin}^{-2}$ for a given tomographic bin $k$ (similar to KiDS-1000), $P^{(m)}(N=0) \approx 0.07$. This means that nearly a tenth of all pixels could be 'masked' in a given realisation of simulations at a higher resolution. To allow for these cases, we decide to also model the mode mixing induced by the sampling variance. However, for the settings discussed in this paper for the main pipeline of forwards simulations where $N_{\mathrm{side}}= 1024$, this effect is not so relevant. Even for the sparsest source bin, S1, with an $n_{\mathrm{gal}}^{(1)} = 0.62 \, \mathrm{\mathrm{arcmin}}^{-2}$, $P^{(m)}(N=0) \approx 0.0005$, so only a negligible fraction of pixels are masked due to random sample variance.

To provide a general overview, the distinguishing factor between the two mixing matrices is how we define the mask $\mathrm{W}(\theta)$ which characterises them. For $ M_{\mu \nu, \ell \ell''} $, $\mathrm{W}(\theta)$ is defined as the mask of the survey footprint, $\Omega_{\mathrm{survey}}$, such that
\begin{equation}
    \mathrm{W}(\theta) = \begin{cases} 1, & \theta \in \Omega_{\mathrm{survey}} \\ 0, & \theta \notin \Omega_{\mathrm{survey}}\end{cases}.
    \label{eq:mm:footprint}
\end{equation}

For $ M^{'(pq)}_{\mu \nu, \ell \ell''}$, rather than defining it using an explicit spatial mask from the pixels masked by each random realisation, $\mathrm{W}^{'\mathit{(p)}}(\theta)$, we can define the expectation value of the mixing matrix due to the random sampling of galaxies directly from the angular power spectrum of $\mathrm{W}^{'\mathit{(p)}}(\theta)$ as
\begin{equation}
\begin{split}
    \frac{\mathcal{W}^{(pq)'}(\ell)}{2\ell+1}= & \; \frac{1}{2\ell+1}\sum_{m=-\ell}^{\ell} w^{(p)}_{\ell m} w^{(q)*}_{\ell m}\\ = &\;
    \frac{4 \pi \sigma^{(pq)}}{N_{\mathrm{pix}}-1} + 4 \pi \Bigg[\mu^{(pq)} - \frac{\sigma^{(pq)}}{N_{\mathrm{pix}}-1} \Bigg] \delta^{\mathrm{K}}_{\ell 0},
    \label{eq:mm:sampling}
    \end{split}
\end{equation}
where $\mu^{(pq)} = P^{(p)}(N>0) P^{(q)}(N>0)$, namely the product of the mean probabilities of success within each tomographic bin (the probability that a given pixel will be populated by at least one galaxy), while $2\sigma^{(pq)2} = P^{(p)2}(N>0)(1-P^{(p)}(N>0))^2+ P^{(q)2}(N>0)(1-P^{(q)}(N>0))^2$, expressly the associated standard deviation of the mean, and $N_{\mathrm{pix}}$ is the number of pixels.

As described in Sect.~\ref{method:fs:salmo}, the galaxies are Poisson sampled according to the  probability of success is given in Eq.~(\ref{eq:method:fs:salmo:position}). As we are interested in the mean over many iterations, we can assume that the matter overdensities are small, $1+b^{(p)}\delta^{(p)} \approx 1$, so we can rewrite the probability as
\begin{equation}
\begin{split}
    P^{(p)}(N>0) =&\; 1 - P^{(p)}(N=0) \\
    =  &\;1 - e^{-\langle N_{m}^{(p)}\rangle (\mathbf{\Theta})}\\
    \approx &\;1 - e^{-n^{(p)}_{\mathrm{gal}} A_{\mathrm{pix}}}\,,
\end{split}
\end{equation}
\noindent where $n^{(p)}_{\mathrm{gal}}$ is the mean galaxy density per tomographic bin $p$ and $A_{\mathrm{pix}}$ is the mean pixel size for a given \texttt{HEALPix} resolution. From this, we can calculate both mixing matrices as shown in Fig.~\ref{fig:mixmat}. In practice, we can see that from Fig.~\ref{fig:mixmat} that the mode mixing $M^{'(pq)}_{\mu \nu, \ell'' \ell'}$ is negligible when compared to $ M_{\mu \nu, \ell \ell''} $ at an $N_{\mathrm{side}}=1024$. However, we still include the term in the signal modelling to allow the flexibility to choose higher resolutions within {KiDS-SBI}. 

With Eqs.~(\ref{eq:mm:footprint}) and (\ref{eq:mm:sampling}) defining the selections which define  $ M_{\mu \nu, \ell \ell''} $ and $M^{'(pq)}_{\mu \nu, \ell'' \ell'}$, respectively, we can compute the matrices as follows \citep{brown2005cosmic, hikage2011shear}:
\begin{equation}
    M_{\mu \nu, \ell \ell''} = \begin{pmatrix} W_{\ell \ell''}^{++} & (W_{\ell \ell''}^{-+} +W_{\ell \ell''}^{+-}) & W_{\ell \ell''}^{--} \\ -W_{\ell \ell''}^{+-} & (W_{\ell \ell''}^{++} - W_{\ell \ell''}^{--}) & W_{\ell \ell''}^{-+} \\ W_{\ell \ell''}^{--} & -(W_{\ell \ell''}^{-+} + W_{\ell \ell''}^{+-}) & W_{\ell \ell''}^{++} \end{pmatrix}\,,
    \label{eq:pcl:mm1}
\end{equation}
\begin{equation}
    M^{'(pq)}_{\mu \nu, \ell'' \ell'} = \begin{pmatrix} W_{\ell'' \ell'}^{'++} & (W_{\ell'' \ell'}^{'-+} +W_{\ell'' \ell'}^{'+-}) & W_{\ell'' \ell'}^{'--} \\ -W_{\ell'' \ell'}^{'+-} & (W_{\ell'' \ell'}^{'++} - W_{\ell'' \ell'}^{'--}) & W_{\ell'' \ell'}^{'-+} \\ W_{\ell'' \ell'}^{'--} & -(W_{\ell'' \ell'}^{'-+} + W_{\ell'' \ell'}^{'+-}) & W_{\ell'' \ell'}^{'++} \end{pmatrix}^{(pq)}\,,
    \label{eq:pcl:mm2}
\end{equation}
\noindent where each of the Wigner symbols is given by
\begin{equation}
     W_{\ell \ell'}^{\pm \pm} = \frac{1}{2\ell+ 1} \sum_{m=-\ell}^{\ell} \sum_{m'=-\ell'}^{\ell'} W_{\ell \ell' m m'}^{\pm} (W_{\ell' \ell m' m}^{\pm} )^{*}\,.
     \label{eq:pcl:wpm}
\end{equation}

Since the shear field is a spin 2 field
\begin{equation}
\begin{split}
    W_{\ell \ell' m m'}^{+} = \frac{1}{2} (_{2}W_{\ell \ell'}^{m m'} + _{-2}W_{\ell \ell'} ^{m m'})\,,\\
    W_{\ell \ell' m m'}^{-} = \frac{\mathrm{i}}{2} (_{2}W_{\ell \ell'}^{m m'} - _{-2}W_{\ell \ell'} ^{m m'})\,,
\end{split}
\end{equation}
\noindent where
\begin{equation}
    _{s}W_{\ell \ell' m m'}^{+} = \int \mathrm{d}\theta \yellmarb{s}{}{\ell' m'}(\theta) \mathrm{W}(\theta) \yellmarb{s}{*}{\ell m}(\theta)\,.
\end{equation}

The latter is difficult to compute, so it is more convenient to compute in phase space such that
\begin{multline}
        \sum_{m=-\ell}^{\ell} \sum_{m'=-\ell'}^{\ell'} \, _{s}W_{\ell \ell'}^{m m'} (_{s'}W_{\ell' \ell}^{m' m})^{*} \\ = \frac{(2\ell+1)(2\ell' + 1)}{4 \pi} \sum_{\ell''=0} ^{\ell''_{\mathrm{max}}} \mathcal{W}_{\ell''} \begin{pmatrix} \ell & \ell' & \ell'' \\ -s & s & 0 \end{pmatrix} \begin{pmatrix} \ell & \ell' & \ell'' \\ -s' & s' & 0 \end{pmatrix}\,,
\end{multline}
\noindent where $s \in \{-2, 2\}$ and
\begin{equation}
    \mathcal{W}_{\ell} = \sum_{m=-\ell}^{\ell} w_{\ell m} w_{\ell m}^{*}\,,
\end{equation}
\noindent where $w_{\ell m}$ are the coefficients of the spherical harmonic transform of the mask, $\mathrm{W}(\theta)$, such that
\begin{equation}
    w_{\ell m} = \int \mathrm{d}\theta \, \mathrm{W}(\theta)\yellmarb{}{*}{\ell m}(\theta)
    \label{eq:pcl:wlm}\,.
\end{equation}

\begin{figure*}
    \centering
    \includegraphics[width=14.5cm]{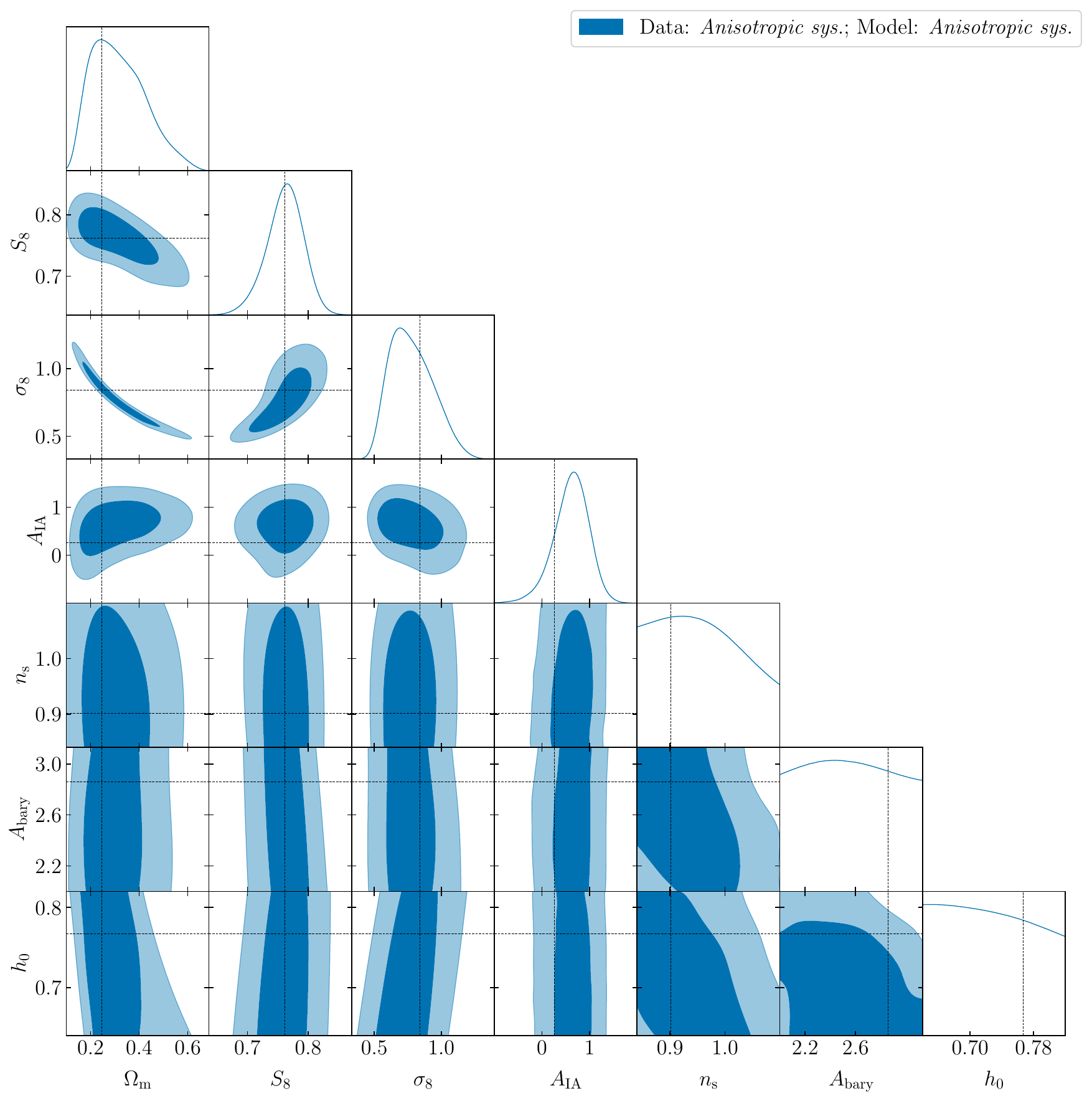}
    \caption{Posterior contours of the seven cosmological and astrophysical parameters which are varied given the anisotropic systematics model within {KiDS-SBI} over the prior space shown in Table~\ref{tab:sbi_priors} when analysing a mock data vector from the same model. The black solid lines indicate the true cosmology of the input mock data vector generated from the anisotropic systematics model while adding noise. All the aforementioned values are shown in Table~\ref{tab:sbi_validation_truth}. These posteriors are obtained from training neural density estimators in \texttt{DELFI} \citep{alsing2019fast} on 18,000 realisations of the forward simulations assuming the anisotropic systematics model, in line with the choices made in \citetalias{lin2022a}. The posterior is obtained from the combined posteriors of six independent conditional Masked Autoregressive Flows (MAFs) each is made up of three to eight Masked Autoencoder for Density Estimations (MADEs) each with two hidden layers of 50 neurons.}
    \label{fig:sbi_validation_posterior}
\end{figure*}

\section{Parameter recovery}\label{validation:paramter_recovery}

Figure~\ref{fig:sbi_validation_posterior} shows the posterior distribution of the seven main parameters varied in the analysis as learnt from 18,000 realisations of the anisotropic systematics model. The data vector used for this posterior is a noisy mock data vector generated with the anisotropic systematics model while assuming the input parameters given in Table~\ref{tab:sbi_priors}. We find that the posteriors from {KiDS-SBI} accurately recover the true input parameters (see Table~\ref{tab:sbi_validation_truth} for the inferred parameter values). As shown in Fig.~\ref{fig:sbi_validation_posterior}, the anisotropic systematics posterior mostly only constrains $S_{8}$ and $A_{\mathrm{IA}}$. At the same time, it becomes apparent that the posterior distributions over $\omega_{\mathrm{b}}$, $n_{\mathrm{s}}$, $A_{\mathrm{bary}}$, and $h_{0}$ are mainly prior driven, namely, the posterior distribution is mostly flat throughout prior space. As the cosmic shear signal is largely degenerate to small changes in these parameters, this is expected.

\begin{table}
    \centering
    \caption{Main inferred cosmological and astrophysical parameters from mocks generated with the anisotropic systematics model.}
    \begin{tabular}{cccc}
        \hline
        Parameter             & Mock truth & Marginal $\pm 1 \sigma$      & MAP $\pm$ PJ-HPD          \\ \hline \\[-0.25cm]
        $S_{8}$               & 0.756      & $0.752^{+0.034}_{-0.029}$    & $0.752^{+0.029}_{-0.031}$ \\[0.15cm]
        $\sigma_{8}$          & 0.706      & $0.720^{+0.093}_{-0.19}$     & $0.613^{+0.096}_{-0.190}$ \\[0.15cm]
        $\Omega_{\mathrm{m}}$ & 0.344      & $0.36^{+0.10}_{-0.14}$       & $0.451^{+0.168}_{-0.140}$ \\[0.15cm]
        $h_0$                 & 0.657      & $< 0.747                   $ & $0.672^{+0.023}_{-0.078}$ \\[0.15cm]
        $\omega_{\mathrm{c}}$ & 0.292      & $0.162^{+0.075}_{-0.048}   $ & $0.182^{+0.060}_{-0.057}$ \\[0.15cm]
        $\omega_{\mathrm{b}}$ & 0.022      & ---                          & $0.022^{+0.002}_{-0.002}$ \\[0.15cm]
        $n_{\mathrm{s}}$      & 1.0        & $< 0.981                   $ & $0.911^{+0.071}_{-0.067}$ \\[0.15cm] \hline \\[-0.25cm]
        $A_{\mathrm{IA}}$     & 0.396      & $0.11^{+0.48}_{-0.31}      $ & $0.266^{+0.068}_{-0.322}$ \\[0.15cm]
        $A_{\mathrm{bary}}$   & 3.133      & ---                          & $2.351^{+0.114}_{-0.250}$
    \end{tabular}
    \tablefoot{For each parameter, we show the underlying true value which was input into the mock data vector, and the inferred value recovered by {KiDS-SBI}. The third column shows the marginal means as well as the upper and lower 68$\%$ confidence intervals, i.e. $1\sigma$, of the marginals. $h_0$, $n_{\mathrm{s}}$, $\omega_{\mathrm{b}}$, and $A_{\mathrm{bary}}$ are not well enough constrained in order to calculate a meaningful marginal parameter estimate. The fourth column shows the maximum a posteriori (MAP), and the uncertainties are defined as the upper and lower 68$\%$ confidence intervals, i.e. $1\sigma$, given by the projected joint highest posterior density, PJ-HPD \citep{robert2007bayesian, joachimi2021kids}. Note: cosmic shear is only expected to significantly constrain the value of $S_{8}$ and $A_{\mathrm{IA}}$.}
    \label{tab:sbi_validation_truth}
\end{table}

In summary, we find that the SBI inference pipeline within {KiDS-SBI} accurately and robustly recovers the underlying posterior distribution of $\Lambda$CDM parameters from two-point statistic measurements of the cosmic shear signal in KiDS-1000 when considering a forward model-based log-normal random fields and biased by relevant systematic effects.

\section{Cosmology dependence of the cosmic shear two-point statistic covariance}\label{appendix:cosmology dependence}

\begin{figure*}
    \centering
    \includegraphics[width=14cm]{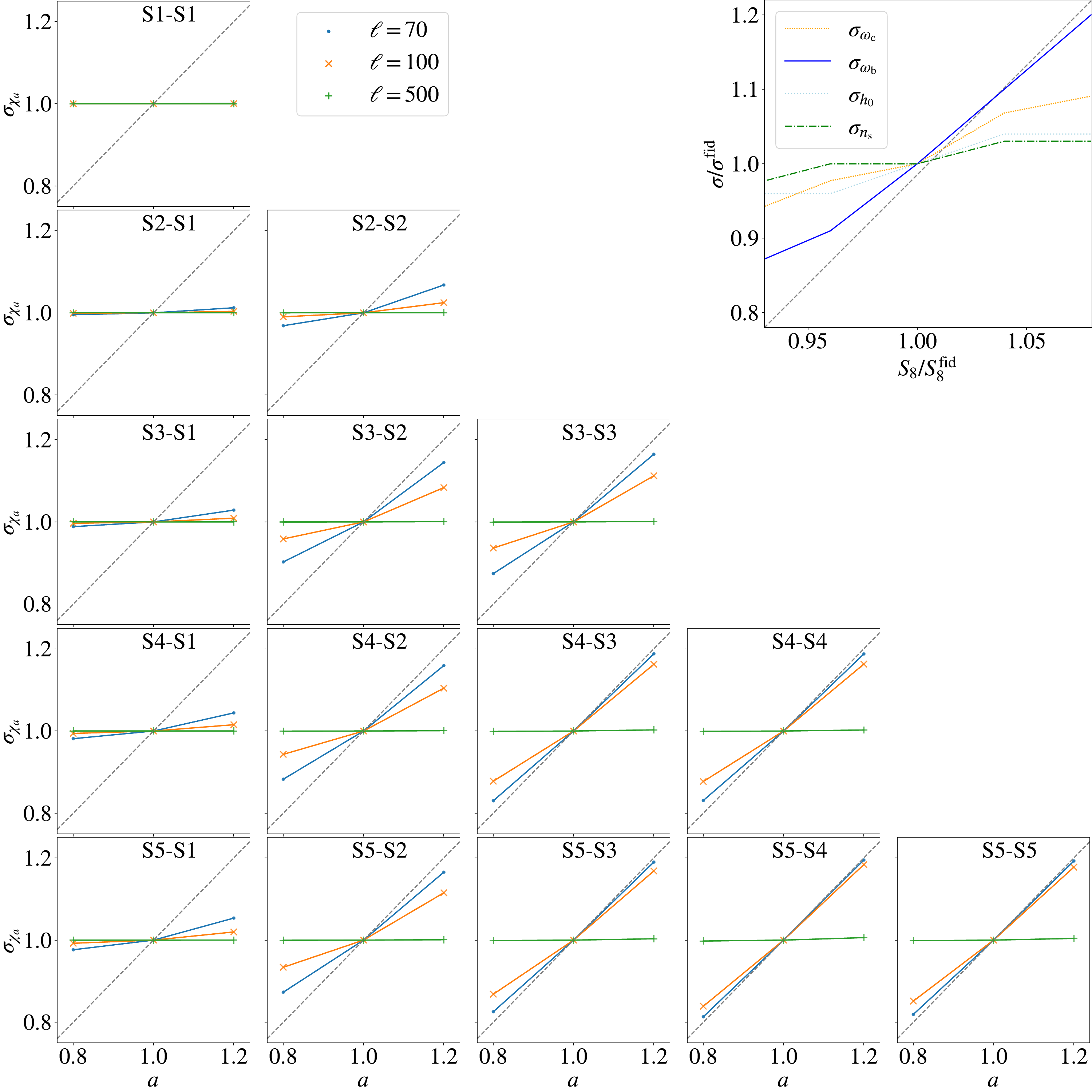}
    \caption{The standard deviation of the analytical likelihood distribution of $\chi_{a}$ (which is given by the goodness-of-fit of the cosmic shear two-point statistic signal scaled by a factor of $a$) as a function of the factor $a$. Each of the panels shows the effect on the shear signal of a different combination of the five KiDS-1000 tomographic bins (S1 to S5). The blue dots represent the $\sigma_{\chi_a}$ values at $\ell =70$ which is just below the scale cuts applied in the anisotropic systematics model. The orange crosses are evaluated at $\ell=100$, while the green plus signs assume $\ell = 500$, where the uncertainty is dominated by the shape noise. The panel in the upper right corner shows how the $1\sigma$ intervals of the likelihood marginals from {KiDS-SBI} vary with respect to the change in $S_8$ relative to $S_{8}^{\mathrm{fid}} = 0.754$ as shown in Fig.~\ref{fig:cosmic_var_likelihood}. The grey dashed line in each panel shows a direct proportionality for reference.}
    \label{fig:cosmic_var_v_sbi_marg}
\end{figure*}

To determine the origin of the observed cosmology dependence in the likelihood in Fig.~\ref{fig:cosmic_var_likelihood}, we investigate whether there is a theoretical justification for the uncertainty in the cosmic shear signal in KiDS-1000 to vary significantly with $S_8$. First, we assume that the likelihood of a full-sky shear angular power spectrum, $C(\ell)$, at a given cosmology, is Gaussian, specifically $\mathcal{L} \propto \mathrm{exp}(-\frac{1}{2} \chi^2)$, where $\chi$ is the goodness of fit given by
\begin{equation}
    \chi^{(p q)} = \big[\hat{C}^{(pq)}(\ell) - C_{\mathrm{fid}}^{(pq)}(\ell)\big]/\sigma_{\mathrm{fid}}^{(p q)}(\ell)\,,
    \label{eq:gof_cl}
\end{equation}
\noindent where $\hat{C}^{(pq)}$ is the observed data vector, $C_{\mathrm{fid}}^{(pq)}$ is the modelled data vector at the fiducial cosmology, and $\sigma_{\mathrm{fid}}^{(p q)}$ is the standard deviation of the data at the fiducial cosmology. Assuming that the underlying shear field is described by a Gaussian random field,  $\sigma_{\mathrm{fid}}^{(p q)}$ is given by
\begin{equation}
     \big( \sigma_{\mathrm{fid}}^{(p q)} \big)^2 = \frac{[\big(C_{\mathrm{fid}}^{(pq)} (\ell) \big]^2  + C_{\mathrm{fid}}^{(pp)}(\ell) \, C_{\mathrm{fid}}^{(q q)} (\ell)}{2 \ell + 1} + \big[ N^{(p q)}(\ell) \big]^2\,,
     \label{eq:sigma_cl}
\end{equation}
\noindent where the first term on the right-hand side defined the contribution due to cosmic variance, while $N^{(p q)}$ is the shape noise contribution for a given tomographic bin combination \citep{joachimi2011constraints}. To assess the impact of varying $S_8$ on such a likelihood distribution, we introduce a toy parameter, $a$, which linearly scales the model's fiducial data vector similar to $S_8$ as follows
\begin{equation}
    C_{\mathrm{fid,} \, a}^{(pq)} (\ell) = a \, C_{\mathrm{fid}}^{(pq)} (\ell)\,.
    \label{eq:cl_a}
\end{equation}

Hence, we can define $\chi^{(p q)}_{a}$ as the goodness-of-fit after rescaling the model by combining Eqs.~(\ref{eq:gof_cl}), (\ref{eq:sigma_cl}) and (\ref{eq:cl_a}) as follows
\begin{equation}
    \chi_a^{(pq)} = \frac{1 - a \pm \chi^{(pq)} \, \sqrt{\frac{1 + \big[R^{(p q)}(\ell) \big]^{-2}}{2 \ell + 1} + \bigg[ \frac{N^{(pq)} (\ell)}{C_{\mathrm{fid}}^{(pq)} (\ell)}\bigg]^2}}{\sqrt{a^2 \, \frac{1 + \big[R^{(p q)}(\ell)\big]^{-2}}{2l + 1} + \bigg[ \frac{N^{(pq)} (\ell)}{C_{\mathrm{fid}}^{(pq)} (\ell)}\bigg]^2}} \,,
\end{equation}
\noindent where $R^{(p q)}(\ell) \equiv C_{\mathrm{fid}}^{(pq)}(\ell)/\sqrt{C_{\mathrm{fid}}^{(pp)}(\ell) \, C_{\mathrm{fid}}^{(qq)} (\ell)}$. By calculating the distribution of $\mathrm{exp}(-\frac{1}{2} \chi_a^2)$ over $\chi$ for different values of $\ell$ and $a$, we can assess the impact of rescaling the shear angular power spectrum directly on the Gaussian likelihood across different angular scales, $\ell$. To facilitate this, we can analytically compute the standard deviation of this distribution, $\sigma_{\chi_a}$, as follows
\begin{equation}
    \sigma_{\chi_a}^2 = 1 +  (a^2 - 1) \Bigg(1 + (2\ell + 1) \, \frac{\big[ N^{(pq)} (\ell)/C_{\mathrm{fid}}^{(pq)} (\ell)\big]^2}{1 + \big[R^{(p q)}(\ell)\big]^{-2}}\Bigg)^{-1}\,.
    \label{eq:sigma_chi_a}
\end{equation}

Applying Eq.~(\ref{eq:sigma_chi_a}) to the five KiDS-1000 tomographic bins, while assuming a shape noise contribution consistent with the values given in Table~\ref{tab:dr4_values}, we obtain Fig.~\ref{fig:cosmic_var_v_sbi_marg}. We find that, as expected, at large angular scales ($\ell = 500$), the likelihood is dominated by the shape noise contribution, so its width does not scale with $a$ under any circumstance. However, at $\ell = 70$ and $\ell =100$, in some tomographic bin combinations where the signal-to-noise ratio is large enough (i.e. the higher redshift bins), the width of the likelihood distribution scales linearly with the scaling applied to the shear angular power spectrum, $a$. This implies that the uncertainty of the KiDS-1000 cosmic shear measurements is dependent on the amplitude of the measurement itself, and therefore the underlying cosmology. In fact, as shown in the upper right panel of Fig.~\ref{fig:cosmic_var_v_sbi_marg}, the broadening of the analytical likelihood with $a$ appears to be generally consistent with the widening observed in the likelihood marginals from the anisotropic systematics model in {KiDS-SBI} shown in Fig.~\ref{fig:cosmic_var_likelihood}.

\addtolength{\tabcolsep}{-0.4em}
\begin{table*}
    \centering
    \caption{All inferred cosmological and astrophysical parameters varied within the anisotropic systematics and the standard isotropic systematics models from the KiDS-1000 gold sample.}
    \begin{tabular}{ccccccccc}
        \hline
          & \multicolumn{2}{c}{SBI: anisotropic sys.} & \multicolumn{2}{c}{SBI: std. isotropic sys.} & \multicolumn{2}{c}{MCMC: anisotropic sys.} & \multicolumn{2}{c}{MCMC: std. isotropic sys.}\\[0.05cm]
        \hline
                                 & Marginal                         & MAP                               & Marginal                          & MAP       & Marginal                          & MAP                           & Marginal                               & MAP \\
                                & $\pm 1\sigma$                     & $\pm$PJ-HPD & $\pm 1\sigma$      &  $\pm$PJ-HPD         & $\pm 1\sigma$                     &  $\pm$PJ-HPD         & $\pm 1\sigma$  &     $\pm$PJ-HPD                \\\hline \\[-0.25cm]
        $S_{8}$                  & $0.731\pm 0.033$                 & $0.743^{+0.015}_{-0.051}$         & $0.772^{+0.038}_{-0.032}$      & $0.780^{+0.020}_{-0.048}$ & $0.725^{+0.034}_{-0.023}$ & $0.731^{+0.023}_{-0.033}$  & $0.764^{+0.031}_{-0.025}    $ &  $0.770 \pm 0.029$\\[0.15cm]
        $\sigma_{8}$             & $0.73^{+0.10}_{-0.21}$           & $0.72^{+0.09}_{-0.20} $        & $0.78^{+0.12}_{-0.23} $         & $0.73^{+0.12}_{-0.20} $  & $0.75^{+0.12}_{-0.18} $  & $0.72^{+0.19}_{-0.11}$  & $0.76^{+0.11}_{-0.14}   $  & $0.74^{+0.14}_{-0.11}$\\[0.15cm]
        $\Omega_{\mathrm{m}}$    & $0.337^{+0.097}_{-0.150}$        & $0.323^{+0.220}_{-0.060} $        & $0.333^{+0.093}_{-0.160} $      & $0.345^{+0.216}_{-0.083}$ & $0.311^{+0.072}_{-0.130} $ & $0.310^{+0.078}_{-0.108}$  & $0.324^{+0.068}_{-0.110}   $ & $0.323^{+0.110}_{-0.086}$\\[0.15cm]
        $h_0$                    & $0.736^{+0.083}_{-0.044}$       & $0.669^{+0.085}_{-0.026} $        & $0.726^{+0.043}_{-0.084} $         & $0.640^{+0.11}_{-0.001}$ & 
        ---  & $0.674^{+0.066}_{-0.005}$ & ---   & $0.673^{+0.071}_{-0.022}$\\[0.15cm]
        $\omega_{\mathrm{c}}$     & $0.158^{+0.084}_{-0.045}   $       & $0.125^{+0.095}_{-0.023} $        &  $0.151^{+0.061}_{-0.089}$     & $0.122^{+0.091}_{-0.027}$ & $0.143^{+0.040}_{-0.073}  $ & $0.118^{+0.069}_{-0.034}$  & $0.147^{+0.039}_{-0.057}   $  & $0.123^{+0.077}_{-0.021}$\\[0.15cm]
        $\omega_{\mathrm{b}}$    &  $0.0224^{+0.0022}_{-0.0033}$     & $0.0190^{+0.004}_{-0.0001} $       &   $0.0225\pm 0.0020 $           & $0.0190^{+0.0046}_{-0.0001} $ & --- & $0.0222^{+0.0016}_{-0.0012}$ & --- & $0.0225 \pm 0.0019$\\[0.15cm]
        $n_{\mathrm{s}}$         & $0.987^{+0.110}_{-0.039}$     & $0.972^{+0.128}_{-0.132}$       & $0.961^{+0.054}_{-0.120}         $  & $0.949_{-0.109}^{+0.151}$ & --- & $0.988^{+0.045}_{-0.105}$  & $0.958^{+0.042}_{-0.110}$ & $0.956^{+0.046}_{-0.108}$\\[0.15cm] \hline \\[-0.25cm]
        $A_{\mathrm{IA}}$        & $0.60^{+0.42}_{-0.32}       $     & $0.63^{+0.36}_{-0.40}$         &  $0.61^{+0.38}_{-0.30}      $ & $0.57^{+0.39}_{-0.29}$ & $0.59^{+0.48}_{-0.36}   $ & $0.73^{+0.35}_{-0.45}$  & $0.64^{+0.42}_{-0.34}        $  & $0.73^{+0.32}_{-0.43}$\\[0.15cm]
        $A_{\mathrm{bary}}$      & $2.60^{+0.52}_{-0.20} $          & $2.66_{-0.66}^{+0.47}  $         & $2.50^{+0.23}_{-0.49}  $           & $2.00^{+1.13}_{-0.001}$ & $> 2.45    $ & $3.02^{+0.09}_{-0.70}$  & $> 2.48  $  & $3.13^{+0.001}_{-0.66}$\\[0.1cm]
        \hline
    \end{tabular}
    \tablefoot{The columns marked with SBI show the constraint obtained from the \texttt{KiDS-SBI} analaysis, while the columns marked with MCMC represent the parameter constraints obtained from a standard MCMC analysis which assumes a Gaussian likelihood with a sample covariance matrix derived from the respective forward model. The second, fourth, sixth and seventh columns show the mean marginals as well as the upper and lower 68$\%$ confidence intervals, i.e. $1\sigma$, of the marginals. The third and fifth columns show the maximum a posteriori (MAP), and the uncertainties are defined as the upper and lower 68$\%$ confidence intervals, i.e. $1\sigma$, given by the projected joint highest posterior density, PJ-HPD \citep{robert2007bayesian, joachimi2021kids}.}
    \label{tab:sbi_full_results}
\end{table*}

\section{Impact of the sampler on the Gaussian likelihood posteriors}\label{appendix:sampler_comparison}
\begin{figure*}
    \centering
    \includegraphics[width=12cm]{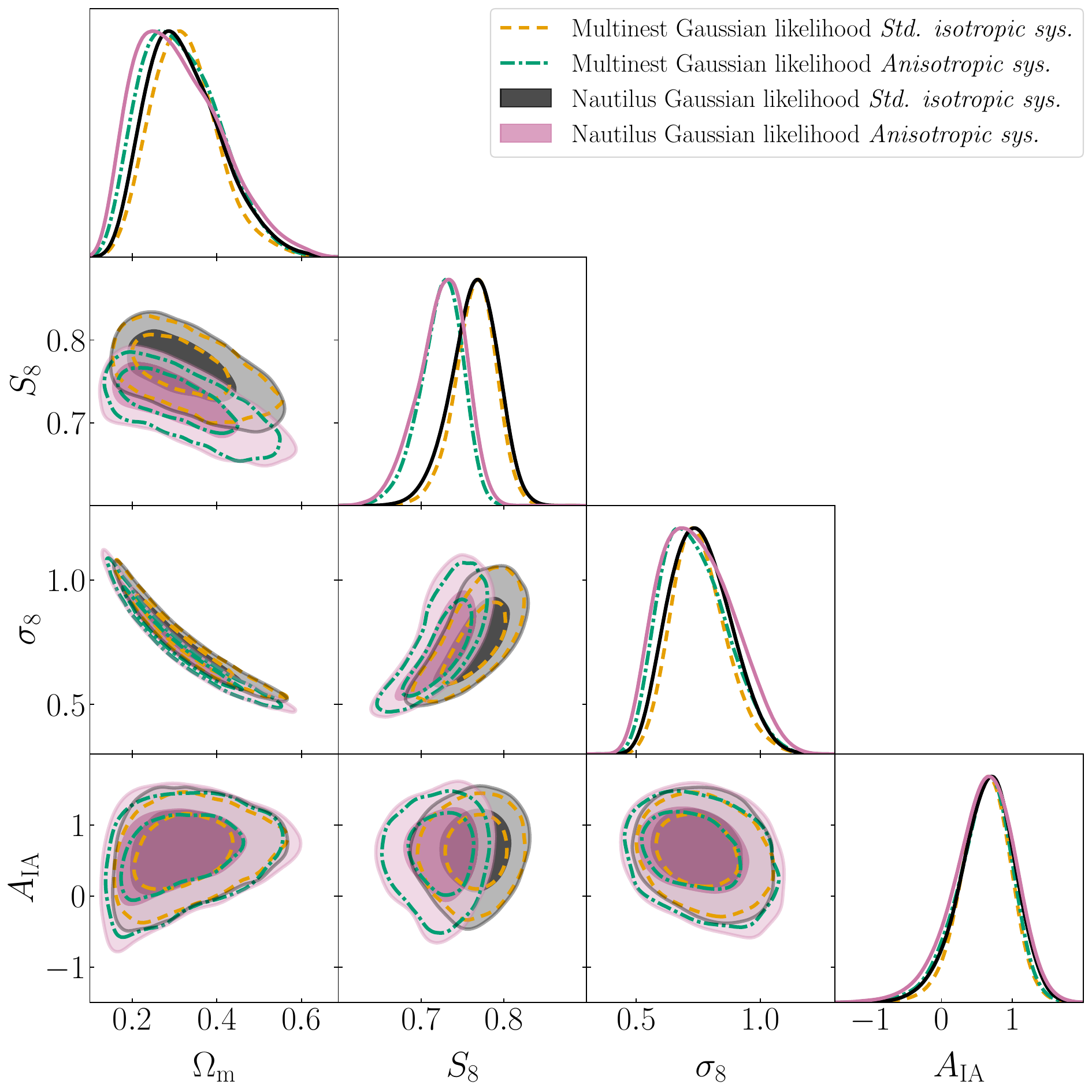}
    \caption{Comparison of posteriors obtained from a standard Gaussian likelihood analysis when using different MCMC samplers for the same noiseless mock data vector. In pink and in grey, we show the posteriors sampled with \texttt{nautilus} \citep{lange2023nautilus} from a likelihood defined by a sample covariance based on mocks from the anisotropic systematics model and the standard isotropic systematics, respectively. In green and orange, we show posteriors sampled from the same two likelihoods as before, but sampling with the \texttt{multinest} sampler \citep{feroz2009multinest} instead. All inference shown is done using the real KiDS data vector.} \label{fig:nautilus_vs_multinest}
\end{figure*}

To show that the Gaussian likelihood constraints shown in Fig.~\ref{fig:sbi_results} and Fig.~\ref{fig:sbi_results_full} obtained from the \texttt{nautilus} sampler \citep{lange2023nautilus} are consistent with previous KiDS-1000 analyses (e.g. \citealt{asgari2021kids, loureiro2021kids}) which used the \texttt{multinest} \citep{feroz2009multinest}, we compare the posteriors obtained from both samplers for the same data vector and covariance matrix in Fig.~\ref{fig:nautilus_vs_multinest}. As has been noted in previous work \citep{lemos2023robust}, we find that \texttt{multinest} consistently gives ${\sim}15\%$ narrower $S_{8}$ posterior marginals than the ones sampled \texttt{nautilus} irrespective of the model assumed to obtain the sample covariance.

\section{Posterior variations with noise realisations}\label{appendix:noise_realisation}
\begin{figure*}
    \centering
    \includegraphics[width=16cm]{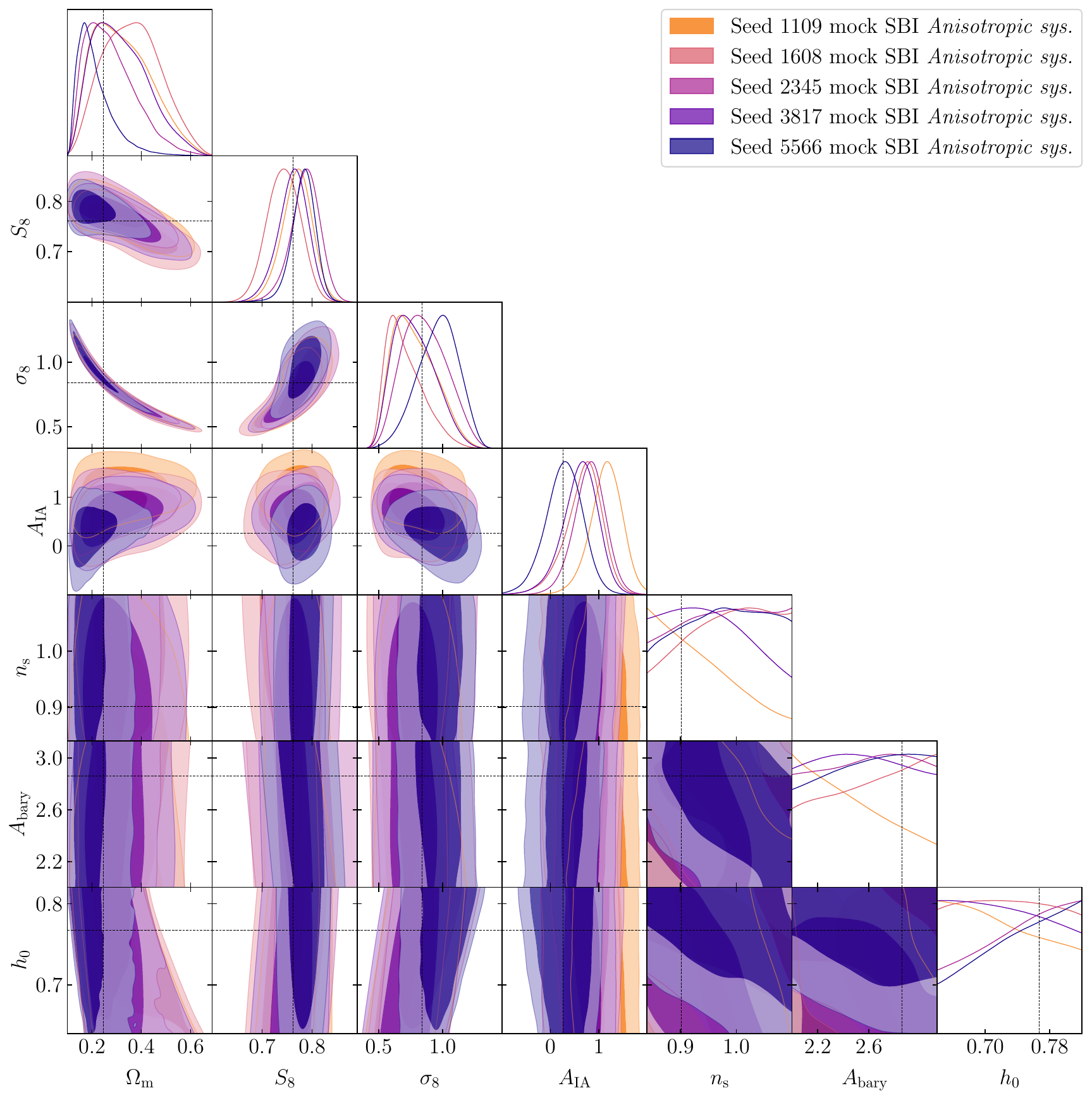}
    \caption{Posterior contours from five different mock data vectors given the anisotropic systematics model within {KiDS-SBI} over the prior space shown in Table~\ref{tab:sbi_priors}. Each data realisation assumes the same input parameters (as given by the black line and shown in Table~\ref{tab:sbi_validation_truth}) while varying the random seeds of the underlying matter fields, galaxy positions, intrinsic galaxy shapes, and shear biases. The posteriors shown in Fig.~\ref{fig:sbi_validation_posterior},\ref{fig:sbi_validation_coverage}, \ref{fig:sbi_validation_vd_impact}, and \ref{fig:sbi_mcmc_v_sbi} are based on the 3817 noise realisation.}
    \label{fig:sbi_validation_seed_comparison}
\end{figure*}

As discussed in Sect.~\ref{method:gof}, different uncertainty realisations of the anisotropic systematics model in {KiDS-SBI} at the same cosmology produces a significant random scatter in the measured final pseudo-Cls. Figure~\ref{fig:sbi_validation_seed_comparison} shows how the choice of noise realisation impacts the inferred posteriors at a given cosmology. We find that overall the variations in the posteriors between uncertainty realisations are small with the mean marginal of $S_{8}$ fluctuating by up to $\pm0.5\sigma$ and the $A_{\mathrm{IA}}$ marginal can fluctuate up to $\pm 0.75\sigma$.

\section{KiDS-SBI posteriors and parameter constraints}\label{appendix:full_posteriors}
\begin{figure*}
    \centering
    \includegraphics[width=16cm]{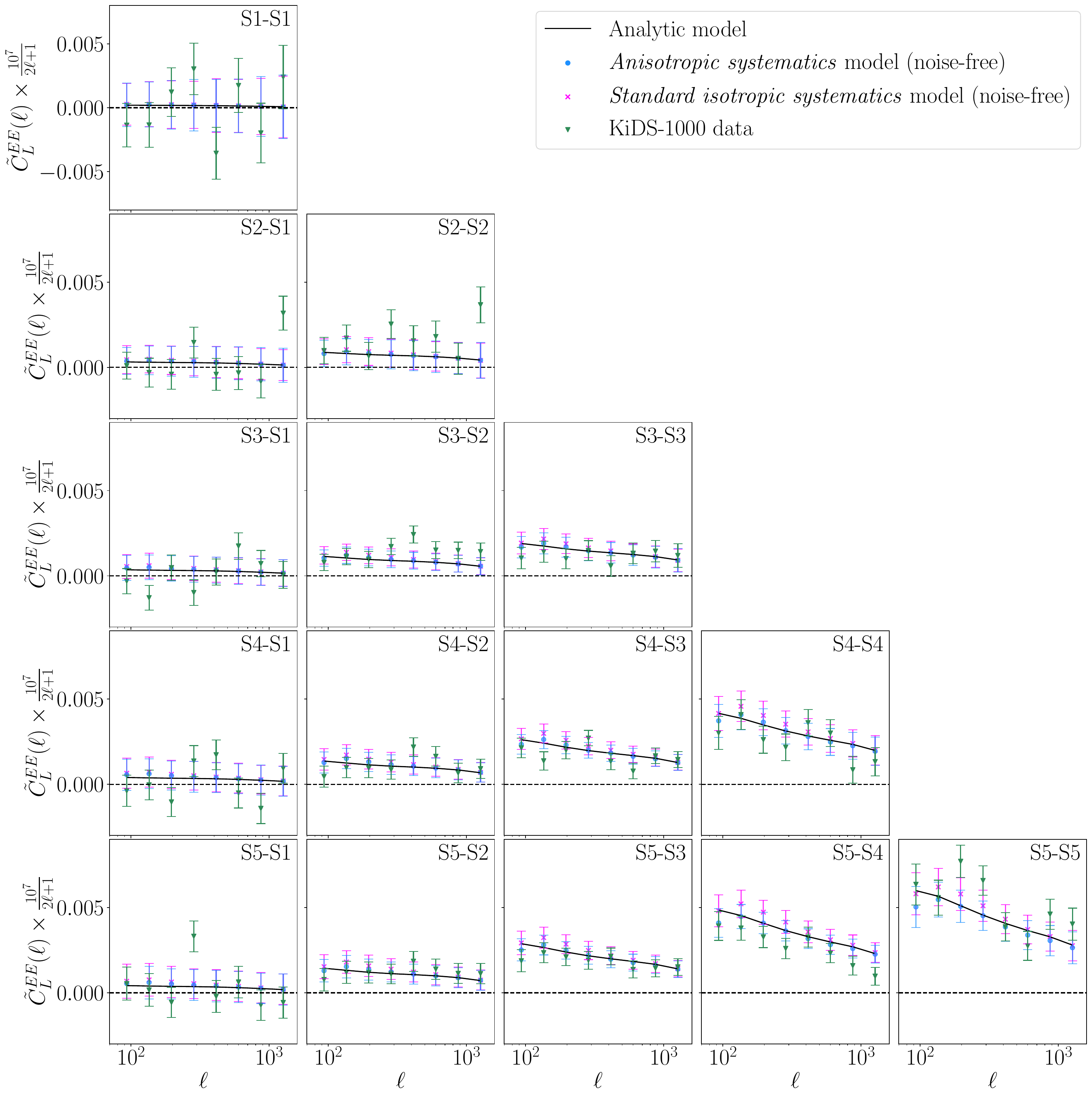}
     \caption{Measured pseudo-Cls, $\tilde{C}^{(pq)}_{\epsilon \epsilon, L}(\bm{\Theta})$, for all combinations for five tomographic bins (S1 to S5). The green triangles show the pseudo-Cls from the KiDS-1000 gold sample data. The blue points show the same for a noise-free realisation single realisation of the anisotropic systematics model in {KiDS-SBI} (see Sect.~\ref{method:sbi}) at the maximum-a-posteriori, MAP (see Table~\ref{tab:sbi_full_results}), while the magenta crosses show the measurements for noise-free realisation single realisation of the standard isotropic systematics model in {KiDS-SBI} (see Sect.~\ref{validation:impact_vd}) at the MAP for this model (see Table~\ref{tab:sbi_full_results}). The uncertainties on the measurements are derived from the covariance matrix described in Sect.~\ref{method:sbi:score}. The solid black line shows the pseudo-Cls, $\tilde{C}_{\epsilon \epsilon}$, as derived from theory (see Appendix~\ref{appendix:signal} for details on this) at the MAP from {KiDS-SBI} based on the anisotropic systematics model. The horizontal dashed line marks the line of zero signal.}
    \label{fig:pcl_measurement}
\end{figure*}

\begin{figure*}
    \centering
    \includegraphics[width=16cm]{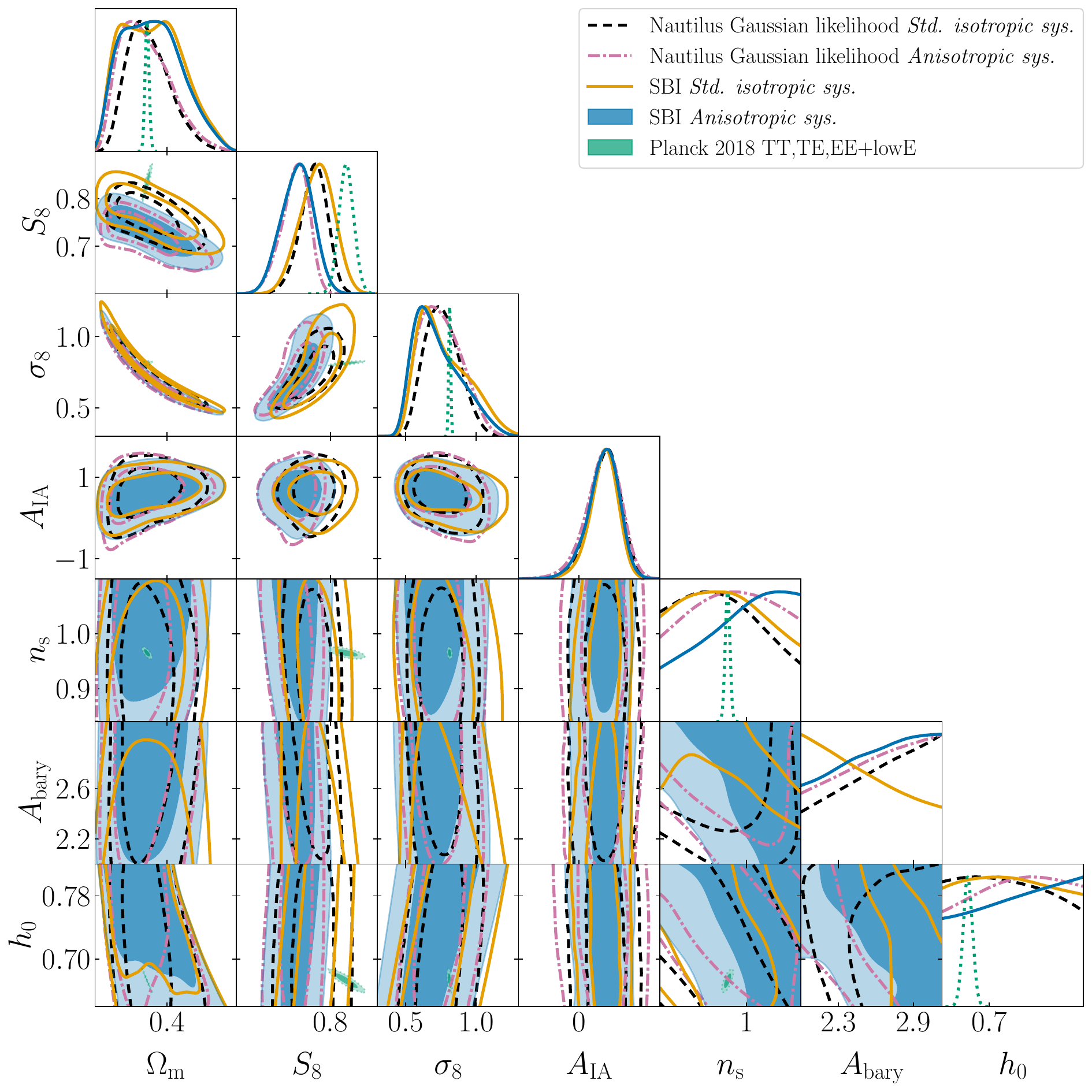}
    \caption{Posterior contours of the main constrained cosmological parameters from the {KiDS-SBI} analysis of the KiDS-1000 cosmic shear data assuming the anisotropic systematics model (in blue) compared against posterior contours from other analyses. In orange, we show the posterior from the same data while assuming the standard isotropic systematics model. In black and pink, we show the posteriors from the equivalent analyses done assuming a Gaussian likelihood for a given model. For clarity purposes, the $\bm{\delta}_{z}$ parameters are marginalised out, while $\omega_{\mathrm{b}}$ and $\omega_{\mathrm{c}}$ are folded into $\Omega_{\mathrm{m}}$. The green contour shows the posterior from the cosmic microwave background constraints from the TT,TE,EE+lowE modes \citep{planck2020planck}. We note that the Planck TT,TE,EE+lowE contours do not have any marginals in $A_{\mathrm{IA}}$ or $A_{\mathrm{bary}}$ as the CMB is not sensitive to the IAs of galaxies or baryonic feedback.}
    \label{fig:sbi_results_full}
\end{figure*}

To expand on the results of the {KiDS-SBI} analysis of the KiDS-1000 gold sample presented in Fig.~\ref{fig:sbi_results} and Table~\ref{tab:sbi_fiducial_results}, we present the full posterior learnt in {KiDS-SBI} to fit the data in Fig.~\ref{fig:pcl_measurement} in  Fig.~\ref{fig:sbi_results_full} for both the anisotropic systematics and standard isotropic systematics models. Table~\ref{tab:sbi_full_results} also provides the parameter best estimates for all cosmological and astrophysical parameters varied in the forward models. As shown in Fig.~\ref{fig:sbi_results_full}, both models leave $A_{\mathrm{bary}}$, $n_{\mathrm{s}}$ and $h_{0}$ unconstrained, while constraining $A_{\mathrm{IA}}$. The posteriors of both models strongly prefer values of $A_{\mathrm{IA}}$ greater than zero and less than 1.5. Despite the wide range of the $A_{\mathrm{IA}}$ priors, the posterior stays within a physical regime for the parameter which indicates a low chance of residual systematics strongly biasing the parameter constraints.

\end{appendix}
\end{document}